\documentclass[12pt,a4paper]{book}
\usepackage[italian]{babel}
\usepackage[dvips]{graphics}
\usepackage{amsmath,amsbsy,epsfig,eucal}
\usepackage{amssymb}
\usepackage{latexsym}
\usepackage{pennames}
\usepackage{graphicx,graphics}
\advance\textheight by \topskip
\makeindex
\newcommand{\decayarrow}{\makebox[0mm][l]%
{\rule{0.33em}{0mm}\rule[0.55ex]{0.044em}{1.55ex}}\rightarrow}

\begin{document}
\selectlanguage{italian}
\begin{titlepage}

\pagestyle{empty}


\def\bw{1}
\ifnum \bw=0
        \def\Red{\color[named]{Black}}
        \def\Cyan{\color[named]{Black}}
        \def\Blue{\color[named]{Black}}
        \def\Green{\color[named]{Black}}
        \def\Magenta{\color[named]{Black}}
        \def\UMagenta{\underline}
        \def\infn{infn-bw.ps}
\else
        \def\Red{\color[named]{Red}}
        \def\Cyan{\color[named]{Cyan}}
        \def\Blue{\color[named]{Blue}}
        \def\Green{\color[named]{Green}}
        \def\Magenta{\color[named]{Magenta}}
        \def\UMagenta{\color[named]{Magenta}}
        \def\infn{infn.ps}
\fi
\begin{center}
{\huge UNIVERSIT\`A DEGLI STUDI DI BARI}
\end{center} 
\begin{center}
{\Large Facolt\`a di Scienze Matematiche Fisiche e Naturali}\\
{\large Corso di Dottorato in Fisica}
\end{center}
\vspace{0.1cm}
\begin{center}
 \includegraphics[scale=1]{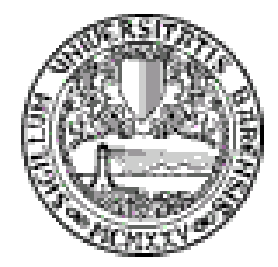}
\end{center}
\vskip 0.2  truecm
\rule{15cm}{0.3mm}
\begin{center}
{\LARGE\sc{Tesi di Dottorato}} \\
\end{center}
\vspace{1cm}
\begin{center}
 {\huge{\bf Produzione di particelle strane in}} \\ \vspace{0.5cm} 
 {\huge{\bf interazioni piombo--piombo e}}\\ \vspace{0.5cm}
 {\huge{\bf protone--piombo a 160 GeV/c per}}\\ \vspace{0.5cm}
 {\huge{\bf nucleone}}
\end{center}
\vspace{2.0cm}
{\hspace{8.0cm} \quad\quad\quad}{\Large Rocco Caliandro} \\
\vskip 0.5  truecm
\vspace{0.4cm}
\rule{15cm}{0.3mm}
\begin{center}
{\Large Undicesimo ciclo}
\end{center}
\end{titlepage}

%
%
\chapter*{{\em Ringraziamenti}}
\vspace{1cm}
Desidero innanzitutto ringraziare il prof. B.~Ghidini per la fiducia
concessami e per l'attenzione prestata al mio operato.
\par
Ringrazio inoltre la dott.ssa R.A.~Fini per la sua pazienza e per
le utilissime discussioni avute con lei riguardo il mio lavoro.
\par
Ringrazio tutto il Gruppo degli ioni pesanti di Bari, per avermi
permesso di lavorare in un clima stimolante, circondato di amicizia
e considerazione, e l'intera Collaborazione WA97, per aver consentito
e sostenuto la mia crescita professionale.
\par
Un saluto particolare va a R.~Loconsole, insostituibile
{\em system manager} e vera anima del Gruppo ed a tutti i miei
colleghi di dottorato, per aver condiviso con me il travagliato
percorso di questi ultimi tre anni.

\vspace{3.5cm}
%
%
\begin{flushright}
``Fatti non foste a viver come bruti,\\
ma per seguir virtute e canoscenza.''\\
{\em Dante, Inferno XXVI, 119-120}
\end{flushright}
%

\pagebreak
\pagenumbering{arabic}\setcounter{page}{1}
%
\pagenumbering{arabic}
\setcounter{page}{0}
\tableofcontents
%
%

%
%
%
\chapter*{Introduzione}
\addcontentsline{toc}{chapter}{Introduzione}

La fisica degli ioni pesanti ad alta energia studia la materia adronica
in condizioni di elevata densit\`a di energia. La teoria pi\`u 
accreditata per la descrizione delle interazioni forti, la
{\em CromoDinamica Quantistica} (QCD), prevede che, in
condizioni estreme, la materia nucleare possa subire una trasformazione 
di fase verso lo stato di {\em Plasma di Quarks e Gluoni} (QGP).
Nel QGP i quarks ed i gluoni non appaiono pi\`u confinati all'interno
degli adroni costituenti il nucleo, ma sono liberi di muoversi entro il volume
del plasma.
\par
Si ritiene che l'intero universo fosse costituito da plasma di
quarks e gluoni pochi istanti dopo il {\em Big Bang} primordiale.
In seguito alla successiva espansione, dopo circa $10^{-5}~sec$ dal
{\em Big Bang}, il plasma si sarebbe adronizzato in materia nucleare,
seguendo una trasformazione di fase inversa a quella suddetta.
Nell'universo attuale lo stato di QGP potrebbe risiedere all'interno
delle stelle di neutroni.
Lo stato di QGP fornisce, inoltre, la possibilit\`a di esplorare le 
interazioni a lunga distanza  tra quarks e gluoni, con particolare
riferimento ai meccanismi di confinamento e rottura di simmetria chirale.
Questi fenomeni, sebbene appartengano alla scala naturale della QCD,
 risultano ancora di difficile comprensione, in quanto cadono nel dominio non 
perturbativo della teoria.
\par
L'estremo interesse suscitato da questa nuova linea di ricerca
si \`e concretizzato, a partire dalla met\`a degli anni '80,
 nella progettazione e
realizzazione di es\-pe\-ri\-menti di interazione tra ioni pesanti agli acceleratori
AGS ({\em Alternate Gradient Synchroton}) di Brookhaven (USA) e 
SPS ({\em Super Proton Synchroton}) del CERN (Ginevra).
Infatti, le condizioni per la formazione dello stato di plasma
potrebbero essere raggiunte
nella regione centrale del sistema interagente creatosi in seguito
alla collisione tra due nuclei pesanti ad alta energia.
La sopravvivenza del plasma 
sarebbe comunque limitata ai primi istanti successivi alla
collisione ($\sim 10^{-22}~sec$).
La sua rivelazione sarebbe, quindi, possibile in maniera indiretta,
 attraverso le
particelle prodotte nello stato finale e cercando segnali caratteristici
dell'avvenuta transizione di fase,
capaci di sopravvivere alla transizione stessa ed alla successiva
evoluzione del sistema collidente.
\par
L'esperimento WA97, inserito in questa linea di ricerca, studia le
collisioni su bersaglio fisso di
nuclei di piombo accelerati fino a 160 GeV/c per nucleone 
nell'SPS del CERN e misura, come  segnale di QGP, la produzione di stranezza.
\`E stato infatti ipotizzato che una delle conseguenze della transizione di fase
di QCD sia l'aumento  della produzione di particelle strane e, in particolare,
di barioni ed antibarioni multi-strani rispetto
alle normali interazioni adroniche.
\par
Il lavoro descritto in questa tesi riguarda l'analisi dei dati raccolti
a partire dal 1995 dall'esperimento WA97, relativi a collisioni piombo-piombo e
a collisioni protone-piombo a 160 GeV/c per nucleone.
\par
Nel primo capitolo verr\`a presentata una introduzione alla fisica del QGP
e delle interazioni nucleari di altissima energia. Verranno inoltre
esaminati i diversi segnali finora proposti quali indicatori della fase
di plasma ed i principali risultati sperimentali finora osservati.
\par
L'esperimento WA97 sar\`a descritto nel secondo capitolo, soffermandosi
in maniera particolare
sull'apparato sperimentale impiegato.
\par
Seguir\`a, nel terzo
capitolo, la descrizione delle procedure usate per identificare le 
particelle strane studiate, isolandole dalla moltitudine di particelle prodotte 
in seguito alle collisioni piombo-piombo e protone-piombo.
\par
Nel quarto capitolo verranno affrontati i problemi relativi all'analisi del
segnale ottenuto, quali le correzioni per l'accettanza geometrica e per
l'efficienza dei programmi di ricostruzione, nonch\`e
 la procedura di estrapolazione
 dei risultati ottenuti.
\par
I risultati sperimentali, riguardanti le distribuzioni di massa trasversa,
i rapporti di produzione e la produzione di particelle strane in funzione
 della centralit\`a, saranno presentati nel quinto capitolo 
e confrontati con i dati esistenti in letteratura.
\par
Nel sesto capitolo l'interpretazione dei risultati sar\`a affrontata
 mediante il loro confronto
con le previsioni di due dei pi\`u accreditati modelli di 
produzione adronica: VENUS ed RQMD.

%
%
\chapter{Il plasma di Quarks e Gluoni e le collisioni nucleari 
ultrarelativistiche}
\index{Il plasma di Quarks e Gluoni e le collisioni nucleari\\ 
ultrarelativistiche} 

\section{Introduzione}
\index{Introduzione}

L'evoluzione scientifica nel campo della fisica delle particelle  \`e
tradizionalmente avvenuta lungo l'asse dell'energia: la possibilit\`a
di accelerare particelle ad energie sempre maggiori ha permesso di
raggiungere di volta in volta le soglie per la creazione di nuove
particelle e per l'innesco di nuovi fenomeni.
Solo recentemente, con la possibilit\`a di accelerare ioni ad energie
ultrarelativistiche ($E~>>~m$) agli acceleratori di Brookhaven e del
CERN, l'evoluzione scientifica ha potuto seguire un altro asse: quello
delle dimensioni del sistema interagente.
Lungo quest'asse, la possibilit\`a di osservare nuovi fenomeni \`e
legata alla creazione di sistemi contenenti un gran numero di costituenti
elementari che interagiscono in maniera collettiva.
Si passa cos\`{\i} dalla fisica delle particelle alla fisica dei 
sistemi di particelle, il cui oggetto di studio
\`e la materia nucleare nelle sue differenti fasi.
\par
Questo nuovo approccio allo studio delle propriet\`a della materia
si avvale di metodi e concetti propri sia della fisica delle particelle
(tradizionalmente denominata ``fisica delle alte energie'') che di 
quella nucleare.
Mentre la prima porta alla descrizione delle interazioni tra oggetti
elementari (adroni o quarks) derivandone le caratteristiche da 
principi primi (teorie di gauge), la seconda studia sistemi estesi
aventi caratteristiche globali, descrivibili mediante modelli 
fenomenologici a base statistica.
L'aspetto microscopico-elementare della fisica delle particelle
e quello macroscopico-fenomenologico della fisica nucleare
hanno portato ad affrontare lo studio delle collisioni tra ioni pesanti
secondo la ``termodinamica di QCD'', in cui stati complessi
costituiti da molte particelle elementari interagenti fortemente sono
descritti in termini di poche variabili macroscopiche.
Le ipotesi necessarie per questo tipo di approccio sono che il 
sistema formatosi in seguito alla collisione tra nuclei sia esteso,
cio\`e le sue dimensioni siano molto maggiori della tipica scala delle
interazioni forti ($~>>~1~fm$) ed  in equilibrio termico,
cio\`e la sua vita media ecceda i tipici tempi di rilassamento ($~>>~1~fm/c$).
In altri termini, \`e necessario che il sistema sia formato da un gran numero
 di particelle e che le loro interazioni siano sufficientemente intense da
 raggiungere l'equilibrio termico entro il tempo di vita del sistema stesso.
\par
Nel seguito del capitolo si vedr\`a come la termodinamica di QCD porti
a prevedere una transizione di fase della materia nucleare in
corrispondenza di una temperatura critica $T_c=150\div200~MeV$ o di una
densit\`a barionica di $0.5\div1.5~nucleoni/fm^3$.
Saranno inoltre descritte le propriet\`a della nuova fase, detta Plasma
di Quarks e Gluoni (QGP), e la possibilit\`a di produrla mediante
 collisioni nucleari di altissima energia.
\par
Seguir\`a una rassegna dei diversi segnali proposti come conferma sperimentale
di un'avvenuta transizione di fase, con particolare riferimento al segnale
di produzione di stranezza, e saranno esposti alcuni dei risultati pi\`u
promettenti ottenuti dagli esperimenti coinvolti in questa linea di ricerca.

\section{Principali caratteristiche della cromodinamica quantistica}
\index{Principali caratteristiche della cromodinamica quantistica}

La prima completa teoria in grado di spiegare le principali caratteristiche
dello spettro barionico in termini di quarks costituenti fu il ``modello
simmetrico a quark'', nel quale i barioni erano considerati come stati legati 
di quarks completamente simmetrici rispetto alle variabili di spin, flavour
e spazio, secondo un gruppo di simmetria $SU(6)\otimes O(3)$ [Gre64].
L'antisimmetria totale della funzione d'onda descrivente tali stati,
necessaria affinch\`e i quarks rispettino la statistica di Fermi-Dirac,
 era garantita dall'introduzione del numero quantico interno di ``colore''.
I quarks erano cos\`{\i} considerati tripletti di colore e si
combinavano negli adroni in modo da formare singoletti di colore,
dato che il colore non \`e una propriet\`a osservata.
\par
In seguito il colore fu interpretato come la carica forte responsabile
delle interazioni tra quarks e si postul\`o l'invarianza di tali
interazioni rispetto a rotazioni locali nello spazio di colore.
Si svilupp\`o cos\`{\i} la teoria quanto-meccanica delle interazioni tra quarks,
 chiamata CromoDinamica Quantistica (QCD), basata sul principio della
invarianza di gauge
per trasformazioni appartenenti al gruppo di simmetria SU(3)
nello spazio del colore.
La richiesta di invarianza di gauge impose la presenza di 8 bosoni vettori di
 massa nulla, detti gluoni, che corrispondevano ai generatori del gruppo
di simmetria e venivano interpretati come mediatori dell'interazione di colore
 tra i quarks. 
\par
\begin{figure}[htb]
\centering
\includegraphics[scale=1.6,clip]
                                  {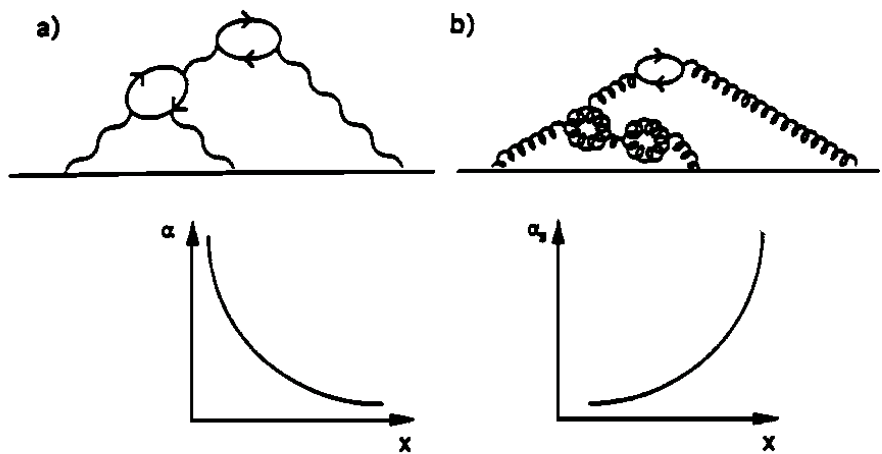}
\caption{\em Differente andamento della costante di accoppiamento
effettiva in QED (a)  e in QCD (b).} 
\label{qed}
\end{figure}

Tale teoria fu sviluppata in analogia con la gi\`a affermata elettrodinamica 
quantistica (QED) che fa uso del gruppo di simmetria U(1), e quindi
di un solo bosone vettore: il fotone.
La principale differenza tra le due teorie risiede nel carattere
non abeliano della QCD e nella conseguente presenza di un termine di
auto-interazione tra i campi gluonici all'interno del suo lagrangiano.
In altri termini, i gluoni, a differenza del fotone che non \`e carico
elettricamente, possiedono una carica di colore, per cui possono interagire
tra loro, oltre che con i quarks.
La possibilit\`a dell'accoppiamento tra gluoni si manifesta in un
diverso andamento della costante di accoppiamento effettiva di QCD
rispetto a quello caratteristico della QED, come mostrato in fig.~\ref{qed}.

Una carica elettrica posta nel vuoto di QED emette e riassorbe continuamente
 fotoni virtuali, i quali  possono produrre temporaneamente coppie $e^+e^-$
che hanno l'effetto di schermare la carica originariamente presente.
L'effetto di schermaggio, simile al fenomeno di polarizzazione del vuoto,
si manifesta in una decrescita della costante di accoppiamento di QED al 
crescere della distanza dalla carica originaria o, equivalentemente, al 
decrescere del momento trasferito nelle interazioni.
In maniera analoga, un quark posto nel vuoto di QCD emette e riassorbe
 continuamente gluoni i quali, per\`o, possono produrre temporaneamente
sia coppie $q\overline{q}$ che coppie $gg$, in virt\`u della loro
auto-interazione. Poich\`e la probabilit\`a di occorrenza di vertici
gluonici \`e maggiore di quelli tra quarks, il contributo dominante
delle coppie $gg$ produrr\`a una distribuzione, nello spazio circostante,
della carica di colore originariamente posseduta dal quark, secondo
un effetto di anti-schermaggio.
La costante di accoppiamento di QCD ($\alpha_s$) mostra quindi una crescita 
all'aumentare della distanza dalla carica di colore originaria o,
equivalentemente, al diminuire del momento trasferito.
L'espressione di $\alpha_s$ in funzione del momento trasferito $q$,
ottenuta in seguito alla procedura di rinormalizzazione e calcolata
 al primo ordine nello sviluppo perturbativo, risulta [Cheng]:

\begin{equation}
\alpha_s(q^2) =
 \frac{4\pi}{(11 - \frac{2}{3}n_f)ln(\frac{q^2}{\Lambda^2})} 
\label{alpha}
\end{equation}

\noindent                
dove $n_f$ indica il numero di flavours e $\Lambda$ \`e il parametro di
scala di QCD, dell'ordine dei decimi di GeV.
Pertanto, purch\`e $n_f$ non superi 16, la costante di accoppiamento
decresce al crescere di $q^2$, 
secondo la propriet\`a nota come ``libert\`a asintotica''.
L'evidenza sperimentale di tale propriet\`a \`e stata ottenuta
attraverso esperimenti di urti altamente inelastici leptone-nucleone, nei
 quali la misura del momento del leptone prima e dopo la collisione
permette di sondare la distribuzione in momento dei quarks costituenti
i nucleoni. Si \`e cos\`{\i} verificato che per grandi momenti trasferiti,
i quarks all'interno degli adroni 
possono essere considerati pressocch\`e liberi.
\par
Al contrario, per piccoli momenti trasferiti ($q\sim\Lambda$), corrispondenti
a distanze superiori ad $1~fm$, la costante di accoppiamento diventa molto 
grande ed il calcolo perturbativo, basato sull'espansione in serie
di potenze della costante di accoppiamento, non \`e pi\`u applicabile.
Questo comportamento viene interpretato con la propriet\`a del
``confinamento'' dei quarks e gluoni all'interno degli adroni, dato che
l'energia necessaria per separare due cariche di colore tende all'infinito
a grandi distanze.
Sperimentalmente non sono mai stati osservati quarks liberi, n\`e creati
in seguito ad interazioni di alta energia innescate da acceleratori o da 
raggi cosmici, n\`e gi\`a esistenti in natura.

\section{La transizione di fase di QCD}
\index{La transizione di fase di QCD}
\label{fasetransition}

Nel paragrafo precedente si \`e visto che i quarks ed i gluoni appaiono
confinati all'interno degli adroni. Se si considera ora un sistema esteso 
costituito di materia nucleare, ossia di nucleoni a loro volta
contenenti quarks e gluoni, ci si pu\`o chiedere se tale sistema possa subire
un cambiamento della propria fase in seguito a specifiche variazioni 
dell'ambiente circostante.
\par
Per la materia ordinaria il fenomeno della transizione di fase \`e frequente 
ed \`e studiato mediante un approccio statistico-termodinamico.
Nel caso specifico di materia nucleare, l'approccio statistico deve essere
formulato all'interno della teoria che descrive le interazioni tra i suoi
costituenti, vale a dire la cromodinamica quantistica.
Sfortunatamente lo studio della transizione di fase riguarda le
interazioni tra quarks e gluoni a grande distanza, per cui cade nel regime
non perturbativo della QCD, l\`{\i} dove si manifesta il fenomeno del
confinamento.
\par
Una trattazione non perturbativa della QCD statistica \`e possibile mediante
la formulazione di una teoria di gauge su un reticolo rappresentante le
coordinate spazio-temporali [Wil74]. La discretizzazione dello spazio-tempo
permette di eliminare le divergenze che nascono nella QCD in seguito 
all'integrazione dei diagrammi di Feynman nella variabile momento
(divergenze ultraviolette). In regime perturbativo tali divergenze sono
eliminate mediante la procedura di normalizzazione, mentre nella QCD su reticolo
il limite inferiore alla distanza tra i punti dello spazio-tempo limita
superiormente la variabile momento, rendendo finiti gli integrali.
D'altro canto, la formulazione su reticolo consente di scrivere la funzione
di partizione della meccanica statistica nella forma di integrale di
cammino, rendendo cos\`{\i} possibile l'utilizzo di metodi Monte Carlo nel
calcolo delle variabili termodinamiche del sistema.
In concreto, le limitazioni nella velocit\`a e memoria dei computers impongono
dei limiti alle dimensioni ed alla spaziatura del reticolo, per cui \`e
necessario ripetere i calcoli per reticoli sempre pi\`u fitti finch\`e
la relazione tra la costante di accoppiamento $\alpha_s$ ed i differenti
passi dei reticoli risulta in accordo con quanto previsto dalla QCD
perturbativa (c.f.r. eq.~\ref{alpha}) [Won94].
\par
\begin{figure}[htb]
\centering
\includegraphics[scale=1.7,angle=1.1,clip]
                                  {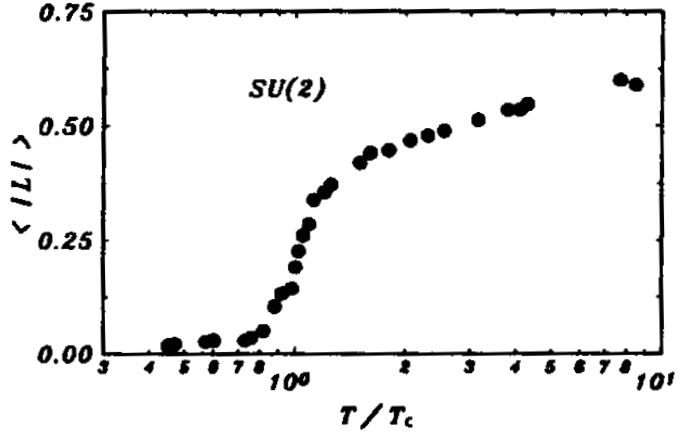}
\caption{\em Andamento del parametro d'ordine $<L>$ in
 funzione della temperatura $T/T_c$ [Sat84].}
\label{fwilson}
\end{figure}

Le transizioni di fase vengono generalmente esaminate per mezzo di
``parametri d'ordine'', vale a dire di variabili termodinamiche che sono 
nulle in una fase del sistema e diverse da zero nell'altra.
Nello studio della transizione di fase di QCD \`e utile usare, come parametro
d'ordine,  il valore di aspettazione dell'operatore di linea di Wilson $<L>$
[Sat84].
Il valore di aspettazione \`e legato all'energia 
 libera $F$ indotta dalla presenza di un quark statico posto
entro un bagno termico gluonico a temperatura $T$ ($<L>\sim e^{-F/T}$).    
In presenza di confinamento, la carica di colore del quark non pu\`o
essere schermata, per cui la sua energia libera \`e infinita e 
$<L>=0$. Nella fase deconfinata, d'altra parte,
il quark interagisce in maniera finita col mezzo gluonico, la sua
energia libera risulta finita e $<L>$ diventa diverso da zero [Kog83].
In fig.~\ref{fwilson} \`e mostrato l'andamento del parametro d'ordine
$<L>$ in funzione della temperature $T$ risultante da un calcolo di QCD su
reticolo.

A bassa temperatura, il valore nullo del parametro indica che i quarks 
si trovano nella usuale fase confinata. Tuttavia, in corrispondenza di un valore
 critico di temperatura $T_c$ (stimato pari a $\sim150~MeV$),
 il parametro subisce una repentina
crescita ed il sistema si porta in una fase deconfinata, non prevista dalla
QCD perturbativa.
\par
\begin{figure}[htb]
\centering
\includegraphics[scale=2.0,clip]
                                  {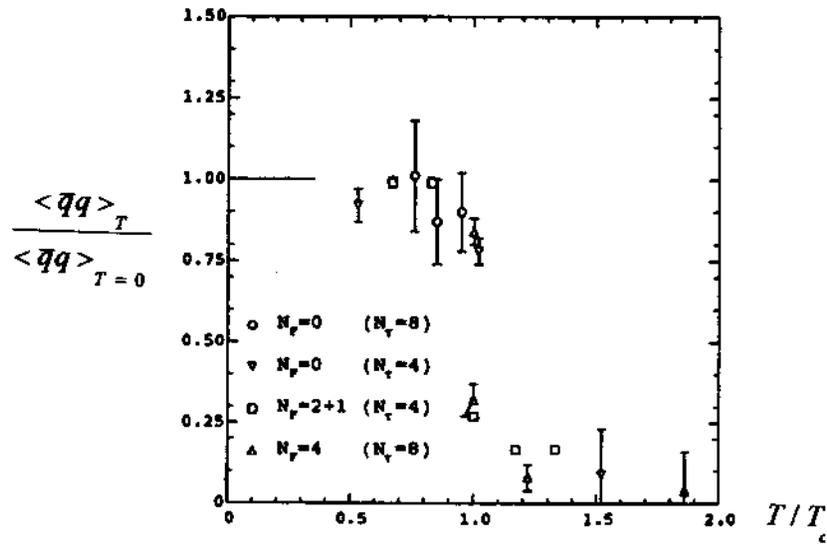}
\caption{\em Andamento del parametro d'ordine $<\overline{q}q>$
             in funzione della
    temperatura $T/T_c$, per reticoli di diverse dimensioni [Wei96].}
\label{fqq}
\end{figure}

Un secondo parametro d'ordine utile per studiare i cambiamenti nelle
proprie\-t\`a fisiche del vuoto di QCD \`e il valore di aspettazione 
$<\overline{q}q>$, dove $q$ \`e il campo dei quarks e $\overline{q}$
quello degli antiquarks. Un valore nullo di tale parametro indica una esatta
simmetria chirale della teoria, mentre un valore non nullo \`e legato
alla rottura spontanea di tale simmetria nello stato di vuoto di QCD.
Poich\`e nel lagrangiano di QCD la simmetria chirale \`e rotta dal termine
di massa dei quarks, del tipo

\[
L_{Massa}(n_f=3)=-(m_u\overline{u}u+m_d\overline{d}d+m_s\overline{s}s)
\]

\noindent  il parametro $<\overline{q}q>$ fornisce anche una
misura della massa effettiva dei quarks.
Simulazioni di QCD su reticolo
 indicano una rapida decrescita del parametro
d'ordine $<\overline{q}q>$ in corrispondenza della temperatura critica
$T_c$ (valutata intorno a $150\div200~MeV$), come mostrato in fig.~\ref{fqq}.

Per temperature inferiori a quella critica, il parametro d'ordine risulta
diverso da zero, i quarks possiedono una massa costituente effettiva
($m_u\sim m_d\sim 300~MeV,\,\,\,\, m_s\sim 500~MeV$)
a causa delle interazioni tra di loro e col vuoto fisico circostante 
e la simmetria
chirale \`e parzialmente rotta; per temperature superiori a quella critica
il parametro d'ordine diventa quasi nullo, le masse dei quarks risultano ridotte
($m_u\sim m_d\sim 5~MeV,\,\,\,\, m_s\sim 150~MeV$), essendo trascurabili
le interazioni col nuovo vuoto perturbativo di QCD, e la simmetria chirale
\`e parzialmente ristabilita.
\par
Pur non essendoci alcuna ragione fondamentale che giustifichi la coincidenza
tra la transizione che porta al deconfinamento e quella che produce un 
ripristino della simmetria chirale, i due fenomeni risultano coincidere
nelle simulazioni della QCD su reticolo.
\par
Riassumendo, la nuova fase della materia nucleare, quale risulta 
dall'analisi delle variabili
termodinamiche calcolate nell'ambito della QCD statistica, \`e costituita
da quarks e gluoni deconfinati ed asintoticamente liberi.
Le loro interazioni hanno una intensit\`a ridotta e ci\`o produce un ripristino
della simmetria chirale, con conseguente riduzione delle masse dei quarks.
Questo stato di plasma di quarks e gluoni (QGP) si forma in corrispondenza
di una temperatura critica compresa tra $150$ e $200~MeV$ e di una
densit\`a di energia di $2\div3~GeV/fm^3$.
\par
\begin{figure}[htb]
\centering
\includegraphics[scale=1.7,clip]
                                  {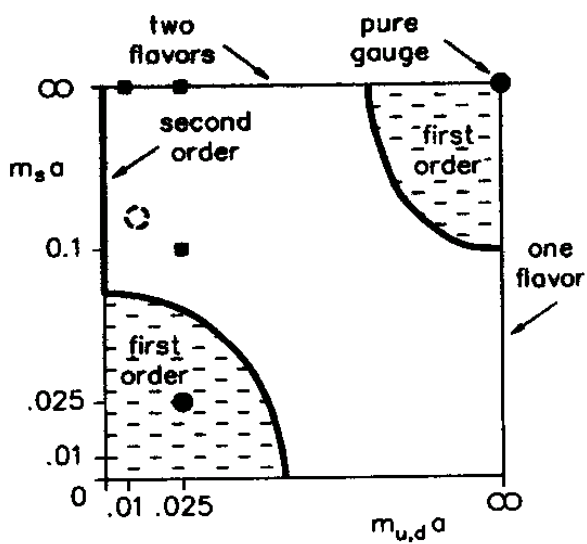}
\caption{\em Le regioni della transizione di fase in funzione della
             massa dei quarks 
($a$ \`e il passo del reticolo usato) [Bro74].}
\label{fordine}
\end{figure}

L'ordine della transizione di fase dipende, cos\`{\i} come il valore della
temperatura critica, dal numero di {\em flavours} e dalle masse dei quarks
usate nelle simulazioni.
La transizione risulta del primo ordine se i calcoli sono effettuati
con masse dei quarks infinite (limite di pura gauge) o nulle (limite chirale).
L'uso di masse realistiche dei quarks 
($m_u\sim m_d\sim 5~MeV, m_s\sim 150~MeV$) comporta una transizione del secondo
ordine, come indicato dal cerchio tratteggiato in fig.~\ref{fordine}.
Tuttavia questo risultato necessita di una  conferma ottenibile utilizzando
 reticoli di dimensioni maggiori.

\section{Il plasma di quarks e gluoni}
\index{Il plasma di quarks e gluoni}

L'analisi di sistemi estesi di materia nucleare pu\`o essere condotta, oltre
che per mezzo della QCD statistica, anche utilizzando modelli fenomenologici
che, partendo da un certo numero di assunzioni fisiche semplificatrici, riescono
a fare previsioni sulle diverse variabili termodinamiche, evitando laboriosi
calcoli numerici.
In questo paragrafo si far\`a uso del ``modello a {\em bag}'' del M.I.T.
[Cho74] per ricavare un'equazione di stato approssimata del QGP.
\par
Nel modello a {\em bag}, gli adroni sono considerati come ``sacche'' di falso
vuoto, in cui quarks di massa nulla e gluoni si propagano liberamente.
L'effetto non perturbativo del confinamento \`e rappresentato 
fenomenologicamente dalla presenza di una pressione esterna $B$ che bilancia
la pressione interna dovuta all'energia cinetica dei quarks.
La pressione $B$ pu\`o essere interpretata come la densit\`a di energia
necessaria per modificare la struttura del vuoto reale di QCD all'interno
 della sacca, dando origine ad una regione di vuoto perturbativo.
La spettroscopia adronica fornisce per $B^{1/4}$ 
valori compresi tra $145~MeV/fm^3$ [Hax80] e $235~MeV/fm^3$ [Has81].
\par
All'interno del modello \`e possibile prevedere l'esistenza di uno stato di 
plasma di quarks e gluoni liberi e deconfinati all'interno di una sacca
estesa di falso vuoto. Le equazioni che regolano tale stato, caratterizzato
dalla temperatura $T$ e dal potenziale chimico dei quarks $\mu$\footnote{
Il potenziale chimico $\mu$ indica l'energia minima necessaria per aggiungere
un quark al sistema per $T=0$. Gli antiquarks hanno potenziale chimico
opposto a quello dei quarks. In generale il valore del potenziale chimico
\`e differente al variare del {\em flavour} del quark considerato.},
 sono:

\begin{equation}
\epsilon = \epsilon_g(T,\mu)+\epsilon_q(T,\mu)+\epsilon_{\overline{q}}(T,\mu)+B
\label{eos1}
\end{equation}
\begin{equation}
P = P_g(T,\mu)+P_q(T,\mu)+P_{\overline{q}}(T,\mu)-B
\label{eos2}
\end{equation}

\noindent dove 
$\epsilon$ e $P$ sono, rispettivamente, la densit\`a di energia e la pressione
dello stato di plasma,
$g$,$q$ e $\overline{q}$ indicano, rispettivamente, i contributi dovuti
al moto dei gluoni, dei quarks e degli antiquarks  e $B$
 indica la densit\`a di energia e la pressione dello stato di falso vuoto.
Il suo segno negativo nell'equazione (\ref{eos2}) \`e interpretabile come
instabilit\`a dello stato di vuoto nella sacca, la quale collassa
se non \`e sostenuta dalla pressione dei costituenti del plasma
che la occupano. Un limite inferiore alla stabilit\`a dello stato di plasma
\`e dunque ottenibile ponendo $P=0$ nella eq.~(\ref{eos2}), per cui i valori
critici di temperatura e potenziale chimico in corrispondenza dei quali ha
luogo la transizione di fase si possono ottenere dall'equazione

\begin{equation}
 P_g(T_c,\mu_c)+P_q(T_c,\mu_c)+P_{\overline{q}}(T_c,\mu_c)=B
\label{eos3}
\end{equation}

Nell'ipotesi che le interazioni tra i costituenti del plasma e le loro
masse siano trascurabili, le rispettive densit\`a di energia risultano:

\begin{equation}
\epsilon_g=n_g\int\frac{d^3p}{(2\pi)^3} \cdot p \cdot
            \left(\frac{1}{e^{p/T}-1}\right)=
          n_g\frac{\pi^2}{30}T^4
\label{ene_g}
\end{equation}
\begin{eqnarray}
\epsilon_q+\epsilon_{\overline{q}} & = & 
n_q\sum_f\int\frac{d^3p}{(2\pi)^3} \cdot p
          \cdot \left(\frac{1}{e^{(p-\mu_f)/T}+1}+\frac{1}{e^{(p+\mu_f)/T}+1}
          \right) \nonumber \\
 & = & 
n_q\sum_f\left(\frac{7\pi^2}{120}T^4+\frac{1}{4}\mu_f^2T^2+\frac{1}{8\pi^2}
\mu_f^4\right)
\label{ene_qq}
\end{eqnarray}

\noindent dove $n_g$ ed $n_q$ indicano rispettivamente il numero di gradi di libert\`a
di colore e {\em spin} associati ai gluoni ed ai quarks: 
$n_g=8\cdot2$ e $n_q=3\cdot2$. La sommatoria \`e invece eseguita sul numero
di {\em flavours} la cui massa pu\`o essere trascurata rispetto alla stato termico 
del plasma.
L'equazione di stato (\ref{eos1}) diventa:

\begin{equation}
\epsilon=\frac{8}{15}\pi^2T^4+3
\sum_f\left(\frac{7}{60}\pi^2T^4+\frac{1}{2}\mu_f^2T^2+\frac{1}{4\pi^2}
\mu_f^4\right)+B
\label{eos}
\end{equation}

\noindent e, considerando che per particelle a massa nulla risulta
 $P=\epsilon/3$, la relazione di stabilit\`a (\ref{eos3}) si scrive:

\begin{equation}
\frac{8}{45}\pi^2T^4+
\sum_f\left(\frac{7}{60}\pi^2T^4+\frac{1}{2}\mu_f^2T^2+\frac{1}{4\pi^2}
\mu_f^4\right)=B
\label{stabile}
\end{equation}

\begin{figure}[htb]
\centering
\includegraphics[scale=1.9,clip]
                                  {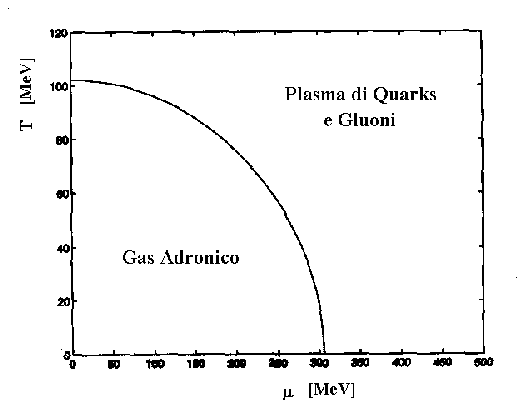}
\caption{\em Tipico risultato ottenuto dall'eq.~(\ref{stabile}), che mostra
la relazione tra i valori critici della temperatura T e del potenziale chimico
$\mu$. Il calcolo \`e stato eseguito con $B^{1/4}=145~MeV/fm^3$ e con due
  flavours ($\mu=\mu_u=\mu_q$).}
\label{diagramma}
\end{figure}

In fig.~\ref{diagramma} \`e mostrata la linea di stabilit\`a (\ref{stabile})
nel piano ($\mu,T$) ottenuta per $B^{1/4}=145~MeV/fm^3$ e considerando due 
{\em flavours} ($u$ e $d$).

\`E interessante notare come la transizione possa avvenire sia per riscaldamento
che per compressione della materia nucleare.
Nel primo caso, la pressione dei costituenti del plasma, necessaria per 
bilanciare la pressione $B$ del vuoto, \`e fornita dal moto termico, nel secondo
essa proviene dalla degenerazione del gas fermionico costituito dai quarks
ed antiquarks compressi.
\par
La densit\`a barionica della materia nucleare \`e legata al valore
del potenziale chimico $\mu$ dalla relazione:

\begin{equation}
\rho_B=\frac{1}{3} (N_q-N_{\overline{q}})=
\frac{n_q}{3}\sum_f\int\frac{d^3p}{(2\pi)^3} 
          \cdot \left(\frac{1}{e^{(p-\mu_f)/T}+1}-\frac{1}{e^{(p+\mu_f)/T}+1}
          \right) 
\label{eq_den} 
\end{equation}

\noindent dove $N_q$ e $N_{\overline{q}}$ sono, rispettivamente, le densit\`a
di quarks ed antiquarks nel plasma e valgono le stesse considerazioni
fatte per l'eq.~(\ref{ene_qq}).
Svolgendo gli integrali, la (\ref{eq_den}) diventa:  

\begin{equation}
\rho_B=\sum_f\left( \frac{\mu_f}{3}T^2+\frac{{\mu_f}^3}{3\pi^2}\right).
\label{eq_den1} 
\end{equation}  

\noindent Sostituendo la (\ref{eq_den1}) nella (\ref{stabile}), si 
ottiene la linea di stabilit\`a nel piano ($\rho_B$, $T$) ed, in particolare,
il valore critico di densit\`a barionica necessario alla formazione dello
stato di QGP per sola compressione (a $T=0$) risulta da $5$ a $10$ volte
maggiore della normale densit\`a barionica della materia nucleare
($\rho_B=0.14~nucleoni/fm^3$).

\section{Collisioni nucleari ultrarelativistiche}
\index{Collisioni nucleari ultrarelativistiche}

\begin{figure}[htb]
\centering
\includegraphics[scale=3.5,clip]
                                  {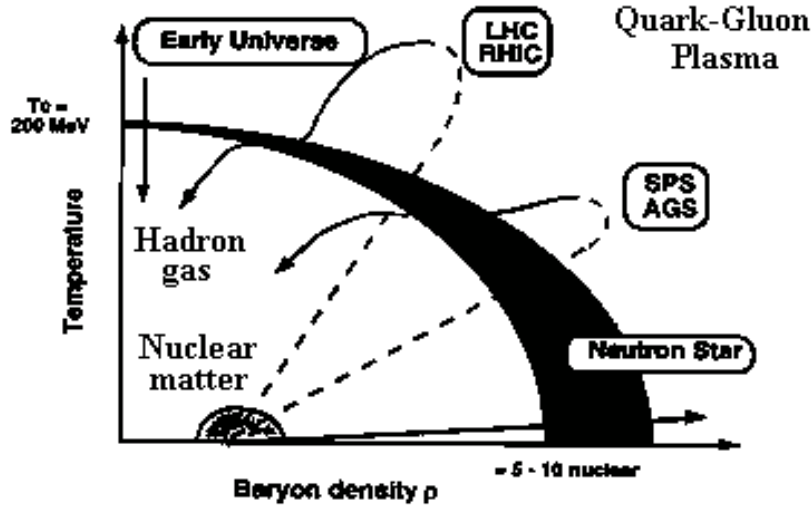}
\caption{\em Diagramma di fase schematico della materia nucleare. Le 
frecce indicano le traiettorie seguite nell'evoluzione dell'Universo,
nelle reazioni tra ioni pesanti agli acceleratori presenti e futuri e
nelle esplosioni di supernova.}
\label{diagramma1}
\end{figure}

Le condizioni estreme di temperatura e densit\`a che, come si \`e visto nel
precedente paragrafo, possono portare alla formazione di uno stato di plasma
di quarks e gluoni possono essere raggiunte in particolari sistemi astrofisici,
quali l'Universo nei primi istanti di vita e la regione pi\`u interna delle
stelle a neutroni, o in seguito a collisioni di alta energia tra ioni pesanti
prodotte nei moderni acceleratori.
Nel diagramma di fase, schematizzato in fig.~\ref{diagramma1} in funzione
della temperatura $T$ e della densit\`a barionica $\rho_B$ della materia 
nucleare, \`e riportato il percorso della transizione di fase che si pensa
possa aver caratterizzato l'Universo dopo circa $10^{-5}~sec$ dal 
{\em Big Bang}.

 Lo studio di transizioni con caratteristiche simili, cio\`e prodotte
per intenso riscaldamento della materia nucleare a bassa densit\`a barionica,
assume estrema importanza in vari aspetti della Cosmologia, in particolare
per quanto riguarda la nucleosintesi, la materia oscura e la struttura a 
larga scala dell'universo [Sch86].
\par
Transizioni di fase che avvengono a bassa temperatura, prodotte per compressione
della materia nucleare, sono invece connesse con problematiche proprie 
dell'Astrofisica. Alle alte densit\`a raggiunte nel nucleo delle stelle
a neutroni i nucleoni possono dissolversi nei loro costituenti, formando 
una plasma di quarks e gluoni freddo e con alta densit\`a barionica.
Inoltre, la dinamica delle esplosioni di supernova, in particolare
quelle dovute ad instabilit\`a di tipo idrodinamico, dipende dalla
compressibilit\`a, e quindi dall'equazione di stato della materia nucleare
[Gle91].
\par
Appare quindi evidente l'importanza di studiare la transizione di fase della materia
nucleare in esperimenti eseguibili in laboratorio e le collisioni
ultrarelativistiche tra nuclei pesanti offrono l'opportunit\`a unica di 
riprodurre e rivelare tale transizione.
Variando l'energia della collisione \`e possibile modificare le condizioni
di temperatura e densit\`a barionica alle quali si manifesta la transizione
di fase, sondando l'intera regione del diagramma di fase.
In fig.~\ref{diagramma1} sono riportati i percorsi tipici ottenibili ai 
presenti sincrotoni (AGS e SPS) ed ai futuri collisionatori (RHIC e LHC).
\par
La particolare caratteristica delle collisioni tra nuclei che permette 
la creazione di stati di materia nucleare ad altissima temperatura e densit\`a
di energia risiede nell'occorrenza di collisioni multiple tra i nucleoni
costituenti.
Infatti, mentre nelle collisioni tra nucleoni i due componenti si 
``attraversano'' depositando solo una piccola frazione della loro energia nella regione
di sovrapposizione (regime di elevata ``trasparenza''), nelle collisioni
tra nuclei ciascun nucleone di un nucleo pu\`o collidere con pi\`u
nucleoni dell'altro nucleo ed ognuna delle singole collisioni produce un
 deposito di energia nella regione di interazione ed un rallentamento dei
due nucleoni interagenti. L'effetto complessivo \`e quello di un parziale
``{\em stopping}'' dei due nuclei e della creazione di una estesa regione
centrale di interazione ad alta densit\`a di energia.
Semplici argomentazioni geometriche portano a stimare,
nel caso di due nuclei di ugual massa $A$,
 una densit\`a di
energia di almeno $\epsilon\sim A^{1/3}~GeV/fm^3$ ed un volume $V\sim A~fm^3$
della regione di interazione, 
mentre tali quantit\`a dipendono in maniera logaritmica 
dall'energia della collisione [Sat86]. Si pu\`o quindi ragionevolmente supporre
che  in collisioni ad alta energia tra nuclei molto pesanti possa essere 
raggiunta la densit\`a di energia necessaria per la formazione di QGP.
Verrebbero cos\`{\i} ad essere
formati sistemi interagenti contenenti un gran numero di particelle
ed aventi dimensioni maggiori della lunghezza di interazione dei suoi 
costituenti. Ci\`o rende possibile l'approccio termodinamico ed il
raggiungimento dell'equilibrio chimico del sistema interagente, cosicch\`e
si possa parlare di fase in senso termodinamico.

\subsection{Dinamica delle collisioni tra nuclei pesanti}
\index{Dinamica delle collisioni tra nuclei pesanti}
\label{par_collisioni}

\begin{figure}[htb]
\centering
\includegraphics[scale=2.6,clip]
                                  {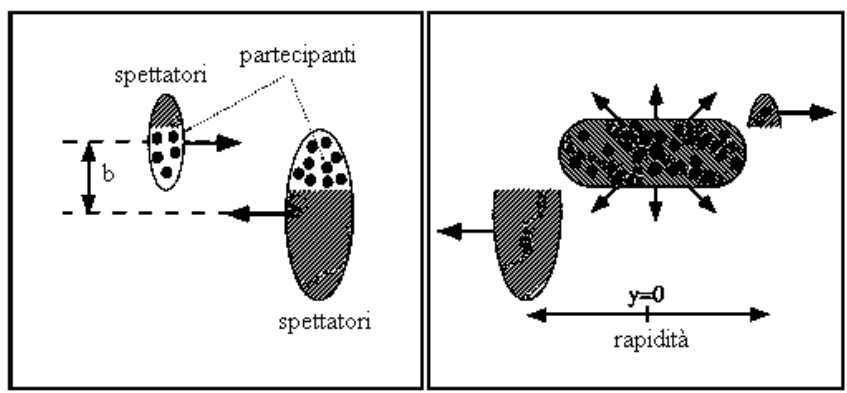}
\caption{\em Rappresentazione schematica di un urto tra ioni pesanti
             con parametro di impatto $b$.}
\label{urto}
\end{figure}

I nuclei sono oggetti estesi, per cui la loro geometria gioca un ruolo
 essenziale nell'urto tra ioni pesanti. In fig.~\ref{urto} \`e illustrata
schematicamente la collisione tra due nuclei asimmetrici nel sistema
del centro di massa.

Il parametro di impatto $b$ separa i nucleoni dei due nuclei in
``partecipanti'', i quali interagiscono mediante collisioni primarie di tipo
nucleone-nucleone, e ``spettatori'', i quali procedono pressocch\`e
indisturbati lungo la loro direzione originale.
L'uso di una semplice rappresentazione geometrica della collisione \`e 
giustificato dal fatto che ad energie molto alte la dimensione
dei nucleoni \`e maggiore della loro lunghezza d'onda Compton ed il raggio
nucleare \`e pi\`u grande della tipica lunghezza di interazione
tra nucleoni ($\sim 1.8~fm$).
\par
Le precedenti argomentazioni geometriche portano a concludere che le
collisioni in cui risulti maggiore il trasferimento della originale energia 
longitudinale in energia di eccitazione della materia nucleare 
interagente  siano quelle centrali, nelle quali il parametro di impatto
\`e nullo ed il numero di nucleoni spettatori risulta minimizzato.
Sperimentalmente la selezione di urti centrali avviene per mezzo di 
osservabili fisiche collegate al numero di nucleoni partecipanti
alla collisione ed alla densit\`a di energia raggiunta nella regione centrale, 
come la molteplicit\`a di particelle secondarie
prodotte e l'energia emessa in direzione trasversa. I primi esperimenti
con nuclei relativistici hanno mostrato che queste variabili sono
strettamente correlate [And92]. In anticorrelazione con l'energia trasversa,
per esperimenti a bersaglio fisso \`e di solito usata anche l'energia residua
misurata in avanti lungo la linea di collisione.
\par
Una variabile adimensionale molto utile per caratterizzare le collisioni tra
nuclei pesanti \`e la rapidit\`a, definita dalla relazione:

\begin{equation}
y=\frac{1}{2}\ln\left(\frac{E+p_L}{E-p_L}\right)
\label{rap}
\end{equation}

\noindent dove $E$ e $p_L$ sono, rispettivamente, l'energia e l'impulso
longitudinale della particella in esame.
Nel sistema del centro di massa (CMS) la regione centrale di interazione 
corrisponde a valori di rapidit\`a intorno allo zero, avendo i suoi
partecipanti perso l'originario impulso longitudinale, mentre i frammenti
provenienti dal bersaglio e dal proiettile assumono valori opposti di 
rapidit\`a, cos\`{\i} come riportato in fig.~\ref{urto}.
I valori di rapidit\`a propri della zona di frammentazione del
bersaglio e del proiettile possono essere ricavati dalle relazioni

\begin{equation}
\left\{ \begin{array}{ccc}
        \sinh y & = & \frac{\textstyle p_L}{\textstyle m_T}\\
        \cosh y & = & \frac{\textstyle E}{\textstyle m_T}
        \end{array}
\right.
\label{b_t.rap}
\end{equation}

\noindent dove $m_T=\sqrt{p_T^2+m^2}$ \`e la massa trasversa della particella
proveniente da tali regioni.
\par
Una notevole propriet\`a della variabile rapidit\`a consiste nell'invarianza 
degli intervalli di rapidit\`a per trasformazioni di Lorentz lungo l'asse
del fascio. Le distribuzioni di rapidit\`a assumono dunque la stessa
forma sia nel sistema del laboratorio che in quello del centro di massa,
ma vengono traslate della quantit\`a

\begin{equation}
y_{CM}=\frac{1}{2}\ln\left(\frac{1+\beta_{CM}}{1-\beta_{CM}}\right)
       \simeq\ln 2\gamma_{CM}
\label{cms.rap}
\end{equation}

\noindent dove $\beta_{CM}$ e $\gamma_{CM}$ sono, rispettivamente,
la velocit\`a ed il fattore di Lorentz del CMS rispetto al sistema del
laboratorio. Il valore $y_{CM}$ \`e comunemente chiamato rapidit\`a del centro
di massa e caratterizza la regione centrale di interazione
nel sistema del laboratorio.
In definitiva, per individuare le particelle provenienti dalla regione 
centrale di interazione, distinguendole dai frammenti di bersagli o proiettile,
si devono selezionare quelle contenute in intervalli di rapidit\`a centrati
sul valore $y_{CM}$, corrispondente
al valore nullo di rapidit\`a nel centro di massa della collisione.
\par
Nel limite ultrarelativistico ($E>>m$) la rapidit\`a pu\`o essere approssimata
dalla pseudorapidit\`a:

\begin{equation}
\eta=\frac{1}{2}\ln\left(\frac{p+p_L}{p-p_L}\right)
\label{pseudorap}
\end{equation}

\noindent usata quando non \`e possibile identificare la  particella.

\subsection{Modelli di collisione}
\index{Modelli di collisione}
\label{parmodel}

Le modalit\`a con le quali si sviluppano le collisioni tra nuclei pesanti
e le caratteristiche fisiche della regione centrale di interazione dipendono
dall'energia fornita ai nuclei interagenti.

\clearpage

\begin{figure}[htb]
\centering
\includegraphics[scale=2.2,clip]
                                  {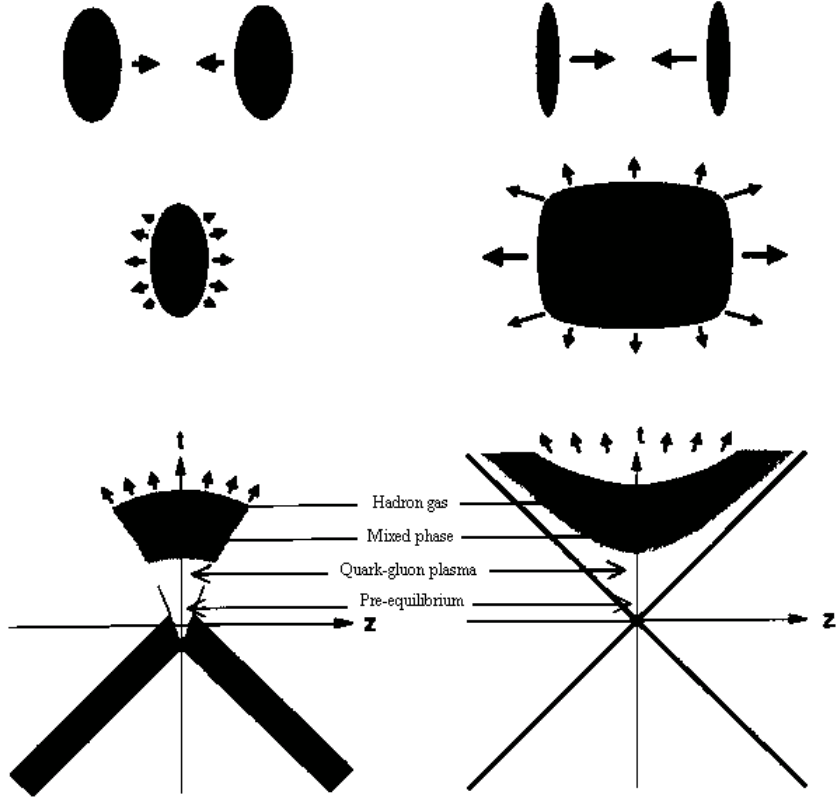}
\caption{\em Rappresentazione schematica della collisione relativistica
tra ioni pesanti nel modello di Fermi-Landau (a sisistra) e quello di 
Bjorken-Mclarran (a destra). L'evoluzione della
collisione \`e mostrata nello spazio (in alto) e nello 
spazio-tempo (in basso).}
\label{modelli}
\end{figure}

In maniera qualitativa, \`e possibile individuare due distinti 
modelli di collisione: uno sviluppato da Fermi e Landau e valido
per energie del sistema del centro di massa dell'ordine delle decine
di $GeV$ per nucleone [Fer50], [Lan53], l'altro valido per energie superiori alle
centinaia di $GeV$ per nucleone e dovuto a Bjorken e Mclerran [Bjo83], [Mcl82].
Il primo modello, mostrato a sinistra in fig.~\ref{modelli}, \`e caratterizzato
da un completo {\em stopping} dei nucleoni costituenti.
Dalle eq.~(\ref{b_t.rap}) si pu\`o calcolare la separazione in rapidit\`a
tra il nucleo proiettile e quello bersaglio alle energie 
caratteristiche di tale modello.
Essa risulta di circa $3\div 4$ unit\`a, dello stesso ordine della perdita media
di rapidit\`a dei nucleoni partecipanti alla collisione\footnote{\`E possibile
valutare la perdita media di rapidit\`a dei nucleoni partecipanti attraverso
l'analisi delle distribuzioni di rapidit\`a dei protoni e degli antiprotoni
prodotti nella collisione (c.f.r. par.~\ref{parrap}).}.
Inoltre la contrazione di Lorentz nel sistema del centro di massa favorisce
l'addensamento longitudinale dei nucleoni costituenti
negli istanti precedenti la collisione.
Di conseguenza, i nucleoni partecipanti perdono completamente la loro energia
originaria e si accumulano nella regione centrale di interazione, formando
uno stato di materia ad alta densit\`a barionica.
Come visto in fig.~\ref{diagramma} per quanto riguarda le collisioni
all'AGS e SPS, la transizione di fase verso lo stato di QGP pu\`o
avvenire in queste condizioni principalmente per compressione.
\par
Il secondo modello, mostrato a destra in fig.~\ref{modelli}, \`e
invece caratterizzato dall'assenza di barioni nella regione centrale
di interazione, e la relativa transizione di fase \`e quella corrispondente
agli acceleratori RHIC e LHC in fig.~\ref{diagramma}.
Alle altissime energie proprie di tale modello, la separazione tra la rapidit\`a
del bersaglio e quella del proiettile supera le 10 unit\`a, per cui i nucleoni
partecipanti non perdono completamente la loro energia in seguito alla
collisione ed il loro moto residuo \`e tale da farli allontanare dalla regione
centrale di interazione. In altre parole, le rapidit\`a delle particelle 
provenienti dalla frammentazione del bersaglio e del proiettile risultano
ben separate anche dopo la collisione.
Di conseguenza, la collisione \`e caratterizzata da una alta ``trasparenza''
dei due nuclei e la materia nucleare nella regione centrale di interazione ha 
un bassissimo contenuto barionico. In questo caso la transizione di fase
avviene per riscaldamento e le condizioni del QGP sono simili a quelle 
proprie dell'Universo dopo il {\em Big Bang}.
\par
All'interno di tale modello la densit\`a di energia iniziale 
raggiungibile nella regione
centrale pu\`o essere stimata dalla formula

\begin{equation}
\epsilon = \frac{<m_T>}{\tau_0 \pi R^2}\left. \frac{dN}{dy} \right|_{y=0}
\label{bio}
\end{equation}

\noindent dove $R$ \`e il raggio dei due nuclei,  $\tau_0$ \`e il tempo
di formazione del plasma, stimato da Bjorken dell'ordine di $1~fm/c$ 
\hspace{0.2cm} e \hspace{0.2cm}
$<m_T>=<\sqrt{m^2+p_T^2}>$ \hspace{0.2cm}
\`e la massa trasversa media delle particelle prodotte.

\subsection{Evoluzione spazio-temporale}
\index{Evoluzione spazio-temporale}

L'evoluzione spazio-temporale della collisione pu\`o essere distinta nelle
fasi il\-lu\-stra\-te in fig.~\ref{modelli} (in basso) ed \`e sostenzialmente analoga
per i due modelli di collisione nucleare prima esaminati.
\par
La prima fase, detta di ``pre-equilibrio'', ha una durata inferiore
ad $1~fm/c$ ed ha luogo durante i primi istanti successivi alla collisione.
Essa \`e caratterizzata dall'occorrenza di interazioni a grande momento 
trasferito ({\em hard}) tra i costituenti partonici dei nucleoni. Le sezioni
d'urto di tali processi, calcolabili attraverso la QCD perturbativa, risultano
molto piccole e rendono trascurabili i loro effetti se confrontati con 
gli effetti delle collisioni tra nucleoni.
Queste avvengono con larga probabilit\`a e con basso momento trasferito 
({\em soft}) e permettono di ridistribuire una frazione dell'originale energia
del fascio nei gradi di libert\`a della materia interagente.
\par
Nella fase successiva i quarks e gli antiquarks si materializzano dal campo
di colore eccitato di QCD. Dopo un tempo di formazione di
circa $1~fm/c$ il sistema interagente raggiunge le condizioni per
la formazione del QGP e l'equilibrio termico pu\`o essere
 raggiunto attraverso le
interazioni tra i suoi costituenti elementari.
La QCD statistica permette di stimare un  libero cammino medio dei quarks
nel plasma di circa $0.5~fm$ alla densit\`a di $2~GeV/fm^3$ [Dan85]:
tenendo conto che le dimensioni trasverse dei nuclei 
($R\sim 7~fm$ per il piombo) sono molto pi\`u grandi di tale cammino medio,
le interazioni nel plasma dovrebbero essere sufficienti a garantire
il raggiungimento dell'equilibrio termico entro il tempo di vita del plasma.
\par
Nella terza fase, detta ``fase mista'', il sistema si espande rapidamente,
principalmente lungo la direzione longitudinale, e la sua temperatura 
diminuisce, raggiungendo il valore critico dopo circa $3\div 5~fm/c$ [Sat90].
Inizia il processo di adronizzazione che porta a riarrangiare i molti gradi di 
libert\`a del QGP nei pochi disponibili nella fase adronica.
Nell'ipotesi che la transizione di fase sia del primo ordine, tale processo
porta ad un rilascio di calore latente ed il sistema permane in una
fase mista di lunga durata (maggiore di $10~fm/c$) in cui sono
contemporaneamente presenti materia adronica e plasma di quarks e gluoni.
\par
Nell'ultima fase il sistema interagente, ormai completamente passato
alla fase adronica, continua la sua espansione che pu\`o essere descritta
in termini idrodinamici come moto ordinato di materia.
Il sistema, assimilabile ad un gas adronico, permane in questa fase per 
decine di $fm/c$, fino al raggiungimento di volumi di interazione dell'ordine
di $10^4\div 10^5~fm^3$.
Quando le tipiche distanze tra le particelle diventano pi\`u grandi del loro
libero cammino medio, le loro interazioni cessano e le
particelle vengono irraggiate
nello spazio circostante. Dopo quest'ultimo stadio, chiamato
{\em freeze-out}, gli effetti della collisione sono osservabili
intercettando parte delle particelle emesse con opportuni rivelatori.

\section{Rivelazione del QGP}
\index{Rivelazione del QGP}

Il sistema creato in seguito a collisioni ultrarelativistiche non \`e statico,
 ma, come si \`e visto nel paragrafo precedente, subisce una rapida evoluzione.
Le osservabili sperimentali corrispondono, in generale, ad un integrale
su tutta la completa evoluzione spazio-temporale della reazione fino al
{\em freeze-out} ed \`e necessario separare i contributi delle diverse fasi
evolutive al segnale rivelato, per poter risalire alle informazioni relative
al primi istanti della collisione.
Tuttavia, un sistema che si evolve attraverso stati di equilibrio per definizione
cancella la sua memoria sugli stati precedenti. In altri termini, 
la complessit\`a del sistema e il gran numero di interazioni tra i suoi 
costituenti che, insieme, giustificano l'utilizzo dell'approccio statistico
allo studio del suo stato e favoriscono il raggiungimento di un
equilibrio termico di tipo locale, al tempo stesso rendono minima
l'informazione proveniente dagli stati precedenti.
Alla base di qualsiasi tentativo di rivelazione di QGP risiede, dunque, 
l'ipotesi che la rapida espansione del sistema interagente impedisca
il raggiungimento di un completo stato di equilibrio in ogni fase della
sua evoluzione, per cui sia preservata qualche traccia della sua storia 
iniziale.
\par
La strategia usata \`e quella di rinunciare alla ricerca di un unico segnale
che permetta l'inequivocabile identificazione della fase di QGP, ma di associare,
per quanto possibile, ogni osservabile ad un particolare stadio o caratteristica
della collisione. In questo modo si cerca di focalizzare l'attenzione su segnali
che si di\-sac\-cop\-piano in differenti momenti dall'espansione del sistema, per
cui contengono informazioni non degradate relative a ciascuno di essi.
\par
Le caratteristiche degli stadi iniziali contenute nei segnali rivelati possono
essere estratte anche per mezzo di modelli termici in grado di
descrivere l'evoluzione della collisione da un dato istante iniziale fino al 
{\em freeze-out}. Partendo dalle informazioni sperimentalmente accessibili,
tali modelli sono in grado di compiere una estrapolazione nel tempo fino
ai primi istanti della collisione, fornendo previsioni circa i parametri
caratteristici del relativo stato del sistema.
\par
Nel seguito del paragrafo verr\`a fornito un quadro dei diversi esperimenti
coinvolti nella fisica degli ioni pesanti e verranno illustrate le
diverse osservabili fisiche proposte, insieme ai pi\`u recenti risultati
 sperimentali ottenuti. Il segnale di produzione di stranezza verr\`a 
analizzato in dettaglio nel prossimo paragrafo.

\subsection{Esperimenti sugli ioni pesanti ultrarelativistici}
\index{Esperimenti sugli ioni pesanti ultrarelativistici}

L'attivit\`a sperimentale riguardante la fisica degli ioni pesanti 
ultrarelativistici si \`e sviluppata a partire dal 1986 con la possibilit\`a
di accelerare ioni $^{16}O$ e $^{32}S$ al sincrotone SPS del CERN (Svizzera)
e ioni $^{28}Si$ al sincrotone a gradiente alternato AGS di Brookhaven (USA).
Gli esperimenti coinvolti in questa prima fase, di carattere esplorativo,
hanno confermato la possibilit\`a di produrre materia nucleare ad alta densit\`a
di energia e temperatura. Nelle collisioni prodotte dal fascio di ioni
$^{32}S$, in particolare, sono state raggiunte temperature di  $\sim 200~MeV$
e densit\`a di energia di $\sim 2~GeV/fm^3$: molto vicini ai valori critici
necessari per la formazione di QGP.
L'informazione sperimentale non ha tuttavia consentito una discriminazione tra 
uno scenario di QGP e uno di normale materia adronica.
\par
Questi risultati preliminari hanno indotto una seconda generazione di
 esperimenti, effettuata all'AGS a partire dal 1993 con fasci di ioni $^{197}Au$
e all'SPS a partire dal 1994 con fasci di ioni $^{208}Pb$.
L'obiettivo di questa seconda fase \`e quello di esplorare le osservabili
del QGP individuate durante la prima fase, in sistemi interagenti pi\`u
complessi, dove le condizioni per la formazione del nuovo stato di materia
sono pi\`u favorevoli.
La tabella \ref{tab1} sintetizza le caratteristiche dei fasci fin qui
utilizzati, insieme alle sigle di alcune tra le principali Collaborazioni
coinvolte.

\begin{table}[htb]
\begin{tabular}{||c|c|c|c|c||} \hline\hline
{ } & {\em Ioni} & {$p_{fascio}/A$ ($GeV/c$)} & {$\sqrt{s_{\scriptstyle NN}}$
 ($GeV$)} & {\em Collaborazioni} \\ \hline\hline
{\em AGS} & {Si} & {14.6} & {5.4} & {E802(E859), E810, E814, E858}
\\ \hline
{\em SPS} & {O, S} & {200} & {19.4} & {NA34,35,36,38;  WA80,85,94}
\\ \hline
{\em AGS} & {Au} & {10.7} & {4.7} & {E866, E877, E878}
\\ \hline
{\em SPS} & {Pb} & {160} & {17.3} & {NA44,45,49,50,52;  WA97,98}
\\ \hline\hline
\end{tabular}
\caption{\em Fasci di ioni ultrarelativistici utilizzati e sigle
delle principali Collaborazioni coinvolte}
\label{tab1}
\end{table}

Tra gli esperimenti di seconda generazione effettuati al CERN, l'esperimento
NA49 \`e stato progettato per essere sensibile, con larga accettanza,
a tutte le osservabili fisiche, consentendo uno studio generale della dinamica
della collisione. La strategia usata dalle altre Collaborazioni \`e, invece,
quella di rivelare in maniera specifica particolari segnali della transizione
di fase.

\subsection{Osservabili globali}
\index{Osservabili globali}

Col nome di ``osservabili globali'' si intende indicare tutte quelle
variabili che forniscono informazioni sulla dinamica delle reazioni nucleari
ultrarelativistiche, per lo pi\`u prescindendo dall'identificazione delle
particelle rivelate. La conoscenza delle condizioni iniziali della collisione
e della sua evoluzione dinamica, ottenibile mediante tali variabili,
costituisce tra l'altro un requisito fondamentale per la selezione
degli eventi e lo studio dei segnali specifici di QGP, esposti in seguito.
Alcune delle osservabili globali,  interpretate 
nell'ambito di modelli teorici, consentono di stimare le variabili
termodinamiche che caratterizzano i vari stadi evolutivi della
collisione, come la temperatura di {\em freeze-out}, il tempo di vita
e le dimensioni del sistema interagente e la densit\`a di energia iniziale.

\subsubsection{Distribuzioni di energia trasversa}
\index{Distribuzioni di energia trasversa}

\begin{figure}[htb]
\centering
\includegraphics[scale=1.7,clip]
                                  {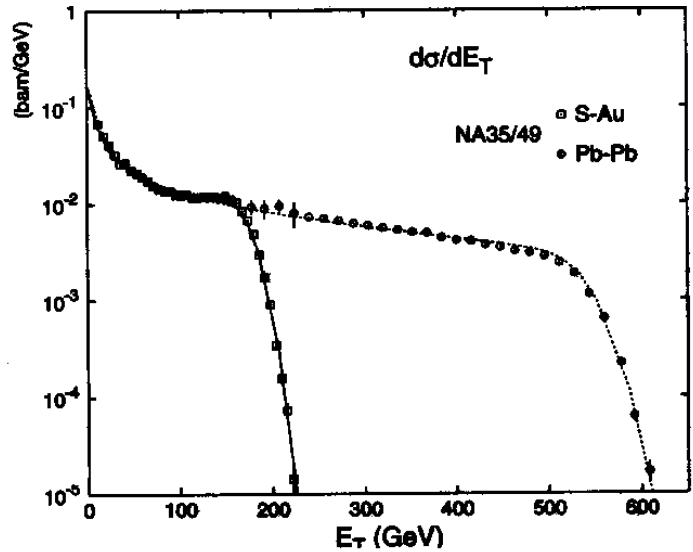}
\caption{\em Spettro di energia trasversa per collisioni S-Au e Pb-Pb
            confrontato col modello di Glauber. La misura \`e eseguita
            nell'intervallo $2.1<\eta<3.4$ [Alb95].}
\label{fet}
\end{figure}

La sezione d'urto differenziale in energia trasversa\footnote{L'energia
trasversa \`e definita come $E_T=\sum_i{E_i\sin\theta_i}$ dove $E_i$ e
$\theta_i$ sono rispettivamente l'energia e l'angolo della particella i-esima rispetto
alla linea di fascio.},
misurata 
dall'esperimento NA35 per interazioni S-Au e dalla Collaborazione
NA49 per interazioni Pb-Pb [Alb95], \`e mostrata in fig.~\ref{fet}.
Ad essa sono sovrapposti i risultati del calcolo condotto utilizzando
il modello di Glauber [Kha97], introdotto in appendice C.

Le distribuzioni di energia trasversa risultano completamente 
interpretabili in termini di geometria della collisione, che \`e 
alla base del modello
di Glauber. A bassi valori di $E_T$ \`e presente il contributo
delle collisioni periferiche; l'appiattimen\-to della  parte centrale
della distribuzione corrisponde ad una
saturazione del numero di nucleoni partecipanti e, infine, dal ginocchio
in poi emerge il contributo di eventi
centrali.
Tale semplice interpretazione geometrica, comune anche ad altre variabili
globali, quali la molteplicit\`a di particelle prodotte e l'energia residua
misurata in avanti lungo la linea del fascio, consente di selezionare
eventi centrali, con grande deposito di energia nella regione centrale
di interazione.
L'estensione dello spettro risulta coprire valori pi\`u alti di energia 
trasversa al crescere della complessit\`a del sistema interagente,
indicando condizioni iniziali pi\`u idonee alla formazione di QGP.
\par
Il valore di energia trasversa corrispondente al ginocchio delle distribuzioni
pu\`o essere utilizzato per stimare la densit\`a di energia raggiunta nella
collisione per messo del modello di Bjorken, esposto nel paragrafo
\ref{parmodel}. La formula (\ref{bio}) pu\`o essere anche scritta come

\begin{equation}
\epsilon = \frac{1}{\tau_0 \pi R^2}\left. \frac{dE_T}{dy} \right|_{max}
\label{bio1}
\end{equation}

\noindent dato che la regione centrale, caratterizzata da valori di rapidit\`a
nel centro di massa intorno allo zero, corrisponde ai valori massimi di energia
trasversa. Inoltre lo spettro in fig.~\ref{fet} \`e stato ricavato
in un ristretto intervallo di pseudorapidit\`a centrato sul valore
di rapidit\`a del centro di massa, per cui il valore di energia trasversa
al ginocchio \`e una buona stima di $\left.\frac{dE_T}{dy}\right|_{max}$.
Per collisioni Pb-Pb si ottiene il valore $\epsilon=2.8~GeV/fm^3$,
da confrontare col valore $\epsilon=2.5~GeV/fm^3$ calcolato per
collisioni S-Au [Fok97].
Tali valori risultano mediati sull'intero volume del sistema interagente
ed \`e stato stimato che la densit\`a di energia per interazioni Pb-Pb pu\`o
raggiungere il valore $\epsilon=3.5~GeV/fm^3$ nella regione pi\`u interna
della sorgente di particelle [Bla96]: valore comparabile con quello critico
predetto dai calcoli di QCD su reticolo.

\subsubsection{Distribuzioni di rapidit\`a}
\index{Distribuzioni di rapidit\`a}
\label{parrap}

\begin{figure}[htb]
\centering
\includegraphics[scale=1.7,clip]
                                  {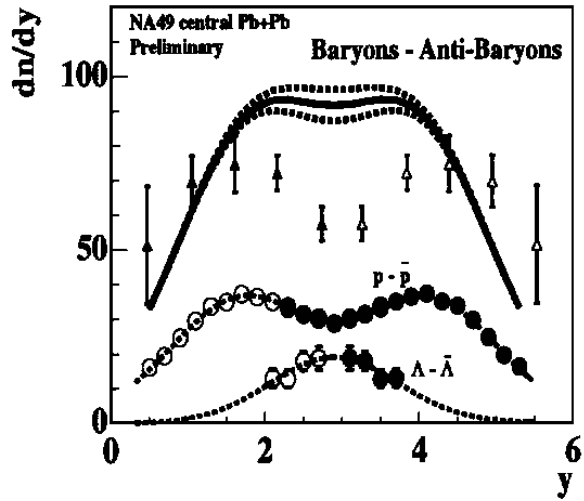}
\caption{\em Distribuzioni di rapidit\`a di $(p-\overline{p})$, di
 ($\Lambda -\overline{\Lambda})$ e del numero barionico netto per
 interazioni Pb-Pb. Per confronto, \`e mostrata  la distribuzione
 del numero barionico netto per collisioni S-S, moltiplicata
 per un fattore 7 (triangoli). I simboli in chiaro rappresentano
 i putni ottenuti per simmetria rispetto al valore centrale di rapidit\`a
 [App97].}
\label{frap}
\end{figure}

Importanti informazioni circa la dinamica della collisione e, in particolare,
le condizioni iniziali del sistema interagente da essa prodotto, possono essere
ottenute dalla misura delle distribuzioni di rapidit\`a delle particelle
nello stato finale. Prima della collisione, tutta l'energia \`e trasportata
dai barioni presenti nei nuclei bersaglio e proiettile; dopo la collisione
il numero barionico netto si distribuisce nello spazio delle fasi
in una maniera che riflette la perdita di energia dei barioni iniziali
durante la collisione. Diventa quindi determinante misurare la distribuzione
di rapidit\`a del numero barionico netto,  non
influenzata dalla produzione delle coppie barione-antibarione che avviene
durante il processo di collisione.
In fig.~\ref{frap} sono mostrate le distribuzioni di rapidit\`a di
$(p-\overline{p})$ e $(\Lambda -\overline{\Lambda})$ in interazioni Pb-Pb
misurate dall'esperimento NA49 [App97]: da queste 
\`e possibile stimare la distribuzione del numero barionico netto mostrata
in grassetto.

\noindent Essa si estende per circa due unit\`a di rapidit\`a intorno
alla rapidit\`a del centro di massa (pari a 2.92 nel laboratorio) 
ed \`e caratterizzata da
una regione centrale piatta e da una rapida decrescita in corrispondenza
della rapidit\`a del bersaglio e del proiettile. 
In fig.~\ref{frap} \`e riportata, per confronto, l'analoga 
distribuzione misurata in collisioni S-S dall'esperimento NA35 [Alb94], la
quale mostra un avvallamento per rapidit\`a centrali e due picchi nelle
vicinanze delle rapidit\`a del bersaglio e del proiettile.
Tutto ci\`o
indica una maggiore decelerazione dei nucleoni partecipanti alle interazioni
Pb-Pb, che quindi subiscono una perdita media di rapidit\`a pi\`u grande che nel
caso di interazioni S-S. In altre parole, in collisioni tra nuclei di piombo
si ha uno {\em stopping} dei nucleoni partecipanti maggiore che nel caso di
interazioni tra nuclei di zolfo, determinando un incremento della densit\`a
di
energia e della densit\`a barionica del sistema interagente successivo alla
collisione.
Queste considerazioni portano a concludere che alle energie proprie
dell'SPS, le collisioni avvengono prevalentemente secondo
il modello di Fermi e Landau descritto nel paragrafo \ref{parmodel}, anche
se il grado di {\em stopping} dipende in larga misura dalla massa dei nuclei
collidenti.

\subsubsection{Distribuzioni di massa trasversa}
\index{Distribuzioni di massa trasversa}
\label{parmassatrasversa}

Lo studio delle distribuzioni di massa trasversa ($m_T=\sqrt{p_T^2+m^2}$)
permette di esaminare la dinamica dell'evoluzione del sistema interagente
formato in seguito agli urti relativistici tra nuclei pesanti. \`E cos\`{\i}
possibile verificare il raggiungimento dell'equilibrio termico locale
del sistema e  assegnare una temperatura al suo stato termico.
\par
Sotto l'ipotesi di raggiungimento dell'equilibrio termico, lo
spettro delle particelle emesse da una singola sorgente stazionaria,
in approssimazione di Boltzmann, \`e parametrizzata da [Sch93]:

\begin{equation}
\frac{d^2N}{dydm_T}\;\sim\;m_T^2\cosh(y-y_{CM})\,e^{-m_T\cosh(y-y_{CM})/T}
\label{temp1}
\end{equation}

\noindent dove $y_{CM}$ \`e la rapidit\`a del centro di massa e $T$ \`e la
temperatura della sorgente.
Per particelle rivelate in un ristretto intervallo di rapidit\`a (tipicamente
$\Delta y<0.5$) l'integrazione della (\ref{temp1})
pu\`o essere approssimata dalla relazione

\begin{equation}
\frac{1}{m_T^2}\frac{dN}{dm_T}\;\sim\;e^{-m_T/T}
\label{temp2}
\end{equation}

\noindent mentre se l'intervallo di integrazione \`e grande ($\Delta y>0.5$)
la (\ref{temp1}) pu\`o essere approssimata dalla relazione

\begin{equation}
\left(\frac{1}{m_T}\right)^{3/2}\frac{dN}{dm_T}\;\sim\;e^{-m_T/T}
\label{temp3}
\end{equation}

\noindent Tale relazione \`e valida anche per integrazioni in piccoli intervalli
di rapidit\`a, se si suppone che l'emissione non avvenga da una singola
sorgente stazionaria, ma da pi\`u sorgenti in moto relativo lungo
la direzione longitudinale con un profilo di velocit\`a Lorentz-invariante.
Mentre il primo scenario \`e caratteristico delle collisioni con {\em stopping}
completo, il secondo caso corrisponde ad una situazione pi\`u realistica
alle energie dell'SPS ed \`e propria di una regione di semi-trasparenza dei
nuclei interagenti [Sch93].
\par
\begin{figure}[htb]
\centering
\includegraphics[scale=1.7,clip]
                                  {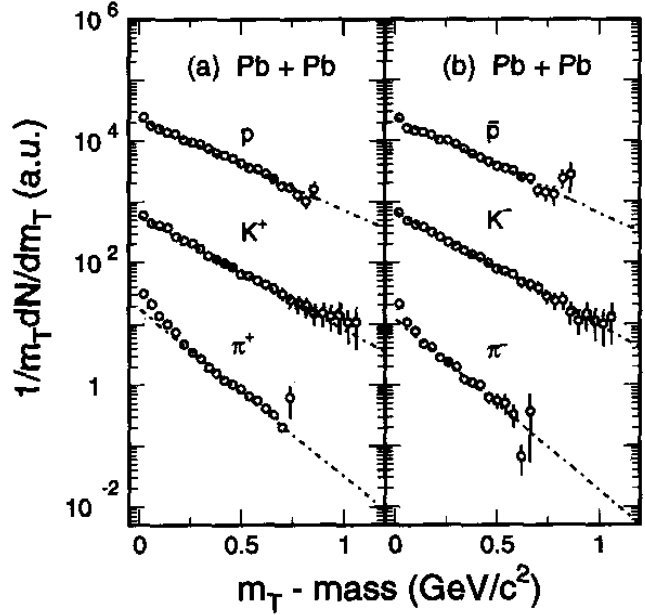}
\caption{\em Distribuzioni di massa trasversa per $p$, $K^+$ e $\pi^+$ (a)
             e le rispettive antiparticelle (b) in collisioni centrali Pb-Pb.
             Le linee tratteggiate rappresentano i fit esponenziali [Bea97].}
\label{ftemp}
\end{figure}

In fig.~\ref{ftemp} sono riportate le distribuzioni di massa trasversa del tipo
$\frac{1}{m_T}\frac{dN}{dm_T}$ per $p$, $K^+$ e $\pi^+$ e le rispettive
antiparticelle identificate in collisioni Pb-Pb dall'esperimento NA44
(per ragioni storiche questo tipo di distribuzione \`e stato preferito
alla forma (\ref{temp2}) in quanto fornisce uno spettro di massa trasversa
invariante per trasformazioni di Lorentz).

Le linee tratteggiate indicano il risultato della 
procedura di {\em best fit} eseguita con la funzione  $e^{-m_T/T}$
[Bea97].
Si pu\`o notare come le distribuzioni siano comples\-sivamente ben
riprodotte da tale funzione, a conferma della natura termica degli
spettri adronici. Ci\`o non \`e tuttavia sufficiente per provare
il raggiungimento dell'equilibrio termico nei primi istanti 
della collisione.
\par
Il parametro $T$ determinato dai {\em fit} delle distribuzioni di massa
 trasversa non pu\`o essere interpretato come temperatura della sorgente
 termica a causa delle distorsioni introdotte dall'espansione del sistema.
Infatti uno stato di equilibrio termico implica la presenza di frequenti
interazioni tra le particelle del sistema. In prossimit\`a della
superficie queste interazioni generano un gradiente di pressione che
d\`a origine ad un flusso trasverso collettivo. D'altra parte, i parametri
$T$ assumono valori inconsistenti con le condizioni presenti al {\em freeze-out},
dato che i liberi cammini medi corrispondenti alle temperature $T$ per le varie
particelle sono molto minori delle dimensioni stimate del sistema interagente.
\par
Il flusso trasverso ha l'effetto di innalzare la temperatura misurata rispetto
a quella presente al {\em freeze-out}, in virt\`u dello spostamento Doppler
introdotto dalla velocit\`a di espansione trasversa, e di fatto i parametri
$T$ misurano sia il contributo dovuto al moto termico (quindi la temperatura
di {\em freeze-out}) che quello dovuto al moto
collettivo (quindi l'energia cinetica  del moto trasverso) all'energia delle particelle
rivelate.
Un'evidenza sperimentale dell'esistenza dell'espansione trasversa \`e fornita
dal sistematico aumento del parametro $T$
con la massa della particella, quindi con la sua energia cinetica trasversa
[Jon96].
\par
Al fine di separare i due contributi nella misura del parametro $T$, i precedenti
modelli termici
sono stati estesi introducendo un'espansione idrodinamica con una velocit\`a 
trasversa comune a tutte le particelle [Sch93], [Kam96].
I parametri liberi di tali modelli sono la temperatura di {\em freeze-out}
$T_f$ e la velocit\`a trasversa $\beta_T$, ma la loro determinazione dal 
confronto con i dati risulta problematica ed ambigua a causa della loro
forte correlazione e della presenza di molteplici minimi della funzione
$\chi^2$ del {\em fit} nello spazio dei parametri. La risoluzione di
tali ambiguit\`a \`e possibile sfruttando informazioni indipendenti
sulla velocit\`a del flusso trasverso.
Queste possono essere ricavate dallo studio della correlazione a due
particelle, nella maniera descritta  qui di seguito.
\par
\begin{figure}[htb]
\centering
\includegraphics[scale=1.8,clip]
                                  {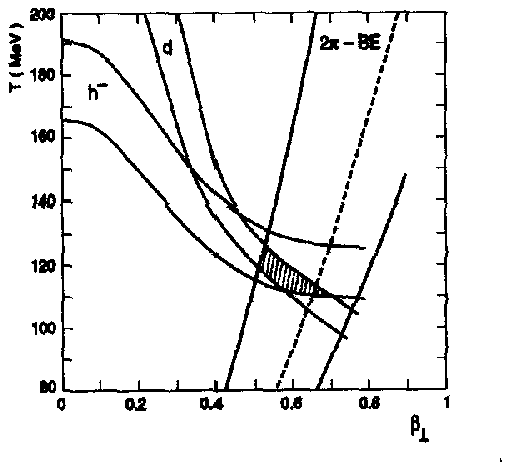}
\caption{\em Temperatura di freeze-out $T_f$ in funzione della
            velocit\`a di flusso trasverso $\beta_T$. Le curve mostrano
         i vincoli provenienti dai fits degli spettri di massa trasversa
           e dall'analisi della correlazione a due particelle [App97].}
\label{fhbt}
\end{figure}

L'interferometria bosonica permette di valutare le dimensioni della regione
di interazione dalla quale i bosoni rivelati sono
stati emessi [Gol60], in analogia con metodo Hambury Brown-Twiss usato in
radio-astronomia per misurare le dimensioni angolari della sorgente 
radio [Han54].
Recentemente il formalismo teorico alla base di tale misura \`e stato
considerevolmente rifinito, introducendo la possibilit\`a di estrarre anche
parametri dinamici, quali la velocit\`a di espansione trasversa e la durata
dell'emissione bosonica [Pra86], [Cha95].
Il sistema interagente \`e descritto come una successione di sorgenti locali
in moto relativo lungo l'asse longitudinale, ognuna caratterizzata da 
 un'estensione spazio-temporale ($R_{//}$, $R_T$, $\tau$) e da una velocit\`a
longitudinale. Tali parametri sono determinati attraverso il 
confronto con i dati sperimentali, eseguito per mezzo di opportune
funzioni di correlazione, e la loro dipendenza dal momento medio della
coppia di bosoni permette di ottenere informazioni sull'espansione del 
sistema [Hei96].
Tali informazioni possono essere appunto 
combinate con quelle provenienti dagli spettri
di massa trasversa per ottenere stime pi\`u precise e non ambigue dei 
parametri $T_f$ e $\beta_T$. In fig.~\ref{fhbt} \`e mostrata la regione
permessa nello spazio dei parametri dopo aver applicato i vincoli provenienti
dal {\em fit} sugli spettri di massa trasversa delle particelle negative
($h^-)$ e del deuterone ($d$) e dall'analisi della correlazione a due particelle
($2\pi-BE$).

I valori permessi indicano una temperatura di {\em freeze-out} $T_f\sim 120~MeV$ 
ed una velocit\`a trasversa $\beta_T\sim 0.55$ per interazioni Pb-Pb.
Il tempo di vita del sistema interagente risulta inoltre di $\sim 8~fm/c$
e le dimensioni della sorgente sono $\sim 6\div 7~fm$, circa uguali al raggio
del nucleo di piombo [Fok97].

\subsection{Segnali elettromagnetici}
\index{Segnali elettromagnetici}

I segnali elettromagnetici sono per molti aspetti ideali per la rivelazione
del QGP in quanto permettono di sondare i primi, pi\`u caldi stadi 
dell'evoluzione del sistema interagente, poich\`e non sono affetti dalle
successive interazioni forti. La loro intensit\`a \`e tuttavia molto bassa
e devono essere distinti dall'abbondante fondo proveniente dai processi
di decadimento adronico.
\par
\begin{figure}[htb]
\centering
\includegraphics[scale=2.0,clip]
                                  {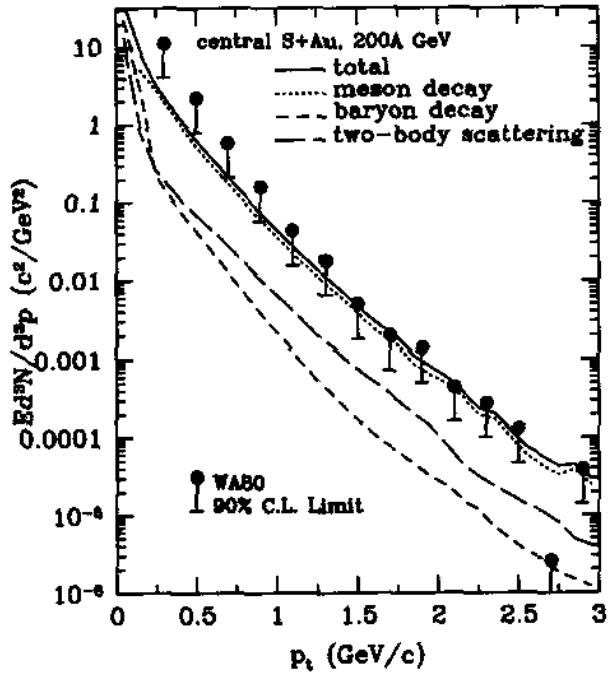}
\caption{\em Spettri di momento trasverso per fotoni prodotti in collisioni
         contrali S-Au, confrontato con il fondo dei decadimenti adronici,
         [Qli97].}
\label{fele0}
\end{figure}

L'emissione di fotoni dallo stato di QGP in equilibrio termico avviene
principalmente attraverso il processo d'urto Compton $gq\rightarrow\gamma q$.
Questo segnale deve competere con i fotoni provenienti dai decadimenti
adronici soprattutto di $\pi^0$ ed $\eta$, e da quelli formati in seguito
a reazioni del tipo $\pi \rho\rightarrow\gamma \rho$ all'interno del
gas adronico termalizzato [Kap93]. Si \`e calcolato che per temperature prossime
a quella critica, la radiazione emessa dal QGP ha intensit\`a e forma spettrale
confrontabile con quella proveniente dal gas adronico [Kap92].
Tuttavia un chiaro segnale di fotoni provenienti dal QGP dovrebbe essere
rivelabile per momenti trasversi compresi tra 2 e 5 $GeV/c$, in seguito alla
formazione di uno stato i\-ni\-ziale di plasma molto caldo [Sri92], [Str94].
Le Collaborazioni NA34 [Ake90], WA80 [Alb91]  e WA98 [Agg96]
 al CERN hanno misurato lo spettro inclusivo
di fotoni in collisioni centrali S-Au e, pi\`u recentemente, Pb-Pb: il confronto
con la predizione del fondo atteso dai decadimenti adronici non ha rivelato
alcun eccesso di fotoni indicante la formazione di QGP.
In fig.~\ref{fele0}, ad esempio, lo spettro di fotoni misurato
in collisioni centrali S-Au dalla Collaborazione WA80 \`e completamente
riprodotto dal fondo calcolato mediante un modello di trasporto 
relativistico.

Poich\`e la produzione di fotoni aumenta fortemente con la temperatura del
 plasma, lo spettro misurato consente di scartare l'ipotesi di formazione di
QGP con temperatura iniziale maggiore di $250~MeV$. Alle energie dei futuri
collisionatori la pi\`u alta temperatura iniziale del sistema interagente
 dovrebbe rendere tale segnale pi\`u significativo.
\par
\begin{figure}[htb]
\centering
\includegraphics[scale=2.7,clip]
                                  {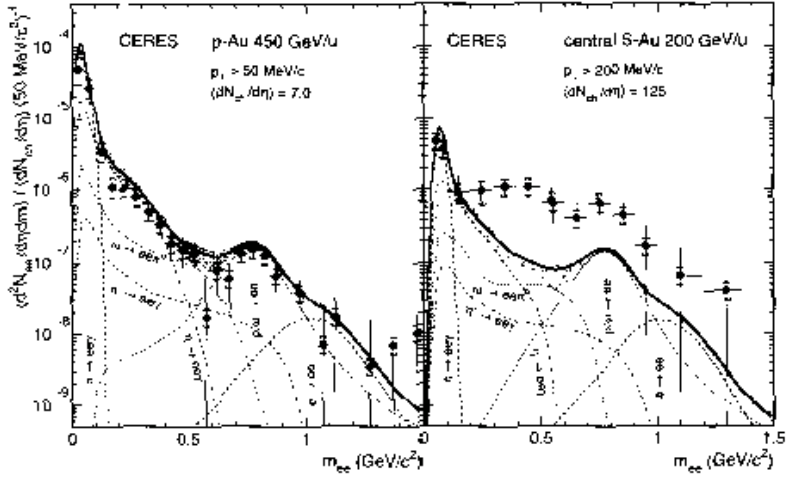}
\caption{\em Spettri di massa invariante per coppie $e^+\,e^-$ in
          $2.1<\eta<2.7$ per interazioni p-Au (a sinistra) e S-Au (a destra)
          [Dre96].}
\label{fele1}
\end{figure}

La produzione di coppie leptoniche in seguito  all'annichilazione
 elettromagnetica di coppie $q\overline{q}$
nel plasma dovrebbe costituire, al
pari della radiazione termica, un segnale non influenzato dalla successiva
adronizzazione e recante informazione circa le condizioni del sistema 
interagente nell'istante della loro formazione.
Il contributo del fondo dovuto ai decadimenti di adroni
\`e determinante per masse invarianti $M_{l^+l^-}<1.5~GeV$, mentre
per $M_{l^+l^-}>5-10~GeV$ diventa dominante il fondo dovuto alle violente
interazioni tra i nucleoni che avvengono nei primi istanti della collisione 
tra nuclei e portano a processi del tipo $q\overline{q}\rightarrow l^+l^-$,
detti di Drell-Yan, o alla produzione di mesoni vettori pesanti che decadono
in coppie leptoniche.
La fig.~\ref{fele1} mostra lo spettro di massa invariante per coppie $e^+e^-$
misurate nell'intervallo di pseudorapidit\`a $2.1<\eta<2.7$
dall'esperimento NA45 in interazioni p-Au (a sinistra) e S-Au (a destra).

\noindent Si pu\`o notare come i decadimenti a due ed a tre corpi dei mesoni
$\pi^0$, $\eta$, $\eta\prime$, $\rho$, $\omega$ e $\phi$, detti ``sorgenti
standard'', riescono a riprodurre sia qualitativamente che quantitativamente
lo spettro misurato in collisioni p-Au.
Nel caso di collisioni S-Au, il contributo delle sorgenti standard, 
calcolato assumendo che l'abbondanza relativa di particelle nello stato finale
sia indipendente dal sistema di collisione, sottostima chiaramente lo spettro
osservato. Questo eccesso di coppie $e^+e^-$ di bassa massa \`e stato
recentemente rivelato anche in collisioni Pb-Au [Aga96] e pu\`o essere cos\`{\i}
quantificato [Dre96]:

\[
\frac{dati}{sorgenti\: standard}=5.0\pm 0.7 (stat.)\pm2.0(sist.).
\]

\begin{figure}[htb]
\centering
\includegraphics[scale=1.85,clip]
                                  {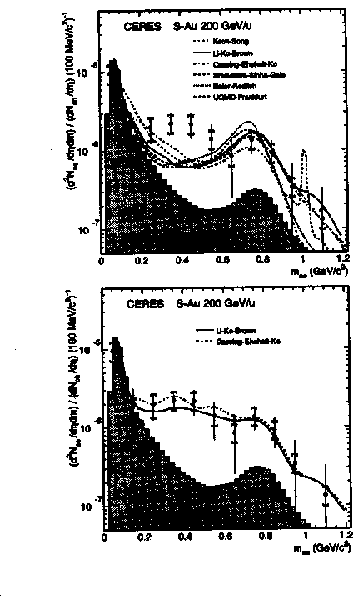}
\caption{\em Confronto tra gli spettri $e^+e^-$ misurati e vari modelli teorici.
I modelli della figura in alto assegnano al mesone $\rho$ la sua massa
nominale, quelli della figura in basso assumono una diminuzione della massa
della $\rho$ [Sto96].} 
\label{fele2}
\end{figure}

In fig.~\ref{fele2} \`e mostrato un confronto tra gli spettri misurati
e vari modelli teorici.

\clearpage

\noindent  I modelli riportati nella figura in alto cercano di spiegare  
l'eccesso di produzione di coppie $e^+e^-$ partendo da differenti
ipotesi, ma con la caratteristica comune di assegnare al mesone $\rho$ la sua
massa nominale. Essi concordano nel riprodurre lo spettro misurato 
in corrispondenza della massa della $\rho$, ma falliscono per massa 
invarianti inferiori. Nella figura in basso, al contrario, lo spettro \`e
interamente riprodotto da modelli che introducono una riduzione della massa
della $\rho$, calcolato a partire dall'andamento del parametro d'ordine
$<q\overline{q}>$ in funzione della densit\`a del mezzo nucleare
(c.f.r. par.~\ref{fasetransition}).
\par
Il mesone $\rho$ ha vita media $\tau =1.3~fm/c$, pi\`u corta di quella delle
altre sorgenti standard e piccola in confronto al tempo di vita del sistema
interagente ($\sim 10\div 20~fm/c$), per cui le coppie leptoniche provenienti
dal suo decadimento hanno una buona probabilit\`a di essere create prima
del {\em freeze-out}. Esse, quindi, forniscono informazioni
sullo stato della materia nucleare in condizioni di alta temperatura e
 densit\`a, favorevoli al parziale ripristino della simmetria chirale ed alla
conseguente riduzione delle masse dei quarks.
\par
L'interpretazione dell'eccesso di produzione di coppie leptoniche di bassa massa
fornisce dunque una indicazione di probabili cambiamenti nelle
caratteristiche fisiche della materia nucleare, in linea con le previste
conseguenze della transizione di fase di QCD.

\subsection{Produzione della $J/\Psi$}
\index{Produzione della $J/\Psi$}

Il mesone $J/\Psi$ \`e costituito da uno stato legato $c\overline{c}$ e
pu\`o essere prodotto nelle collisioni nucleari in seguito a processi
di fusione del tipo $gg\rightarrow c\overline{c}$ e 
$q\overline{q}\rightarrow c\overline{c}$. In ipotesi di formazione di QGP,
l'alta densit\`a di cariche di colore produce uno schermaggio della forza
di colore, quantificato dal raggio di schermaggio.
Quando tale raggio, inversamente proporzionale alla densit\`a di carica 
di colore e quindi alla densit\`a di energia del sistema interagente, diventa
minore della lunghezza tipica di legame della $J/\Psi$ ($\sim0.5~fm$), lo
stato $c\overline{c}$ si dissolve ed i suoi quarks appaiono dopo 
l'adronizzazione in mesoni ``charmati'' [Mat89].
Simulazioni su reticolo mostrano che tale condizione dovrebbe essere 
soddisfatta per temperature del sistema di poco superiori a quella critica
($T>1.2\,T_c$), in corrispondenza del completo deconfinamento
della materia nucleare. [Kan86], [Bla91].
Gli stati eccitati del sistema $c\overline{c}$, quali  $\Psi\prime$ e 
$\chi_c$, possono essere dissociati pi\`u facilmente e dovrebbero essere
soppressi non appena la temperatura raggiunge il valore critico.

\begin{figure}[htb]
\centering
\includegraphics[scale=1.6,clip]
                                  {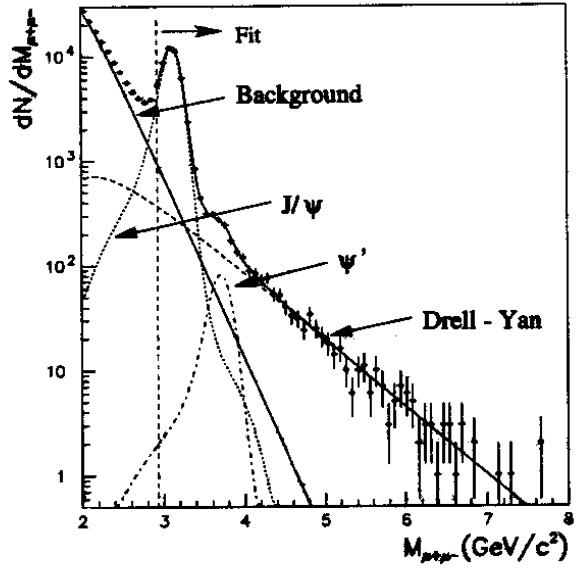}
\caption{\em Spettro di massa invariante di coppie muoniche [Abr96].}
\label{fsupp1}
\end{figure}

La produzione di mesoni vettori $J/\Psi$ e $\Psi\prime$ \`e stata studiata
attraverso il loro
decadimento in coppie $\mu^+\mu^-$  
dagli esperimenti NA38  e NA50 in vari sistemi di collisione.
In fig.~\ref{fsupp1} \`e mostrato lo spettro di massa invariante delle coppie
muoniche rivelate in interazioni Pb-Pb.
Esso \`e interpretato come sovrapposizione dei contributi relativi
al decadimento della $J/\Psi$ e della $\Psi\prime$,  
ai processi di Drell-Yan e ad
 eventi di fondo costituiti da coppie muoniche provenienti da decadimenti
simultanei di pioni e/o kaoni.
\`E cos\`{\i} possibile misurare la sezione d'urto per i processi di Drell-Yan
e per la produzione di $J/\Psi$ e $\Psi\prime$ nel canale di 
decadimento osservato, indicate rispettivamente con 
$\sigma(DY)$, $B_{\mu \mu}\sigma (J/\Psi)$ e
$B_{\mu \mu}\sigma (\Psi\prime)$
($B_{\mu \mu}$ \`e la frazione di decadimento nel canale $\mu \mu$).
La sezione d'urto $\sigma(DY)$ risulta proporzionale al numero di nucleoni
partecipanti alla collisione, essendo i processi di Drell-Yan provenienti
dalle singole collisioni primarie tra nucleoni, per cui pu\`o essere
utilizzata come elemento di normalizzazione nel valutare la produzione di 
$J/\Psi$ in diversi sistemi di collisione.

\begin{figure}[htb]
\centering
\includegraphics[scale=1.8,angle=359.5,clip]
                                  {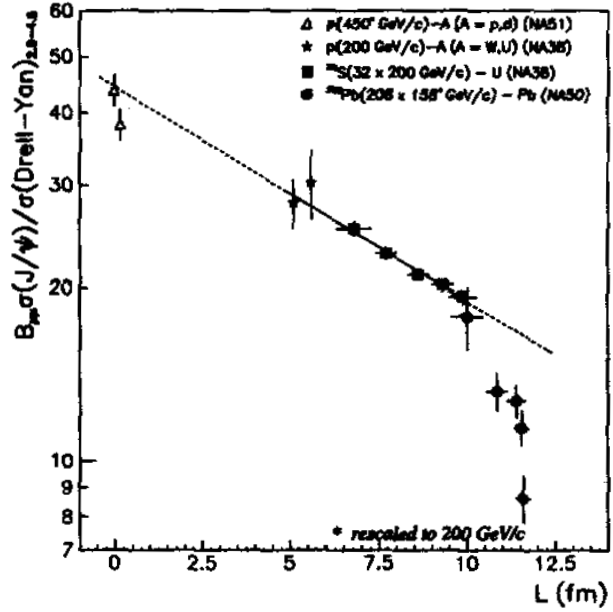}
\caption{\em  Rapporto della sezione d'urto di produzione del mesone
          $J/\Psi$ rispetto a quella di Drell-Yan in funzione
          del libero cammino medio $L$ [Abr96].}
\label{fsupp2}
\end{figure}

In fig.~\ref{fsupp2} \`e riportata la quantit\`a 
$\frac{B_{\mu \mu}\sigma(J/\Psi)}{\sigma(DY)} $
relativa a differenti sistemi interagenti in funzione del 
cammino medio $L$ dello stato $c\overline{c}$ all'interno
della materia nucleare.
Questa \`e una variabile calcolabile
in maniera geometrica che esprime la centralit\`a della collisione.
Si pu\`o notare una decrescita nella produzione di $J/\Psi$ relativamente
agli eventi di fondo in funzione delle dimensioni del sistema
interagente, interpretata in termini di assorbimento di uno stato 
pre-risonante $c\overline{c}$ nella ordinaria materia nucleare.
La produzione della $J/\Psi$ pu\`o infatti essere soppressa anche in presenza
dei gas adronico, mediante processi anelastici del tipo
$\pi+J/\Psi\rightarrow D\overline{D}X$ innescati da adroni co-moventi di
sufficiente energia [Gav88]. L'andamento mostrato in fig.~\ref{fsupp2}
con la retta tratteggiata \`e ottenibile
assumendo una dipendenza del tipo 
$ \sigma (J/\Psi)\;\sim\; \exp(-\rho_0 \sigma_{ass} L) $
 considerando una sezione
d'urto di assorbimento  $\sigma_{ass}=6.2\pm 0.7~mb$ 
(risultato del {\em best fit} ai dati sperimentali)
ed una densit\`a nucleare $\rho_0=0.138/fm^3$.
Tuttavia i punti relativi alle collisioni Pb-Pb rivelano un assorbimento anomalo
nella produzione della $J/\Psi$ rispetto al normale assorbimento nucleare,
quantificabile come

\[
\frac{\left(\frac{\sigma(J/\Psi)}{\sigma(DY)}\right)_{misurato}}
     {\left(\frac{\sigma(J/\Psi)}{\sigma(DY)}\right)_{ass.\: nucl.}}
=0.71\pm0.03
\]

\`E interessante notare che l'assorbimento 
\`e comparabile con  il modello di assorbimento adronico
per le collisioni pi\`u periferiche, mentre se ne discosta per quelle
 pi\`u centrali, aumentando con la centralit\`a.
 I valori di cammino
medio ai quali si manifesta la soppressione anomala sono quelli $L>10~fm$,
corrispondenti a collisioni Pb-Pb con parametro di impatto $b<8~fm$.
\par
\begin{figure}[htb]
\centering
\includegraphics[scale=1.6,clip]
                                  {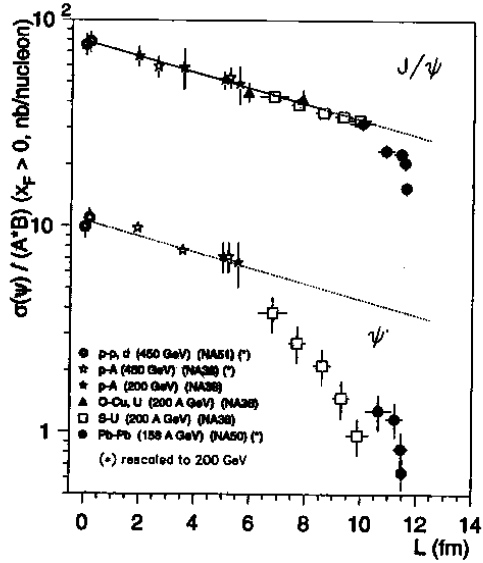}
\caption{\em  Sezioni d'urto di produzione dei mesoni $\Psi\prime$ e
          $J/\Psi$ in funzione
          del cammino medio $L$ [Lou96].}
\label{fsupp3}
\end{figure}

I risultati concernenti la produzione del mesone $\Psi\prime$
in funzione del cammino medio nucleare
sono mostrati in fig.~\ref{fsupp3} e confrontati con quelli relativi alla
$J/\Psi$. La risonanza $\Psi\prime$ viene assorbita in maniera anomala gi\`a
per collisioni S-U ed in corrispondenza di liberi cammini medi $L>6~fm$,
coerentemente con la minore energia di legame di questo stato $c\overline{c}$.

\par
La rivelazione della soppressione anomala delle risonanze $J/\Psi$ e 
$\Psi\prime$ ha innescato una intensa attivit\`a teorica che al momento
pu\`o essere cos\`{\i} sintetizzata: i precedenti modelli di assorbimento nucleare
non sono in grado di predire la soppressione anomala, per cui
alcuni autori [Kha97], [Bla96] hanno attribuito il fenomeno alla
formazione di QGP; altri [Cap97], [Gav97] hanno sviluppato modelli pi\`u conservativi che
riescono a riprodurre la soppressione anomala in collisioni Pb-Pb per mezzo
delle interazioni con gli adroni co-moventi, mostrando tuttavia
un disaccordo con i precedenti risultati relativi a nuclei pi\`u leggeri.

\subsection{Fenomeni nuovi}
\index{Fenomeni nuovi}

La formazione del QGP potrebbe essere associata all'apparenza di fenomeni
completamente nuovi e la loro rivelazione fornirebbe una prova
univoca della sua esistenza.
Storicamente, la proposta di cercare uno stato di QGP in collisioni nucleari
nacque in seguito alla rivelazione, nelle interazioni tra raggi cosmici,
di eventi caratterizzati da un rapporto di carica dei pioni
 $N_{\pi^0}/N_{\pi}$
notevolmente differente da $1/3$ [Lat80].
Questi eventi, conosciuti come ``eventi di Centauro'', potrebbero
essere originati dal decadimento di stati di vuoto ``disorientato'',
nei quali la simmetria chirale sia temporaneamente ristabilita. 
Quindi una nuova osservazione di eventi con queste caratteristiche in
 esperimenti di ioni pesanti fornirebbe una verifica diretta dell'avvenuta
transizione di fase chirale.
\par
L'oggetto esotico pi\`u probabile che potrebbe essere formato da uno stato di
QGP \`e la materia di quark strani. Essa conterrebbe stati metastabili
di materia avente approssimativamente lo stesso numero di quarks 
$u$, $d$ ed $s$, detti ``{\em strangelets}'', formati in seguito al 
raffreddamento del plasma ed alla conseguente distillazione del suo
contenuto di stranezza [Liu84].
A causa della consistente presenza di quarks $s$, il rapporto massa/carica
di tali stati dovrebbe essere molto grande ($\frac{m}{|Z|}>10~GeV$)
e questa caratteristica pu\`o essere usata come segnale per una
loro ricerca sperimentale.
\par
L'esperimento NA52 al CERN non ha finora trovato alcuna particella
con tempo di vita $t_{lab}>1.2~\mu s$ e $\frac{m}{|Z|}>5~GeV$
in collisioni Pb-Pb [App96]. Il limite superiore per la produzione
di {\em strangelets} \`e attualmente di 

\[
E\left.\frac{d^3\sigma}{dp^3}\right|_{p_T=0}<50~\frac{nbarn}{GeV^2}c^3.
\]

\section{La produzione di stranezza come possibile segnale di QGP}
\index{La produzione di stranezza come possibile segnale di QGP}
\label{par_s}

Uno dei primi strumenti proposti per l'analisi della formazione di plasma
nelle collisioni nucleari \`e stato lo studio della produzione
di stranezza, data la sua connessione 
con la dinamica del processo in esame [Raf82], [Raf86].

\begin{figure}[htb]
\centering
\includegraphics[scale=1.6,clip]
                                  {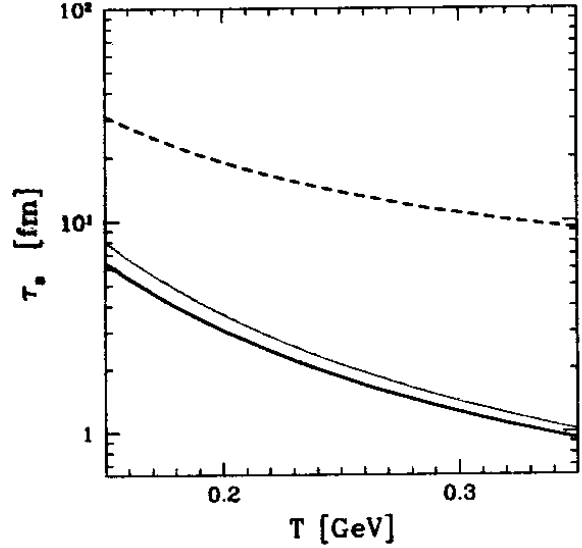}
\caption{\em  Costante di tempo di rilassamento chimico nel QGP in funzione
              della sua temperatura. Il contributo dei quarks (linea
         tratteggiata), dei gluoni (linea sottile) e totale (linea spessa)
          sono calcolati per $m_s=160~MeV$ e $\alpha_s=0.6$ [Raf96].}
\label{fstra1}
\end{figure}

In ipotesi di formazione di QGP, infatti, la produzione di 
stranezza pu\`o essere
 calcolata nell'ambito della QCD perturbativa ed avviene per mezzo della
fusione di gluoni ($gg\rightarrow s\overline{s}$), dell'annichilazione
di quarks ($q\overline{q}\rightarrow s\overline{s}$) e del
decadimento di stati massivi di eccitazione collettiva del QGP, detti
``plasmoni'' ($g\rightarrow s\overline{s}$).
Pur essendo le sezioni d'urto dei tre processi confrontabili, il maggior
numero di gradi di libert\`a a disposizione dei gluoni
rende dominante il loro contributo alla produzione di stranezza.
Si pu\`o dimostrare [Raf96] che 
la densit\`a di quarks strani nel plasma ($\rho_s$)
raggiunge l'equilibrio chimico in maniera asintotica, secondo la legge:

\[
\rho_s(t) \sim \rho_s(\infty)\,\,(1-2 e^{-t/\tau_s}) \:\:\:\:\:\:
{\textstyle per}\:\:\: t>\tau_s
\]

\noindent
dove $\rho_s(\infty)$ indica la densit\`a di quarks strani all'equilibrio e
$\tau_s$, detta costante di tempo di rilassamento chimico, fornisce
una stima della scala di tempo necessaria per la saturazione 
della produzione di stranezza.
In fig.~\ref{fstra1} \`e mostrato l'andamento di tale costante 
in funzione della temperatura del QGP.
 Il contributo dei processi gluonici (linea sottile) risulta
dominante rispetto a quello proveniente dall'annichilazione dei quarks (linea
tratteggiata) e la somma dei contributi (linea spessa) indica che l'equilibrio
chimico nella produzione di quarks strani \`e raggiunto entro $\sim3~fm/c$
se la temperatura del plasma si mantiene al di sopra dei $200~MeV$.
Il calcolo della costante di tempo di rilassamento dipende dai valori usati
per la massa del quark $s$ e per la costante di accoppiamento $\alpha_s$,
ma in ogni caso, essa risulta confrontabile col tempo di vita del QGP.
La saturazione della produzione di stranezza risulta quindi possibile in uno
scenario di QGP.
\par
Il parziale ripristino della simmetria chirale contribuisce a favorire
la produzione di stranezza nel QGP, riducendo la soglia di produzione di
coppie $s\overline{s}$. Inoltre la produzione di quarks $u$ e $d$ risulta
sfavorita in regioni ad alta densit\`a barionica, quali quelle create in
 seguito  a collisioni all'SPS. Infatti la grande abbondanza di quarks $u$
e $d$ inizialmente presenti si traduce in un alto valore del
potenziale bariochimico $\mu_B$ del plasma ed in una riduzione di
un fattore $e^{-\mu_B/3T}$ dell'ulteriore produzione di quarks leggeri.
Questi meccanismi fanno s\`{\i} che in un plasma chimicamente
equilibrato l'abbondanza di quarks $\overline{u}$, $\overline{d}$
e $\overline{s}$ sia simile e, di conseguenza, sia fortemente
favorita la produzione di antibarioni strani e multi-strani
rispetto a quanto avviene nelle normali interazioni adroniche.
\par
In assenza di transizione di fase, la produzione di stranezza avviene per 
mezzo di interazioni secondarie nella fase di gas adronico successiva alla
collisione tra nuclei.
Le tipiche reazioni coinvolte sono quelle di produzione associata, del tipo:

\[
\pi\,\pi\rightarrow K\,\overline{K} ,\hspace{0.5cm} \pi\,N\rightarrow K\,Y ,
\hspace{0.5cm} N\,N\rightarrow N\,K\,Y    \hspace{0.5cm} (Y=\Lambda\,\, o\,\, \Sigma) 
\]

\noindent ma tali reazioni hanno una soglia molto alta
rispetto alla produzione di quarks $s$ e $\overline{s}$ nelle 
interazioni elementari, specialmente per quanto
riguarda la produzione di barioni multi-strani ed ancora di pi\`u per 
anti-barioni multi-strani.
Per esempio, la produzione diretta di un'antiomega
 ($\overline{\Omega}(\overline{s}\,\overline{s}\,\overline{s})$) pu\`o avvenire
mediante la reazione $\pi\,\pi\rightarrow \Omega\,\overline{\Omega}$
con una soglia di $3.3~GeV$.
\par
Dal punto di vista teorico, l'alta soglia di produzione di stranezza
in interazioni adroniche \`e riconducibile
alla rottura di simmetria chirale ad al conseguente innalzamento della massa
del quark $s$ nello stato confinato.
\par
\begin{figure}[htb]
\centering
\includegraphics[scale=1.6,clip]
                                  {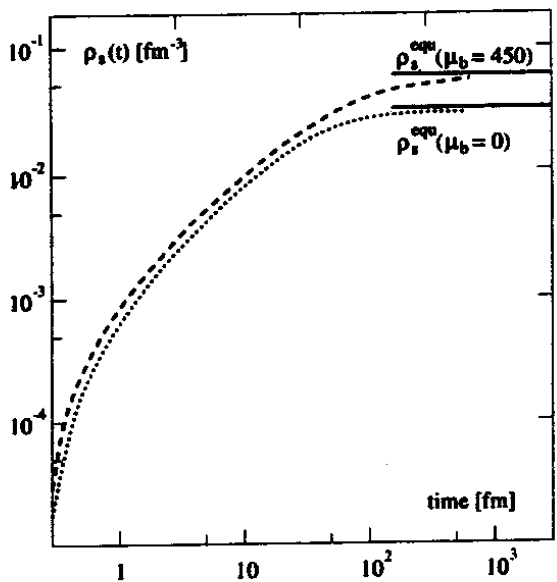}
\caption{\em  Produzione di stranezza in funzione del tempo in un gas
         adronico a $T=160~MeV$. Sono mostrati i risultati per due valori
         del potenziale chimico barionico ($\mu_B=0$ e $450~MeV$) [Koc86].}
\label{fstra2}
\end{figure}

La produzione di anti-barioni strani in gas adronico pu\`o anche avvenire
mediante interazioni successive, ciascuna delle quali con soglia pi\`u bassa
della reazione diretta. D'altra parte \`e stato dimostrato che, in un gas adronico
ad alta densit\`a in equilibrio chimico la produzione di stranezza
\`e confrontabile con quella relativa al plasma [Cle91].
Diventa quindi cruciale valutare il tempo
in cui l'equilibrio chimico di stranezza pu\`o essere raggiunto in interazioni
adroniche.
In fig.~\ref{fstra2} \`e mostrato l'andamento della produzione di stranezza
in funzione del tempo per un gas adronico avente temperatura $T=160~MeV$ e 
potenziale chimico barionico $\mu_B=0~MeV$ (linea punteggiata)
e $\mu_B=450~MeV$ (linea tratteggiata).

\noindent La saturazione nella produzione di stranezza \`e raggiunta dopo
tempi dell'ordine di $100~fm/c$, molto pi\`u lunghi rispetto
al tempo di vita del sistema interagente.
\par
In definitiva, il segnale di QGP consiste in un incremento nella produzione
di stranezza in reazioni tra nuclei, nelle quali \`e possibile la
transizione di fase di QCD, rispetto a quelle
tra nucleoni, nelle quali intervengono soltanto processi di produzione adronica.
L'incremento atteso risulta maggiore per i barioni ed antibarioni multi-strani
e, in generale, dovrebbe crescere col contenuto di stranezza della particella.
La rivelazione di questo segnale fornirebbe una considerevole prova
a favore dell'avvenuta transizione di fase verso uno stato a parziale
ripristino di simmetria chirale.

\begin{figure}[htb]
\centering
\includegraphics[scale=0.5,clip]
                                  {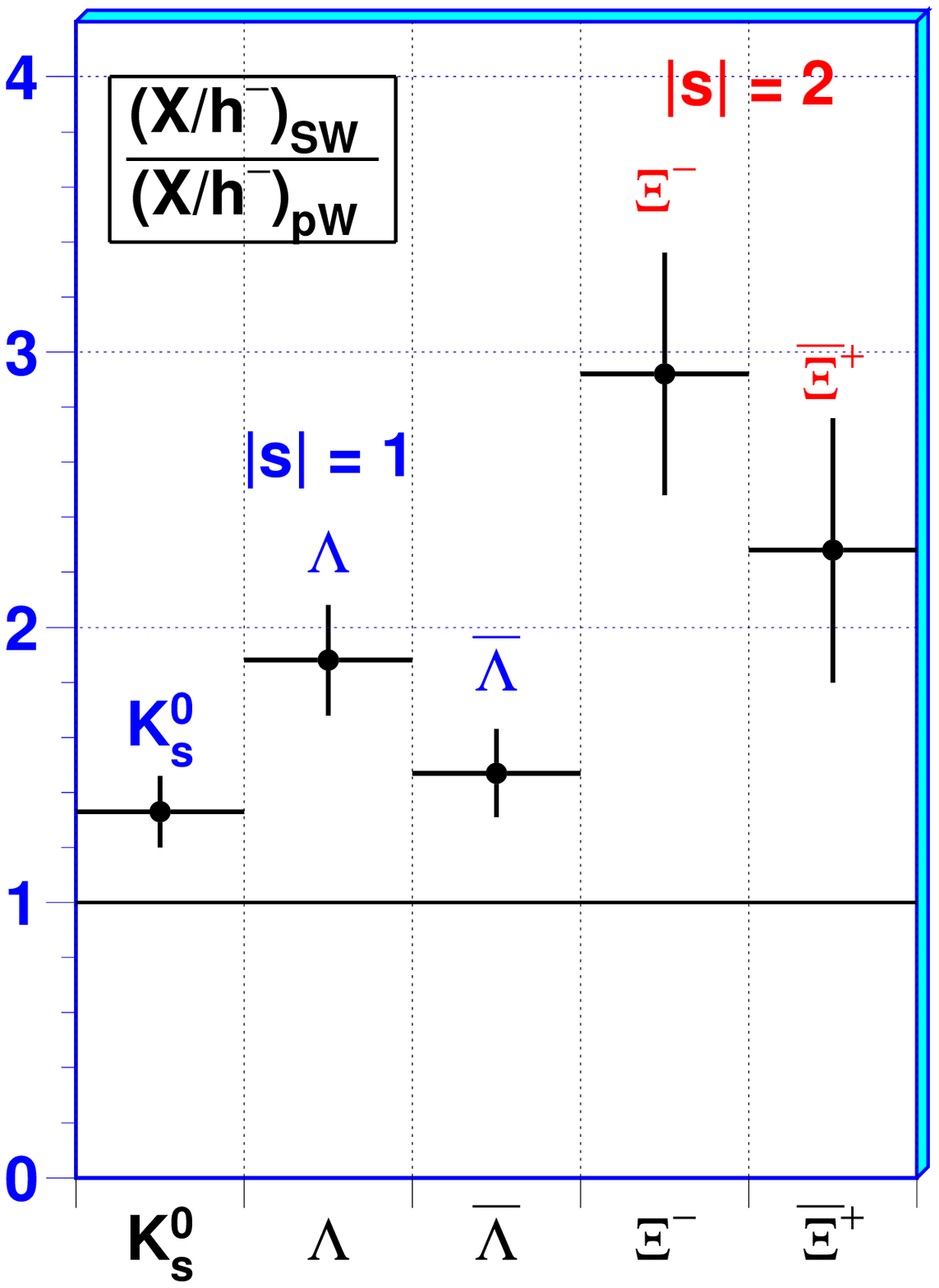}
\caption{\em  Incremento nella produzione di particelle strane in interazioni
             S-W rispetto a collisioni p-W. Le particelle negative ($h^-$)
             prodotte nella collisione sono usate come elemento di 
             normalizzazione [Eva94], [Dib95].}
\label{fig1}
\end{figure}

Gli esperimenti WA85 e WA94 effettuati al CERN con fasci di protoni e ioni
$^{32}S$ hanno fornito incoraggianti risultati a riguardo. In fig.~\ref{fig1}
\`e mostrato l'incremento della produzione di particelle strane in funzione
del loro contenuto di stranezza misurato da WA85 [Eva94], [Dib95].
L'incremento \`e stato valutato utilizzando come elemento
di normalizzazione le particelle negative ($h^-$) prodotte nelle collisioni,
in maniera da svincolarsi dall'ef\-fetto dovuto alla sovrapposizione 
di pi\`u nucleoni incidenti.
La produzione relativa di particelle strane rispetto a quella delle particelle
negative risulta in effetti maggiore in interazioni tra nuclei
che in interazioni protone-nucleo e l'aumento cresce col contenuto di stranezza.
\par
Questi risultati hanno suggerito l'opportunit\`a di estendere lo studio all'intero intervallo
di stranezza (da $|s|=0$ a $|s|=3$), in sistemi pi\`u complessi (Pb-Pb)
e con un notevole incremento della statistica accumulata. Questo studio
\`e stato effettuato dall'esperimento WA97  e sar\`a esposto nei capitoli 
successivi.

 
%
%
%
%
%
%
\chapter{L'esperimento WA97}
\index{L'esperimento WA97} 

\section{Introduzione}
\index{Introduzione}

L'esperimento WA97 si propone di studiare la produzione di particelle strane
in interazioni piombo-piombo (Pb-Pb), protone-piombo (p-Pb) e
protone-berillio (p-Be)  a $158~GeV/c$ per nucleone [WA97p].
Esso utilizza fasci di ioni $^{208}Pb$ e  protoni accelerati dal
sincrotone SPS del CERN ed incidenti su un bersaglio fisso.
La rivelazione di particelle strane avviene nello spettrometro OMEGA, situato
nella ``{\em West area}'' del CERN (da cui la sigla dell'esperimento).
\par
Nell'ambito del programma di studio delle interazioni tra nuclei pesanti
di altissima
energia portato avanti al CERN, l'esperimento WA97 rappresenta la naturale
evoluzione dei precedenti esperimenti WA85 e WA94.
Questi hanno studiato rispettivamente
interazioni zolfo-tungsteno (protone-tungsteno) e zolfo-zolfo (protone-zolfo)
nello spettrometro OMEGA, utilizzando un sistema di camere proporzionali
multifili per la ricostruzione dei decadimenti delle particelle strane.
\par
Al fine di aumentare il volume e la densit\`a di energia della regione centrale
di interazione, si \`e dunque passati a studiare collisioni tra nuclei di
piombo, molto pi\`u massivi dei nuclei impiegati in precedenza.
E' stato inoltre previsto un notevole incremento del
campione  statistico da accumulare,
consentendo lo studio di particelle strane molto rare come gli iperoni $\PgOm$
e $\PagOp$.
Per far fronte alla alta molte\-plicit\`a di particelle cariche presenti nella
regione centrale di rapidit\`a, 
lo spettrometro OMEGA
\`e stato dotato di un telescopio di rivelatori 
al silicio di altissima risoluzione
spaziale, che costituisce la caratteristica peculiare dell'esperimento WA97.
\par
Per quanto esposto nel paragrafo \ref{par_s}, risulta importante, oltre
allo studio delle caratteristiche proprie delle interazioni nucleo-nucleo,
il confronto con la produzione di particelle strane nelle normali interazioni
adroniche. L'esperimento WA97 prevede, quindi, anche lo studio dell'interazione
tra un fascio di protoni di $158~GeV/c$ con un bersaglio di piombo 
e uno di berillio (quest'ultimo inteso come la migliore approssimazione
possibile dell'interazione protone-nucleone).
In queste interazioni non \`e prevista la formazione di plasma ed il loro
studio dovrebbe contentire di apprezzare gli eventuali incrementi nella
produzione di stranezza prodottisi nell'interazione piombo-piombo,
al di l\`a di quelli dovuti semplicemente al numero di nucleoni
partecipanti alla collisione.
\par
Come illustrato nel capitolo I,
la ricerca di segnali di plasma 
deve essere effettuata nella regione centrale del sistema interagente, dove
l'elevato deposito di energia dei nucleoni collidenti potrebbe aver
determinato le condizioni per una transizione di fase. La ricerca di particelle
strane nell'esperimento WA97 avviene, dunque, in una regione 
centrale di rapidit\`a, lontano dai valori estremi di rapidit\`a 
caratterizzanti le regioni di frammentazione del bersaglio e del proiettile.
Il valore centrale di rapidit\`a corrispondente ad interazioni tra sistemi
simmetrici a $158~GeV/c$ \`e $y_{CM}=2.92$ nel sistema del laboratorio
(c.f.r. eq.~\ref{cms.rap}) e l'intervallo di rapidit\`a coperto 
dall'esperimento WA97 \`e centrato su tale valore e si estende per
circa una unit\`a di rapidit\`a.
Nello studio delle interazioni p-Pb e p-Be la rivelazione avviene nello
stesso intervallo di rapidit\`a ed il valore $y_{CM}=2.92$ corrisponde alla
rapidit\`a del centro di massa del sistema nucleone-nucleone.
\par
Da un punto di vista sperimentale, limitare la rivelazione ad una ristretta
regione dello spazio delle fasi permette di escludere le numerose particelle
provenienti dai nucleoni spettatori. Ci\`o consente di migliorare l'efficienza
di ricostruzione delle tracce rivelate, di ridurre i tempi morti nella 
registrazione degli eventi ed i tempi di analisi dei dati.
Tuttavia la scelta di una finestra cinematica cos\`{\i} ridotta rende
alquanto difficoltosa l'interpretazione dei risultati per mezzo
di modelli teorici ed il loro confronto con quelli di altri esperimenti.
\par
L'esperimento WA97 ha raccolto dati per la prima volta nell'anno 1994,
veri\-ficando il funzionamento del nuovo telescopio di silicio in condizioni
di elevata molteplicit\`a di tracce presenti in seguito a collisioni tra
nuclei di piombo.
Durante il 1995   sono stati raccolti circa 287 milioni 
di eventi p-Pb e 123 milioni di eventi Pb-Pb per l'analisi fisica.
 Nell'anno successivo, una ulteriore presa dati di eventi Pb-Pb ha 
aggiunto altri
117 milioni di eventi. Si \`e inoltre eseguita una presa dati con
bersaglio di berillio raggiungendo
219 milioni di eventi p-Be.
\par
Nel seguito del capitolo sar\`a descritto l'apparato nelle configurazioni
usate per lo studio di interazioni Pb-Pb e p-Pb; i dati relativi all'interazione
p-Be non saranno analizzati in questa tesi.

\section{Il fascio}
\index{Il fascio}
\label{par_fascio}

L'esperimento WA97 utilizza un fascio di ioni piombo completamente ionizzati
($Pb^{82+}$) accelerati fino a $158~GeV/c$ per nucleone dal complesso di macchine
acceleratrici del CERN, avente come stadio finale il sincrotone SPS.
Il fascio primario estratto ha una intensit\`a tipica di $10^{6}$ ioni per
ciclo ed \`e convogliato nella linea di fascio H1 [Pla83] prima di essere
focalizzato sul bersaglio di piombo. Il ciclo completo di accelerazione
dell'SPS dura $19.2~sec$ e la durata dell'estrazione \`e di $4.8~sec$.
\par
Per lo studio dell'interazione protone-nucleo viene utilizzato un fascio 
secondario, cio\`e costituito da protoni non direttamente accelerati dall'SPS,
 in quanto quest'ultimo \`e stato progettato per accelerare protoni
fino a $450~GeV/c$.
Il fascio primario di protoni, subito dopo la sua estrazione dall'SPS,
interagisce con 5 bersagli di berillio, per uno spessore
complessivo di $4~cm$. Le particelle secondarie positive aventi momento
di $158~GeV/c$ ($\pm 0.5\%$) sono selezionate ed introdotte nella
linea di fascio H1 per mezzo di collimatori e magneti bipolari; esse sono
infine foca\-lizzate sul bersaglio per mezzo di magneti quadrupolari.
L'intensit\`a del fascio primario estratto di protoni \`e di 
circa $10^{9}$ protoni 
per ciclo, alla quale corrisponde una intensit\`a del fascio
secondario di circa $10^6$ protoni per ciclo. La durata del ciclo di
accelerazione per i protoni \`e di $14.4~sec$ e la durata dell'estrazione \`e
di $2~sec$.
\par
La maggior parte delle particelle secondarie selezionate \`e costituita
da protoni, ma esiste una cospicua contaminazione dovuta principalmente
ai pioni, che deve essere eliminata. A tale scopo si utilizzano due
rivelatori \v{C}erenkov differenziali a gas, denominati CEDAR, posti lungo
la linea di fascio H1. La regolazione della pressione del gas consente
di ottimizzare la loro efficienza di identificazione di
 protoni ($\simeq 100\%$).
Tali rivelatori sono inclusi nella logica di trigger per scartare interazioni
prodotte da particelle diverse dai protoni (c.f.r. par.~\ref{partrigger}).
La reiezione operata dai CEDAR corrisponde a circa il $33\%$ delle particelle
incidenti.

\section{Il bersaglio}
\index{Il bersaglio}

Il bersaglio \`e caratterizzato dalla massa atomica del materiale impiegato
e dal suo spessore.
Per l'esperimento WA97 \`e stato scelto un bersaglio di piombo
($A=207.19$) per ottenere un sistema simmetrico nella collisione.
Tale simmetria pu\`o essere sfruttata per estendere l'intervallo di rapidit\`a
coperto dall'esperimento, dato che le regioni di accettanza possono essere
simmetrizzate rispetto al valore di rapidit\`a del centro di massa.
D'altra parte, l'alto numero atomico del piombo consente il raggiungimento
di alte densit\`a di energia 
e di estesi volumi di interazione.
Il bersaglio di piombo \`e stato usato anche col fascio di protoni,
per minimizzare gli elementi di discordanza nel confronto con
l'interazione nucleo-nucleo.
\par
Per quanto riguarda la scelta dello spessore del bersaglio, essa rappresenta un
compromesso tra la necessit\`a di aumentare la frequenza delle interazioni
e quella di minimizzare la probabilit\`a di interazioni multiple
e di conversioni di fotoni al suo interno.
La probabilit\`a che una particella abbia una interazione all'interno di un
bersaglio di spessore $L$ \`e

\begin{equation}
P(L)=1-e^{-L/\lambda_I}
\label{probcoll}
\end{equation}

\noindent dove $\lambda_I$ \`e la lunghezza di interazione, 
che per l'urto Pb-Pb vale $3.99~cm$, mentre per l'urto p-Pb vale $14.08~cm$.
La probabilit\`a di conversione di fotoni in $e^+\,e^-$
all'interno del bersaglio ha una espressione analoga alla (\ref{probcoll}),
in cui al posto di $\lambda_I$ si introduce la lunghezza di radiazione $X_0$,
 pari a $0.56~cm$ per il piombo.
\par
Nell'esperimento WA97, per lo studio dell'interazioni Pb-Pb, \`e stato scelto
uno spessore pari a $0.4~mm$, corrispondente a $1\%$ di lunghezza di 
interazione
ed a $0.071$ lunghezze di radiazione.
Per l'interazione p-Pb, il problema delle collisioni multiple e della 
conversione di fotoni \`e meno rilevante e la necessit\`a
di aumentare la frequenza delle interazioni ha portato ad adottare
un bersaglio di  $1.1~cm$ di spessore, corrispondente all'$8\%$ di lunghezza
di interazione ed a $2.01$ lunghezze di radiazione.

\section{L'apparato sperimentale}
\index{L'apparato sperimentale}

\begin{figure}[htb]
\centering
\includegraphics[scale=0.56,clip]
                                 {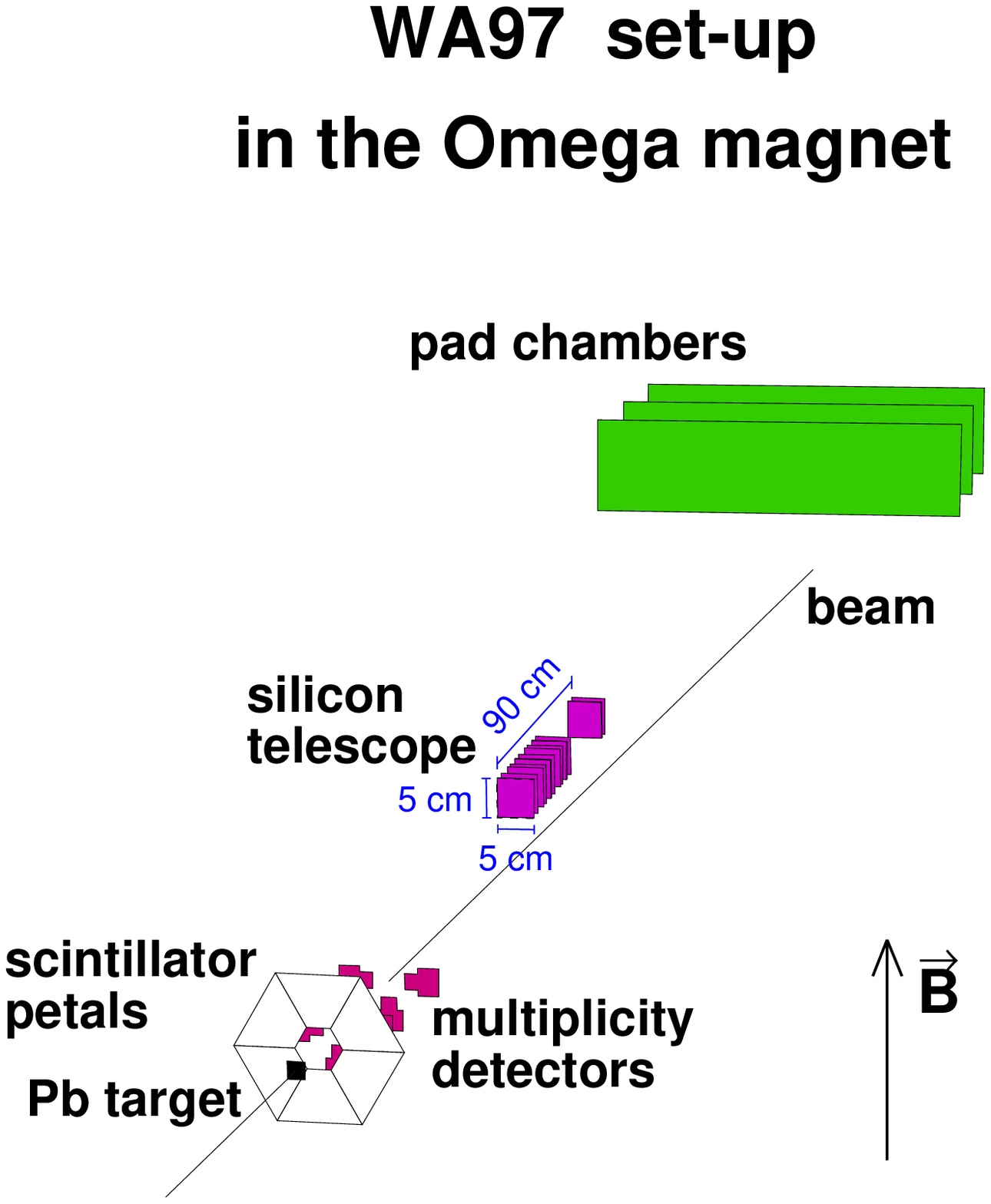}
\caption{\em L'insieme dei rivelatori usati nell'esperimento WA97.}
\label{setup}
\end{figure}

L'insieme dei rivelatori utilizzati nell'esperimento WA97 \`e mostrato
in fig.~\ref{setup}. Esso consiste di  una corona
di scintillatori a ``petali'', di due piani di rivelatori di molteplicit\`a
(microstrip al silicio),
del telescopio al silicio, a sua volta contenente piani di rivelatori
a {\em pixel} e piani di microstrip al silicio,
 e di rivelatori a multifili a catodo segmentato (camere a 
{\em pad}).
Tali rivelatori sono stati utilizzati sia nello studio delle
interazioni Pb-Pb che in quello delle interazioni p-Pb, sebbene
l'impiego di alcuni di essi risulti funzionale solo nel caso di collisioni tra
 nuclei di piombo. 
Il bersaglio e il telescopio al silicio sono posti entro un campo magnetico
uniforme generato dal magnete OMEGA del CERN.
Nel seguito, il magnete ed i rivelatori verranno analizzati singolarmente,
mentre la descrizione del trigger e dei connessi contatori di fascio
\`e contenuta nel successivo paragrafo, diversificata per le due interazioni.

\subsection{Il magnete}
\index{Il magnete}

\begin{figure}[htb]
\centering
\includegraphics[scale=1.6,clip]
                                 {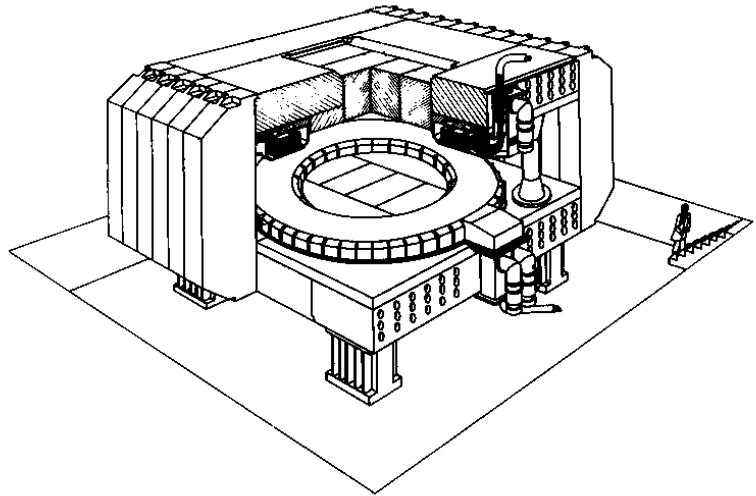}
\caption{\em Magnete dello spettrometro OMEGA.}
\label{figomega}
\end{figure}

Il magnete OMEGA \`e stato originariamente concepito come parte di uno
spettrometro, capace di rivelare
con grande accettanza angolare
eventi ad alta moltepli\-cit\`a, fornendo una misura accurata dei parametri
cinematici delle particelle cariche che l'attraversano [Beu84].
Si tratta di un magnete di grande apertura, illustrato
in fig.~\ref{figomega}, le cui espansioni polari, di $6\times 6~m^2$,
sono poste a distanza di circa $1.5~m$ tra loro.
Due bobine superconduttrici con un diametro di $3~m$ generano un campo
magnetico sostanzialmente uniforme nella regione centrale, che 
raggiunge un valore massimo di circa $1.8~T$.

\begin{figure}[ht]
\centering
\includegraphics[scale=0.85,bb=29 153 685 530,angle=90.,clip]
                                 {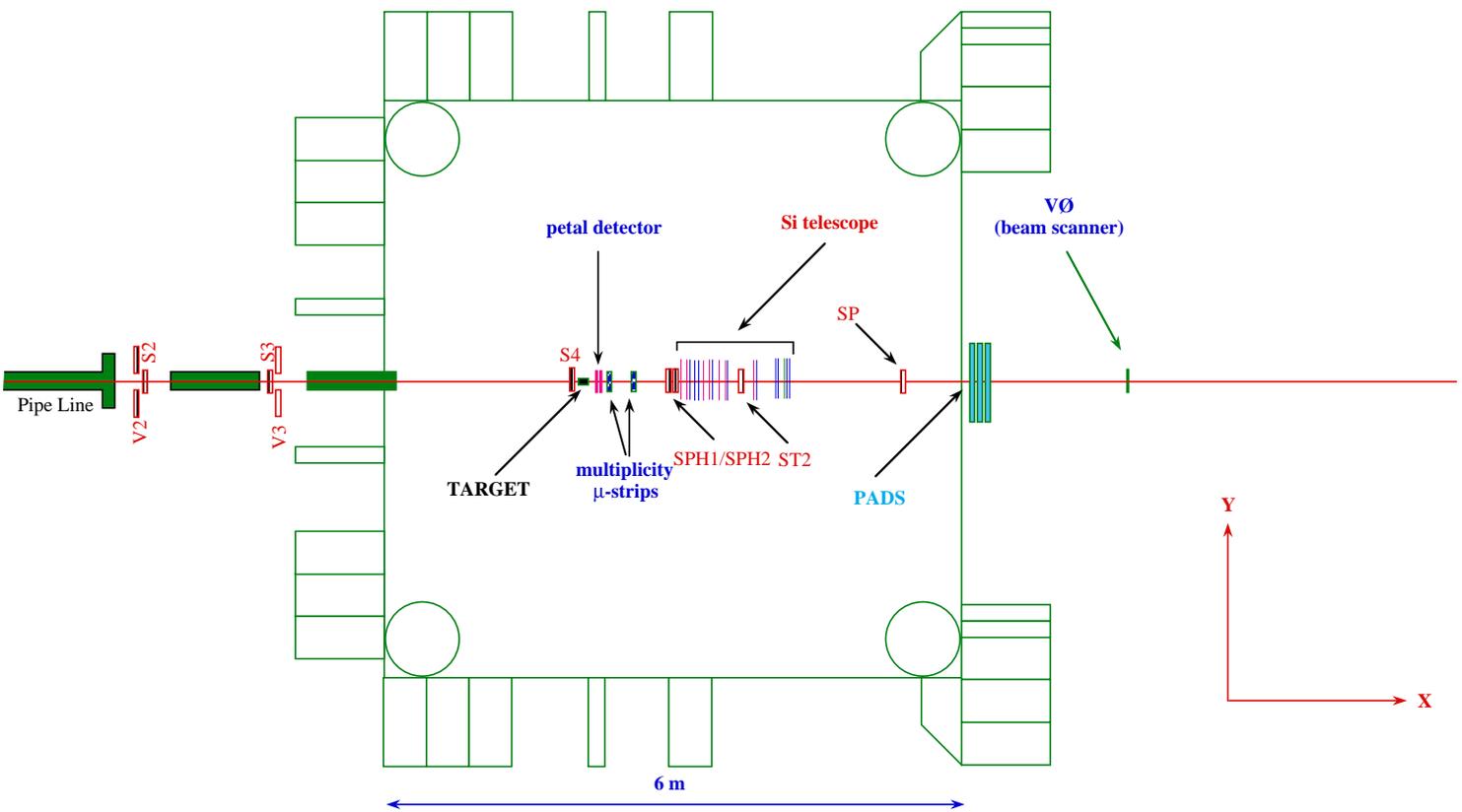}
\caption{\em Schema dell'apparato sperimentale di WA97 usato per lo studio di
            interazioni p-Pb.}
\label{setup_pPb}
\end{figure}

All'interno dello spettrometro possono essere configurati apparati sperimentali
anche molto diversi tra loro.  L'apparato 
 usato nell'esperimento WA97 per lo studio di interazioni p-Pb  
 \`e mostrato
in fig.~\ref{setup_pPb}; esso risulta solo leggermente modificato 
nello studio di interazioni Pb-Pb.

\clearpage

\clearpage

Il sistema di riferimento adottato ha l'origine nel centro
del magnete,  l'asse $x$ diretto orizzontalmente secondo il verso
del fascio incidente, l'asse $z$ diretto verso l'alto perpendicolarmente
alle espansioni polari e l'asse $y$ tale da formare una terna destrorsa.
Pertanto il campo magnetico \`e diretto approssimativamente
secondo l'asse $z$ e le traiettorie delle particelle cariche vengono
incurvate essenzialmente nel piano $xy$.
Il verso del campo magnetico \`e determinato dal verso della corrente nelle
bobine; sono stati raccolti dati con i due versi del campo,
al fine di evidenziare eventuali errori sistematici compiuti nella misura
della mappa del campo magnetico e di verificare l'allineamento dei 
rivelatori, dato che
l'apparato sperimentale \`e simmetrico rispetto al piano $xz$.

\subsection{Gli scintillatori a petali}
\index{Gli scintillatori a petali}
\label{par_petals}

Nella parte immediatamente posteriore al bersaglio, a 
$10~cm$ da questo, sei scintillatori
trapezoidali (``petali'')  circondano la linea del fascio, formando
un'apertura centrale esagonale (c.f.r. fig.~\ref{setup}).
Il loro compito \`e quello di selezionare eventi ad alta molteplicit\`a,
dovuti ad urti centrali tra nuclei.
Come gi\`a discusso nel capitolo precedente (par.~\ref{par_collisioni}),
la selezione di eventi centrali permette di isolare le
collisioni in cui \`e massimo il numero di nucleoni partecipanti, 
caratterizzate da una grande quantit\`a di energia rilasciata: in questo tipo
di collisioni risulta pi\`u probabile la formazione di QGP.
La condizione di trigger consiste nel richiedere che almeno 5 scintillatori
su 6 forniscano un segnale in coincidenza. La soglia al di sopra della
quale i singoli scintillatori forniscono un segnale \`e stata scelta
considerando che, in corrispondenza di un evento centrale, ognuno di essi
\`e attraversato da circa $20\div30$ particelle.
La percentuale di eventi selezionati con questa condizione corrisponde a circa
il $40\%$ della sezione d'urto totale anelastica relativa a collisioni Pb-Pb
[Ant95].

\subsection{I rivelatori di molteplicit\`a}
\index{I rivelatori di molteplicit\`a}

\begin{figure}[htb]
\centering
\includegraphics[scale=1.6,clip]
                                 {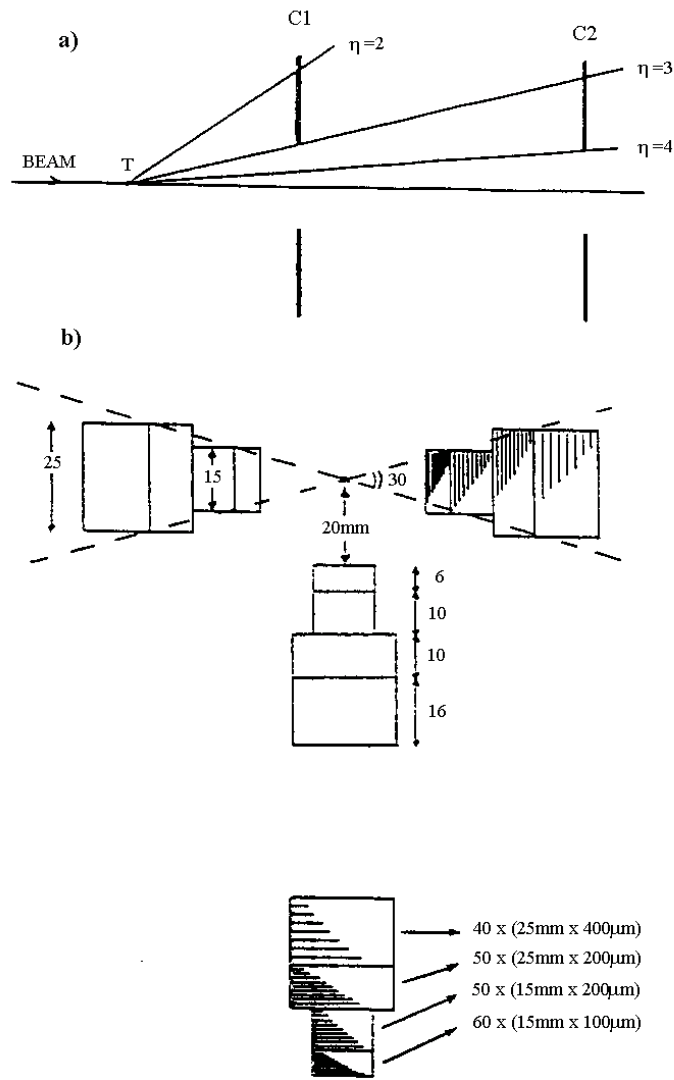}
\caption{\em Rivelatori di molteplicit\`a: a) le stazioni identiche C1 e C2 coprono
 un intervallo di pseudorapidit\`a $2<\eta<4$; b) struttura delle singole
 stazioni [WA97p].}
\label{figmicro}
\end{figure}

La molteplicit\`a di particelle cariche \`e campionata per mezzo di due
 stazioni identiche di rivelatori a microstrip di silicio.
Esse sono posizionate lungo la linea di fascio, dopo i rivelatori a petali
(c.f.r. fig.~\ref{setup}) e ciascuna di esse \`e costituita da tre piani 
di microstrip 
montati su altrettanti bracci di un'intelaiatura vuota al centro, 
in corrispondenza della zona di passaggio del fascio.
Come mostrato in fig.~\ref{figmicro}, ognuno dei 6 piani di silicio
\`e formato da microstrip di differente lunghezza (da $15$ a $25~mm$) e
passo (da $100$ a $400~\mu m$), per un totale di $200$ canali,
disposti in modo tale da garantire un'occupazione costante di
particelle nell'ipotesi di distribuzione piatta di pseudorapidit\`a.
Tale ipotesi risulta ragionevole in prossimit\`a della regione 
centrale di rapidit\`a e l'occupazione risultante rimane minore di $0.4$
particelle per strip, anche per gli eventi pi\`u centrali [WA97p].
\par
La disposizione dei piani permette di coprire le regioni di pseudorapidit\`a
$2\lesssim\eta\lesssim 3$ (stazione 1) e $3\lesssim \eta\lesssim 4$ (stazione 2), con una accettanza media
di circa $27\%$ in tale intervallo. Il numero medio di particelle cariche
registrate per evento dalle due stazioni di microstrip \`e
maggiore di $400$ e non dipende in maniera significativa dall'azione del campo
magnetico presente nell'apparato.
La misura di molteplicit\`a fornita da tali rivelatori consente di studiare 
in maniera dettagliata la dipendenza dalla centralit\`a della produzione
delle particelle ricostruite dal telescopio.

\clearpage

\subsection{Il telescopio}
\index{Il telescopio}

Il telescopio
\`e costituito da 7 piani di rivelatori a {\em pixel} e 10 piani di
 microstrip al silicio, descritti in dettaglio pi\`u avanti.
Esso rappresenta il dispositivo primario per la 
ricostruzione spaziale delle tracce cariche e consente di
ricostruire il decadimento delle particelle strane in un ambiente ad alta
molteplicit\`a.
Infatti, in questo esperimento, l'identificazione
avviene tramite la ricostruzione di decadimenti contenenti
solo particelle cariche nello stato finale, quali \hspace{0.2cm}
$\PgL \rightarrow \Pp + \Pgpm$,  \hspace{0.2cm}
$\PKzS \rightarrow \Pgpp + \Pgpm$, \hspace{0.2cm}
$\PgXm \rightarrow \PgL + \Pgpm$, \hspace{0.2cm}
$\PgOm \rightarrow \PgL + \PKm$ \hspace{0.2cm}
e quelli corrispondenti alle rispettive antiparticelle.
\par
\begin{figure}[htb]
\centering
\includegraphics[scale=0.66,bb=135 503 534 695,clip]
                                 {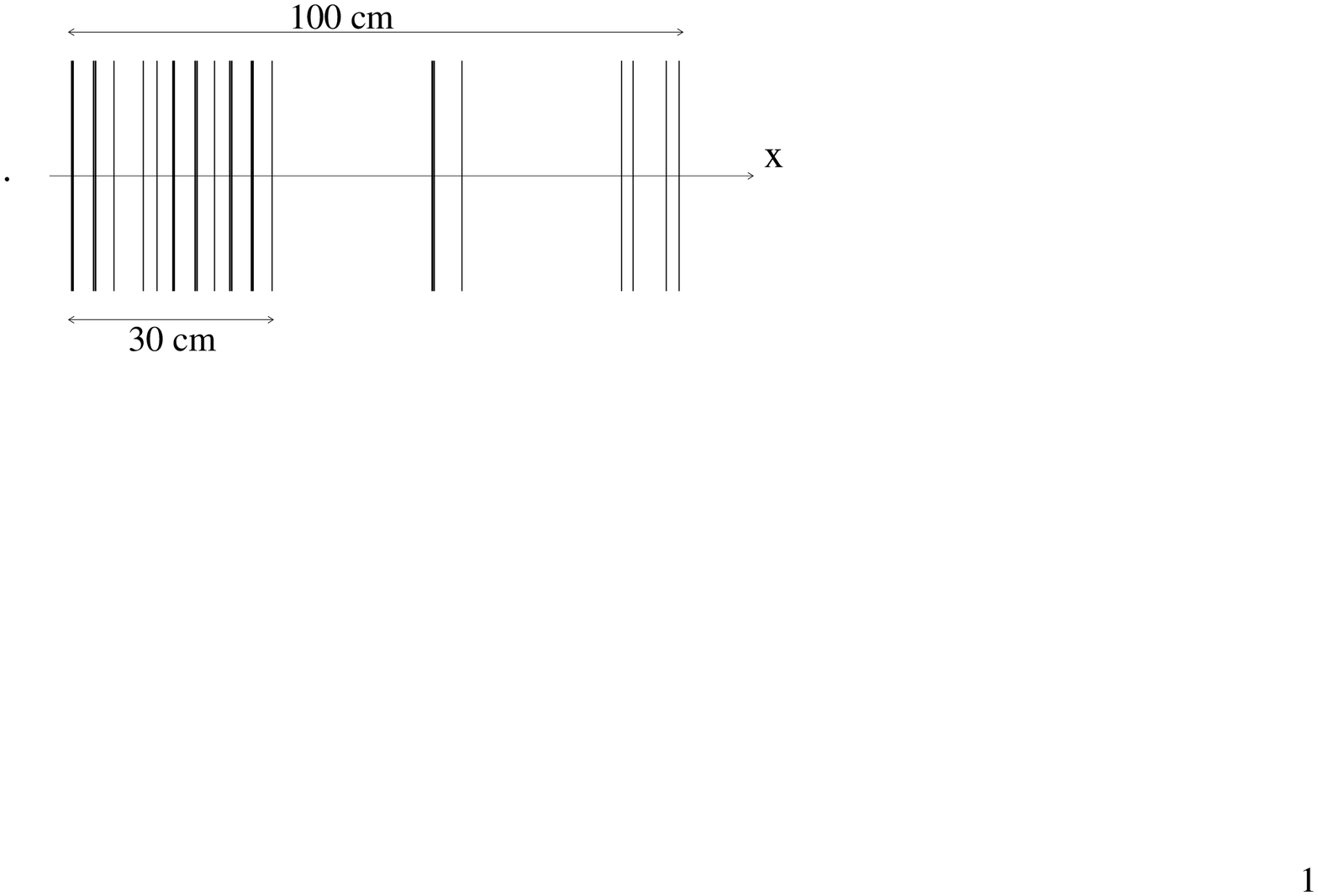}
\caption{\em Rappresentazione schematica del telescopio al silicio. I piani
  di pixel sono evidenziati in grassetto, mentre quelli di
          microstrip sono riportati con linee tratteggiate.}
\label{figtele}
\end{figure}

Il telescopio \`e posto leggermente al di sopra
della linea di fascio e inclinato in modo da puntare al bersaglio.
L'angolo di inclinazione, di $48~mrad$, \`e stato scelto in modo
da accettare particelle strane 
nell'intervallo di rapidit\`a $|y-y_{CM}|\leq0.5$ e con momento trasverso
$p_T\ge 0.3~GeV/c$.
L'area sensibile di ciascun piano ha una sezione di $5\times5~cm^2$ e la 
loro disposizione \`e mostrata in fig.~\ref{figtele}.
L'intero telescopio \`e lungo circa $1~m$.

\begin{figure}[htb]
\centering
\includegraphics[scale=0.65,bb=73 520 558 685,clip]
                                 {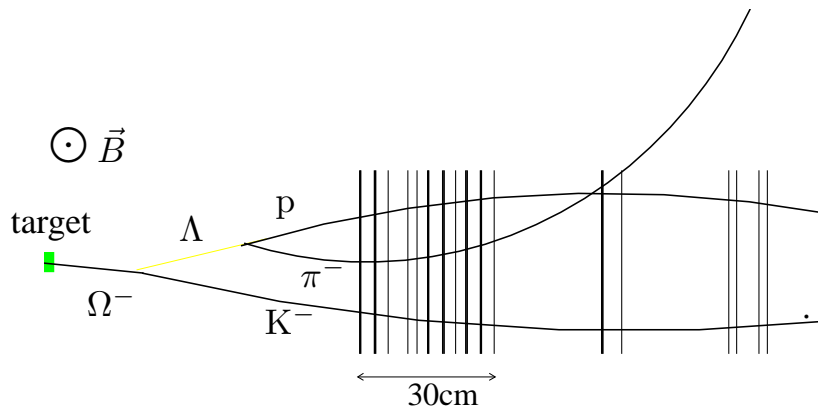}
\caption{\em Tipico decadimento di una $\PgOm$ ricostruibile nel
telescopio.}
\label{figdecay}
\end{figure}

Per la ricostruzione delle tracce vengono inizialmente utilizzati i piani
contenuti nella prima parte (``compatta'') del telescopio, dove sono 
disposti in maniera ravvicinata 6 piani di {\em pixel} e 5 piani di microstrip,
per una lunghezza totale di $30~cm$. I rimanenti piani del telescopio,
insieme alle camere a {\em pad} descritte in seguito, hanno il compito
di migliorare la precisione nella misura del momento delle particelle
veloci.
In fig.~\ref{figdecay} \`e mostrato un tipico decadimento $\PgOm$ che 
\`e possibile ricostruire nel telescopio.

\vspace{1cm}

\par
Per quanto riguarda la distanza del telescopio dal bersaglio, la scelta
operata rappresenta un compromesso tra la necessit\`a di garantire la 
rivelazione di un numero cospicuo di particelle strane nell'arco di
tempo destinato all'acquisizione dati e quella di rendere minimo il numero
di tracce che attraversano il telescopio, consentendo un'efficiente
 ricostruzione. Studi effettuati con simulazioni Monte Carlo hanno suggerito
la scelta di porre il telescopio a $60~cm$ dal bersaglio
per interazioni Pb-Pb. Una distanza minore avrebbe consentito una
maggiore accettanza per le particelle strane studiate, aumentando tuttavia il
fondo presente nel campione di dati.
Nello studio di interazioni p-Pb, come si vedr\`a nel paragrafo \ref{partrigger},
\`e presente un trigger che permette di rigettare eventi con meno di due tracce
nel telescopio. Questo consente di scegliere la collocazione del
telescopio in modo da rendere massimo il numero di particelle strane per 
evento ``triggerato''.
Le simulazioni effettuate hanno mostrato che tale numero \`e piccolo 
a grandi distanze, dato l'esiguo numero di particelle strane rivelate, ma 
anche a piccole distanze, a causa dell'aumento del numero di eventi selezionati.
Esso risulta massimo in corrispondenza della distanza di $90~cm$ dal bersaglio,
che \`e stata cos\`{\i} adottata nella presa dati p-Pb.

\subsubsection{I rivelatori a pixel}
\index{I rivelatori a pixel}
\label{parpixel}

\begin{figure}[htb]
\centering
\includegraphics[scale=2.,angle=1.5,clip]
                                 {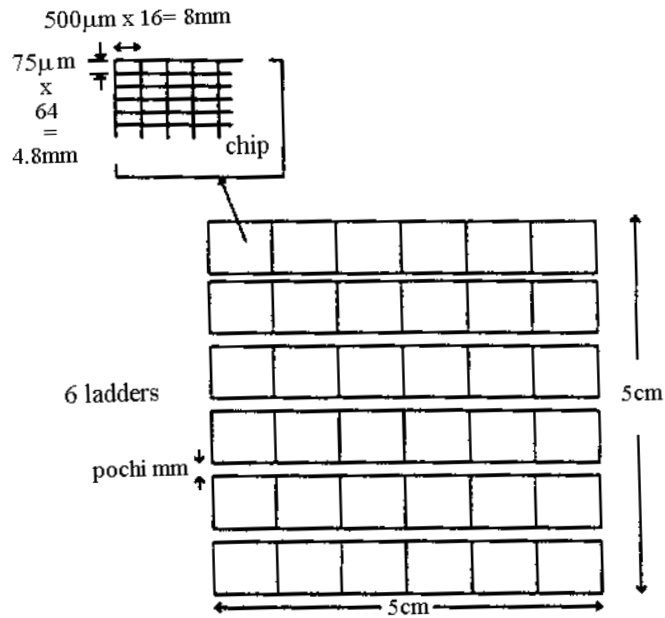}
\caption{\em Illustrazione schematica di un array, costituito da 6 ladder
formati da 6 chip adiacenti. La figura in alto visualizza in dettaglio il
singolo chip con la sua segmentazione a pixel.}
\label{figpixel1}
\end{figure}

L'unit\`a sensibile di questo nuovo tipo di rivelatori, denominata appunto
{\em pixel}, \`e un diodo rettangolare formato per impiantazione ionica
su un substrato di silicio di tipo n ad alta resistivit\`a ($>5~k\Omega cm$).
Queste celle elementari, di dimensioni $500\times75\mu m^2$, con uno spessore
di $300~\mu m$, sono allineate in righe e colonne a comporre strutture 
bidimensionali di rivelazione pi\`u estese [Hei94].
Una matrice di $64$ righe e $16$ colonne realizza un {\em chip}
e un segmento di sei {\em chip} a righe adiacenti
costituisce un {\em ladder}, composto da 64 righe e 96 colonne di {\em pixel},
cos\`{\i} come mostrato in fig.~\ref{figpixel1}.

\begin{figure}[htb]
\centering
\includegraphics[scale=1.5,clip]
                                 {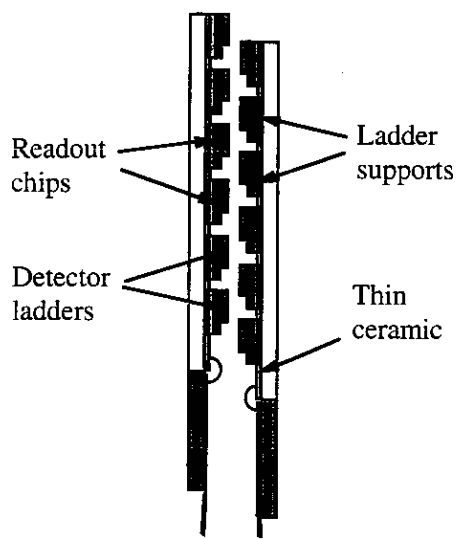}
\caption{\em Sezione di un piano logico di un rivelatore a pixel e dettaglio
della struttura di supporto e di lettura.}
\label{figpixel2}
\end{figure}

A loro volta, 6 {\em ladder} paralleli alloggiano su di un'unica lastra di ceramica
spessa $630~\mu m$, spaziati di alcuni millimetri, indispensabili
per le connessioni elettroniche, a formare una struttura piana rigida
detta {\em array}.
Infine, due {\em array} identici sono appaiati in una singola impalcatura, in cui
si dispongono specularmente, slittati l'uno rispetto all'altro in modo da
evitare che gli interstizi tra due {\em ladder} costituiscano delle
discontinuit\`a di rivelazione (c.f.r. fig.~\ref{figpixel2}).
Una coppia di piani fisici cos\`{\i} congegnata rappresenta un unico piano
``logico'' di rivelatore, che consiste di 72576 {\em pixel} concentrati in
un'area di $5\times5~cm^2$. Il telescopio di WA97 contiene 7 di questi piani
logici, per cui il numero totale di {\em pixel} \`e di circa mezzo milione.
L'elettronica di lettura \`e affidata ad un {\em microchip} saldato 
su ciascun {\em chip}
del rivelatore e frazionato in modo che a ciascun {\em pixel} sia associata
una cella circuitale contenente un preamplificatore a minimo rumore,
un comparatore ed unit\`a di ritardo e di memoria atte a registrare i segnali
in concomitanza con la selezione operata dalla logica di trigger.

I piani di {\em pixel} sono disposti nel telescopio in modo che presentino la
frammentazione in $75\mu m$ alternativamente lungo l'asse $y$ (piani~$y$)
e lungo l'asse $z$ (piani~$z$), per garantire una uguale risoluzione
spaziale nelle due direzioni trasverse. Essi sono in grado di fornire
con grande accuratezza le coordinate dei punti di impatto delle particelle
con ciascun piano, permettendo la ricostruzione delle relative tracce cariche.

\subsubsection{I rivelatori a microstrip}
\index{I rivelatori a microstrip}

I 10 piani di microstrip al silicio sono disposti secondo quanto mostrato in 
fig.~\ref{figtele} ed hanno uno spessore di $300~\mu m$ ed un passo di
 $50~\mu m$.
Sette di questi piani sono orientati in modo che le rispettive strip siano
disposte verticalmente, in modo da fornire la misura della coordinata
$y$ della traccia (la curvatura \`e nel piano $xy$). 
I rimanenti tre hanno le strip disposte orizzontalmente,
per la misura della coordinata $z$.
Sebbene si tratti di dispositivi con elementi sensibili
unidimensionali, tali rivelatori consentono di affinare ulteriormente la
precisione con cui vengono ricostruite le tracce cariche nel telescopio,
avendo essi un passo di $50~\mu m$ a fronte dei $75~\mu m$ dei {\em pixel}.

\subsection{Le camere a pad}
\index{Le camere a pad}

Il sistema di tracciamento \`e completato da un tripletto di camere 
proporzionali a fili col
catodo segmentato in celle ({\em pad}), collocate appena fuori del magnete
OMEGA, in modo da coprire l'angolo
solido individuato dal telescopio.
L'area sensibile delle camere misura $80\times 25~cm^2$ e le dimensioni delle
singole celle sono di $4\times 12~mm^2$.
Una risoluzione spaziale di $\sigma_y=0.5~mm$ e $\sigma_z=1.5~mm$ pu\`o
essere raggiunta determinando il centroide della distribuzione di cariche
per ogni traccia, usando anche le celle adiacenti a quella colpita dalla
particella. 
Alle camere a {\em pad} compete la funzione di ottimizzare le prestazioni del 
telescopio nel caso di tracce molto veloci, incrementandone la risoluzione in
impulso e l'efficienza di tracciamento a grande distanza dal bersaglio.

\section{Selezione degli eventi}
\index{Selezione degli eventi}
 
In esperimenti che studiano particolari categorie di particelle
\`e importante e\-strarre le interazioni in cui esse sono contenute dagli eventi
di fondo, imponendo dei criteri sperimentali di selezione, detti
``condizioni di trigger''.
Nell'esperimento WA97 la selezione cambia a seconda che si studi l'interazione
Pb-Pb o quella p-Pb.

\subsection{Condizioni di trigger in eventi p-Pb}
\index{Condizioni di trigger in eventi p-Pb}
\label{partrigger}

Un primo livello di trigger, detto ``trigger di fascio'', permette di selezionare
collisioni prodotte da particelle incidenti ben focalizzate sul bersaglio
e sufficientemente distinte una dall'altra. Esso fa uso dei contatori
di fascio S2, V2, S3 e V0, mostrati in fig.~\ref{setup_pPb}.
S2 ed S3 sono scintillatori plastici posti anteriormente al bersaglio lungo
la linea di fascio e destinati al conteggio delle particelle che 
percorrono la direzione attesa di collimazione del fascio sul bersaglio.
Gli scintillatori V3 e V4, posti anch'essi prima del bersaglio, hanno una sagoma
anulare del diametro di qualche centimetro, sufficiente per permettere
il passaggio delle particelle proiettile: essi consentono di rivelare 
ed antiselezionare la componente divergente
del fascio.
Il trigger di fascio \`e definito dall'equazione logica:
\[
BEAM=S2 \cdot S3 \cdot \overline{V2} \cdot \overline{V3} 
\]
\noindent La richiesta di coincidenza tra i segnali
di S2 ed S3 o di V2 e V3
riduce la probabilit\`a di  selezionare del rumore.
\par
\begin{figure}[htb]
\centering
\includegraphics[scale=1.7,clip]
                                 {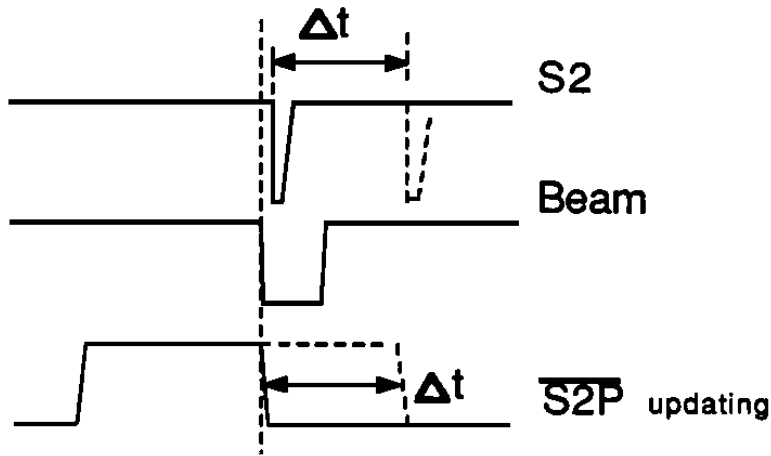}
\caption{\em Segnali che concorrono a produrre la ``protezione passato-futuro''.
I segnali in linea tratteggiata sono generati da una seconda
particella proiettile.}
\label{fig_CB}
\end{figure}

Un secondo livello di trigger di fascio fornisce la cosiddetta
``protezione passato-futuro'', scartando quegli eventi nei quali due
particelle proiettile molto vicine tra loro provocano due diverse
interazioni nel bersaglio. A tal fine, il segnale proveniente dal contatore
S2 \`e inviato in un discriminatore operante in {\em updating mode}.
L'arrivo  di un secondo segnale proveniente da S2 entro un 
tempo $\Delta t$ determina un prolungamento del segnale di uscita del
discriminatore, denominato S2p in fig.~\ref{fig_CB}.
Finch\`e il discriminatore \`e attivo, non viene consentito al
sistema di esaminare ed acquisire nuove interazioni.
Un ``fascio pulito'' \`e dunque definito secondo l'equazione
\[
CB=BEAM \cdot S2 \cdot \overline{S2p}
\] 
\noindent ponendo in anticoincidenza il segnale del discriminatore.
La durata della protezione \`e stata fissata a $\Delta t=20~nsec$. 

Successivamente, un trigger di interazione impone delle condizioni sulla 
qualit\`a degli eventi da accettare. Eventi con almeno una traccia nel
telescopio possono essere selezionati utilizzando gli scintillatori
SPH1, SPH2, ST2, SP e V0, mostrati in fig.~\ref{setup_pPb}, secondo
l'equazione:

\[
TEL\,INT=SPH1 \cdot SPH2 \cdot ST2 \cdot SP \cdot \overline{V0}.
\]

\begin{figure}[htb]
\centering
\includegraphics[scale=1.5,clip]
                                 {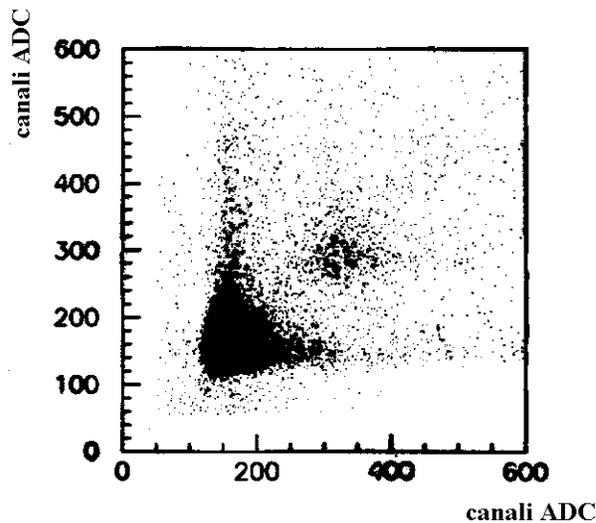}
\caption{\em Correlazione tra i segnali provenienti dagli scintillatori
 SPH1 ed SPH2.}
\label{fig_counter}
\end{figure}

\noindent Gli scintillatori SPH1 ed SPH2 sono posti all'inizio del telescopio
ed hanno le stesse dimensioni dei piani di rivelatore in esso presenti.
La correlazione tra le altezze d'impulso dei due rivelatori \`e mostrata
in fig.~\ref{fig_counter}.

\noindent In essa \`e ben visibile la zona di addensamento dovuta al
passaggio di una particella in SPH1 ed SPH2, ma \`e anche presente una 
seconda regione di addensamento, relativa al passaggio
di due particelle in ciascuno degli scintillatori.
Regolando opportunamente le soglie di SPH1 ed SPH2 e ponendo i loro segnali
in coincidenza, \`e dunque possibile selezionare eventi in cui almeno una
o almeno due tracce entrano nel telescopio.
Richiedendo poi la coincidenza col segnale proveniente dallo scintillatore 
ST2, posto dopo la parte compatta del telescopio, e da SP, distante
circa 1 metro dal telescopio ed avente un'area di $26\times12~cm^2$,
si impone che almeno una traccia attraversi l'intero telescopio.
Infine, ponendo in anticoincidenza il segnale proveniente dallo scintillatore
 V0, posto dietro il telescopio lungo la linea di fascio, ci si assicura che
il protone incidente abbia interagito in qualche maniera all'interno dello
spettrometro.
\par
Dal punto di vista fisico, tale trigger di interazione favorisce
la rivelazione di eventi nei quali il protone incidente abbia interagito
con un nucleo del bersaglio ed almeno una  particella carica 
attraversi l'intero telescopio. Regolando
le soglie degli scintillatori SPH1 ed SPH2 \`e anche possibile richiedere che
almeno due particelle cariche entrino nel telescopio.
Quest'ultima condizione favorisce la rivelazione di particelle
strane che si presentano con due tracce cariche ($\PgL$ e $\PKzS$) o con
tre tracce cariche ($\PgXm$ e $\PgOm$) ed aumenta quindi il
numero di particelle strane per evento selezionato.
\par
I successivi livelli del trigger di interazione sono i seguenti:
\begin{itemize}
\item $ INT\,\overline{DT}=BEAM \cdot \overline{DT} \cdot TEL\,INT $
\end{itemize}
\noindent che  assicura il rigetto di successive interazioni 
      qualora i livelli pi\`u
      alti del trigger stiano ancora analizzando l'evento precedente. Ci\`o
      si ottiene ponendo in anticoincidenza il segnale di ``tempo morto''
      ($DT$) proveniente da un ``OR'' logico al cui ingresso sono posti
       i segnali provenienti dai livelli di trigger pi\`u alti e 
	dal sistema di acquisizione dati.
\begin{itemize}
\item $ CU0=CB \cdot INT\,\overline{DT} $ 
\end{itemize}
\noindent in cui viene richiesta la coincidenza 
	tra il segnale di interazione e quello di ``fascio pulito''.
\begin{itemize}
\item $ CU1=CU0 \cdot CEDARS $
\end{itemize}
\noindent   in	cui viene
	aggiunto nella coincidenza  il segnale  proveniente dai $CEDARS$ 
	(c.f.r. par.~\ref{par_fascio}) per selezionare interazioni prodotte
	unicamente da protoni incidenti.

\subsection{Condizioni di trigger in eventi Pb-Pb}
\index{Condizioni di trigger in eventi Pb-Pb}
\label{partrigger1}

Nell'apparato sperimentale per lo studio di interazioni Pb-Pb, gli scintillatori
S2, S3 e V0 sono stati sostituiti da altrettanti contatori \v{C}erenkov al quarzo,
appositamente preposti alla discriminazione degli ioni all'ingresso ed
all'uscita dello spettrometro. Infatti, i valori elevati di carica propri
del fascio di ioni possono indurre fenomeni di saturazione sugli scintillatori
 di fascio, pregiudicando la loro efficacia e la linearit\`a della loro
risposta. Invece i \v{C}erenkov al quarzo offrono appropriati requisiti di
resistenza a fasci di considerevole intensit\`a e permettono l'identificazione
degli ioni piombo di adeguata energia, dato che l'altezza dell'impulso ai
relativi fotomoltiplicatori dipende, oltre che dalla velocit\`a
della particella incidente, dal quadrato della sua carica.
\par
La definizione del trigger di fascio \`e analoga alla precedente, mentre per
il trigger di interazione sono usate le seguenti definizioni:
\begin{itemize}
\item $ TARG\,INT=PETALS \cdot \overline{V0} $
\end{itemize}
\noindent  che definisce un'interazione nel 
	bersaglio ($\overline{V0}$)  e seleziona eventi a grande 
moltepli\-cit\`a
         per mezzo dei  petali di scintillatore 
        (c.f.r. par.~\ref{par_petals}). Ci\`o permette di selezionare urti
	 centrali tra nuclei di piombo.
\begin{itemize}
\item $ INT\,\overline{DT}=BEAM \cdot \overline{DT} \cdot TARG\,INT $
\end{itemize}
          \noindent dove la definizione di interazione tiene conto del 
          tempo morto dei rivelatori.
\begin{itemize}
\item $ CU1=CB \cdot INT\,\overline{DT} $
\end{itemize}
        \noindent dove la protezione 
	``passato-futuro'' \`e inclusa nella selezione	degli eventi.

\section{Il campione di dati}
\index{Il campione di dati}

Gli eventi che soddisfano tutte le condizioni di trigger imposte vengono
registrati su nastri magnetici, per mezzo di un sistema dedicato di
acquisizione dati. Questo sistema legge le informazioni provenienti
dai vari rivelatori e dai contatori di fascio, le compatta e le scrive in
formato EPIO [Gro81] (leggibile direttamente con ogni computer).
Si ottiene cos\`{\i} una 
registrazione dei dati allo stato grezzo ({\em raw data}) nella quale
 i vari eventi costituiscono dei {\em record} logici allineati
all'interno di uno stesso {\em record} fisico ed aventi tutti la medesima 
struttura. Il cuore del {\em record} logico risiede nel
blocco ROMULUS, che contiene le informazioni provenienti dai rivelatori
ed ha una struttura ad albero tale da renderle facilmente accessibili.
\par
L'analisi presentata in questa tesi si riferisce ai dati raccolti a partire 
dal 1995 relativi all'interazione Pb-Pb (200 milioni di eventi)
e p-Pb (120 milioni di eventi) con le seguenti specifiche:
\begin{description}
 \item Pb-Pb:
  \begin{itemize}
   \item bersaglio di spessore pari all'$1\%$ di lunghezza di interazione
   \item telescopio distante $60~cm$ dal bersaglio
   \item trigger di centralit\`a (mediante i petali, come spiegato sopra)
  \end{itemize}
 \item p-Pb:
  \begin{itemize}
   \item bersaglio di spessore pari all'$8\%$ di lunghezza di interazione
   \item telescopio distante $90~cm$ dal bersaglio
   \item trigger a due tracce, ossia richiedente almeno 
         due tracce entranti nel telescopio,
	 con almeno una traccia che l'attraversa per intero.
  \end{itemize}
\end{description}

Verranno inoltre analizzati i dati relativi all'interazione p-Pb
registrati con la richiesta di almeno una traccia nel telescopio
(trigger ad una traccia).


%
%
%
%
%
%
\chapter{Ricostruzione e selezione delle particelle strane}
\index{Ricostruzione e selezione delle particelle strane}

\section{Introduzione}
\index{Introduzione}

Come gi\`a accennato nel capitolo 2,
l'apparato sperimentale di WA97 consente l'identificazione 
delle particelle strane attraverso la ricostruzione dei loro
 decadimenti carichi. 
La misura dei vettori momento delle particelle cariche che attraversano il 
telescopio, effettuata mediante la ricostruzione delle loro tracce nel 
campo magnetico, permette di assegnare una massa ai vertici di
decadimento ricostruiti, facendo delle ipotesi sulla massa delle particelle
rivelate. Nel caso di decadimento a due corpi l'assegnazione avviene mediante
la relazione

\begin{equation}
M^2=m_1^2+m_2^2+2\sqrt{ (m_1^2+p_1^2)(m_2^2+p_2^2)} - 2\,\vec{p_1}\cdot \vec{p_2}
\label{massainvariante}
\end{equation}

\noindent  
dove gli indici 1 e 2 indicano le variabili relative alle due tracce di 
decadimento, $m_1$ ed $m_2$ sono le masse ad esse assegnate e $M$ \`e la
massa invariante della particella che decade, relativa alla topologia di
decadimento ipotizzata.
\par
L'identificazione risulta particolarmente favorita per
le particelle strane che decadono in maniera debole
le quali, a causa del lungo tempo di vita
media, hanno una buona probabilit\`a di avere il vertice di decadimento
ben distinto dal vertice primario di interazione.
Nel caso di particelle strane neutre, i decadimenti carichi formano delle
configurazioni
denominate $V^0$, intendendo con ci\`o una coppia di tracce di curvatura
opposta nel campo magnetico che si diramano da uno stesso punto di cuspide
distinto dal vertice primario di interazione.
In particolare, possono essere ricostruiti i seguenti decadimenti di particelle
strane neutre, riportati con le relative probabilit\`a di
 decadimento ({\em Branching Ratios}) [Par96]:

\begin{center}
  \[
   \begin{array}{rll}
   \PKzS \rightarrow & \Pgpp + \Pgpm       & \mathrm{BR}=68.6\pm 0.3\% \\
   \PgL  \rightarrow & \Pp + \Pgpm         & \mathrm{BR}=63.9\pm 0.5\% \\
   \PagL \rightarrow & \Pap + \Pgpp        & \mathrm{BR}=63.9\pm 0.5\% \\
   \end{array}
  \]
\end{center}
\vspace{0.2cm}
\`E inoltre possibile identificare particelle multi-strane
attraverso i loro decadimenti a ``cascata''. A questo proposito le $V^0$
ricostruite vengono combinate con un'altra traccia carica dell'evento
per ottenere i decadimenti

\begin{center}
  \[
   \begin{array}{rll}
   \PgXm \rightarrow & \PgL + \Pgpm            & \mathrm{BR}=99.89\pm 0.04\% \\
                     & \decayarrow \Pp + \Pgpm &                     \\
                     &                         &                     \\
   \PgOm \rightarrow & \PgL + \PKm             & \mathrm{BR}=67.8\pm 0.7\%  \\
                     & \decayarrow \Pp + \Pgpm &                     \\
   \end{array}
  \]
\end{center}

\noindent
o quelli corrispondenti alle rispettive antiparticelle.
Un tipico decadimento $\PgOm$ che \`e possibile ricostruire nel telescopio
\`e stato mostrato in fig.~\ref{figdecay}.
\par
In questo capitolo verranno descritte le procedure utilizzate per
ricostruire ed identificare sia le particelle strane, corrispondenti a
topologie $V^0$, che quelle multi-strane, corrispondenti a topologie a cascata.
L'analisi sar\`a descritta in dettaglio per i dati relativi all'interazione
p-Pb, nei quali la moderata presenza di eventi di fondo consente una
migliore definizione dei criteri di selezione. Alla fine del capitolo 
sar\`a discussa l'applicazione di tali criteri ai dati Pb-Pb e saranno
mostrati i relativi segnali.

\section{Ricostruzione delle $V^0$}
\index{Ricostruzione delle $V^0$}

La ricostruzione delle tracce delle particelle
cariche a partire dagli impatti sui piani del telescopio
\`e operata dal programma
ORHION [Las94], sviluppato appositamente per questo esperimento.
Le traiettorie delle particelle sono approssimate da eliche cilindriche, 
rappresentate da un arco di circonferenza nel piano $xy$
(ortogonale alla componente principale del campo) e da un arco di
sinusoide nel piano $xz$, a sua volta approssimabile con un segmento di
retta sulla lunghezza del telescopio. 
Come passo successivo, il codice STRIPV0, contenuto nel programma ORHION,
 ricostruisce i vertici di decadimento di tipo $V^0$ 
come intersezione di due tracce di carica opposta estrapolate
all'indietro.
Se un evento ha almeno una coppia di tracce di segno opposto, STRIPV0
insegue all'indietro tutte le sue tracce fino alla posizione nominale del 
bersaglio. Il programma, poi, associa tra di loro in maniera
combinatoria le tracce di segno opposto e calcola la minima distanza tra 
le due traiettorie nello spazio; le coppie per le quali tale distanza
\`e maggiore di una distanza massima, posta a $1~mm$ per collisioni
Pb-Pb ed a $1~cm$ per collisioni p-Pb, vengono rigettate.
Le coordinate del vertice $V^0$ sono valutate come valori medi delle
coordinate dei punti di minima distanza tra le tracce associate, mentre il
vettore impulso della $V^0$ \`e dato dalla somma vettoriale degli impulsi
delle sue tracce, calcolati nei punti di minima distanza.
\par
\begin{figure}[htb]
\centering
\includegraphics[scale=0.4,clip]
                                {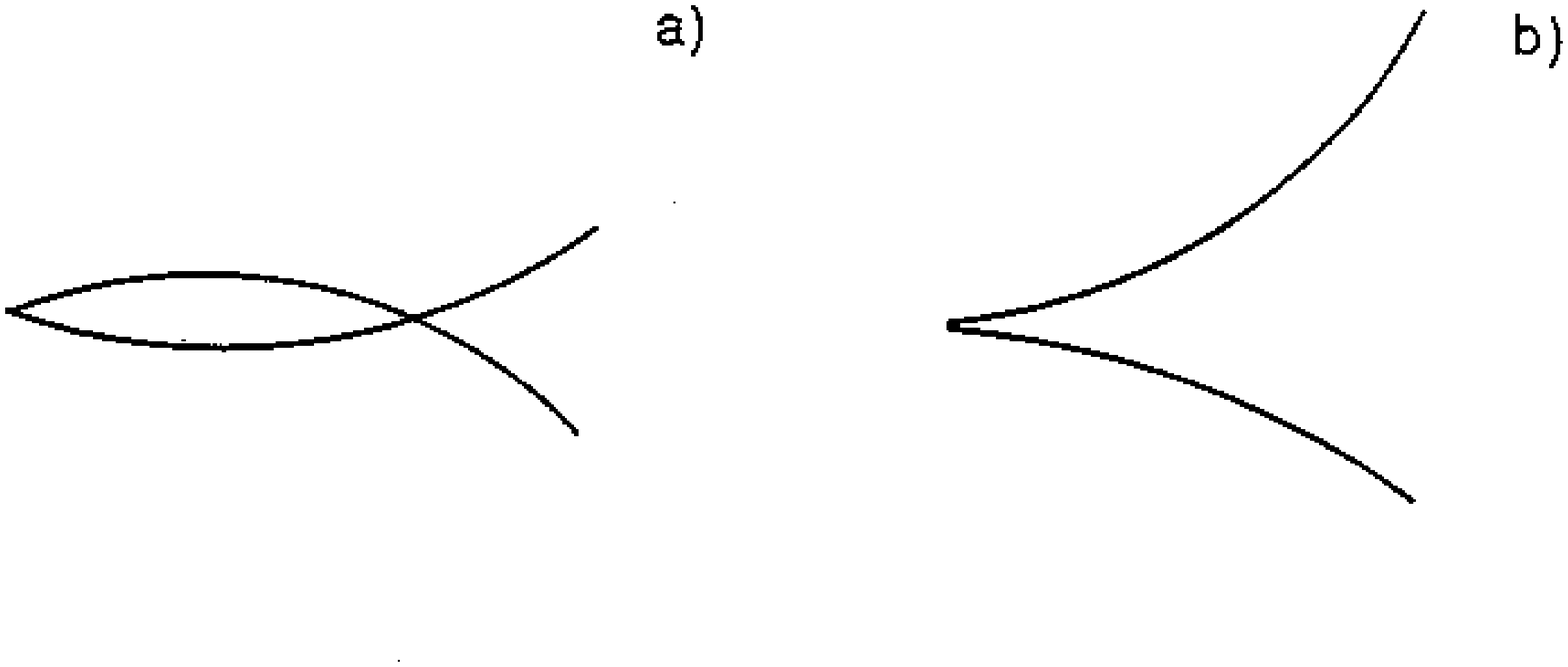}
\caption{\em Possibili configurazioni di decadimento delle $V^0$:
         a) ``cowboy'', b) ``sailor''.}
\label{figcowsai}
\end{figure}

A seconda che le tracce associate si intersechino o meno nel piano $xy$ 
in un ulteriore punto in avanti rispetto al vertice ricostruito, la $V^0$
sar\`a denominata ``cowboy'' o ``sailor''.
I due casi corrispondono alle due diverse geometrie di decadimento mostrate
rispettivamente in fig.~\ref{figcowsai}a e \ref{figcowsai}b ed hanno la
stessa probabilit\`a di occorrenza.

\noindent
La rivelazione di vertici cowboy risulta, tuttavia, favorita
rispetto a quelli sailor, a causa della piccola sezione trasversa del
telescopio. Essi corrispondono al $70\%$ dei vertici $V^0$ ricostruiti.
 D'altra parte, l'errore sulla determinazione dell'angolo
di apertura $\phi$ della $V^0$, determinato dagli inevitabili errori
di misura delle tracce, risulta maggiore per la configurazione sailor che per
quella cowboy. Questo errore \`e legato all'errore sulla massa della $V^0$
 dalla relazione

\[
dM=\frac{p_1p_2}{M}\,\sin{\phi}\,\,d\phi
\]

\noindent 
ottenuta differenziando la (\ref{massainvariante}).
Risulta pertanto conveniente considerare nell'ana\-lisi i soli
vertici cowboy, pi\`u numerosi e meglio determinati. 
\par
In appendice A viene mostrato che, per quanto riguarda la configurazione 
cowboy, la distanza $L$ del secondo punto di intersezione nel piano $xy$ dal
vertice della $V^0$ \`e una quantit\`a invariante
per trasformazioni di Lorentz. Essa \`e quindi
caratteristica del tipo di particella che decade ed assume come valore
massimo $L_{MAX}\sim 40~cm$ per $\PgL$ e $\PagL$ e 
$L_{MAX}\sim 80~cm$ per $\PKzS$ nel campo magnetico di OMEGA.
In appendice A si dimostra, inoltre, che la distanza trasversa
nel piano $xy$ tra le due tracce di decadimento lungo il percorso $L$
\`e piuttosto piccola, dell'ordine del centimetro
 per $\PKzS$ e $\PgL$. La rivelazione di tali particelle \`e dunque
possibile pur utilizzando un rivelatore a limitata sezione trasversa,
quale il telescopio adoperato in WA97, di cui si pu\`o sfruttare tutta la
lunghezza per la misura delle traiettorie e delle particelle di decadimento.
\par
Un'ulteriore selezione imposta dal programma STRIPV0 sui vertici $V^0$ 
(al solo fine di ridurre le combinazioni spurie)
corrisponde alla richiesta che il momento trasverso $q_T$ delle tracce
di decadimento rispetto alla linea di volo della particella madre
sia minore di $0.4~GeV/c$. Tale limite \`e ben al di sopra del massimo
valore che $q_T$ pu\`o assumere per $\PKzS$, $\PgL$ e $\PagL$, 
dato dalla relazione

\begin{equation}
q_{T_{MAX}}=p^\star=\frac{1}{2M}\,\sqrt{M^2-(m_1+m_2)^2}\:\sqrt{M^2-(m_1-m_2)^2}
\label{momentotrasverso}
\end{equation}

\noindent
dove $p^\star$ \`e il momento delle particelle di decadimento nel 
centro di massa, $M$ \`e la massa della particella madre considerata
e $m_1$ e $m_2$ sono le masse dei suoi prodotti di decadimento.
La (\ref{momentotrasverso}) fornisce $q_{T_{MAX}}=0.206~GeV/c$ per i $\PKzS$
e $q_{T_{MAX}}=0.101~GeV/c$ per le $\PgL$ e $\PagL$.

\begin{figure}[htb]
\centering
\includegraphics[scale=0.7,clip]
                                {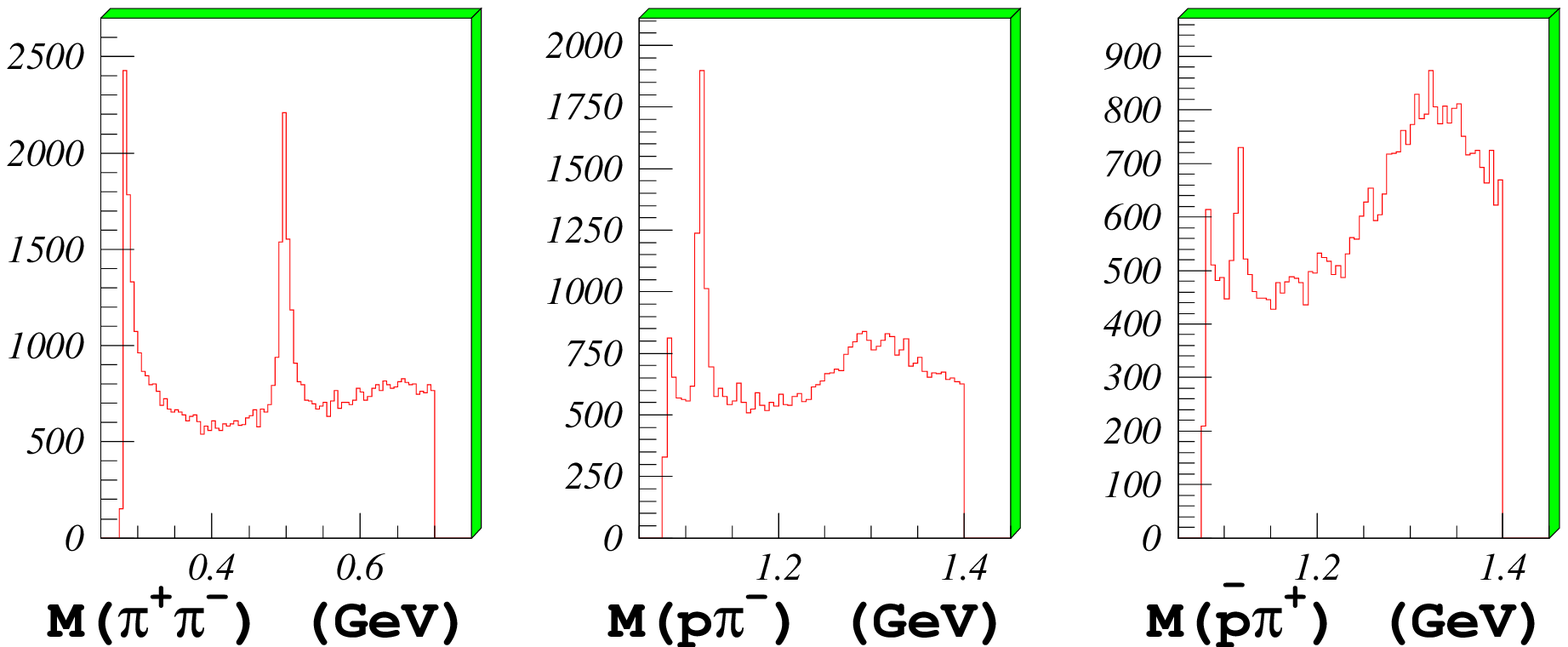}
\caption{\em Spettri di massa invariante
$M(\Pgpp ,\Pgpm)$ (a sinistra), $M(\Pp,\Pgpm)$ (al centro)
 e $M(\Pap ,\Pgpp)$ (a destra) per un campione di
$V^0$ candidate ricostruite da STRIPV0.}
\label{figmasse1}
\end{figure}

In fig.~\ref{figmasse1} sono mostrati gli spettri di massa invariante
$M(\Pgpp ,\Pgpm)$, $M(\Pp,\Pgpm)$ e $M(\Pap ,\Pgpp)$ per i vertici
$V^0$ ricostruiti e selezionati da STRIPV0; il campione in figura
 corrisponde 
all'1\% dell'intera statistica disponibile. Gli spettri sono stati
ottenuti applicando l'eq.~(\ref{massainvariante}) e le particelle
in parentesi indicano le masse assegnate alle tracce di decadimento.
I segnali di $\PKzS$, $\PgL$ e $\PagL$ risultano evidenti nei rispettivi
spettri, in corrispondenza della loro massa nominale, 
in misura statisticamente significativa rispetto al fondo dovuto a false $V^0$.
Queste sono attribuibili a casuali intersezioni geometriche di tracce presenti
nell'evento, interpretate come $V^0$ a causa di errori di misura e di
 estrapolazione, oppure ad errate combinazioni tra le tracce dell'evento
(pi\`u $V^0$ candidate possono avere una traccia in comune).
Le distribuzioni, inoltre, mostrano altri picchi dovuti ad un'errata
assegnazione delle masse alle particelle di decadimento,
 detti ``riflessi''. Il lavoro di analisi eseguito sui vertici $V^0$
ricostruiti da STRIPV0 ha lo scopo di isolare il segnale fisico di
$\PKzS$, $\PgL$ e $\PagL$ dagli eventi di fondo e dai riflessi.

\section{Selezione delle $V^0$}
\index{Selezione delle $V^0$}

Per identificare le particelle
$\PKzS$, $\PgL$ e $\PagL$ \`e necessario minimizzare la contaminazione
geometrica, dovuta alle false $V^0$, e quella cinematica, dovuta
alla presenza dei riflessi, presente nel campione di $V^0$ candidate
fornito da STRIPV0.
Ci\`o pu\`o essere ottenuto operando una serie di criteri di selezione,
di natura sia geometrica che cinematica.
La scelta di tali criteri \`e stato oggetto di uno studio sistematico, 
condotto valutando di volta in volta il loro effetto sul rapporto 
segnale/fondo caratteristico degli spettri di massa delle particelle
da identificare.
I risultati di questo studio saranno esposti
 nel paragrafo~\ref{taglianalisi}, mentre nei due paragrafi
successivi i criteri di selezione saranno esaminati uno per volta
e la loro scelta sar\`a giustificata a posteriori,
analizzando di volta in volta i campioni di $V^0$ candidate ottenuti
applicando tutti i tagli selezionati tranne quello da studiare.
\par
Il campione di dati utilizzato consiste di $120000$ eventi p-Pb raccolti
col trigger di interazione a due 
tracce, pari all'1\% dell'intera statistica accumulata.

\subsection{Selezioni generali}
\index{Selezioni generali}

Si considerano dapprima le selezioni comuni alle particelle
$\PKzS$, $\PgL$ e $\PagL$, idonee ad eliminare la contaminazione
geometrica del campione di $V^0$ candidate e parte di quella
cinematica, dovuta alla presenza dei $\gamma$.

\begin{figure}[htb]
\centering
\includegraphics[scale=0.6,clip]
                                {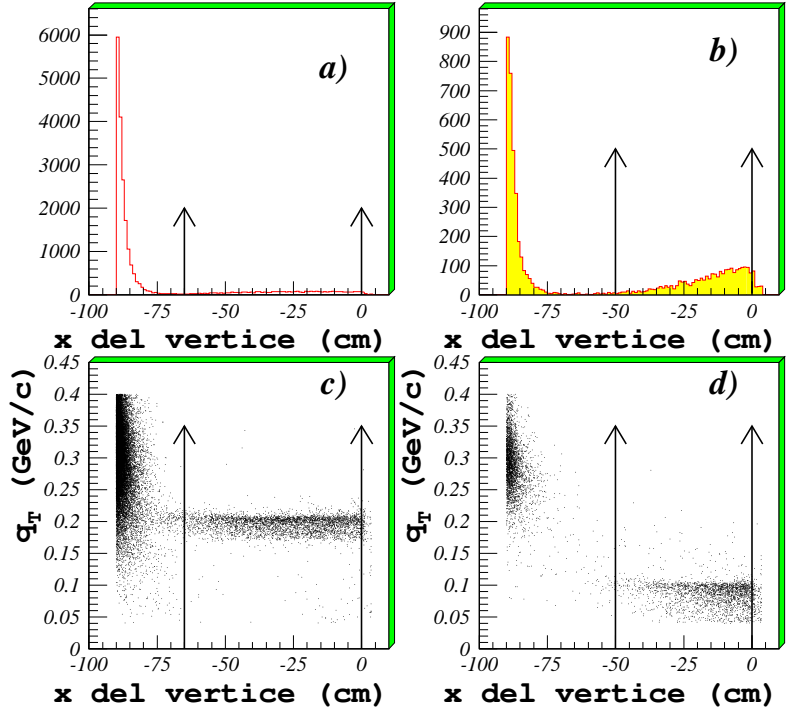}
\caption{\em Distribuzione della posizione lungo l'asse $x$ del vertice di
decadimento per un campione di candidate 
$\PKzS$ (a) e $\PgL$ (b) (si veda il testo per la selezione delle
candidate). Correlazione della posizione del vertice con
la variabile $q_T$ per gli stessi campioni di $\PKzS$ (c) e $\PgL$ (d).}
\label{figvtxfig}
\end{figure}

1) \hspace{0.5cm}
In fig.~\ref{figvtxfig} \`e mostrata la distribuzione della posizione
del vertice delle $V^0$ lungo l'asse $x$ del sistema di riferimento
 di OMEGA per
un campione di candidate $\PKzS$ (fig.~\ref{figvtxfig}a) e uno di candidate
 $\PgL$ (fig.~\ref{figvtxfig}b).
Questi campioni sono stati ottenuti applicando tutti gli altri tagli scelti
per isolare i segnali di $\PKzS$ e $\PgL$ (descritti in seguito), tranne
quello della posizione del vertice di decadimento. 
Nel sistema di ascisse in figura, il centro del 
primo piano del telescopio \`e posto
a $x=3.5~cm$ (il piano \`e inclinato), mentre il bersaglio si 
trova a $x=-90~cm$.
Le frecce indicano i tagli operati su tali distribuzioni:
le candidate $\PKzS$ sono accettate se decadono nell'intervallo $[-65, 0]~cm$,
mentre per le candidate $\PgL$ l'intervallo \`e $[-50, 0]~cm$.
Il limite superiore coincide con la posizione del primo degli 
scintillatori SPH1 ed SPH2, posti innanzi al telescopio, ed \`e imposto per
eliminare le false $V^0$ create 
utilizzando
tracce prodotte nell'interazione col materiale di tali scintillatori.
La scelta del limite inferiore ha, invece, l'obiettivo di eliminare
i vertici di decadimento posti nelle vicinanze del bersaglio.
Questi, infatti, costituiscono in prevalenza delle false $V^0$,
formate associando due tracce provenienti dal bersaglio
che apparentemente si incrociano al di fuori di esso
a causa degli errori di misura.
La scelta del taglio \`e stata affinata osservando le correlazioni
tra  la posizione del vertice lungo l'asse $x$ ed il  momento
trasverso  $q_T$ ad esso associato per i due campioni, mostrate
rispettivamente  in fig.~\ref{figvtxfig}c 
e \ref{figvtxfig}d. I decadimenti delle false $V^0$ 
si accumulano in corrispondenza della posizione del bersaglio
e ad alti momenti trasversi, mentre le particelle reali si
distribuiscono tra il bersaglio ed il telescopio,
accumulandosi in corrispondenza del massimo momento trasverso
a loro consentito.
 La necessit\`a di richiedere una regione
fiduciale pi\`u estesa nel caso dei $\PKzS$ \`e dovuta alla loro
minore vita media rispetto a quella delle $\PgL$, e quindi
alla loro maggiore probabilit\`a di decadere nelle vicinanze del bersaglio.
I tagli per le $\PagL$, se non altrimenti
specificato, devono intendersi uguali a quelli delle $\PgL$.

\begin{figure}[htb]
\centering
\includegraphics[scale=0.75,clip]
                                {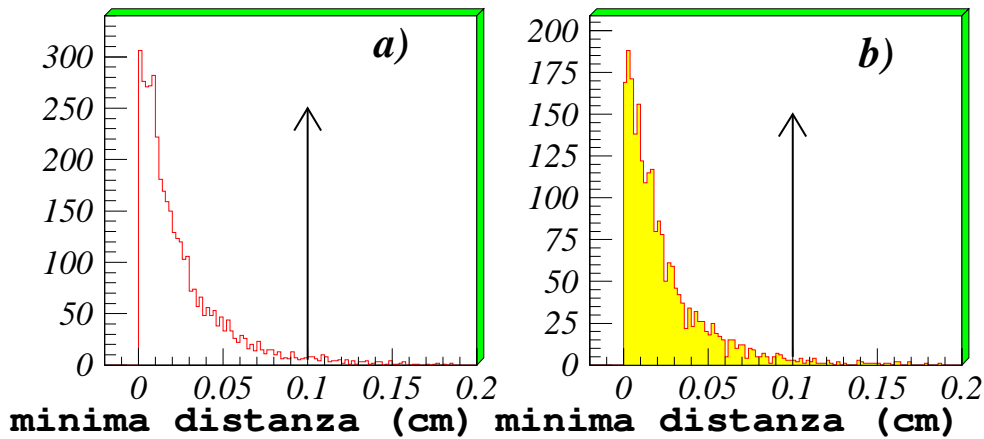}
\caption{\em Distribuzione della minima distanza tra le tracce della $V^0$
per un campione di candidate $\PKzS$ (a) e $\PgL$ (b).}
\label{figclosefig}
\end{figure}

2) \hspace{0.5cm}
In fig.~\ref{figclosefig} \`e mostrata la distribuzione della minima distanza
tra le due tracce che formano la $V^0$ per i campioni di candidate $\PKzS$
(\ref{figclosefig}a) e $\PgL$ (\ref{figclosefig}b), ottenuti applicando tutti i
rispettivi rimanenti criteri di selezione.
Il taglio indicato dalla freccia \`e stato scelto in modo da eliminare le $V^0$
con tracce di decadimento aventi minima distanza maggiore di $1~mm$. Ci\`o
riduce la contaminazione dovuta a false $V^0$ generate 
dall'associazione di due tracce non correlate.

\begin{figure}[htb]
\centering
\includegraphics[scale=0.75,clip]
                                {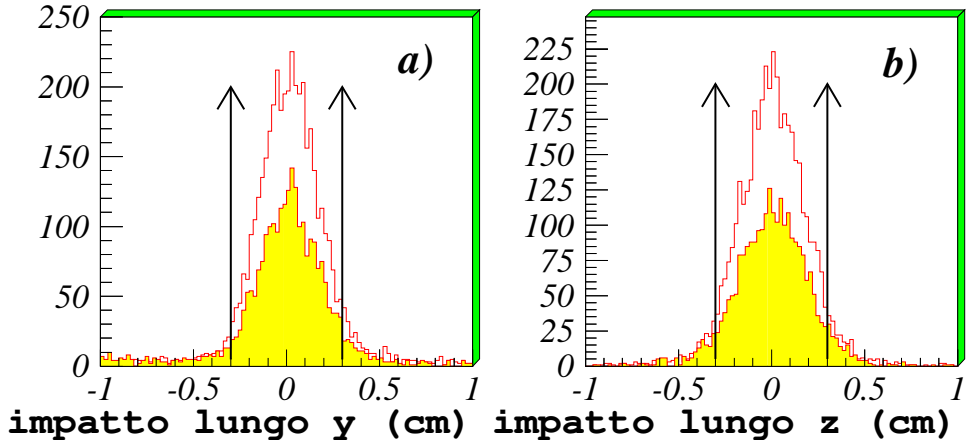}
\caption{\em Distribuzione della proiezione del parametro di impatto
lungo l'asse $y$ (a) e l'asse $z$ (b)
per un campione di candidate $\PKzS$ (in chiaro) e $\PgL$ (in grigio).}
\label{figbybzfig}
\end{figure}

3) \hspace{0.5cm}
Un'altra variabile utile per selezionare le $V^0$ reali \`e il loro parametro
di impatto, vale a dire la distanza, nel piano del bersaglio, tra
il vertice primario di interazione e la linea
di volo della $V^0$.
La posizione del vertice primario di interazione \`e stata campionata ogni
$12000$ eventi circa, calcolando il valor medio
degli impatti di tutte le candidate $V^0$ in essi contenuti; essa
\`e nota con una incertezza intorno al $20\%$.
In fig.~\ref{figbybzfig} sono mostrate le distribuzioni della proiezione
del parametro di impatto lungo l'asse $y$ (fig.~\ref{figbybzfig}a) e
l'asse $z$ (fig.~\ref{figbybzfig}b) per i campioni di candidate $\PKzS$
(in chiaro) e $\PgL$ (in grigio) ottenuti rilasciando la selezione 
su tale variabile.
In entrambi i casi il taglio operato, indicato con le frecce, richiede che
le proiezioni del parametro di impatto non siano maggiori di $3~mm$.
Esso consente di selezionare le $V^0$ candidate provenienti dal vertice
primario di interazione, individuando intorno ad esso una regione fiduciale
quadrata di lato $6~mm$, necessaria per tener conto degli errori di
 estrapolazione delle tracce di decadimento inseguite all'indietro e della
incertezza nella posizione del vertice primario stesso.

\begin{figure}[htb]
\centering
\includegraphics[scale=0.75,clip]
                                {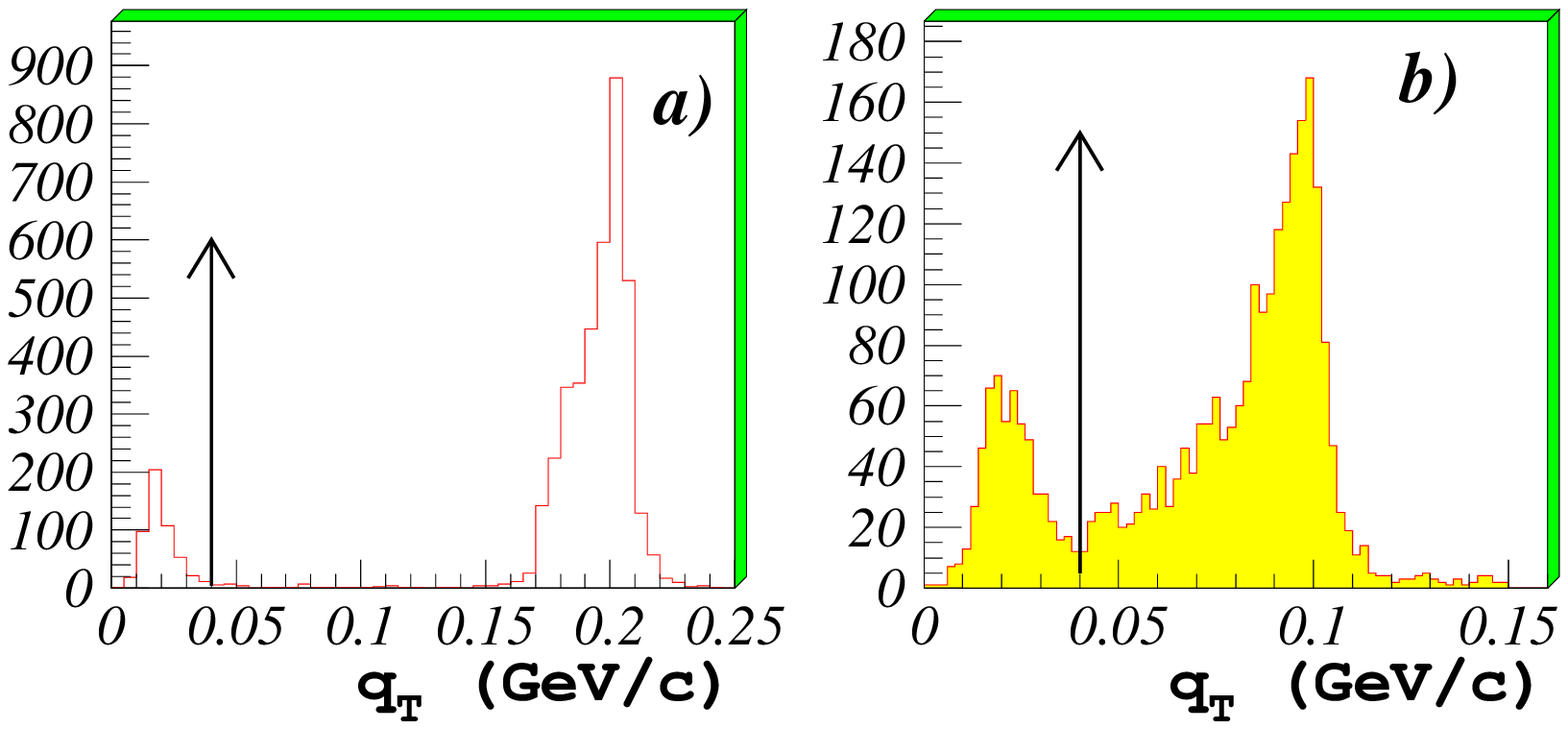}
\caption{\em Distribuzione del momento trasverso $q_T$
per un campione di candidate $\PKzS$ (a) e $\PgL$ (b).}
\label{figqtfig}
\end{figure}

4) \hspace{0.5cm}
In fig.~\ref{figqtfig} sono riportate le distribuzioni del momento trasverso
$q_T$ per il campione di candidate $\PKzS$ (fig.~\ref{figqtfig}a)
e $\PgL$ (fig.~\ref{figqtfig}b) ottenuti al solito applicando tutti gli altri
tagli.
 I picchi a bassi $q_T$ sono dovuti prevalentemente alla contaminazione
di $\gamma$ che convertono in coppie $e^+ e^-$, ricostruiti come $V^0$ dal
programma STRIPV0. Tale contaminazione, di carattere cinematico, pu\`o
essere eliminata selezionando vertici aventi $q_T > 40~MeV/c$, secondo
il taglio mostrato dalle frecce in figura. I vertici veri 
$\PKzS$ e $\PgL$ scartati
con questo taglio sono del resto molto ridotti, dato che la distribuzione 
prevista nella variabile momento trasverso per decadimenti reali, ottenuta
assumendo l'isotropia del decadimento nel sistema del centro di massa,
risulta:

\begin{equation}
\frac{dN}{dq_T}=\frac{q_T}{2p^\star\sqrt{p^{\star^2}-q_T^2}}.
\label{distribuzioneqt}
\end{equation}

\noindent
Essa raggiunge il massimo valore in corrispondenza del momento $p^\star$ (picchi
principali in fig.~\ref{figqtfig}), ma si riduce a zero per bassi momenti
trasversi, l\'{\i} dove \`e stato operato il taglio.

\vspace{1cm}
5) \hspace{0.5cm}
Come ultimo taglio di carattere generale si \`e imposto che le tracce
di decadimento delle $V^0$ selezionate soddisfino
le condizioni di trigger presenti durante la presa dati.
Il trigger a due tracce richiedeva il passaggio di almeno
 due particelle
per SPH1 e SPH2, e di almeno una particella per ST2 e SP
 (c.f.r. par.~\ref{partrigger}).
Tale condizione \`e stata riprodotta in fase di analisi imponendo che entrambe
le tracce di decadimento attraversino la parte compatta del telescopio e che
almeno una delle due attraversi gli scintillatori ST2 ed SP.
 Si sono di
conseguenza selezionati gli eventi in cui il trigger \`e stato
determinato unicamente dalla $V^0$, scartando quelli per i quali 
\`e stata determinante la presenza di tracce diverse da quelle di decadimento.
Questo taglio non serve a ridurre le contaminazioni del campione
di $V^0$ candidate, ma lo rendono omogeneo e consentono l'applicazione
della procedura di correzione per accettanza, descritta nel capitolo successivo.

\subsection{Identificazione di $K^0_S$, $\Lambda$ e $\overline{\Lambda}$}
\index{Identificazione di $K^0_S$, $\Lambda$ e $\overline{\Lambda}$}

Una volta operati i tagli comuni, la selezione pu\`o essere
ulteriormente affinata per distinguere i vari tipi di particelle
considerando la distribuzione nello spazio delle fasi dei decadimenti
ricostruiti.

\begin{figure}[htb]
\centering
\includegraphics[scale=0.76,clip]
                                {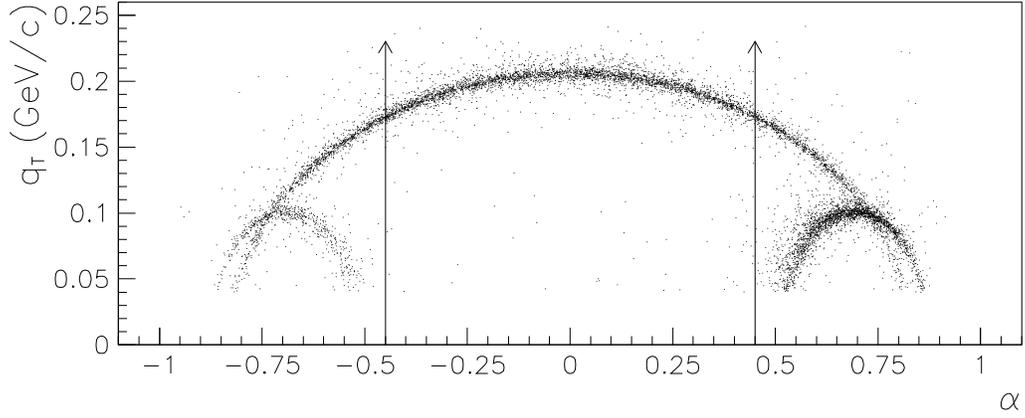}
\caption{\em Distribuzione dei decadimenti $V^0$ selezionati con i tagli comuni
nel piano ($q_T$, $\alpha$).}
\label{figarmenfig}
\end{figure}

\noindent
In fig.~\ref{figarmenfig} \`e mostrata tale distribuzione utilizzando
come variabili il momento trasverso $q_T$ di ciascun decadimento
e la relativa variabile di Armenteros [Pod54]
ottenuta combinando linearmente i momenti longitudinali delle due tracce
di decadimento:

\begin{equation}
\alpha=\frac{q_{L_1} - q_{L_2}}{q_{L_1} + q_{L_2}}.
\label{variabilearmenteros}
\end{equation}

\begin{figure}[htb]
\centering
\includegraphics[scale=0.75,clip]
                                {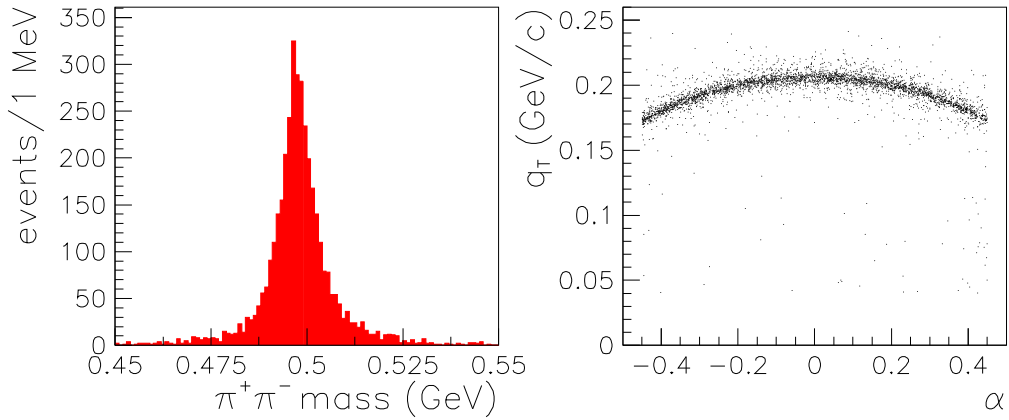}
\caption{\em Spettro di massa invariante e distribuzione nel piano ($q_T$, $\alpha$)
per un campione di $\PKzS$ identificati.}
\label{figkapfig}
\end{figure}

\noindent
Si pu\`o dimostrare [Pod54] che, nello spazio delle fasi cos\`{\i}
determinato, i punti corrispondenti ad un dato schema di decadimento si
posizionano lungo l'ellisse di equazione

\begin{equation}
\left(\frac{q_T}{p^\star} \right)^2 +
\left(\frac{\alpha-\bar{\alpha}}{\alpha_L} \right)^2=1
\label{ellissearmenteros}
\end{equation}

\noindent con

\begin{equation}
\bar{\alpha}=\frac{m_1^2-m_2^2}{M}, \:\:\:\:\:\:\: \alpha_L\sim\frac{2p^\star}{M}
\label{ell2}
\end{equation}

\noindent
dove la seconda delle relazioni (\ref{ell2}) \`e valida
sotto l'ipotesi che i decadimenti avvengano ad energie relativistiche.
La forma dell'ellisse dipende, dunque,
solamente dalle masse
delle particelle coinvolte nel decadimento, mentre la posizione del 
punto rappresentativo del decadimento lungo l'ellisse
dipende dall'angolo di decadimento nel centro di massa (dalla
relazione $q_T=p^\star\,\sin{\theta^*}$).
Risulta quindi agevole distinguere le varie specie di particelle neutre
presenti nel campione di candidate $V^0$: l'ellisse pi\`u grande
in fig.~\ref{figarmenfig} corrisponde al decadimento (simmetrico) dei $\PKzS$,
mentre le due ellissi pi\`u piccole sono dovute ai decadimenti (asimmetrici)
delle $\PgL$ e $\PagL$.
Il taglio nella variabile $\alpha$ indicato dalle frecce in figura consente
di operare una prima distinzione tra queste tre topologie di decadimento,
riducendo la contaminazione di tipo cinematico presente nel campione
di $V^0$ candidate. Il campione di $\PKzS$ cos\`{\i} ottenuto \`e mostrato in
fig.~\ref{figkapfig}, insieme al suo spettro di massa invariante:
l'ambiguit\`a con le $\PgL$ e $\PagL$ \`e stata completamente eliminata.

\begin{figure}[htb]
\centering
\includegraphics[scale=0.75,clip]
                                {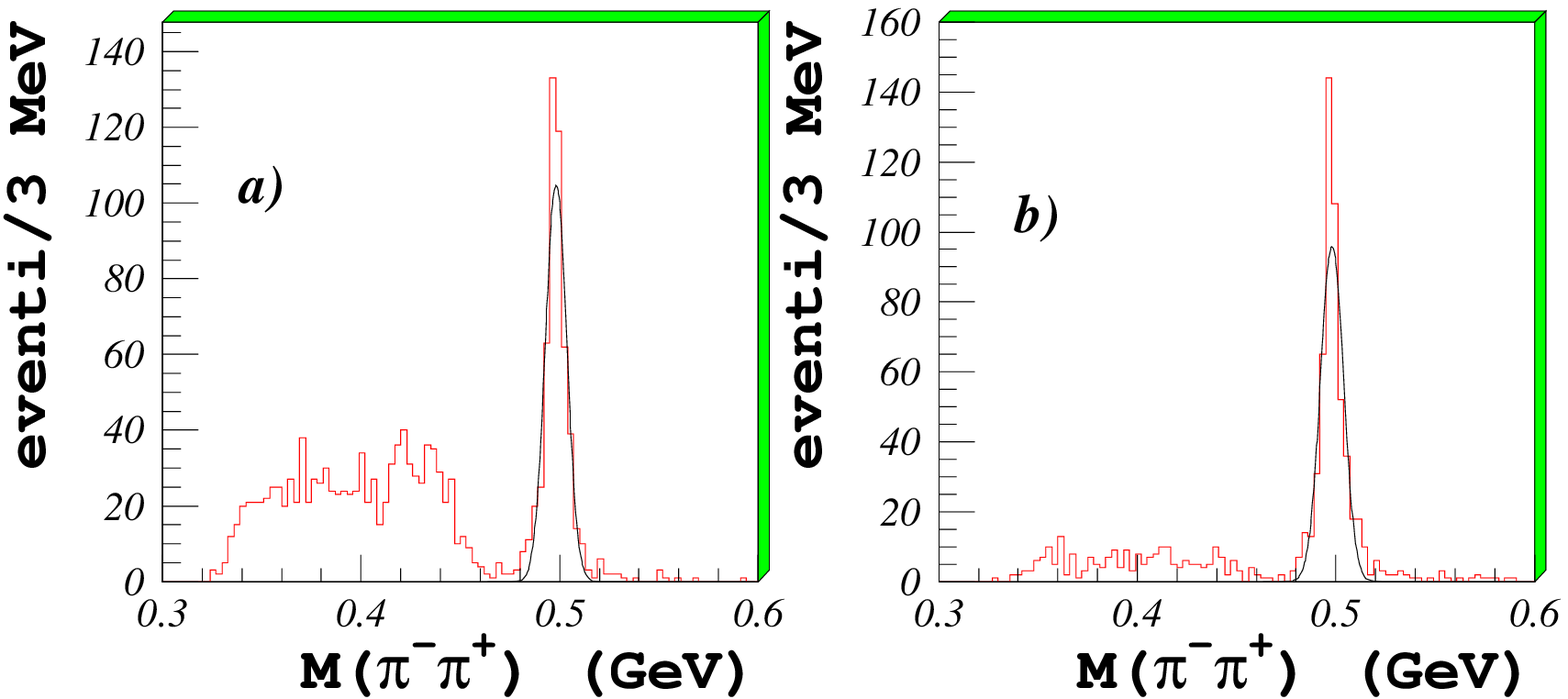}
\caption{\em Spettri di massa invariante $M(\Pgpp, \Pgpm)$ relativi ad un
campione di $\PgL$ (a) e $\PagL$ (b).} 
\label{figriflfig}
\end{figure}

\noindent
Per i campioni di $\PgL$ e $\PagL$, invece, \`e necessario eliminare
l'ambiguit\`a residua dovuta ai $\PKzS$ situati nella regione
di intersezione delle ellissi. 
Si \`e allora considerata la regione dello spazio delle fasi
individuata dal taglio $0.45\le |\alpha |\le 0.65$, nella quale
le ellissi delle
$\PgL$ e $\PagL$ risultano ben separate da quella dei $\PKzS$ 
(si veda la fig.~\ref{figarmenfig})
e in tale regione si sono studiati
gli spettri
di massa invariante $M(\Pgpp, \Pgpm)$ relativi ai campioni di $\PgL$
e $\PagL$ candidate, mostrati rispettivamente in fig.~\ref{figriflfig}a
e~\ref{figriflfig}b.

\noindent
In entrambi gli spettri il picco dovuto ai $\PKzS$ presenti nel campione,
situato in corrispondenza della loro massa nominale, risulta chiaramente
distinto dal riflesso del segnale di $\PgL$ ($\PagL$), distribuito su tutto lo
spettro, ma concentrato nella regione a masse invarianti minori.
Il picco dei $\PKzS$ \`e stato approssimato con una gaussiana di larghezza
$\sigma=5.7~MeV$ e dai campioni di $\PgL$ e $\PagL$
sono stati rigettati gli eventi distanti meno di $3\,\sigma$ dal valore nominale
della massa dei $\PKzS$.

\begin{figure}[htb]
\centering
\includegraphics[scale=0.75,clip]
                                {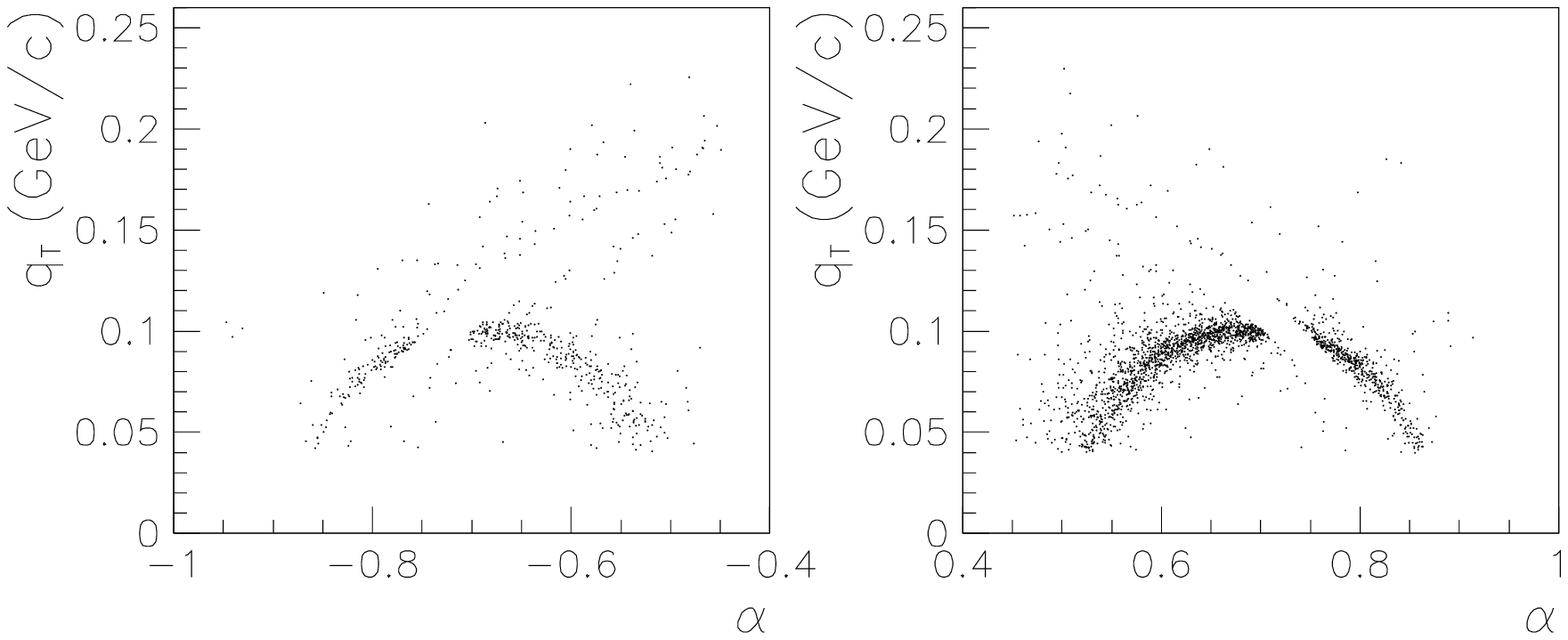}
\caption{\em Distribuzioni nel piano ($q_T$, $\alpha$) per due
campioni anambigui di $\PagL$  e $\PgL$.} 
\label{figrela3}
\end{figure}

\begin{figure}[htb]
\centering
\includegraphics[scale=0.75,clip]
                                 {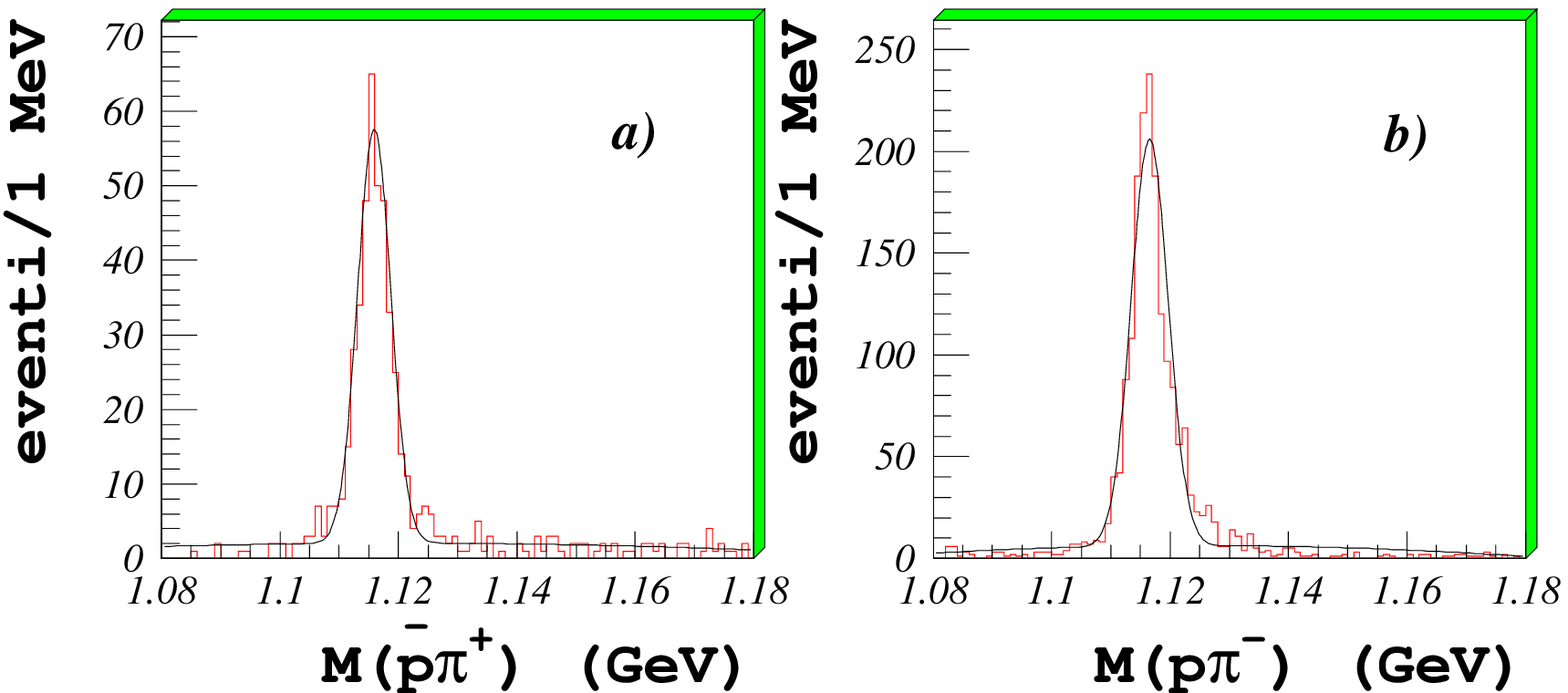}
\caption{\em Spettri di massa invariante $M(\Pap , \Pgpp)$ e $M(\Pp , \Pgpm)$
relativi a due campioni anambigui di $\PagL$ (a)  e $\PgL$ (b).} 
\label{figrela4}
\end{figure}

\noindent
L'effetto del taglio nello spazio delle fasi per i campioni di $\PgL$ e $\PagL$
\`e visibile in fig.~\ref{figrela3}: si pu\`o stimare che esso comporta la
perdita di circa il $20\%$ dei vertici $\PgL$ e $\PagL$ reali.
In fig.~\ref{figrela4} sono mostrati i rispettivi spettri di massa invariante.

\subsection{Fit degli spettri di massa}
\index{Fit degli spettri di massa}

Per valutare la qualit\`a dei segnali di $\PKzS$, $\PgL$ e $\PagL$ selezionati
e stimare il numero di vertici reali (segnale) rispetto al fondo dovuto a
contaminazioni di tipo cinematico e geometrico si \`e eseguito
un {\em best fit} dei rispettivi spettri di massa invariante,
utilizzando una gaussiana per descrivere il segnale ed un
polinomio di secondo grado per descrivere il fondo\footnote{ \`E noto
da numerosi altri studi su $\PKzS$ e $\PgL$ che le distribuzioni
di massa non corrispondono a semplici gaussiane (basti osservare, per
esempio, che gli eventi non hanno tutti la stessa precisione
di misura). Ci\`o \`e riflesso nel $\chi^2$ dei {\em best fit}, che 
non \`e ottimale. Tuttavia, l'approssimazione \`e sufficiente per gli scopi
qui prefissi.}.
I risultati dei {\em fit}, eseguiti col programma MINUIT [Jam94], sono riportati
in tab.~\ref{tabfit}. \`E interessante notare come i valori centrali della
gaussiana ($\mu$) siano in completo accordo con i valori nominali delle 
rispettive particelle, mentre la larghezza della gaussiana ($\sigma$)
fornisce una stima della risoluzione in massa degli spettri e risulta compresa
tra $3$ e $5~MeV$.

\begin{table}[htb]
\centering
\caption{\em Risultati dei fit sugli spettri di massa invariante dei
             $\PKzS$, $\PgL$ e $\PagL$ identificati in interazioni p-Pb,
$\mu$ \`e il valore centrale e $\sigma$ \`e la larghezza  della gaussiana.}
\begin{tabular}{|c|c|c|c|}
\hline
            & $\mu (MeV)$     &  $\sigma (MeV)$   & $\chi^2/ndf$  \\
\hline
\PKzS       &  $497.4\pm 0.1$ & $4.95\pm 0.10$    & $2.35$ \\
\PgL        & $1116.5\pm 0.1$ & $3.1\pm 0.1  $    & $3.20$ \\
\PagL       & $1115.9\pm 0.1$ & $2.8\pm 0.1  $    & $0.59$ \\
\hline
\end{tabular}
\label{tabfit}
\end{table}

\noindent
La procedura di {\em fit} permette, inoltre, di ottimizzare la scelta degli intervalli
 di massa da considerare per isolare le $V^0$ reali: essa \`e stata
effettuata in relazione alla conseguente diminuzione del segnale di ciascun
campione (inteso come l'area della gaussiana al di sopra del polinomio)
e cercando di rendere minimo il contributo del fondo (inteso come
l'area del polinomio).
Sono stati cos\`{\i} selezionati un intervallo di $30~MeV$ per il
campione di $\PKzS$ ed uno di $15~MeV$ per i campioni di $\PgL$ e $\PagL$,
centrati sui rispettivi valori nominali di massa.
Il contributo del fondo in tali intervalli di massa per i campioni di
$\PKzS$, $\PgL$ e $\PagL$ risulta rispettivamente del $6.8\%$, $2.8\%$ e 
$6.8\%$.
Nel primo caso il fondo \`e dovuto esclusivamente alle contaminazioni di tipo
geometrico, mentre negli ultimi due casi esso \`e in gran parte
attribuibile all'ambiguit\`a residua dei $\PKzS$.

\subsection{Analisi dei tagli}
\index{Analisi dei tagli}
\label{taglianalisi}

Riassumendo, per identificare i $\PKzS$, $\PgL$ e $\PagL$ in interazioni
p-Pb sono stati effettuati i seguenti criteri di selezione:
\begin{description}

\item[a)] il vertice della $V^0$ deve avere una topologia di tipo cowboy;

\item[b)] il vertice della $V^0$ deve trovarsi entro una regione fiduciale situata
immediatamente prima del telescopio. La regione si estende per $50~cm$ nel
caso di $\PgL$ e $\PagL$ e per $65~cm$ nel caso di $\PKzS$;

\item[c)] la minima distanza tra le tracce di decadimento della $V^0$ non deve
essere maggiore di $1~mm$;

\item[d)] il parametro di impatto della $V^0$ deve trovarsi entro un quadrato
di lato $6~mm$ centrato sul vertice primario di interazione;

\item[e)]il momento trasverso $q_T$ della $V^0$ deve essere maggiore di 
$0.04~GeV/c$;

\item[f)] la variabile di Armenteros deve assumere i valori $|\alpha|\leq 0.45$
per i $\PKzS$, $\alpha > 0.45$ per le $\PgL$ e $\alpha < -0.45$ per le $\PagL$;

\item[g)] nei campioni di $\PgL$ e $\PagL$ devono essere rigettati gli eventi
che, nello spettro di massa $M(\Pgpp , \Pgpm)$ distano meno di $17~MeV$
dal valore nominale di massa di $\PKzS$;

\item[h)] entrambe le tracce di decadimento devono attraversare la parte 
compatta del telescopio 
ed almeno una deve attraversare gli scintillatori ST2 ed SP;

\item[i)] la massa invariante della $V^0$ deve trovarsi in un intervallo
di $30~MeV$ per i $\PKzS$ e $15~MeV$ per le $\PgL$ e $\PagL$, centrato sui
rispettivi valori nominali di massa.

\end{description}
\par
\begin{figure}[htb]
\centering
\includegraphics[scale=0.75,clip]
                                {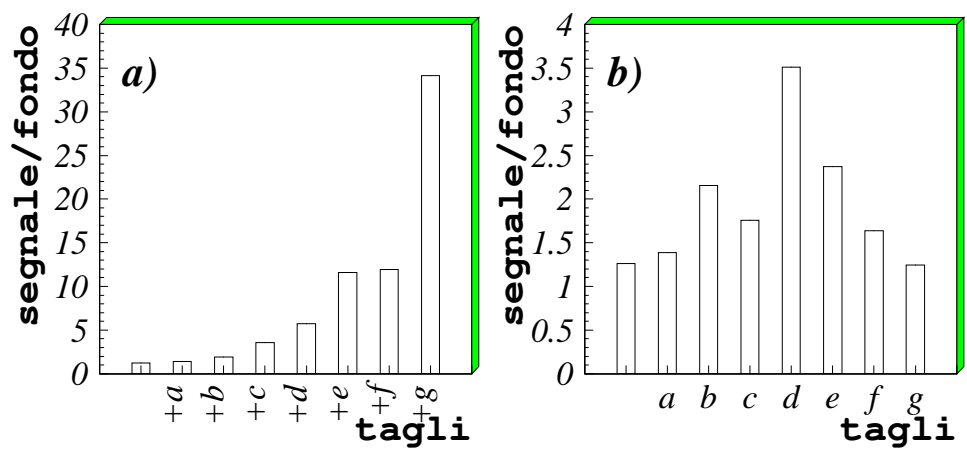}
\caption{\em  Rapporto segnale/fondo negli spettri di massa
del campione di $\PgL$ ottenuti applicando i tagli dell'analisi in
maniera sequenziale (a) e separata (b).}
\label{figtaglifig}
\end{figure}

Il miglioramento nel rapporto tra segnale e fondo apportato dai tagli esaminati
pu\`o essere valutato applicando la procedura di {\em best fit}, discussa nel
paragrafo precedente, agli spettri di massa ottenuti nei differenti stadi
della selezione. L'analisi \`e stata eseguita per il campione di $\PgL$,
relativamente ai tagli compresi tra i punti a) e g) ed
 all'interno dell'intervallo
di massa riportato al punto i).
Il rapporto segnale/fondo risultante dai {\em fit} \`e stato calcolato dapprima
applicando i tagli in successione (fig.~\ref{figtaglifig}a) e poi uno
separatamente dall'altro (fig.~\ref{figtaglifig}b). 

\noindent
Dalla fig.~\ref{figtaglifig}a si deduce che l'applicazione in sequenza dei tagli
porta ad un costante miglioramento del rapporto segnale/fondo, il quale passa
da $1.3$ per il campione iniziale 
di candidate $V^0$ a $34.1$ per il campione di $\PgL$ 
identificate. Al tempo stesso, si osserva una decrescita del contributo del
fondo nel campione, calcolabile con la formula

\[
fondo (\%)=\frac{100}{1+\frac{segnale}{fondo}}.
\]

\noindent
All'inizio della selezione esso risulta del $44.2\%$, mentre alla fine diventa
del $2.8\%$, come riportato in precedenza.
\par
D'altra parte, dall'analisi di fig.~\ref{figtaglifig}b \`e possibile
valutare il miglioramento della qualit\`a del campione introdotto dai
singoli tagli, indipendentemente dall'ordine con cui essi sono operati.
I tagli pi\`u efficaci nell'eliminare il fondo presente nel campione di
$V^0$ candidate risultano quelli sul parametro di impatto (taglio d) ) e sul
momento trasverso $q_T$ (taglio e) ), seguiti dal taglio sulla posizione 
del vertice di
 decadimento (taglio b) ). Essi consentono di eliminare gran parte delle false
$V^0$ formate da tracce provenienti dal bersaglio (tagli b) e d) ) e
la contaminazione dei numerosi $\gamma$ presenti nel campione (taglio e) ).
Si osservi come l'azione del taglio g), riguardante l'eliminazione della
contaminazione dei $\PKzS$, 
appaia del tutto trascurabile in 
fig.~\ref{figtaglifig}b, ma  
risulti molto vantaggiosa per ridurre il fondo
 alla fine della catena dei tagli,
una volta che tutta la contaminazione 
geometrica \`e stata eliminata
(si veda fig.~\ref{figtaglifig}a).

\section{Ricostruzione e selezione di $\Xi$ e $\Omega$}
\index{Ricostruzione e selezione di $\Xi$ e $\Omega$}
\label{partaglixiom}

Per la ricostruzione dei decadimenti a cascata delle particelle multi-strane,
gli eventi nei quali il programma STRIPV0 ha ricostruito almeno una $V^0$ 
candidata sono esaminati da un ulteriore programma. Questo combina la
$V^0$ con una terza traccia carica dell'evento, per formare una candidata
cascata, secondo lo schema di decadimento illustrato in fig.~\ref{figdecay}.
Alcuni tagli preliminari sono operati in fase di ricostruzione del decadimento.
Per i vertici $V^0$ si richiede, in aggiunta ai tagli fatti da STRIPV0,
che  le tracce di decadimento abbiano minima distanza minore di $0.5~mm$,
 coordinata $x$ compresa nell'intervallo $[-80, 0]~cm$,
momento trasverso $q_T > 0.02~GeV/c$, proiezione del parametro di impatto
lungo l'asse $y$ maggiore di $1.5~cm$ (in quanto il vertice $V^0$ appartenente
al vertice cascata non deve puntare al vertice primario di interazione)
e topologia cowboy.
Poich\`e sia nel caso di $\Xi$ che in quello di $\Omega$ \`e necessario
che il vertice $V^0$ sia identificato come una $\PgL$ ($\PagL$), si richiede
inoltre che la sua massa invariante non si discosti da quella nominale delle
$\PgL$ per pi\`u di $15~MeV$.
Per quanto riguarda i vertici cascata, si richiede che la minima distanza tra
la linea di volo della $V^0$ e la traccia carica sia minore di $0.5~mm$,
la proiezione del parametro di impatto lungo l'asse $y$ non sia maggiore di
$1.5~cm$ e la topologia di decadimento sia di tipo cowboy.

\begin{figure}[htb]
\centering
\includegraphics[scale=0.7,clip]
                                {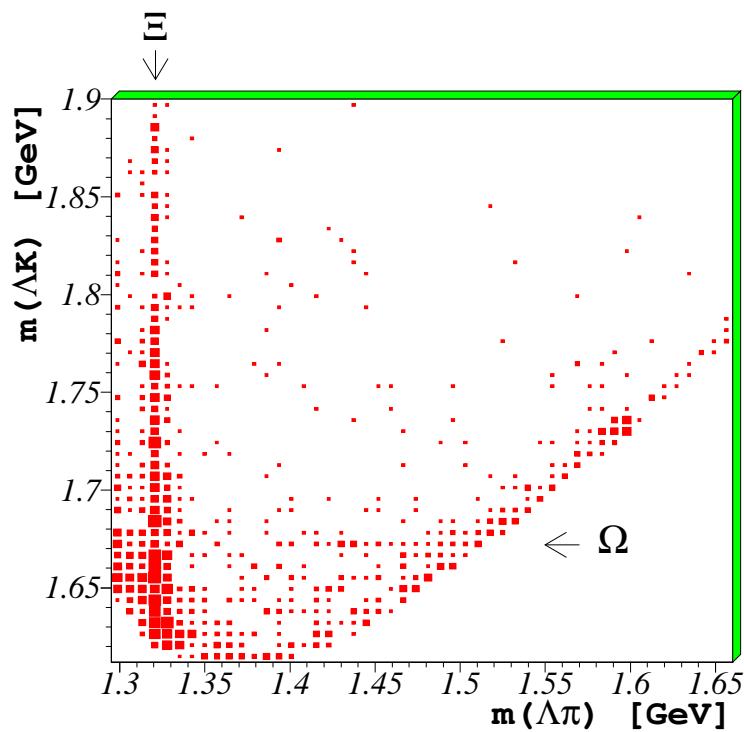}
\caption{\em Correlazione tra lo spettro di massa
massa invariante $M(\PgL , \pi)$ e quello
$M(\PgL , K)$ per le candidate cascate ricostruite in interazioni p-Pb.}
\label{figcorrfig}
\end{figure}

Applicando il programma di ricostruzione delle cascate all'intera statistica
p-Pb relativa al trigger a due tracce
 si ottengono $2534$ candidate cascate, per le quali la 
correlazione tra la massa invariante $M(\PgL , \pi)$ e quella
$M(\PgL , K)$ \`e mostrata in fig.~\ref{figcorrfig}.
Il segnale di $\Xi$, in corrispondenza del valore nominale
$M_\Xi =1.321~GeV$ nello spettro di massa invariante  $M(\PgL , \pi)$,
risulta gi\`a ben visibile e distinto dalle regioni di accumulazione di punti
poste ai bordi delle regioni permesse dello spazio delle fasi. Il segnale
di $\Omega$ dovrebbe trovarsi in corrispondenza del valore nominale
$M_\Omega =1.672~GeV$ nello spettro di massa $M(\PgL , K)$, ma
risulta ancora poco distinguibile dal fondo.
\par
Per estrarre il segnale di $\Xi$ ed $\Omega$ si sono applicati
ulteriori criteri di selezione, scelti in base ad una analisi
analoga a quella presentata per l'identificazione delle particelle strane
neutre. I tagli finali operati sono i seguenti:

\begin{itemize}

\item il vertice della cascata deve trovarsi entro una regione fiduciale situata
subito prima del telescopio. La regione si estende per $80~cm$ nel
caso delle $\Xi$  e per $60~cm$ nel caso delle $\Omega$ (la cui vita media
\`e circa la met\`a di quella delle $\Xi$);

\item i vertici $V^0$ devono seguire i vertici cascata, con una distanza di
almeno $2.5~cm$. la distanza \`e uguale per $\Xi$ e $\Omega$ perch\`e essa
dipende dalla vita media della $V^0$, che in entrambi i casi \`e una $\Lambda$;

\item il parametro di impatto della cascata deve trovarsi entro un quadrato
di lato $1~cm$ per le $\Xi$ e $0.8~cm$ per le $\Omega$,
centrato sul vertice primario di interazione;

\item le tracce di decadimento cariche della cascata devono 
soddisfare la condizione di trigger a due tracce presente durante la presa dati.
Allora si \`e imposto che tutte le tre tracce attraversino la parte 
compatta ed almeno una di loro colpisca gli scintillatori ST2 ed SP;

\end{itemize}
 
\begin{figure}[htb]
\centering
\includegraphics[scale=0.7,bb= 0 0 398 395,clip]
                                {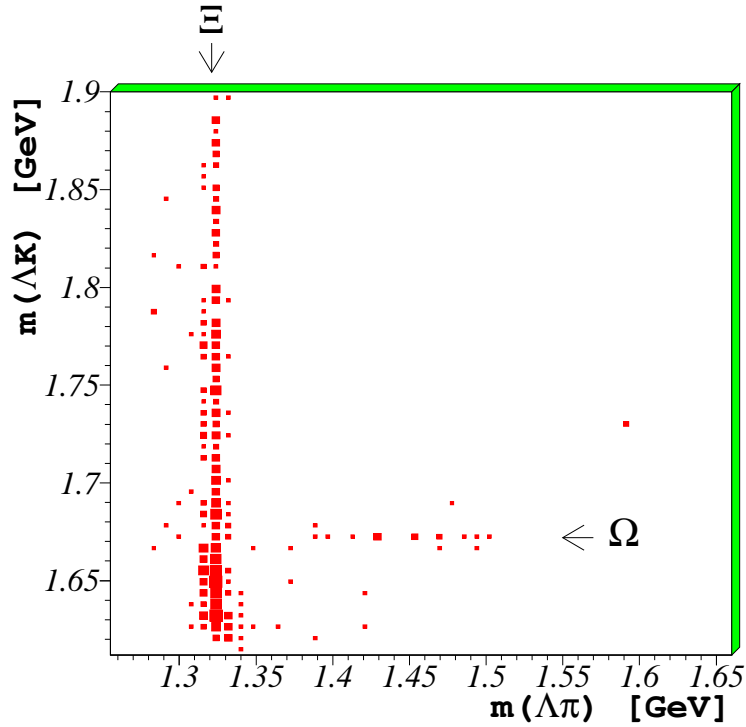}
\caption{\em Correlazione tra lo spettro di massa
invariante $M(\PgL , \pi)$ e quello
$M(\PgL , K)$ per le candidate cascate selezionate nell'analisi
di eventi p-Pb.}
\label{figpad1}
\end{figure}

\noindent
La correlazione corrispondente a quella di fig.~\ref{figcorrfig},
ottenuta dopo aver applicato i suddetti criteri di selezione, \`e mostrata in
in fig.~\ref{figpad1}: il segnale di $\Omega$ appare ora chiaramente visibile
e gran parte del fondo dovuto ai falsi vertici cascata ricostruiti \`e stato
eliminato.

\begin{figure}[htb]
\centering
\includegraphics[scale=0.75,clip]
                                {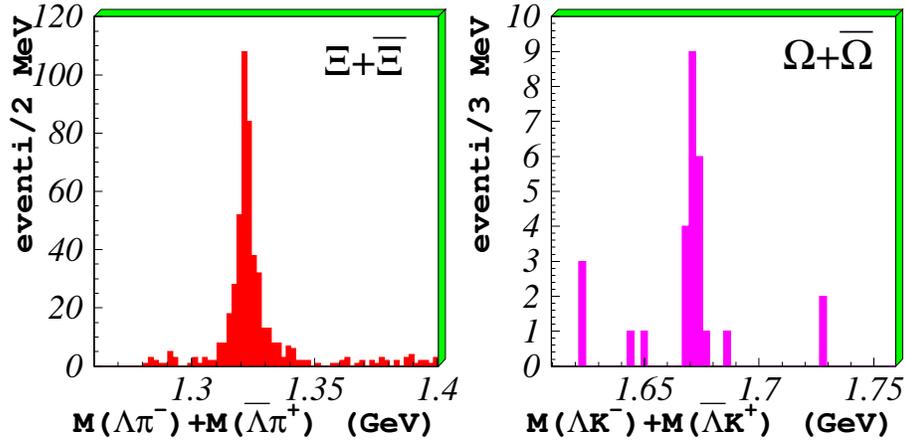}
\caption{\em Spettri di massa
invariante $M(\PgL , \pi)$ e 
$M(\PgL , K)$ per i campioni di $\Xi$ e $\Omega$ identificati
in eventi p-Pb.}
\label{figxiompfig}
\end{figure}

Rimane ora da eliminare la contaminazione cinematica tra i segnali di 
$\Xi$ ed $\Omega$.
Con riferimento alla fig.~\ref{figpad1}, per isolare le $\Xi$ non ambigue si
 sono scartati gli eventi che nello spettro di massa $M(\PgL ,K)$  sono
compresi nell'intervallo $[1.667,\, 1.677~GeV]$, mentre per isolare le
$\Omega$ non ambigue si sono scartati gli eventi per i quali la
massa $M(\PgL , \pi)$ risulta minore di $1.35~GeV$.
Gli spettri di massa finali sono mostrati in fig.~\ref{figxiompfig}
e gli intervalli di massa scelti per isolare il segnale fisico hanno
un'ampiezza di $30~MeV$ per le $\Xi$ e di $15~MeV$ per le $\Omega$, centrati
sui rispettivi valori nominali di massa.

\noindent
Si pu\`o notare come anche per le particelle multi-strane il contributo 
del fondo compreso nel segnale fisico sia trascurabile.

\section{Selezione delle particelle strane in interazioni Pb-Pb}
\index{Selezione delle particelle strane in interazioni Pb-Pb}

La ricostruzione dei decadimenti di particelle strane in interazioni Pb-Pb
\`e stata eseguita adoperando criteri analoghi a quelli  usati per
l'interazione p-Pb. L'identificazione dei segnali fisici a partire
dall'insieme di vertici ricostruiti risulta per\`o pi\`u problematica che nel
caso precedente, a causa dell'aumento della molteplicit\`a media di tracce
nell'evento.
La massiccia presenza del fondo dovuto alla ricostruzione di falsi vertici
rende necessari tagli pi\`u stringenti in fase di ana\-lisi.
D'altra parte, la maggiore intensit\`a del segnale fisico permette una migliore
definizione dei criteri di selezione e, soprattutto, l'alta molteplicit\`a
delle tracce ricostruite consente di determinare
la posizione del vertice primario di interazione con una buona precisione
evento per evento.
Il taglio sul parametro di impatto risulta di conseguenza molto pi\`u
efficiente che nel caso dell'interazione p-Pb.
\par
I criteri di selezione usati per identificare i $\PKzS$, $\PgL$ e $\PagL$
sono stati studiati con le stesse tecniche usate per l'interazione p-Pb
e sono stati applicati sulle stesse variabili del decadimento.
In riferimento all'elenco riportato nel paragrafo~\ref{taglianalisi},
vengono ora evidenziate solo le differenze introdotte nello studio
degli eventi Pb-Pb:
\begin{description}

\item[b)] la regione fiduciale si estende per $50~cm$ per
$\PKzS$,  $\PgL$ e $\PagL$ (si ricordi che per questo 
campione il telesopio \`e distante $60~cm$ dal bersaglio);

\item[c)] la minima distanza tra le tracce di decadimento della $V^0$ non deve
essere maggiore di $0.5~mm$;

\item[d)] la proiezione lungo l'asse $y$
del parametro di impatto della $V^0$ deve essere minore di $3~mm$, mentre
la sua proiezione lungo l'asse $z$ deve essere minore di $1.5~mm$; 

\item[e)]il momento trasverso $q_T$ della $V^0$ deve essere maggiore di 
$0.03~GeV/c$;

\item[g)] nei campioni di $\PgL$ e $\PagL$ devono essere rigettati gli eventi
che, nello spettro di massa $M(\Pgpp , \Pgpm)$, distano meno di $25~MeV$
dal valore nominale di massa del $\PKzS$;

\item[h)] entrambe le tracce di decadimento devono attraversare la parte 
compatta del telescopio (nel trigger per il campione Pb-Pb 
non sono richieste condizioni
sugli scintillatori che precedono o seguono il telescopio);

\end{description}

Per quanto riguarda l'identificazione di $\Xi$ e $\Omega$,
 le differenze rispetto
ai tagli riportati nel paragrafo~\ref{partaglixiom} sono:
\begin{itemize}
\item la regione fiduciale si estende per $60~cm$ per 
      entrambe le particelle;
\item per isolare le $\Omega$ non ambigue si sono scartati 
      gli eventi che hanno massa minore di $1.38~GeV$ nello 
      spettro $M(\PgL , \pi)$.
\end{itemize}

\begin{figure}[htb]
\centering
\includegraphics[scale=0.7,bb=0 14 454 387,clip]
                                {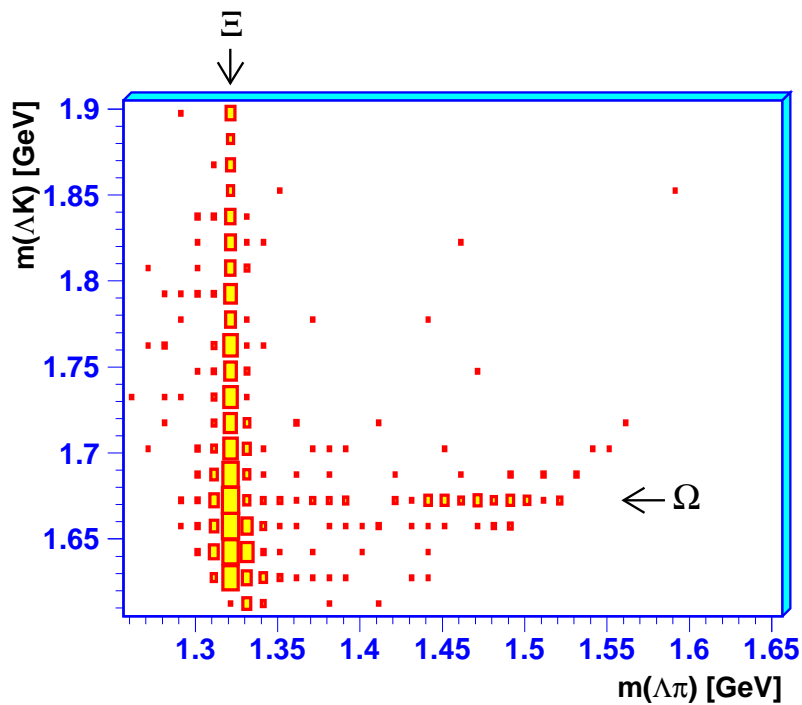}
\caption{\em Correlazione tra lo spettro di massa
invariante $M(\PgL , \pi)$ e quello
$M(\PgL , K)$ per le candidate cascate selezionate nell'analisi
di eventi Pb-Pb.}
\label{figpad2}
\end{figure}

\noindent
In fig.~\ref{figpad2} \`e riportata la correlazione analoga a quella
mostrata in fig.~\ref{figpad1} per eventi p-Pb.
Si pu\`o notare la maggiore intensit\`a dei segnali di $\Xi$ e $\Omega$,
che risultano comunque ben distinti dal fondo residuo presente nel
campione.

\begin{figure}[htb]
\centering
\includegraphics[scale=0.7,clip]
                                {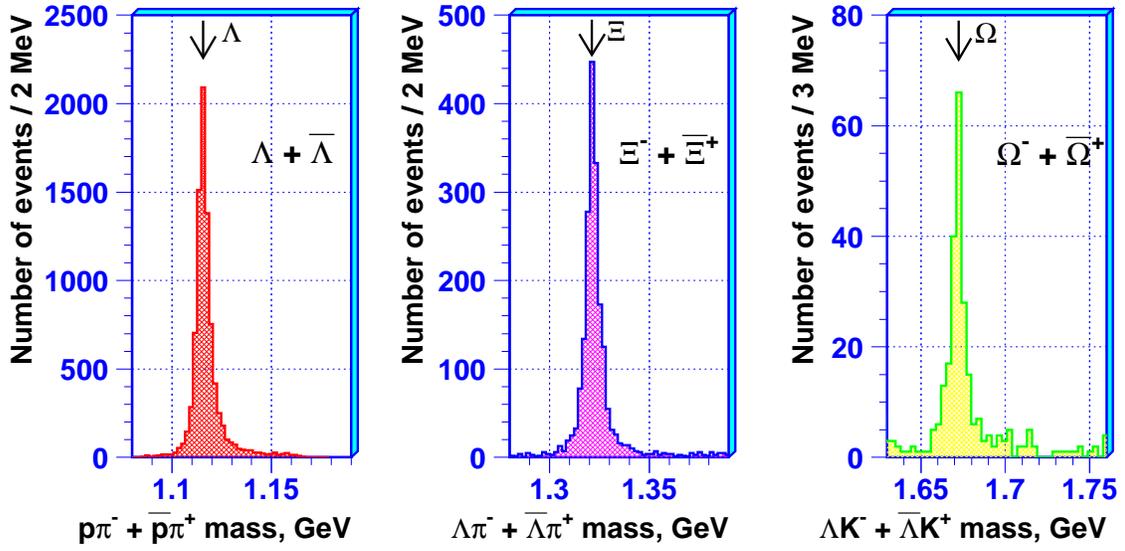}
\caption{\em Spettri di massa
per  campioni di $\PgL$, $\Xi$ e $\Omega$ identificati in eventi Pb-Pb.}
\label{figpad3}
\end{figure}

\begin{figure}[ht]
\centering
\includegraphics[height=7cm,width=13cm]
                                {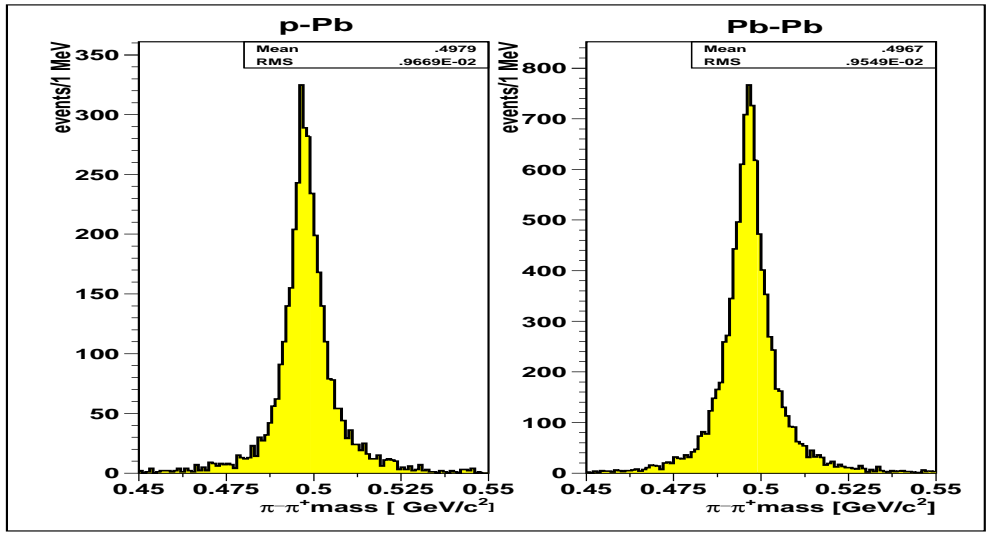}
\caption{\em Spettri di massa
per  campioni di $\PKzS$ identificati
in eventi p-Pb e Pb-Pb.}
\label{figpad4}
\end{figure}

\nopagebreak \noindent
In fig.~\ref{figpad3} sono riportati gli spettri di massa finali per un
 campione di $\PgL$, $\Xi$ ed $\Omega$ identificate in eventi Pb-Pb.
La risoluzione in massa \`e di circa $3~MeV$ e la presenza di fondo risulta
trascurabile. In fig.~\ref{figpad4} sono invece confrontati gli spettri
di massa dei $\PKzS$ identificati in eventi p-Pb e Pb-Pb.


%
%
%
%
%
%
\chapter{Analisi dei dati}
\index{Analisi dei dati}

\section{Introduzione}
\index{Introduzione}

Questo capitolo riguarda lo studio dei segnali fisici di particelle strane
ottenuti applicando i criteri selettivi discussi e giustificati nel
capitolo precedente.
Verranno riportati  il numero di particelle strane identificate 
analizzando l'intera statistica p-Pb e Pb-Pb accumulata dall'esperimento
WA97 e le regioni cinematiche scelte per lo studio dei differenti segnali.
I valori sono stati corretti per l'accettanza geometrica dell'apparato 
sperimentale, per
 l'efficienza dei rivelatori e dei programmi di ricostruzione
e, infine,  per i tagli introdotti nella procedura di
identificazione dei diversi tipi di particella.
Le propriet\`a fisiche dei segnali sono state studiate all'interno delle
relative regioni cinematiche; in esse \`e stato calcolato il tasso di
 produzione
di particelle strane ed \`e stata applicata una procedura di {\em best fit}
che ha permesso di determinare le distribuzioni di massa trasversa ed i loro
parametri caratteristici.
Questi sono stati poi utilizzati per estrapolare il calcolo del tasso
di produzione
di stranezza ad una regione cinematica comune ai diversi tipi di 
particella e pi\`u direttamente confrontabile con quella di altri esperimenti.
Le possibili sorgenti di errori sistematici introdotte nelle varie fasi
di analisi del segnale saranno evidenziate di volta in volta e 
saranno illustrate le procedure che hanno portato al calcolo dei 
relativi contributi.

\section{Caratteristiche dei segnali fisici}
\index{Caratteristiche dei segnali fisici}

I numeri di particelle strane presenti nel campione di dati registrato 
dall'esperimento WA97 ed identificate per mezzo dei criteri di selezione
esposti nel capitolo precedente sono riassunti in tab.~\ref{tabstat1}.

\begin{table}[ht]
\centering
\caption{\em Statistica delle particelle strane contenute nell'intero
             campione di dati relativi all'interazione p-Pb e Pb-Pb.}
\begin{tabular}{|c|c|c|}
\hline
            & p-Pb      & Pb-Pb  \\
\hline
\PgOm       &  $\hphantom{000}15$  &  $\hphantom{000}363$ \\
\PagOp      &  $\hphantom{0000}4$  &  $\hphantom{000}145$ \\
\PgXm       &  $\hphantom{00}275$  &  $\hphantom{00}3365$ \\
\PagXp      &  $\hphantom{00}101$  &  $\hphantom{000}848$ \\
\PgL        &  $106180           $ &  $653800$            \\
\PagL       &  $\hphantom{0}26125$ &  $100350$            \\
\PKzS       &  $208085$            &  $750000$            \\
\hline
\end{tabular}
\label{tabstat1}
\end{table}

\noindent
Per quanto riguarda l'interazione p-Pb, l'analisi \`e stata eseguita su tutti
i $110~M$ eventi registrati,  scartando tuttavia gli eventi nei quali si
sono verificate anomalie in fase di presa dati.
Per l'interazione Pb-Pb, invece, l'analisi si riferisce ad un
sotto-campione di $77~M$ eventi, pari al $40\%$ dell'intera
statistica registrata (l'analisi del resto del campione \`e ancora in corso).
 I numeri riportati in tab.~\ref{tabstat1},
relativi all'intera statistica accumulata, sono 
stati stimati sulla base delle particelle identificate in tale sotto-campione.

\begin{figure}[htb]
\centering
\includegraphics[scale=0.45,clip]%
                                {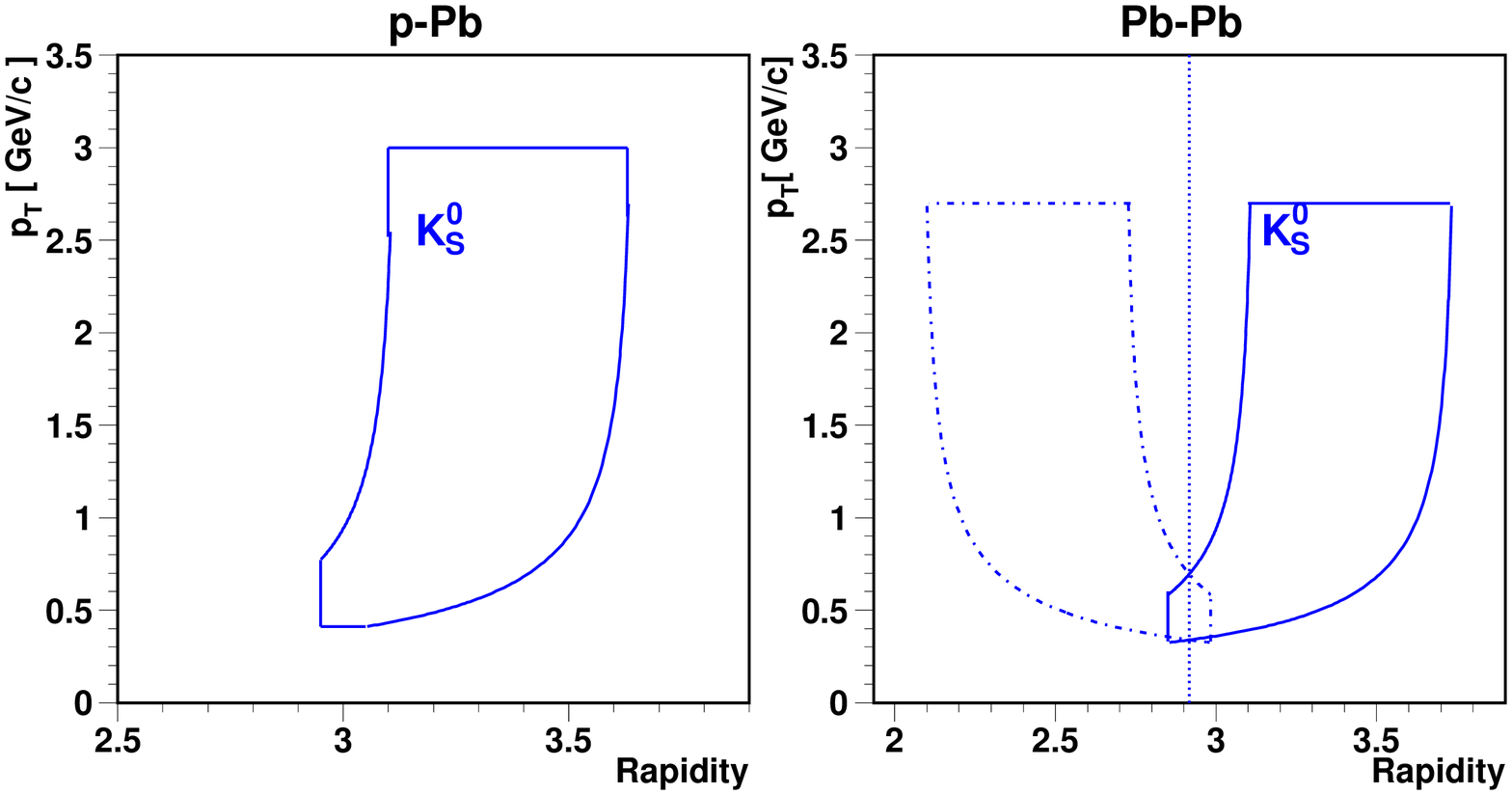}
 \includegraphics[scale=0.62,bb=8 1 262 240,clip]%
                                {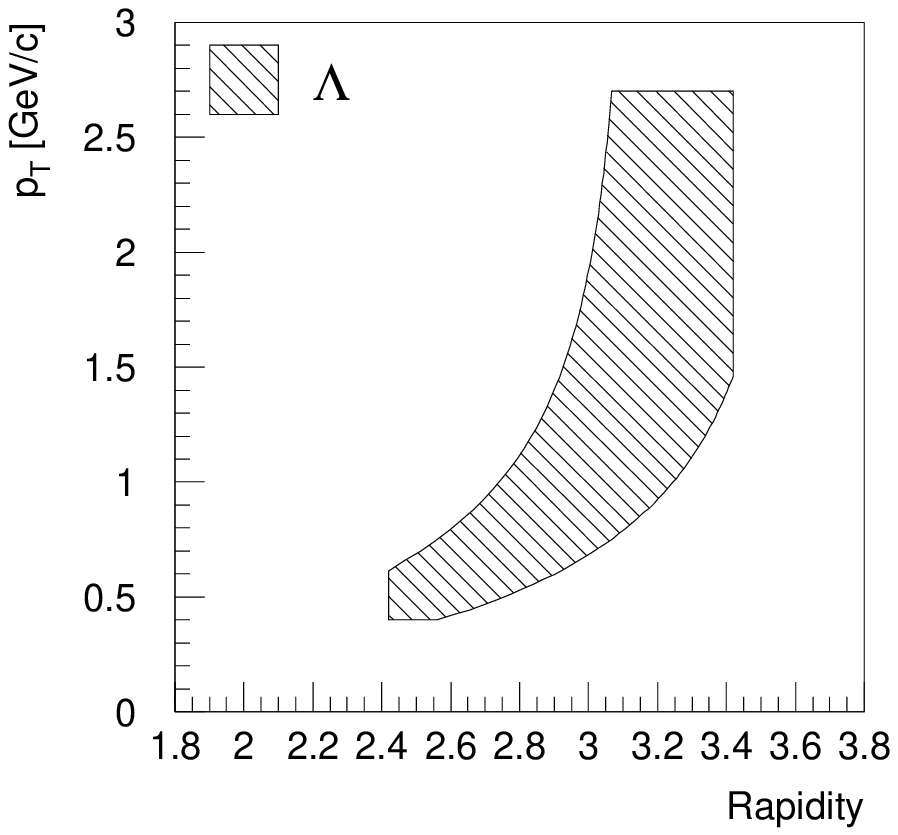}
 \includegraphics[scale=0.62,bb=8 1 262 240,clip]%
                                {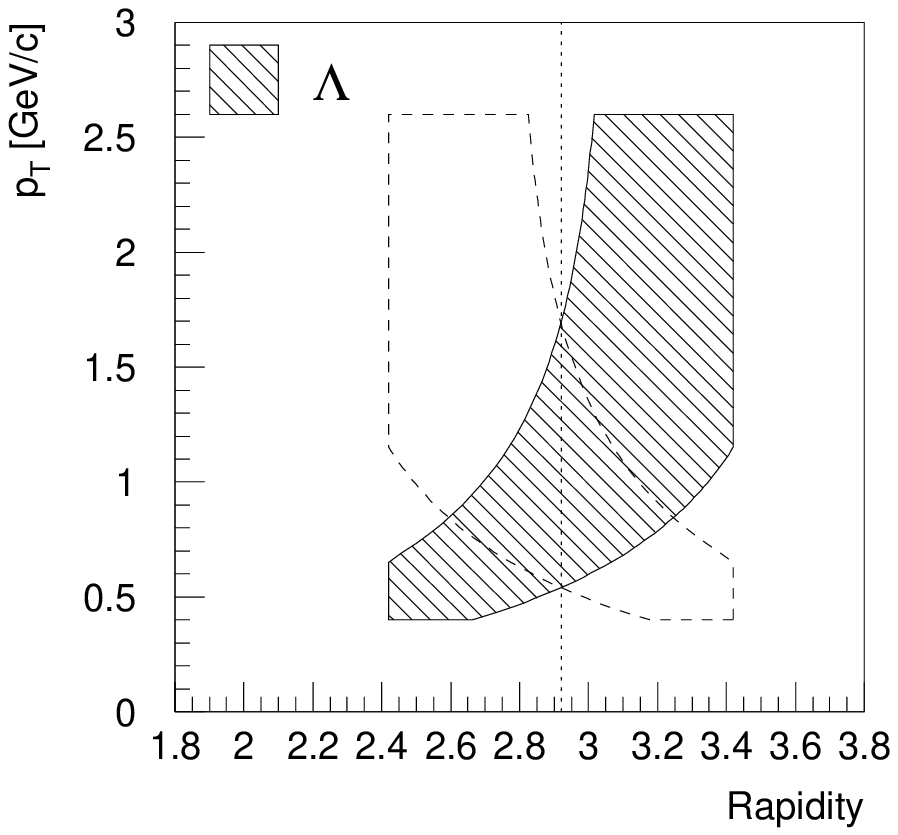}
 \includegraphics[scale=0.62,bb=8 1 262 240,clip]%
                                {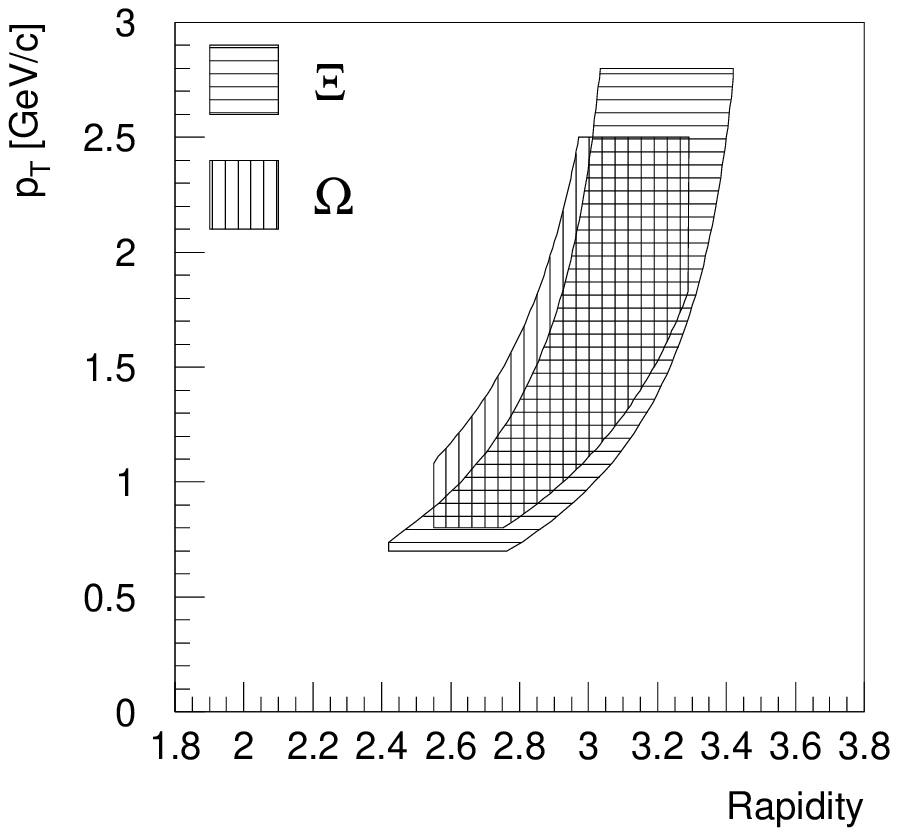}
 \includegraphics[scale=0.62,bb=8 1 262 240,clip]%
                                {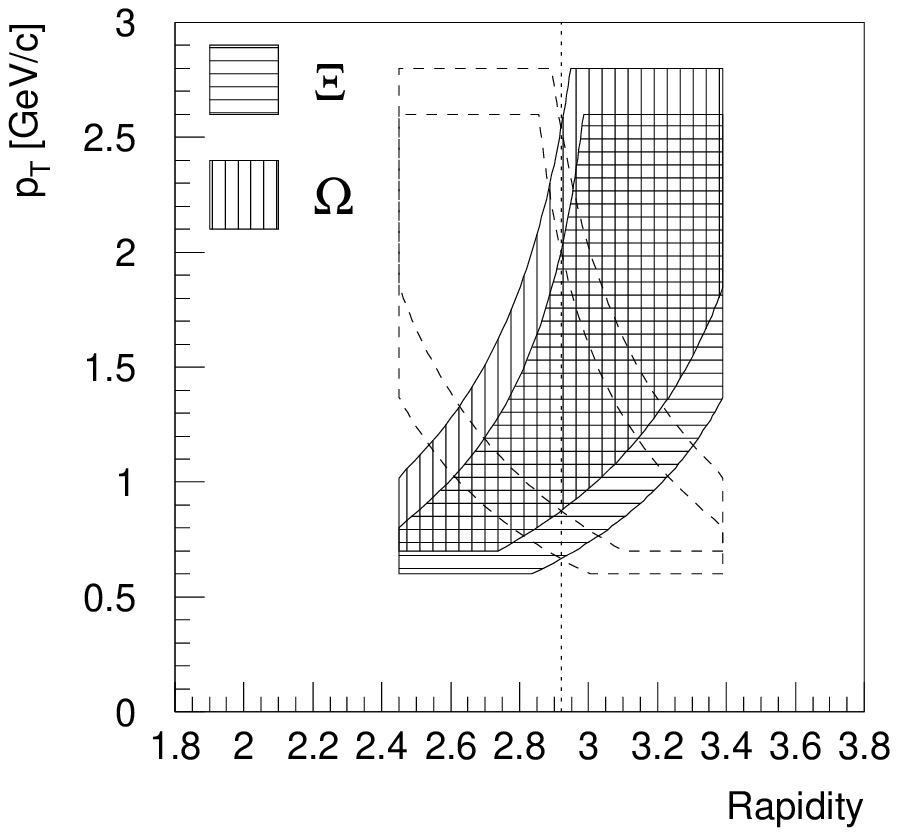}
    
\caption{\em Regioni di accettanza per  $\PKzS$, $\PgL$, $\Xi$ e $\Omega$
in interazioni p-Pb (a sinistra) e Pb-Pb (a destra). Per le collisioni
Pb-Pb sono mostrate in tratteggio
le regioni cinematiche ottenute per riflessione rispetto
al valore centrale di rapidit\`a, indicato da una linea verticale.}
\label{figoverlap}
\end{figure}

\clearpage

Le regioni cinematiche scelte per lo studio delle propriet\`a fisiche delle
particelle strane sono riportate in fig.~\ref{figoverlap},
nel piano individuato dalle variabili rapidit\`a ($y$) e momento trasverso
rispetto alla direzione del fascio ($p_T$).
A sinistra sono riportate le regioni di $\Omega$, $\Xi$, $\PgL$ e $\PKzS$
relative all'interazione p-Pb e a destra quelle relative all'interazione
Pb-Pb; esse sono uguali sia per le particelle che per le antiparticelle,
grazie alla simmetria dell'apparato sperimentale per inversione del
verso del campo magnetico.
La scelta delle regioni di spazio delle fasi nel piano $(p_T , y)$
\`e stata fatta considerando i valori delle correzioni calcolate
per ogni particella identificata, secondo quanto sar\`a esposto nel
prossimo paragrafo. \`E stato infatti richiesto che il valore della
correzione relativo alle particelle situate all'interno delle regioni 
cinematiche selezionate non superi di pi\`u di un ordine di grandezza
il minimo valore calcolato.
Ci\`o assicura una accettanza sufficientemente
uniforme all'interno delle regioni scelte
ed evita che valori di correzione troppo alti, corrispondenti a 
a basse accettanze, possano introdurre anomalie nei risultati finali, quali,
ad esempio, la distorsione in un punto specifico di una distribuzione di massa
trasversa.
I limiti delle regioni cinematiche sono stati individuati considerando
fissati valori delle variabili $p_T$ e $y$ e, per quanto riguarda
i confini curvi, fissati angoli $\theta$ tra la linea di fascio e la direzione
della particella strana identificata.
Tali curve, valide sotto l'ipotesi
di traiettoria rettilinea della particella strana,
sono individuate nel piano $(p_T , y)$ dall'equazione

\begin{equation}
p_T=\frac{m\: \sinh{y}\:\tan{\theta}}{\sqrt{1-\sinh^2 y\:\tan^2 \theta}}
\label{eqcont}
\end{equation}

\noindent ottenuta usando la prima delle eq.~\ref{b_t.rap} e indicando
con
$m$ la massa della particella identificata.
Ci\`o significa che l'accettanza delle particelle strane di un dato tipo 
dipende soprattutto dalla loro direzione di emissione piuttosto
che dalle direzioni ed impulsi dei loro prodotti di decadimento e
che l'azione del campo magnetico sulla traiettoria delle $\Xi$ ed $\Omega$ 
risulta trascurabile.
\par
Le regioni cinematiche relative all'interazione Pb-Pb possono essere 
simmetrizzate per riflessione intorno al valore centrale di rapidit\`a, grazie
alla simmetria del sistema di collisione. Le regioni riflesse sono indicate
con linee tratteggiate in fig.~\ref{figoverlap} e permettono di
estendere la regione di spazio delle fasi coperta dall'esperimento.
Tuttavia nell'analisi mostrata in questa tesi la simmetria del sistema
non \`e stata usata per questo scopo: essa \`e servita unicamente per verificare
il calcolo delle correzioni per accettanza ed efficienza.

\subsection{Segnali aggiuntivi}
\index{Segnali aggiuntivi}
\label{par_neg}

Per valutare l'incremento di produzione di stranezza nel passaggio 
dall'interazione p-Pb a quella Pb-Pb si \`e soliti impiegare,
come elemento di normalizzazione, il numero di particelle negative presenti
nell'evento, in modo da svincolarsi dalla dipendenza dalla sezione
d'urto del processo e, quindi, dal naturale incremento dovuto alla
sovrapposizione di pi\`u nucleoni incidenti.
La scelta del segno delle particelle \`e giustificata dal fatto che le negative
sono certamente prodotte nell'interazione, mentre tra le positive possono essere
contenuti anche i protoni spettatori dei nuclei interagenti.
Accanto ai segnali di particelle strane, si \`e dunque isolato un segnale
di particelle negative (indicate col simbolo $h^-$) provenienti da una
ristretta regione intorno al vertice primario di interazione.
Non potendo identificare le particelle che attraversano l'apparato sperimentale
di WA97, a tutte le tracce negative cos\`{\i} selezionate \`e stata
assegnata la massa nominale del $\Pgpm$, la particella negativa pi\`u 
frequentemente prodotta in tali interazioni.

\begin{figure}[htb]
\centering
\includegraphics[scale=0.45,clip]
                                {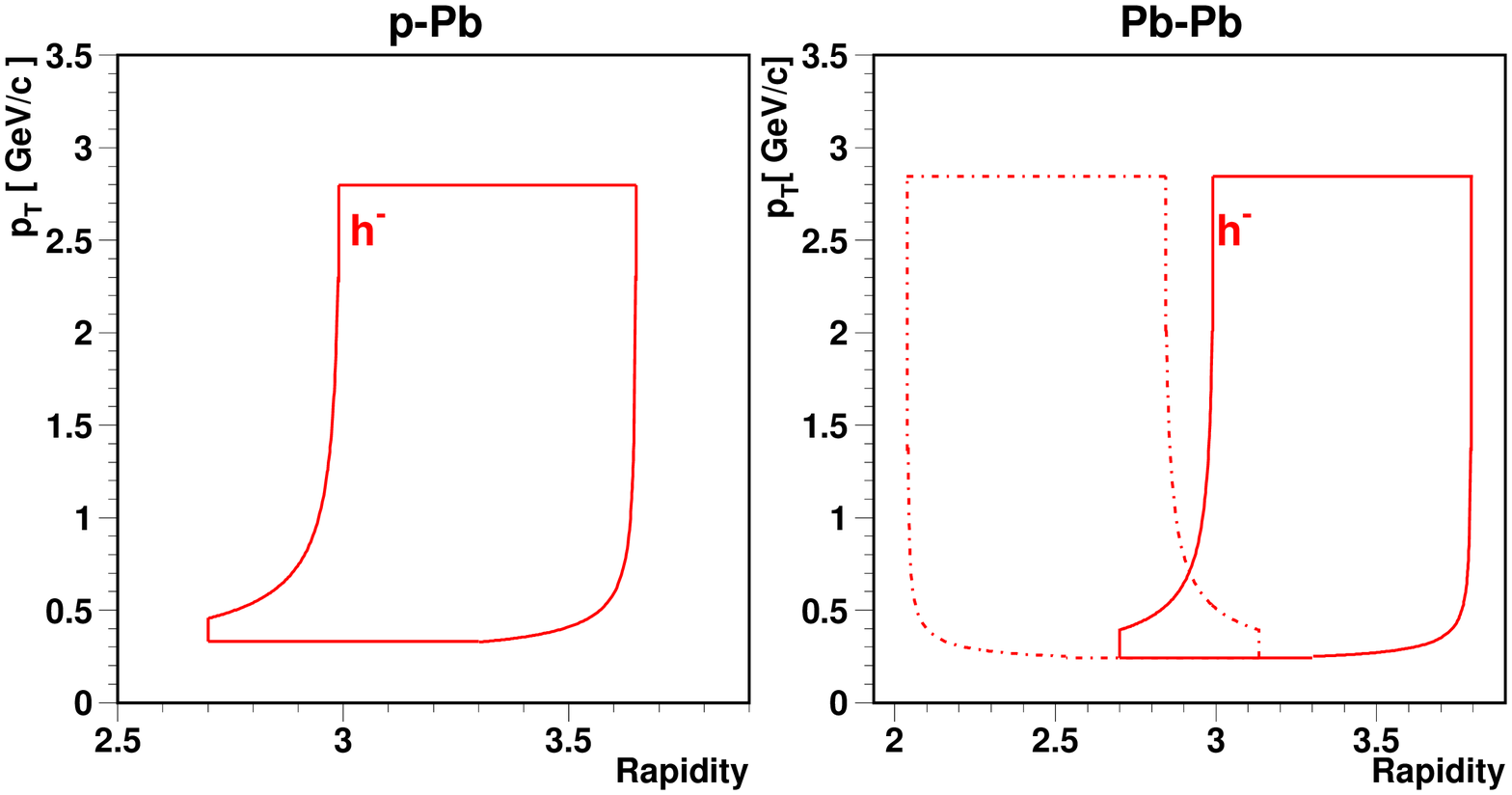}
    
\caption{\em Regioni di accettanza per  particelle negative  ($h^-$)
in interazioni p-Pb (a sinistra) e Pb-Pb (a destra). Per le collisioni
Pb-Pb \`e mostrata in tratteggio
la regione cinematica ottenuta per riflessione rispetto
al valore centrale di rapidit\`a.}
\label{figoverlap1}
\end{figure}

Le regioni di accettanza utilizzate per lo studio delle negative in eventi
p-Pb e Pb-Pb sono mostrate in fig.~\ref{figoverlap1}.
In tal caso la traiettoria delle particelle selezionate risente dell'azione
del campo magnetico ed i contorni curvi nel piano $(p_T, y)$ sono
individuati mediante formule empiriche diverse dalla (\ref{eqcont}).
\par
Le particelle negative relative all'interazione p-Pb sono state
selezionate all'in\-terno degli eventi registrati con la richiesta
che almeno una traccia attraversi il telescopio (trigger ad una traccia,
secondo quanto riferito nel par.~\ref{partrigger}).
Al fine di verificare la consistenza tra
la produzione di particelle strane calcolata in eventi con trigger di
interazione a due tracce e quella calcolata con trigger di interazione
ad una traccia, si \`e isolato un segnale aggiuntivo di $\PgL$ 
dagli eventi p-Pb con trigger ad una traccia.
I tagli utilizzati sono gli stessi di quelli descritti nel
par.~\ref{taglianalisi}, ad eccezione del taglio riportato al punto h),
al posto del quale si \`e richiesto che le tracce di decadimento attraversino
solo la parte compatta del telescopio.

\section{Correzione per accettanza ed efficienza}
\index{Correzione per accettanza ed efficienza}

Per poter valutare la produzione di diversi tipi di particelle \`e
necessario conoscere il modo in cui la particolare geometria dei
rivelatori limita la loro misura (accettanza geometrica) e l'efficienza
con la quale esse sono rivelate e ricostruite.
Il calcolo dell'accettanza assume un'importanza cruciale nell'esperimento
WA97, caratterizzato da un apparato di altissima risoluzione, ma
disposto in modo da coprire una porzione molto limitata dello spazio
delle fasi permesso alle particelle nello stato finale.
La precisione richiesta nella misura del tasso di produzione di particelle
strane, d'altra parte, impone che la loro efficienza
 di rivelazione e ricostruzione sia nota con grande accuratezza.
\`E stato dunque sviluppato un metodo di calcolo basato su una
simulazione Monte Carlo che consente di associare ad ognuna delle
particelle identificate una correzione che contiene  un contributo
dovuto all'accettanza geometrica ed uno dovuto all'efficienza
di rivelazione e ricostruzione.
La procedura consiste nel generare, per ogni particella identificata,
particelle Monte Carlo dello stesso tipo  aventi momento trasverso,
rapidit\`a e posizione del vertice primario di interazione uguali alle
corrispondenti variabili misurate.
Il decadimento di tali particelle viene simulato facendo variare la sua
posizione e orientazione e i prodotti di decadimento sono tracciati
all'interno dell'apparato sperimentale per mezzo del programma
GEANT [Brun]. Per velocizzare il calcolo, le particelle sono 
generate in modo che l'angolo $\phi$ di azimuth 
che il loro momento trasverso $p_T$ forma
con il piano $xy$ del sistema di OMEGA sia compreso tra $40^\circ$ e
$140^\circ$, in modo da includere l'angolo solido sotteso dal telescopio.
Durante la generazione vengono registrate tutte le informazioni relative
alle particelle simulate le cui tracce di decadimento attraversano la parte
compatta del telescopio; la generazione corrispondente ad ogni particella
reale termina quando vengono raggiunte $2500$ particelle registrate
(l'errore sulla correzione \`e cos\`{\i} reso indipendente
dalla accettanza della particella sotto esame).
All'interno di GEANT \`e stata inserita la struttura a {\em pixel} dei piani del
telescopio ed ai {\em pixel} interessati dal passaggio dei prodotti carichi
di decadimento sono assegnati degli {\em hits} elettronici.
L'efficienza dei {\em pixel}, misurata per ogni {\em chip} di lettura [Hel98]
(c.f.r. par.~\ref{parpixel}), \`e stata utilizzata in tale
assegnazione. 
Per eventi p-Pb, essa introduce una diminuzione del $25\%$
nell'efficienza di ricostruzione. 
\par
\begin{figure}[htb]
\centering
\includegraphics[scale=0.65,clip]
                                {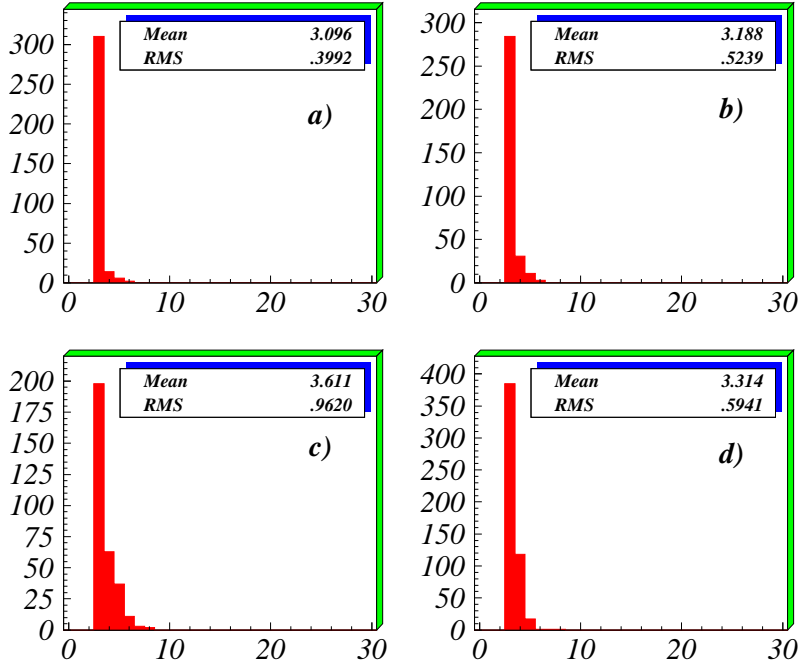}
    
\caption{\em Molteplicit\`a di tracce per un campione di $\Xi$ 
relativo all'interazione p-Pb impiantato
su eventi di fondo acquisiti con trigger di fascio (a), trigger ad una traccia
(b) e trigger a due tracce (c), confrontate con la molteplicit\`a di tracce
degli eventi reali contenenti le $\Xi$ identificate (d).}
\label{figtestfig}
\end{figure}

Per riprodurre al meglio la situazione reale in cui ogni particella strana
\`e accompagnata da altre tracce e dal rumore elettronico dei rivelatori,
gli {\em hits} elettronici cos\`{\i} simulati da GEANT per 
ogni particella 
vengono impiantati entro eventi reali campionati sull'intera statistica
 accumulata, detti eventi di fondo, costruendo cos\`{\i} degli eventi
``ibridi'', composti cio\`e da {\em hits} di tracce reali e {\em hits}
 di tracce simulate.
Per quanto riguarda l'interazione Pb-Pb, gli eventi di fondo sono 
stati scelti in modo che  la loro molteplicit\`a di {\em hits}
nei piani di {\em pixel} del telescopio sia compatibile con quella
dell'evento contenente la particella da correggere.
Questo criterio \`e verificato in media una volta ogni $100$
eventi di fondo ed ha lo scopo di ricreare nell'evento ibrido la
stessa situazione fisica presente in quello originale.
Per l'interazione p-Pb, invece, gli eventi di fondo sono stati scelti in 
base al livello di trigger col quale sono stati acquisiti.
Il livello di trigger da usare \`e stato scelto utilizzando la
seguente procedura:
\begin{itemize}
\item un campione di $\Xi$ \`e stato impiantato su eventi di fondo acquisiti
con trigger di fascio, con trigger ad una traccia e con trigger a
due tracce; 

\item le molteplicit\`a di tracce risultanti, mostrate rispettivamente in
fig.~\ref{figtestfig}a,b e c, sono state confrontate con la molteplicit\`a
di tracce in eventi reali contenenti le $\Xi$ identificate
nell'interazione p-Pb, mostrate in fig.~\ref{figtestfig}d;

\item gli eventi ibridi che meglio riproducono la molteplicit\`a di tracce
degli eventi reali sono quelli relativi al trigger a due tracce, per
cui nel calcolo della correzione sono stati utilizzati eventi di fondo
acquisiti con tale trigger.

\end{itemize}

\noindent L'introduzione degli eventi di fondo ha l'effetto di ridurre 
dell'$8\%$ l'efficienza di ricostruzione in interazioni p-Pb.
\par
Gli eventi ibridi sono poi processati con ORHION e con gli stessi programmi
di analisi usati per gli eventi reali,
al fine di ricostruire ed identificare le
particelle strane in essi presenti. Vengono cos\`{\i} calcolati,
per ogni particella da correggere, il
numero di particelle Monte Carlo generate ($gen$), quello delle particelle
Monte Carlo generate che soddisfano i tagli dell'analisi (${gen}_{tagli}$)
e quello delle particelle ricostruite negli eventi ibridi che soddisfano
i tagli dell'analisi (${ricostr}_{tagli}$). La correzione \`e data da:


\begin{equation}
w    =  \frac{1}{acc\:\: eff}, 
\label{eqpeso}
\end{equation}

\[
acc  =  \frac{\Delta\phi}{360^\circ}\:\frac{{gen}_{tagli}}{gen}
\;\;\;\;\;\;\;\;\;
eff  =  \frac{{ricostr}_{tagli}}{{gen}_{tagli}}
\]

\noindent dove $\Delta\phi$ corrisponde all'intervallo di generazione
usato (al di fuori di quest'angolo nessuna particella \`e accettata) e
$acc$ ed $eff$ sono i contributi dovuti rispettivamente all'accettanza
geometrica e all'efficienza di rivelazione e ricostruzione.
Poich\`e la simulazione \`e compiuta per mezzo di generatori di
numeri casuali, gli errori sui contributi di accettanza ed efficienza
sono dati dalle relazioni binomiali:

\begin{equation}
\delta acc =\sqrt{\frac{ acc\:(\frac{\Delta\phi}{360^\circ} - acc) }{gen}}
\;\;\;\;\;\;\;\;\;
\delta eff =\sqrt{\frac{eff\,(1 - eff )}{{gen}_{tagli}}}
\label{eqerrpeso}
\end{equation}

\noindent I valori medi della correzione globale e 
dell'efficienza di ricostruzione per i segnali studiati
in interazioni p-Pb e Pb-Pb
sono riportati in tab.~\ref{tabpesi}.

\begin{table}[ht]
\centering
\caption{\em Valori medi della correzione globale 
 e dell'efficienza di ricostruzione per i segnali studiati
in interazioni p-Pb e Pb-Pb.}
\begin{tabular}{|c|cc|cc|}
\hline
         &\multicolumn{2}{|c|}{ p-Pb}&\multicolumn{2}{|c|}{ Pb-Pb }  \\
\hline
         &    $<w>$       &$<{eff}>(\%)$&   $<w>$      &$<{eff}>(\%)$ \\
\hline
$\Omega$ &  $6384\pm 1766$&$39\pm 9$&$10396\pm 900$&$26\pm 2  $\\
$\Xi$    &  $7040\pm 431$ &$38\pm 2$&$11523\pm336$ &$23.1 \pm 0.6 $\\
$\PgL$   &  $584\pm 18$   &$64\pm 2$&$744\pm 17 $  &$ 38.8\pm 0.9 $\\
$\PKzS$  &  $396\pm 15$   &$71\pm 2$&$719\pm 33$   &$35\pm 1  $\\
$h^-$    &  $22.6\pm 0.7$ &$65\pm 2$&$16.9\pm0.5$  &$65\pm 2  $\\
\hline
\end{tabular}
\label{tabpesi}
\end{table}
 
\noindent
I valori di efficienza per interazioni Pb-Pb sono pi\`u bassi di quelli
 calcolati in interazioni p-Pb poich\`e l'alta molteplicit\`a
di tracce  rende pi\`u difficoltosa la ricostruzione dei decadimenti
ed i tagli dell'analisi sono pi\`u stringenti.
L'aumento di circa un ordine di grandezza nelle correzioni per le
particelle multi-strane rispetto a quelle per le particelle strane, inoltre,
\`e dovuta alla minore accettanza dell'apparato per i decadimenti
di tipo cascata rispetto a quelli di tipo $V^0$.
\par
Nel calcolo della correzione per le particelle negative, l'identificazione
 della traccia corrispondente (se esiste)
a quella simulata, nell'evento ibrido ricostruito, 
\`e stata eseguita utilizzando
come parametro di decisione la pi\`u piccola tra le ``distanze'' $d_j$,
definite dalla relazione

\begin{center}
\[
d_j=\frac{1}{N_p}\:\sum_{i}
        \sqrt{(y_{ij}-\overline{y}_i)^2 + (z_{ij}-\overline{z}_i)^2} 
\]
\end{center}

\noindent
dove $N_p$ \`e il numero di piani del telescopio in cui c'\`e stato
l'impatto registrato della traccia, ($\overline{y}_i$,
$\overline{z}_i$) sono le coordinate del punto di impatto della traccia
simulata sul piano $i-$esimo e
($y_{ij}$, $z_{ij}$) sono le coordinate
 della  traccia $j-$esima dell'evento impiantato sul piano $i-$esimo.
La quantit\`a $d_j$ viene calcolata confrontando la traccia simulata
con ciascuna delle tracce ricostruite dello stesso segno e scegliendo il valore
pi\`u piccolo.
\par
La necessit\`a di generare e processare migliaia di eventi per ogni particella
rivelata rende il calcolo dell'accettanza ed efficienza molto dispendioso 
in termini di tempo di esecuzione al computer, per cui \`e stato
finora possibile correggere solo una frazione delle particelle
identificate, gi\`a riportate in tab.~\ref{tabstat1}.
Esse sono state campionate in modo uniforme 
tra tutte quelle disponibili, in modo da
essere comunque rappresentative dell'intera statistica raccolta.
Il numero di particelle corrette \`e riportato 
in tab.~\ref{tabstat2}, insieme
alla corrispondente percentuale sul totale delle particelle identificate.
La correzione delle particelle  rimanenti \`e tuttora in corso;
nel contempo sono in via di sviluppo
metodi di correzione pi\`u veloci, adatti per
campioni ad alta statistica quali quelli delle $h^-$, delle $\Lambda$
e dei $\PKzS$ [Lie98], [Calia98].

\begin{table}[ht]
\centering
\caption{Statistica delle particelle corrette in interazioni p-Pb e Pb-Pb
e corrispondenti percentuali sul totale delle particelle identificate.}
  \begin{tabular}{|c|cc|cc|}
   \hline
            & \multicolumn{2}{|c|}{p-Pb} & \multicolumn{2}{|c|}{Pb-Pb} \\ 
   \cline{2-5}
            & Corrette & \% sul totale & Corrette & \% sul totale \\
   \hline 
   \PgOm    &\hphantom{00}15 & 100\% &\hphantom{0}145 & 40\%\\
   \PagOp   &\hphantom{000}4 & 100\% &\hphantom{00}58 & 40\%\\
   \PgXm    &\hphantom{0}275 & 100\% &         1346   & 40\%\\
   \PagXp   &\hphantom{0}101 & 100\% &\hphantom{0}339 & 40\%\\
   \PgL \hphantom{$^-$} & 1056 & 1\%      & 1868   & 0.3\%\\
   \PagL\hphantom{$^-$} &\hphantom{0}243& 1\%&\hphantom{0}669 & 0.6\%\\
   \PKzS    &1093 & 0.5\%          &\hphantom{0}750 & 0.1\%   \\
    $h^-$   &1500 & 0.002\% &               1800 & 0.0003\%\\
   \hline
  \end{tabular}
   \label{tabstat2}
 \end{table}

\section{Calcolo della produzione di particelle strane}
\index{Calcolo della produzione di particelle strane}
\label{parprod}

Una volta determinata la correzione da apportare ad ogni particella
identificata, il numero di particelle strane prodotte pu\`o essere
calcolato all'interno delle regioni cinematiche selezionate,
sommando tali correzioni e normalizzando il risultato al
numero di eventi raccolti.
Per l'interazione Pb-Pb, si \`e calcolato il tasso di
produzione di particelle strane 
per evento selezionato, utilizzando la formula

\begin{equation}
P_{Pb}=\frac{\sum_{i}\: w_i}{BR}\:\frac{1}{\sum CU1}
\label{eqcrosspb}
\end{equation}

\noindent
dove la sommatoria delle correzioni \`e estesa a tutte le particelle
di un dato tipo che cadono all'interno delle corrispondenti regioni
cinematiche, $BR$ indica la probabilit\`a di decadimento della
particella considerata nel canale osservato e l'elemento di normalizzazione
consiste nella somma degli eventi selezionati dal
livello di trigger pi\`u alto (c.f.r. par.~\ref{partrigger1}).
L'errore su tale quantit\`a \`e calcolato come

\begin{equation}
\delta P_{Pb}=\frac{\sum_{i}\:\sqrt{ w_i^2+(\delta w_i)^2 }}{BR}
\:\frac{1}{\sum CU1}
\label{eqerrcrosspb}
\end{equation}

\noindent
sommando in quadratura l'errore statistico legato al numero di particelle
presenti e l'errore sulla correzione dato dalla (\ref{eqerrpeso});
le incertezze legate alla probabilit\`a di decadimento e al numero di
eventi selezionati risultano trascurabili.
\par
La produzione di particelle strane calcolata secondo la (\ref{eqcrosspb})
\`e relativa ai soli urti centrali tra nuclei di piombo,
selezionati mediante il trigger di interazione.
La frazione di sezione d'urto anelastica $\sigma_I$ cui essi corrispondono
\`e stata calcolata tramite la relazione

\begin{equation}
\frac{\sigma_{trig}}{\sigma_I}=\frac{\lambda_I}{L}\: 
\left[
<\frac{CU1}{BEAM\,\overline{DT}}>\:<\frac{INT\,\overline{DT}}{CU1}>%
 \right]
\label{eqpercross}
\end{equation}

\noindent
dove $L$ \`e lo spessore del bersaglio ed il termine tra parentesi quadre
contiene le medie, fatte su tutta la presa dati, dei
conteggi dei corrispondenti livelli di trigger e misura la frequenza
di interazione.
Il primo fattore rappresenta il rapporto tra il numero di eventi selezionati
($CU1$) e quello degli ioni del fascio
contati escludendo il tempo morto di rivelazione ($BEAM\,\overline{DT}$);
il secondo fattore tiene conto della protezione passato-futuro ($CB$),
essendo $CU1$  il risultato della coincidenza tra $CB$ e 
$INT\,\overline{DT}$ (c.f.r. par.~\ref{partrigger1}).
Dalla (\ref{eqpercross}) risulta che la sezione d'urto selezionata 
$\sigma_{trig}$ corrisponde a circa il $40\%$ della sezione
d'urto totale anelastica $\sigma_I$.
\par
Per quanto riguarda l'interazione p-Pb, la produzione di particelle
strane deve essere calcolata in modo da tenere conto del particolare
trigger di interazione usato; infatti il suo effetto \`e proprio
quello di aumentare artificialmente il numero di tali particelle per evento.
Si \`e allora usata la formula:

\begin{equation}
P_{p}=\frac{\sum_{i}\: w_i}{BR}\:\frac{1}{\sum BEAM\,\overline{DT}}
\:<\frac{CU0}{CU1}>\:<\frac{INT\,\overline{DT}}{CU1}>
\:\frac{1}{(1-e^{-L/\lambda_I})}
\label{eqcrossp}
\end{equation}

\noindent
che fornisce il numero di particelle strane per interazione.
Il fattore di normalizzazione $\sum\: BEAM\,\overline{DT}$, la cui somma
\`e estesa a tutto il periodo di acquisizione dati,
conta le particelle incidenti tenendo conto del tempo morto; il termine
$<\frac{CU0}{CU1}>$ tiene conto del contributo dei CEDARS, cio\`e
del fatto che non tutte le particelle del fascio sono protoni 
($CU1$ \`e infatti 
il risultato della coincidenza tra i segnali $CU0$ e $CEDARS$); il termine
$<\frac{INT\,\overline{DT}}{CU1}>$ tiene conto della protezione passato-futuro
e l'ultimo fattore rappresenta la probabilit\`a di interazione del protone
all'interno del bersaglio di piombo (si veda l'eq.~(\ref{probcoll})).
I valori dei rapporti $<\frac{CU0}{CU1}>$ e $<\frac{INT\,\overline{DT}}{CU1}>$
sono rispettivamente $0.67\pm 0.02$ e $0.83\pm0.02$, per cui
l'effetto dei CEDARS \`e quello di eliminare il $33\%$ delle particelle
di fascio e l'effetto della protezione passato-futuro \`e quello di eliminarne
il $17\%$.
L'errore sulla quantit\`a $P_p$ \`e stato calcolato in analogia con la 
(\ref{eqerrcrosspb}), considerando trascurabili le incertezze
sui fattori di normalizzazione e sulle probabilit\`a di decadimento
e interazione.
\par
Per avere una stima dell'inefficienza del trigger di interazione applicato
nella presa dati p-Pb, la quantit\`a $P_p$ \`e stata calcolata per le
 $\PgL$ identificate  in eventi con trigger ad una traccia ($1TT$) e per quelle
identificate in eventi con trigger a due tracce ($2TT$). 
La produzione di $\PgL$ nei due livelli di trigger differisce della quantit\`a

\begin{equation}
A=\frac{(P_{p})_{1TT}}{(P_{p})_{2TT}}=1.17\pm 0.10
\label{eqfactrig}
\end{equation}

\noindent
La quantit\`a $(1-1/A)$ \`e interpretabile come inefficienza del trigger
a due tracce rispetto a quello ad una traccia, di livello inferiore al
precedente; le produzioni calcolate per le particelle strane
sono state cos\`{\i} moltiplicate per il fattore $A$ per includere 
tale inefficienza.

\subsection{Errori sistematici nella determinazione del numero di $\Lambda$}
\index{Errori sistematici nella determinazione del numero di $\Lambda$}
\label{par_feed}

Nel capitolo precedente si \`e visto come, combinando le tracce dell'evento
ed applicando una serie di criteri selettivi su variabili geometriche e 
cinematiche, sia possibile ricostruire i decadimenti e 
identificare le particelle strane nell'apparato dell'esperimento WA97.
Alcune ambiguit\`a rimangono tuttavia insolute nei campioni
di particelle strane identificate; esse sono peraltro comuni a tutti gli
esperimenti che misurano questo tipo di segnale.
\par
In primo luogo, la procedura di identificazione non consente di distinguere
le $\PgL$ prodotte direttamente nelle collisioni primarie
da quelle provenienti dal decadimento elettromagnetico
$\Sigma^0 \rightarrow \PgL +\Pgg$.
 La vita media di tale
decadimento ($\tau\sim 10^{-20}~sec$) rende infatti
il suo vertice di
decadimento praticamente indistinguibile dal vertice primario di interazione.
Nello studio delle propriet\`a fisiche del campione di $\PgL$ (e  $\PagL$)
misurato bisogner\`a ricordare che esso in realt\`a \`e un miscuglio di
$\PgL$ e $\Sigma^0$
e, nel confronto di tali propriet\`a con le
previsioni di modelli teorici, bisogner\`a includere in quest'ultimi
 il contributo dovuto al decadimento delle $\Sigma^0$.
\par
In secondo luogo, le $\PgL$ identificate includono quelle provenienti dal
 decadimento debole di barioni pi\`u pesanti, quali $\PgXm$, $\PagXz$ e
$\PgOm$.
Il contributo di tali decadimenti, chiamato {\em feed-down}, pu\`o essere
stimato grazie alla possibilit\`a di ricostruire i decadimenti deboli
delle particelle madri
nell'apparato di WA97. Il calcolo \`e stato fatto solo per il
contributo di $\PgXm$ e $\PagXz$ (quello pi\`u consistente) 
in interazioni Pb-Pb, in base a considerazioni prettamente geometriche.
Il metodo utilizzato \`e cos\`{\i} riassumibile:
\begin{itemize}
\item si sono generati, mediante simulazione Monte Carlo, i decadimenti
di $\PgXm$, $\PagXz$ e $\PgL$ (primarie) all'interno della regione cinematica
$2<y<4$ e $0.1< p_T <2.0~GeV/c$, contenente quella coperta dal telescopio,
e con una distribuzione di momento trasverso avente caratteristiche simili
a quella misurata (per le $\PagXz$ essa si \`e assunta uguale a quella delle
$\PgXm$);

\item i prodotti carichi di decadimento sono stati tracciati all'interno 
dell'apparato sperimentale ed ai vertici $\PgL$ sono stati applicati
i tagli dell'analisi riguardanti la topologia di decadimento, la posizione
lungo l'asse $x$, il parametro di impatto e il passaggio attraverso la
parte compatta del telescopio;

\item sono state calcolate le accettanze per $\PgXm$, $\PagXz$ e $\PgL$,
denominate rispettivamente $acc_{\:\PgXm}$, $acc_{\:\PagXz}$ e $acc_{\:\PgL}$.
I loro valori sono $acc_{\:\PgXm}=0.024\pm0.001\%$,
$acc_{\:\PagXz}=0.015\pm0.001\%$ e $acc_{\:\PgL}=0.080\pm0.001\%$.

\end{itemize}

\noindent
Le $\PgL$ provenienti dal {\em feed-down} ($\PgL_{feed}$) possono
essere stimate dalla relazione

\begin{equation}
\PgL_{feed}=\frac{1}{acc_{\:\PgL}}\:\:(\PgXm\, acc_{\:\PgXm}\:+\:\PagXz
\,acc_{\:\PagXz}).
\end{equation}

\noindent
Assumendo uguale la produzione di $\PgXm$ e $\PagXz$, la frazione di 
$\PgL_{feed}$ rispetto a quelle prodotte direttamente nell'interazione primaria
($\PgL_{vere}$) \`e data da

\begin{equation}
\frac{\PgL_{feed}}{\PgL_{vere}}=
\frac{1}{acc_{\:\PgL}}\:\frac{\PgXm}{\PgL_{vere}}\:(acc_{\:\PgXm}\:+\: 
acc_{\:\PagXz}).
\label{eqfeed1}
\end{equation}

\noindent
Approssimando il rapporto $\frac{\PgXm}{\PgL_{vere}}$ con quello misurato
$\frac{\PgXm}{\PgL_{oss.}}\sim 0.1$ e sostituendo i valori noti
delle accettanze nella (\ref{eqfeed1}) si ottiene
$\frac{\PgL_{feed}}{\PgL_{vere}}=0.05$.
Analogamente per le $\PagL$, utilizzando il rapporto misurato
$\frac{\PagXp}{\PagL_{oss.}}\sim 0.2$, si ottiene
$\frac{\PagL_{feed}}{\PagL_{vere}}=0.10$.
L'effetto del {\em feed-down}, interpretabile come errore sistematico
nella misura della produzione di $\PgL$ e $\PagL$, risulta dunque
del $5\%$ nel caso di $\PgL$ e del $10\%$ nel caso di $\PagL$, per quanto
riguarda l'interazione Pb-Pb.
\par
Nel caso di collisioni p-Pb, la 
produzione di $\Xi$ relativa a quella delle $\PgL$ \`e 
minore che nel caso Pb-Pb e i valori di accettanza non sono
eccessivamente influenzati dalla diversa disposizione del telescopio.
L'effetto del {\em feed-down}, stimato a partire 
dai valori di accettanza calcolati 
e dai rapporti $\frac{\PgXm}{\PgL_{oss.}}$ e
$\frac{\PagXp}{\PagL_{oss.}}$ misurati in questo tipo di collisioni, 
risulta dell'ordine del
$3\%$ nel caso di $\PgL$ e del $5\%$ nel caso di $\PagL$.

\section{Distribuzioni di massa trasversa}
\index{Distribuzioni di massa trasversa}
\label{par_rapid}

Uno degli strumenti pi\`u comuni usati per studiare le collisioni ad alta
energia \`e la distribuzione di massa trasversa, secondo quanto esposto nel
par.~\ref{parmassatrasversa}.
Il moto trasverso \`e infatti generato durante la collisione, per cui 
fornisce informazioni relative alla sua dinamica.
Nell'esperimento WA97 le distribuzioni di massa trasversa
 sono state studiate per
mezzo di una procedura di {\em best fit} applicata agli spettri di
momento corrispondenti
a ciascun tipo di particella identificata, all'interno delle rispettive regioni
cinematiche.
La funzione usata \`e 

\begin{equation}
\frac{d^2N}{dm_T\,dy}=f(y)\; m^\alpha_T\: e^{-m_T/T}
\label{eqlikely}
\end{equation}

\noindent
nella quale $f(y)$ descrive la distribuzione di rapidit\`a. Il parametro  
$T$ \`e interpretato, nell'ambito di un modello termico, come la
temperatura apparente della sorgente che emette le particelle osservate.
Per facilitare il confronto dei risultati con quelli di altri esperimenti,
sono stati usati per il parametro $\alpha$ i valori $1$ e $\frac{3}{2}$,
corrispondenti rispettivamente alle distribuzioni di massa trasversa
date dalle equazioni (\ref{temp2}) e (\ref{temp3}).
Scopo della procedura di {\em fit} \`e quello di determinare il parametro
$T$ ed eventuali parametri che descrivono la distribuzione di rapi\-dit\`a.
Esso \`e stato perseguito applicando il metodo della
``massima verosimiglianza'' ({\em maximum likelihood}), col quale
\`e possibile evitare le distorsioni nelle distribuzioni introdotte dalla 
particolare forma delle regioni di accettanza.
\par
\begin{figure}[htb]
\centering
\includegraphics[scale=2,angle=0.5,clip]
                                {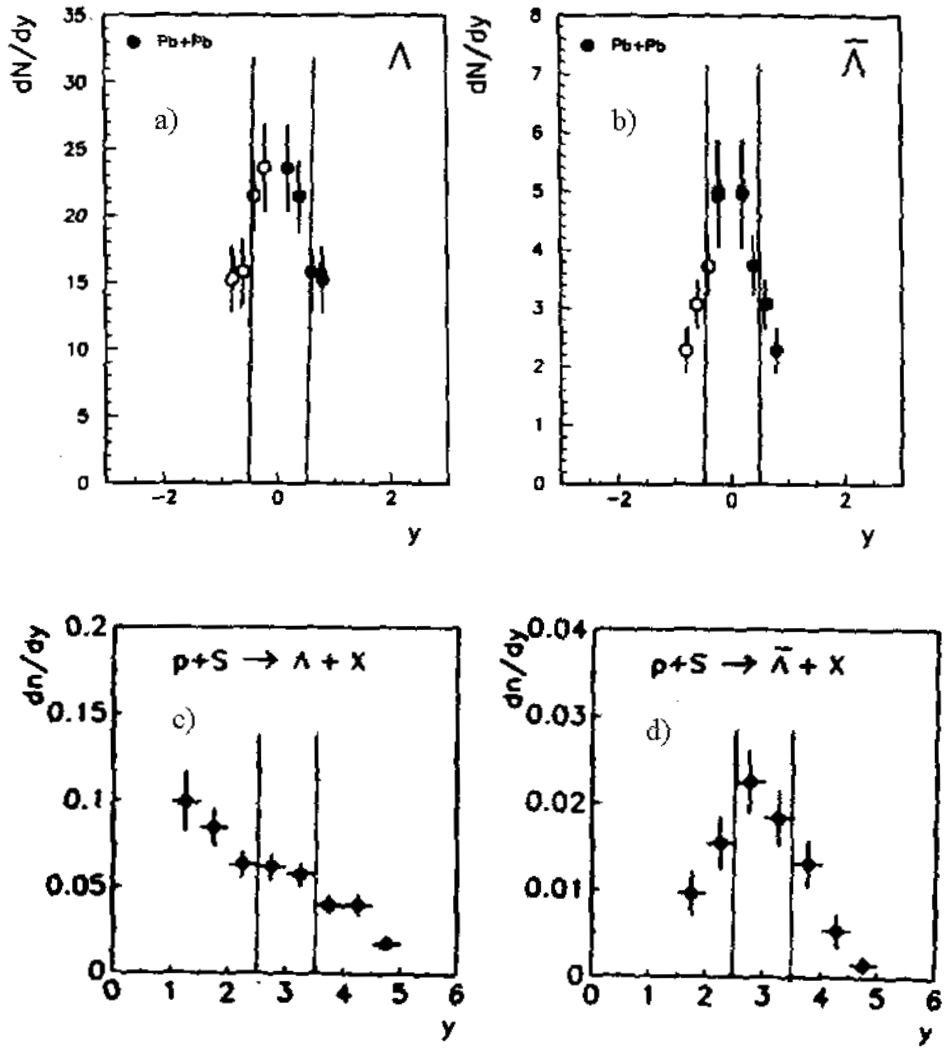}
    
\caption{\em Distribuzioni di rapidit\`a per $\PgL$ e $\PagL$ misurate
in interazioni Pb-Pb (rispettivamente a e b) [Rol97] e p-S
(rispettivamente c e d) [Alb94]. Per le collisioni
Pb-Pb i pallini bianchi indicano i punti ottenuti per riflessione
rispetto
al valore centrale di rapidit\`a.
I limiti dell'intervallo di rivelazione dell'esperimento
WA97 sono  indicati da  linee verticali.}
\label{figrapfig}
\end{figure}

A causa della limitata statistica, nella presente analisi le distribuzioni
di rapi\-dit\`a sono state assunte piatte all'interno dell'intervallo 
coperto dall'esperimento, che per ognuna delle particelle esaminate non supera
una unit\`a di rapidit\`a intorno alla rapidit\`a del centro di massa.
La funzione  $f(y)$ \`e stata dunque  posta costante nei {\em fit} e gli errori
sistematici introdotti da tale assunzione sono stati stimati considerando
le distribuzioni di rapidit\`a misurate in collisioni p-S, p-Au
[Alb94], [Bam89], [Fok95] e Pb-Pb [Rol97].
In fig.~\ref{figrapfig}, ad esempio, sono mostrate le distribuzioni di 
rapidit\`a di $\PgL$ e $\PagL$ misurate in interazioni Pb-Pb 
nell'esperimento NA49 (fig.~\ref{figrapfig}a e b) e in interazioni p-S
nell'esperimento NA35 (fig.~\ref{figrapfig}c e d); 
l'intervallo di rapidit\`a coperto 
dall'esperimento WA97 \`e indicato dalle linee verticali.
Nel primo caso la simmetria del sistema consente di simmetrizzare le
distribuzioni di rapidit\`a intorno alla rapidit\`a del centro di massa
(pallini bianchi) ed esse mostrano entrambe un picco pronunciato all'interno
dell'intervallo di rivelazione.
Nel secondo caso, invece, la forma delle distribuzioni di rapidit\`a
appare molto differente per le due specie: la distribuzione delle $\PgL$
mostra un picco molto accentuato in corrispondenza della regione di frammentazione
del bersaglio, mentre la $\PagL$ risulta prodotta pi\`u centralmente.
In entrambi i casi si osserva una consistente discrepanza con l'ipotesi
di distribuzione piatta.
Si \`e allora supposto un andamento lineare della distribuzione di rapidit\`a,
secondo la funzione

\begin{equation}
f(y)=1+K\:(y-y_{CM})
\label{eqraplin}
\end{equation} 

\noindent
e le pendenze $K$ corrispondenti a ciascun tipo di particella sono state
valutate dalle distribuzioni riportate in [Bam89], [Fok95] e [Rol97].
La temperatura apparente \`e stata allora determinata nei {\em fit} sia supponendo
la rapidit\`a piatta ($T(f(y)=cost)$) che utilizzando l'eq.~(\ref{eqraplin})
con $K$ uguale a quello sperimentale ($T(f(y)\neq cost$) e gli errori
sono stati calcolati con l'espressione

\begin{equation}
\epsilon(\%)=\left(\frac{T(f(y)\neq cost)}{T(f(y)= cost)}-1
\right)\times 100
\end{equation} 

\noindent
Essi sono riportati in tab.~\ref{tabrap}, sia per le interazioni p-Pb che per
quelle Pb-Pb.

\clearpage

In base a considerazioni legate al contenuto di quarks di valenza
delle particelle, la pendenza $K$ delle $\PgXm$ \`e stata posta uguale
a quella delle $\PgL$, quella delle $\PagXp$ e $\Omega$ \`e stata posta
uguale a quella delle $\PagL$.
L'errore sistematico nella misura della temperatura apparente rimane, dunque,
al di sotto del $10\%$ per tutte le particelle e, come
preannunciato nell'analisi delle distribuzioni in fig.~\ref{figrapfig},
risulta maggiore per le collisioni p-Pb che per quelle Pb-Pb.

\begin{table}[ht]
\centering
\caption{\em Stime degli errori sistematici sul calcolo delle
temperature apparenti dovuti all'ipotesi di distribuzione piatta
in rapidit\`a.}
\begin{tabular}{|c|c|c|}
\hline
                 & p-Pb & Pb-Pb \\
\hline
$\PgOm + \PagOp$ &3\%   &   1\%  \\    
$\PgXm$          &8\%   &  $<1$\% \\
$\PagXp$         &3\%   &  $<1$\% \\ 
$\PgL$           &10\%  &  $<1$\% \\ 
$\PagL$          &5\%   &  $<1$\% \\ 
$\PKzS$          &2\%   &   2\%  \\ 
$h^-$            &2\%   &   1\%  \\ 
\hline
\end{tabular}
\label{tabrap}
\end{table}

I risultati della procedura di {\em fit} sono riassunti in tab.~\ref{tabtemp},
nella quale sono  riportate in $MeV$ le temperature apparenti
determinate per le particelle identificate in collisioni p-Pb e Pb-Pb,
utilizzando entrambi i valori del parametro $\alpha$ nella funzione
(\ref{eqlikely}).

\begin{table}[ht]
\caption{Temperature apparenti $T$ espresse in $MeV$ calcolate in collisioni
p-Pb e Pb-Pb, per entrambi i valori del parametro $\alpha$.} 
\centering
\begin{tabular}{|c|cc|cc|}
\hline
           &\multicolumn{2}{|c|}{ p-Pb}&\multicolumn{2}{|c|}{ Pb-Pb } \\ 
\cline{2-5}
Particelle    & $\alpha= 3/2$ & $\alpha= 1$ & $\alpha = 3/2$  & $\alpha = 1$\\
\hline
$h^-$         &  $163\pm 4$ & $185\pm 5$   & $183\pm 6$ & $213\pm 8$  \\
$\PKzS$       & $197\pm 5$& $217\pm 6$ &$210 \pm 7$ & $232\pm 9$  \\
$\PgL$        & $184\pm 5$ & $196\pm 6$   & $266\pm 7 $ & $289\pm 8$  \\
$\PagL$       & $172\pm 10$ & $183\pm 11$   & $264\pm 11$ & $287\pm 13$  \\
$\PgXm$       & $222\pm 12$ & $235\pm 14$  & $266\pm 8$  & $286\pm 9$  \\
$\PagXp$      & $211\pm 19$ & $224\pm 21$ & $265 \pm 15$& $284\pm 17$  \\
$\PgOm+\PagOp$& $312\pm 86$ &$334\pm 99$ &$238\pm 17$&$251\pm 19$ \\
\hline
\end{tabular}
\label{tabtemp}
\end{table}

\noindent
A causa della statistica limitata, i dati delle $\PagOp$ sono stati uniti 
a quelli delle $\PgOm$; gli errori riportati sono solo statistici e non 
includono i suddetti errori sistematici dovuti all'assunzione di rapidit\`a piatta.
Si pu\`o osservare, in generale, che le temperature apparenti 
calcolate in interazioni p-Pb sono 
pi\`u basse di quelle relative ad interazioni Pb-Pb;
gli errori non consentono invece di trarre alcuna conclusione circa
le temperature apparenti delle $\Omega$.
Una estesa discussione sui risultati dello studio delle distribuzioni
di massa trasversa sar\`a affrontata nel prossimo capitolo.

\section{Estrapolazione della produzione di particelle strane}
\index{Estrapolazione della produzione di particelle strane}

Il calcolo della  produzione di particelle strane descritto
nel par.~\ref{parprod} si riferisce alle
diverse regioni cinematiche individuate per ogni
particella studiata, mostrate in fig.~\ref{figoverlap} e \ref{figoverlap1}.
Tuttavia, la particolare forma di tali regioni rende difficile il confronto
dei risultati con quelli forniti da altri esperimenti. Il confronto
tra la produzione delle diverse particelle all'interno dell'esperimento
WA97 risulta parimenti problematico a causa della diversit\`a delle
relative regioni cinematiche.
La porzione di spazio delle fasi comune a tutte le finestre cinematiche
mostrate in fig.~\ref{figoverlap} e \ref{figoverlap1} ha una estensione
molto limitata ed il calcolo della produzione di particelle al suo interno
sarebbe affetto da grossi errori statistici.
Per ottenere risultati meglio definiti e maggiormente confrontabili si \`e
cos\`{\i} deciso di estrapolare la produzione di particelle ad una
regione cinematica comune sia ai due tipi di collisione che alle
diverse particelle studiate.
Tale regione, scelta in modo da contenere tutte le finestre cinematiche
delle fig.~\ref{figoverlap} e \ref{figoverlap1}, \`e definita
dagli intervalli $|y-y_{CM}|\,<\,0.5$ e $p_T\,>\,0~GeV/c$.
Per l'estrapolazione si sono usati i risultati della procedura di 
{\em best fit} usata per studiare le distribuzioni di massa trasversa e
la produzione di particelle estrapolata \`e calcolata mediante la formula


\begin{equation}
\overline{P} =  P \:\times\: S 
             =  P \:\times\: 
\frac{\int_{y_{CM} -0.5}^{y_{CM} +0.5}\int_0^\infty \left(\frac{d^2N}{dy\,dp_T}
 \right)\;\;dy\:dp_T}
{\int\int_{reg. cinematica} \left(\frac{d^2N}{dy\,dp_T}
 \right)\;\;dy\:dp_T}
\label{eqextra}
\end{equation}

\noindent
dove $P$ \`e la produzione di particelle data dalla relazione
(\ref{eqcrosspb}) o (\ref{eqcrossp}) a seconda che si tratti di eventi Pb-Pb o p-Pb,
lo spettro di momento $\frac{d^2N}{dy\,dp_T}$ \`e determinato tramite il
metodo di massima verosimiglianza, utilizzando la funzione (\ref{eqlikely}),
e l'integrale a denominatore \`e eseguito all'interno della regione
cinematica della particella considerata.
L'errore sulla produzione estrapolata \`e determinato dall'equazione

\begin{equation}
\delta\overline{P}=\sqrt{S^2\,\delta P^2\:+\:P^2\,\delta S^2}
\label{eqerrextra}
\end{equation}

\noindent
sommando in quadratura l'errore $\delta P$ dato dalla (\ref{eqerrcrosspb})
con quello $\delta S=\frac{dS}{dT}\delta T$ dovuto 
al fattore $S$ di estrapolazione.
$\delta T$ indica l'errore nella determinazione del parametro $T$ e la
derivata $\frac{dS}{dT}$ \`e stata calcolata
facendo variare di una quantit\`a molto piccola,
 nell'espressione (\ref{eqlikely}), la temperatura
apparente risultante dal {\em fit}.
I risultati della procedura di estrapolazione sono riassunti 
in tab.~\ref{tabextra}, nella quale sono
riportati solo gli errori statistici determinati dalla (\ref{eqerrextra}).

\begin{table}[ht]
\centering
\caption{\em Produzione di particelle estrapolata ($\equiv$ numero di 
particelle per evento), relativa   all'interazione p-Pb e a quella Pb-Pb.}
\begin{tabular}{|c|c|c|}
\hline
            &              p-Pb         &               Pb-Pb      \\
\hline
\PgOm       &  $(2.2\pm 0.8)\,10^{-4}  $&  $0.15\pm 0.02          $\\
\PagOp      &  $(4.7\pm 2.5)\,10^{-5}  $&  $(5.7\pm 1.0)\,10^{-2} $\\
\PgXm       &  $(3.04\pm 0.23)\,10^{-3}$&  $0.81\pm 0.03          $\\
\PagXp      &  $(1.17\pm 0.1)\,10^{-3} $&  $0.202\pm 0.01         $\\
\PgL        &  $(5.97\pm 0.22)\,10^{-2}$&  $8.12\pm 0.22          $\\
\PagL       &  $(1.52\pm 0.1)\,10^{-2} $&  $1.08\pm 0.05          $\\
\PKzS       &  $(9.79\pm 0.43)\,10^{-2}$&  $12.8\pm 0.6           $\\
$h^-$       &  $1.19\pm 0.05           $&  $81\pm 2               $\\
\hline
\end{tabular}
\label{tabextra}
\end{table}

\noindent
Nella colonna relativa alle interazioni p-Pb \`e riportata la produzione di 
particelle per interazione, estrapolata a partire dalla (\ref{eqcrossp}),
mentre in quella relativa alle interazioni Pb-Pb \`e riportata la produzione
di particelle per evento selezionato, estrapolata a partire dalla
(\ref{eqcrosspb}).
Tali produzioni saranno analizzate e confrontate nel prossimo capitolo.

\subsection{Calcolo degli errori sistematici di estrapolazione}
\index{Calcolo degli errori sistematici di estrapolazione}

Per calcolare gli errori sistematici connessi alla procedura di
estrapolazione si \`e utilizzato il generatore di eventi Monte Carlo
VENUS [Wer93], le cui caratteristiche verranno esaminate nell'ultimo capitolo.
Il metodo consiste nel generare collisioni p-Pb e Pb-Pb e nel contare
le particelle prodotte all'interno della regione
$|y-y_{CM}|\,<\,0.5$ e $p_T\,>\,0~GeV/c$ usata per l'estrapolazione
in eventi reali.
Il risultato del conteggio, chiamato $P_{VENUS}$, viene confrontato
con la quantit\`a $\overline{P}_{VENUS}$, ottenuta contando
le particelle generate all'interno delle regioni cinematiche
mostrate in fig.~\ref{figoverlap} e \ref{figoverlap1} ed estrapolando poi
il risultato con lo stesso procedimento usato per i dati reali.
L'errore sistematico di estrapolazione \`e dato dalla relazione

\begin{equation}
\epsilon_{extrap.}(\%)=\left(
\frac{\overline{P}_{VENUS}}{P_{VENUS}}-1 \right)\times 100
\label{eqerrextrasis}
\end{equation}

\begin{figure}[htb]
\centering
\includegraphics[scale=0.57,clip]%
                                {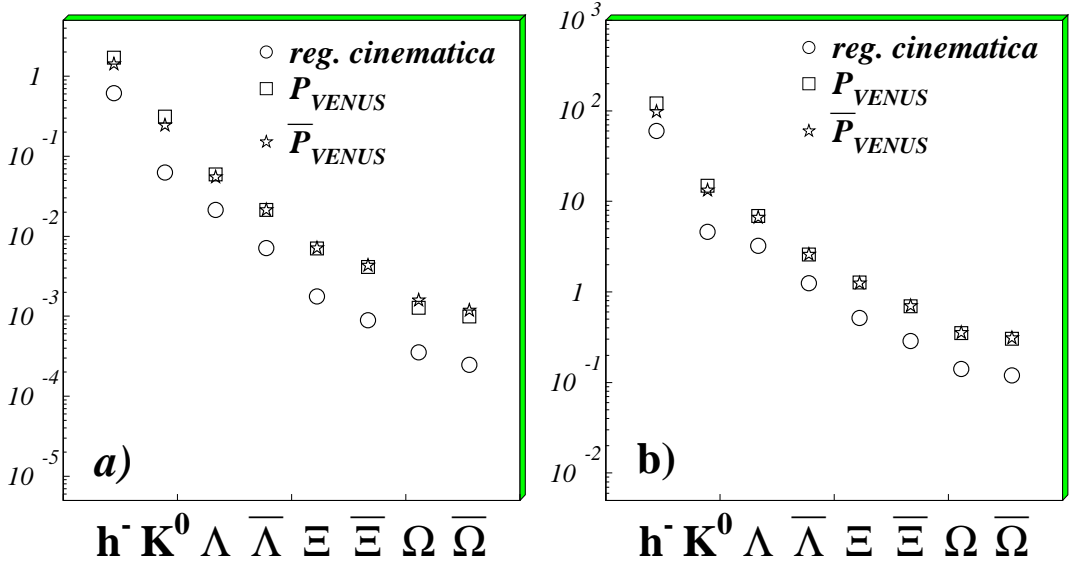}
    
\caption{\em Conteggi di particelle generate da VENUS all'interno
delle regioni cinematiche delle varie particelle (cerchi) e
nella regione di estrapolazione (quadrati), relativi
alle interazioni p-Pb (a) e Pb-Pb (b). Le stelle indicano la quantit\`a
$\overline{P}_{VENUS}$ calcolata applicando ai dati del Monte Carlo
la stessa procedura di estrapolazione usata per i dati reali.} 
\label{figextrafig}
\end{figure}

\noindent
Esso contiene un contributo dovuto all'estrapolazione nella variabile
$p_T$, che tiene conto dell'errata determinazione del parametro $T$ di 
temperatura apparente, e un contributo dovuto all'estrapolazione nella 
variabile $y$, che tiene conto delle incertezze introdotte dall'ipotesi
di distribuzione piatta di rapidit\`a.
In fig.~\ref{figextrafig}, 
i conteggi delle particelle simulate da VENUS nelle regioni
cinematiche determinate dall'accettanza del telescopio
per le varie particelle (cerchi) e nella regione di
estrapolazione (quadrati) sono confrontati con la quantit\`a 
$\overline{P}_{VENUS}$ (stelle), nella generazione di eventi
p-Pb (fig.~\ref{figextrafig}a) e Pb-Pb (fig.~\ref{figextrafig}b).

\noindent
In tab.~\ref{taberrextra} sono riportati i corrispondenti errori
sistematici di estrapolazione ottenuti mediante la (\ref{eqerrextrasis}),
che possono essere interpretati 
graficamente come la distanza tra le stelle ed i quadrati
contenuti in fig.~\ref{figextrafig}. L'incertezza con la quale sono
determinati dipende, naturalmente, dal numero di particelle generate.

\begin{table}[ht]
\centering
\caption{\em  Errori sistematici di estrapolazione relativi
             all'interazione p-Pb e Pb-Pb.}
\begin{tabular}{|c|c|c|}
\hline
            &            p-Pb      &               Pb-Pb  \\
\hline
$h^-$       &  $-17\pm 2\%$&  $-19\pm 2\%$\\
\PKzS       &  $-22\pm 2\%$&  $-11\pm 2\%$\\
\PgL        &  $-7\pm 2\% $&  $-3\pm 2\%$\\
\PagL       &  $-0.4\pm 3\%$& $-0.3\pm 2\%$\\
\PgXm       &  $1.1\pm 6\%$& $-1.8\pm 3\%$\\
\PagXp      &  $4.7\pm 8\%$&  $0.5\pm 4\%$\\
\PgOm       &  $24\pm 16\%$&  $0.4\pm 5\%$\\
\PagOp      &  $19\pm 18\%$&  $2\pm 6\%$\\
\hline
\end{tabular}
\label{taberrextra}
\end{table}

\noindent
In entrambi i tipi di collisione si hanno errori sistematici superiori
al $10\%$ per $h^-$ e $\PKzS$ e ci\`o \`e dovuto principalmente al
contributo dell'errore sull'estrapolazione in momento trasverso.
La distribuzione in massa trasversa di tali particelle, infatti, si discosta
dall'andamento esponenziale a causa di un eccesso di eventi a basse $m_T$,
al di sotto della regione misurata dall'esperimento WA97.
Di conseguenza, l'estrapolazione compiuta fino a momenti trasversi
molto piccoli, utilizzando la pendenza della distribuzione di massa trasversa
misurata all'interno della regione cinematica di rivelazione, introduce una
sottostima della reale produzione di tali particelle.
\par
Per le altre particelle, gli errori di estrapolazione sono piccoli e
compatibili con zero nel caso di interazioni Pb-Pb, 
mentre per quelle p-Pb, pur restando compatibili con zero a causa dei
grandi errori statistici,
 essi variano con continuit\`a in funzione della
massa della particella studiata, passando da valori negativi per le
particelle pi\`u leggere a valori positivi per quelle pi\`u pesanti.
Questo effetto \`e attribuibile all'estrapolazione in rapidit\`a
 ed alla particolare forma delle regioni cinematiche.
Si ricordi, infatti, che 
la distribuzione in rapidit\`a \`e stata assunta piatta, mentre in
realt\`a per interazioni p-Pb si \`e visto che essa ha un andamento
crescente passando dalla regione di frammentazione del proiettile
a quella del bersaglio.
D'altra parte, dall'esame delle regioni cinematiche
mostrate in fig.~\ref{figoverlap} e \ref{figoverlap1} si osserva
che la loro estremit\`a inferiore (a basso $p_T$),
 nella quale cade il maggior numero
di  particelle a causa dell'andamento esponenziale in $m_T$,
si sposta verso valori di rapidit\`a sempre pi\`u bassi al
crescere della massa della particella.
Di conseguenza, per le particelle pi\`u pesanti la rivelazione avviene in
gran parte in corrispondenza di rapidit\`a minori di $y_{CM}$, per le quali
la distribuzione reale di rapidit\`a dovrebbe essere al di sopra di quella
ipotizzata: la produzione estrapolata \`e cos\`{\i} sovrastimata
rispetto a quella effettiva.
Per le particelle pi\`u leggere, invece,
 la rivelazione avviene principalmente in corrispondenza di rapidit\`a
maggiori di $y_{CM}$, l\`{\i} dove la distribuzione di rapidit\`a effettiva
dovrebbe cadere al di sotto di quella ipotizzata: la produzione
estrapolata viene cos\`{\i} ad essere sottostimata.
Il maggiore errore di estrapolazione dei $\PKzS$ rispetto a quello delle
$h^-$ negli eventi p-Pb pu\`o essere spiegato in virt\`u di tale ragionamento,
considerando che la loro regione cinematica copre un intervallo di rapidit\`a
completamente in avanti rispetto a $y_{CM}$.
Per gli eventi Pb-Pb, invece, la simmetria delle distribuzioni di rapidit\`a
rende trascurabile
il contributo all'errore 
sistematico dovuto all'estrapolazione in tale variabile.
\par
\`E opportuno osservare che l'obiettivo dell'esperimento WA97 \`e il
confronto tra la produzione di stranezza tra interazioni p-Pb e Pb-Pb, per cui
i risultati finali, riportati nel prossimo capitolo, saranno espressi
come rapporto tra la produzione delle particelle strane identificate 
nei due tipi di interazione. Il contributo degli errori sistematici
di estrapolazione risulta, dunque, ridotto per compensazione,
soprattutto per quanto riguarda le  $h^-$, $\PKzS$ e $\PgL$ 
(c.f.r. tab.~\ref{taberrextra}).


%
%
%
%
%
%
\chapter{Risultati sperimentali e discussione}
\index{Risultati sperimentali e discussione}

\section{Introduzione}
\index{Introduzione}

I risultati finali dell'esperimento WA97, ottenuti a partire dal lavoro 
di analisi descritto nelle pagine precedenti e dalla applicazione delle
opportune correzioni sui dati, verranno presentati e discussi in questo
capitolo.
Essi riguardano lo studio delle distribuzioni di massa trasversa
 e l'analisi della
produzione di particelle nella regione di estrapolazione.
Le temperature apparenti determinate dalle distribuzioni di massa 
trasversa
saranno esaminate in funzione della massa della
particella a cui si riferiscono ed in funzione del sistema di collisione,
utilizzando i risultati dei precedenti esperimenti eseguiti nello spettrometro
OMEGA (WA85 e WA94), nonch\`e dei pi\`u
recenti esperimenti che utilizzano fasci di piombo (NA44 ed NA49).
Successivamente verranno presentati i rapporti di produzione delle diverse
particelle studiate e sar\`a discussa la loro variazione in funzione
del sistema di collisione, in particolare nel passaggio da interazioni
di tipo protone-nucleo a quelle di tipo nucleo-nucleo.
Il risultato principale dell'esperimento WA97 riguarda tuttavia lo studio
della produzione di stranezza in funzione della centralit\`a della 
collisione.
Questo studio sar\`a esposto in dettaglio, descrivendo il modello sviluppato
per il calcolo della centralit\`a a partire 
dalla molteplicit\`a di particelle cariche
misurata e considerando le indicazioni che esso fornisce circa
il meccanismo di produzione di stranezza nella collisione.
\par
Versioni preliminari di tali risultati, succedutesi nel corso degli ultimi
due anni di dottorato, sono state oggetto di relazioni presentate
al ``{\em Relativistic Heavy Ion School-Workshop}'' tenutosi a Praga dal
1 al 5 Settembre 1997 [Cal97], ai ``{\em XXXIIIrd Rencontres de Moriond}''
 tenutisi a Les Arcs (Francia) dal 21 al 28 Marzo 1998 [Cal98] ed alla
 ``{\em 6th Conference on Strangeness in Quark Matter}''
 tenutasi a Padova dal 20 
al 24 Luglio 1998 [Cali98].

\section{Temperature apparenti}
\index{Temperature apparenti}
\label{par_tempris}

\begin{figure}[htb]
\centering
\includegraphics[scale=0.58,bb=0 20 305 333,clip]%
                                {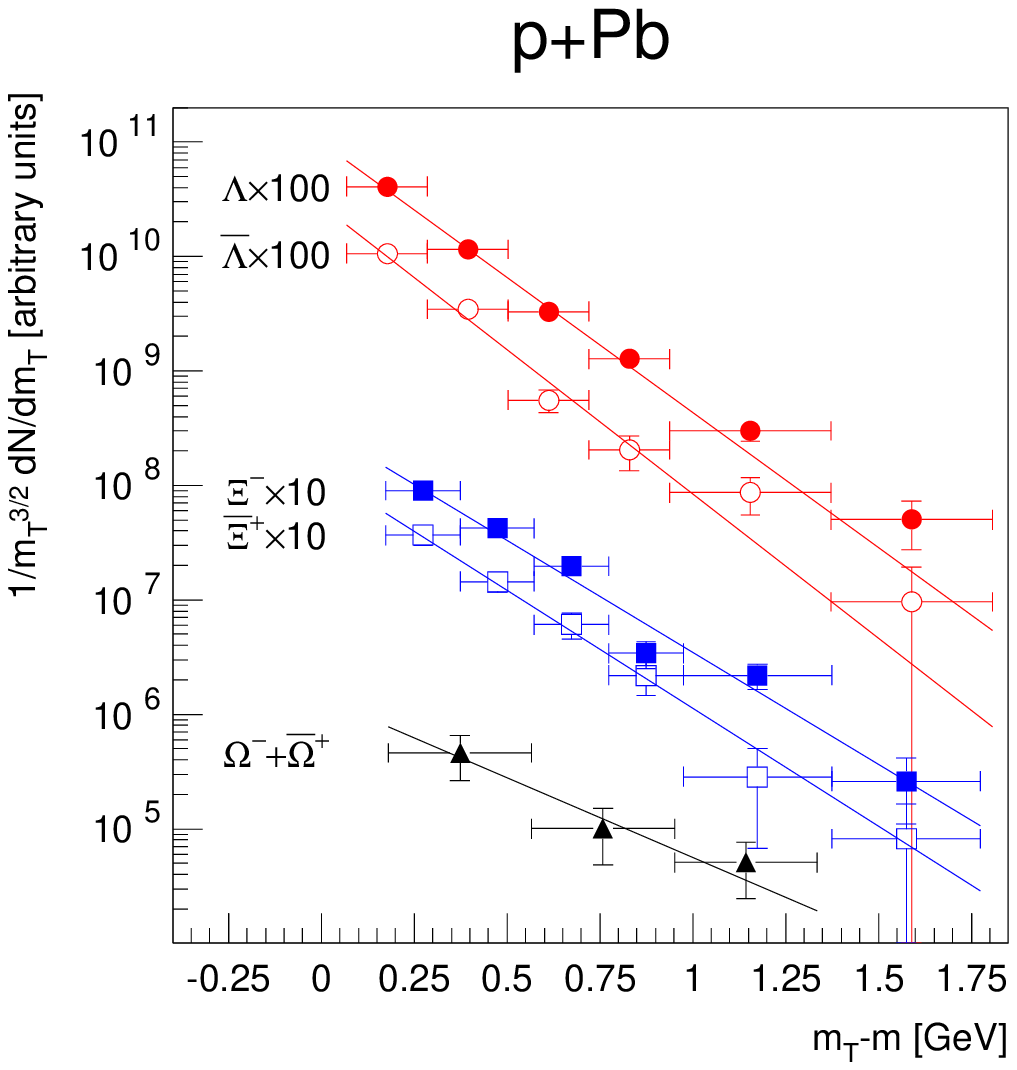}
\includegraphics[scale=0.58,bb=0 20 305 333,clip]%
                                {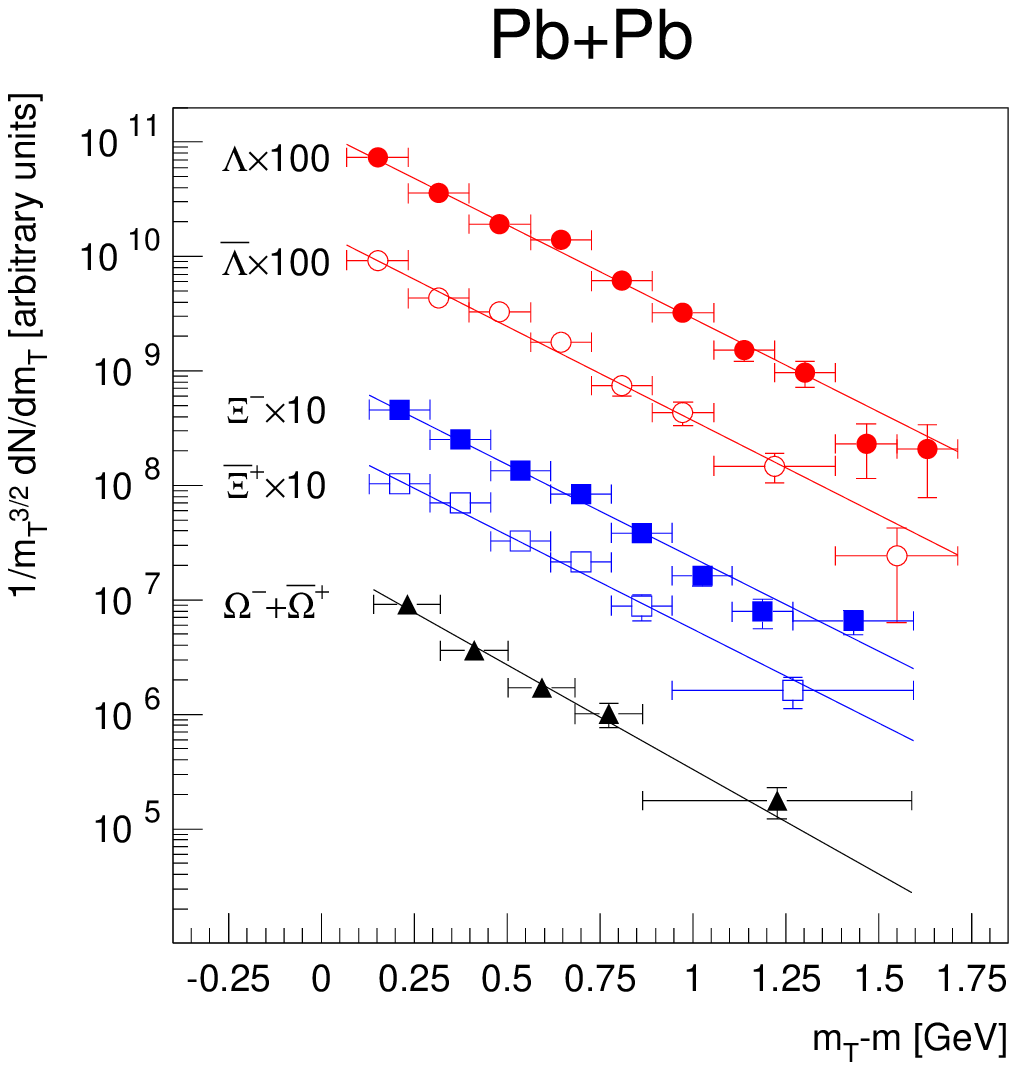}
\caption{\em Distribuzioni di massa trasversa di $\PgL$, $\PagL$, $\PgXm$,
$\PagXp$ e $\PgOm + \PagOp$ in interazioni p-Pb (a sinistra) e Pb-Pb (a destra)
con i risultati dei fit sovraimposti.}
\label{figmtfig}
\end{figure}

Lo studio degli spettri di momento delle particelle prodotte in collisioni
tra nuclei pesanti fornisce indicazioni circa il raggiungimento dell'equilibrio
termico locale nella sorgente. 
Le temperature apparenti, determinate con la procedura di {\em best fit} 
descritta nel capitolo precedente,
applicata all'interno delle regioni di accettanza per ciascun tipo di
particella identificata, sono riportate in tab.~\ref{tabtemp}. 
In fig.~\ref{figmtfig} sono mostrate le distribuzioni in massa 
trasversa di $\PgL$, $\PagL$, $\PgXm$, $\PagXp$ e $\PgOm + \PagOp$
per interazioni p-Pb e Pb-Pb; le rette sovrapposte sono il risultato
dei {\em best fit} corrispondenti alla funzione
(\ref{eqlikely}) e le temperature
apparenti rappresentano il reciproco della pendenza di tali rette.
Si pu\`o notare come le interazioni Pb-Pb siano
caratterizzate da temperature apparenti simili per $\PgL$, $\PagL$, $\PgXm$
e $\PagXp$, mentre per le $\PgOm +\PagOp$ esse assumono valori leggermente
inferiori. In generale, si osservano temperature apparenti simili per 
particelle ed antiparticelle in entrambi i tipi di interazione.

\par
Come gi\`a accennato nel par.~\ref{parmassatrasversa}, la temperatura
apparente fornisce una misura sia della temperatura della sorgente
termica nell'istante del {\em freeze-out} che del flusso trasverso che si
sviluppa in seguito alla collisione a causa delle interazioni tra le particelle
secondarie.
L'espansione trasversa, in particolare, produce un appiattimento delle 
distribuzioni di massa trasversa, quindi la temperatura apparente risulta
sempre maggiore di quella presente al {\em freeze-out}. Per alte
masse trasverse, la temperatura apparente $T$ \`e legata a quella di 
 {\em freeze-out} $T_f$ dalla relazione [Hei98]

\begin{equation}
T=T_f\;\sqrt{\frac{c+<v_T>}{c-<v_T>}}
\label{eqdoppler}
\end{equation}

\noindent
che indica lo spostamento Doppler introdotto dalla velocit\`a di
 flusso trasverso $v_T$.
Per basse masse trasverse, tipicamente per $m_T - m_0 < m_0$, dove
$m_0$ \`e la massa della particella esaminata, l'espansione trasversa
introduce una curvatura concava degli spettri di massa trasversa, tanto
pi\`u accentuata quanto maggiore \`e la massa $m_0$.
La temperatura apparente pu\`o essere parametrizzata qualitativamente
dalla relazione

\begin{equation}
T=T_f\: +\: m_0\, <v_T>^2
\label{eqxu}
\end{equation}

\noindent
nella quale si sommano i contributi dovuti all'energia termica ed a
 quella cinetica del moto trasverso [Bea97].
Nell'espansione idrodinamica della materia interagente, infatti, tutte
le particelle si muovono con la stessa velocit\`a trasversa. Poich\`e
classicamente l'energia cinetica collettiva dipende dalla massa della
 particella, si ha che particelle di massa maggiore trasportano energia 
pi\`u elevata, in accordo con la (\ref{eqxu}).
Osservando gli spettri in fig.~\ref{figmtfig} si pu\`o notare che essi
si estendono in prevalenza nella regione a bassa massa trasversa ed il maggior
numero di particelle presenti in tale regione contribuisce in maniera
determinante alla misura del parametro $T$ tramite i {\em fit}. Una parametrizzazione
delle temperature apparenti secondo la (\ref{eqxu}) sar\`a dunque pi\`u 
utile nell'interpretazione dei risultati dell'esperimento WA97.
\par
\begin{figure}[htb]
\centering
\includegraphics[scale=0.3,bb=0 25 453 510,clip]%
                                {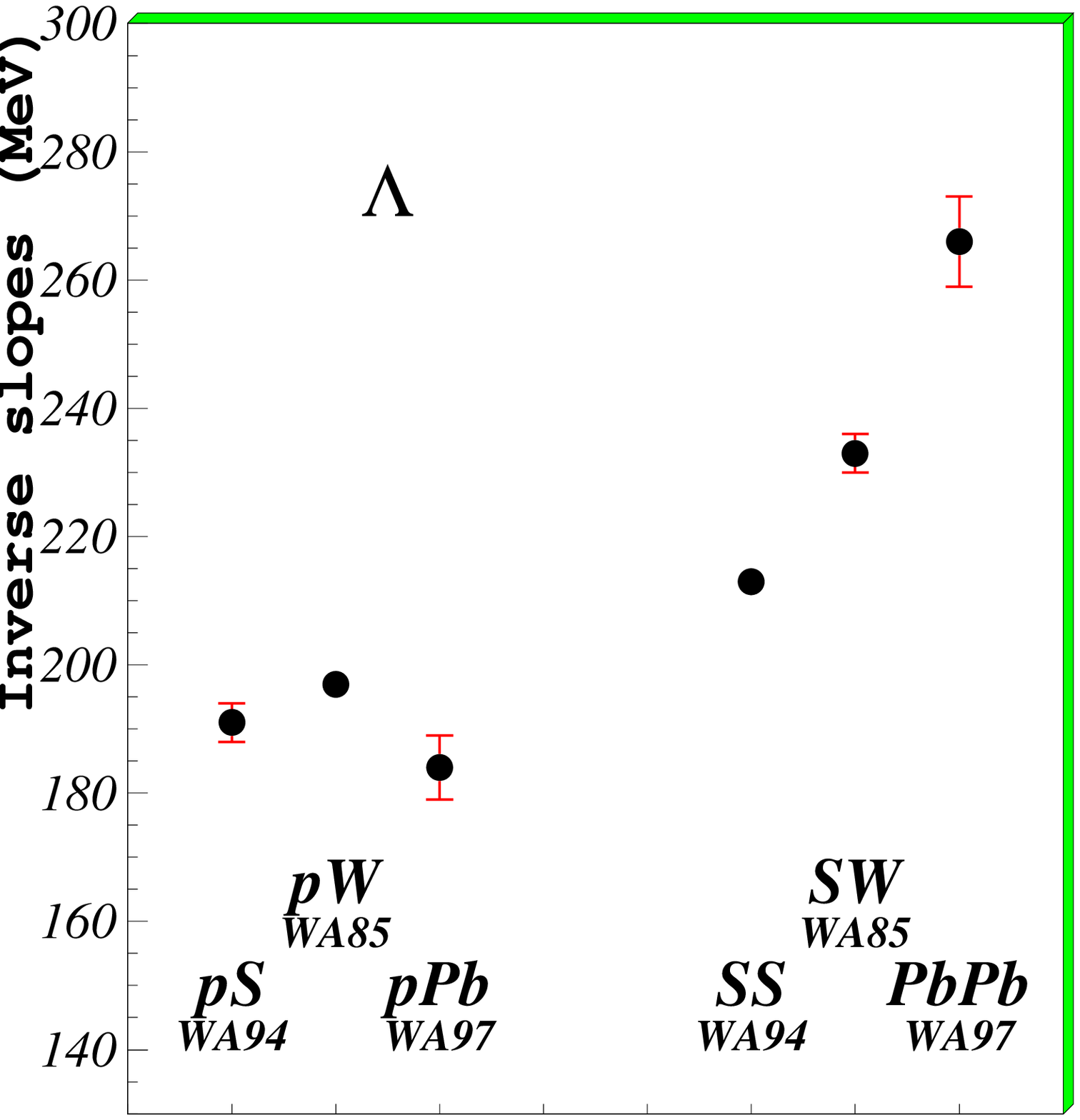}
\includegraphics[scale=0.3,bb=0 25 453 510,clip]%
                                {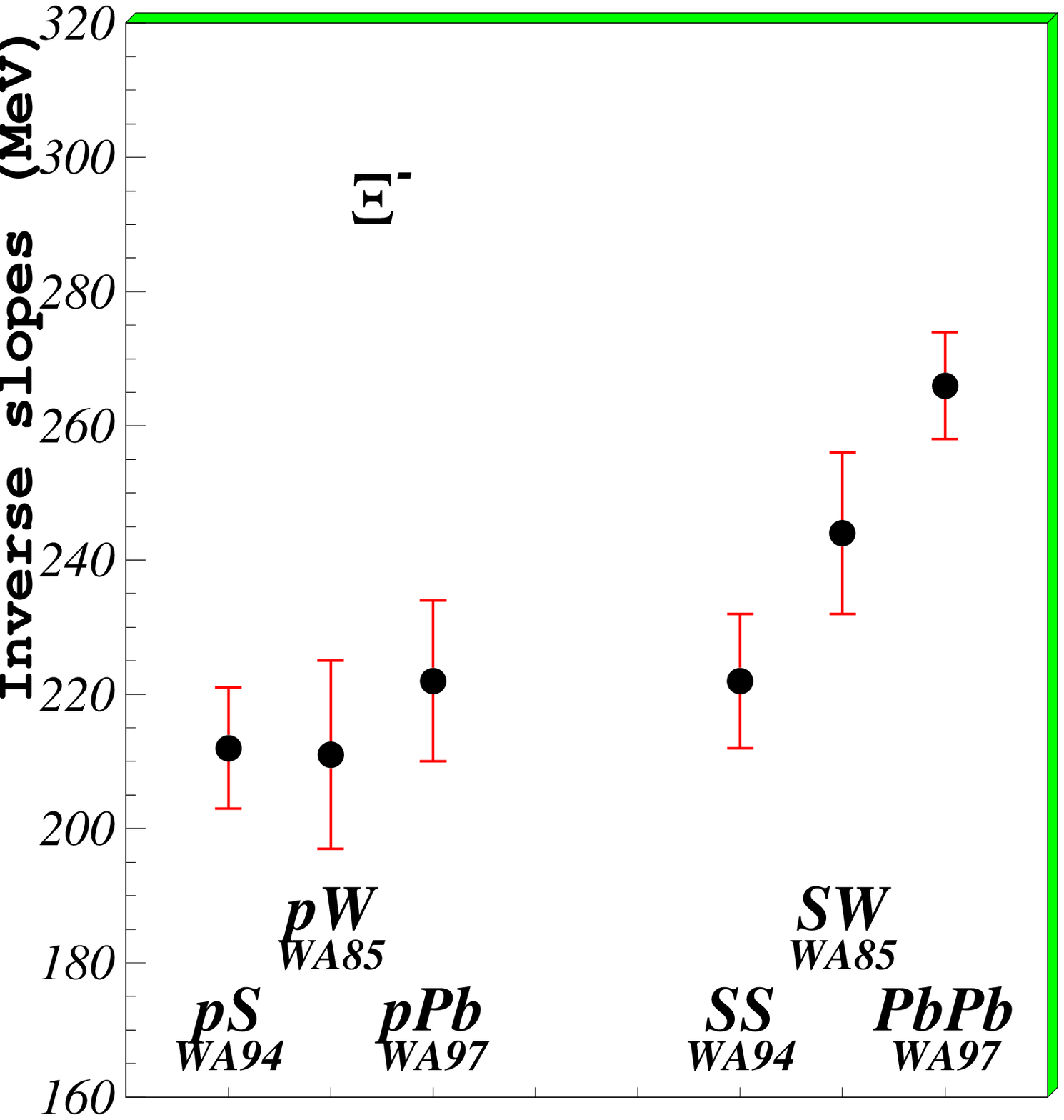}
\includegraphics[scale=0.3,bb=0 25 453 510,clip]%
                                {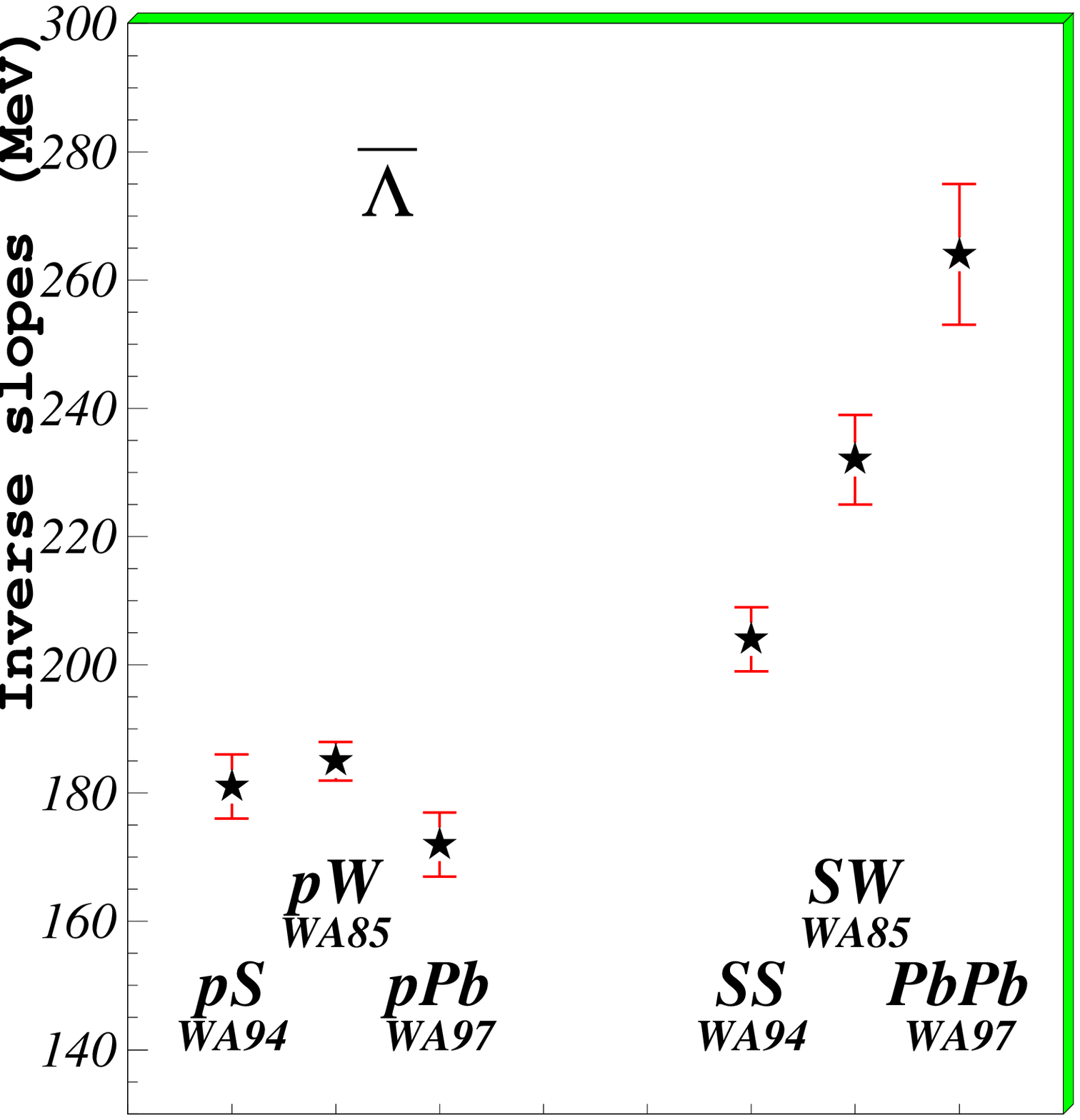}
\includegraphics[scale=0.3,bb=0 25 453 510,clip]%
                                {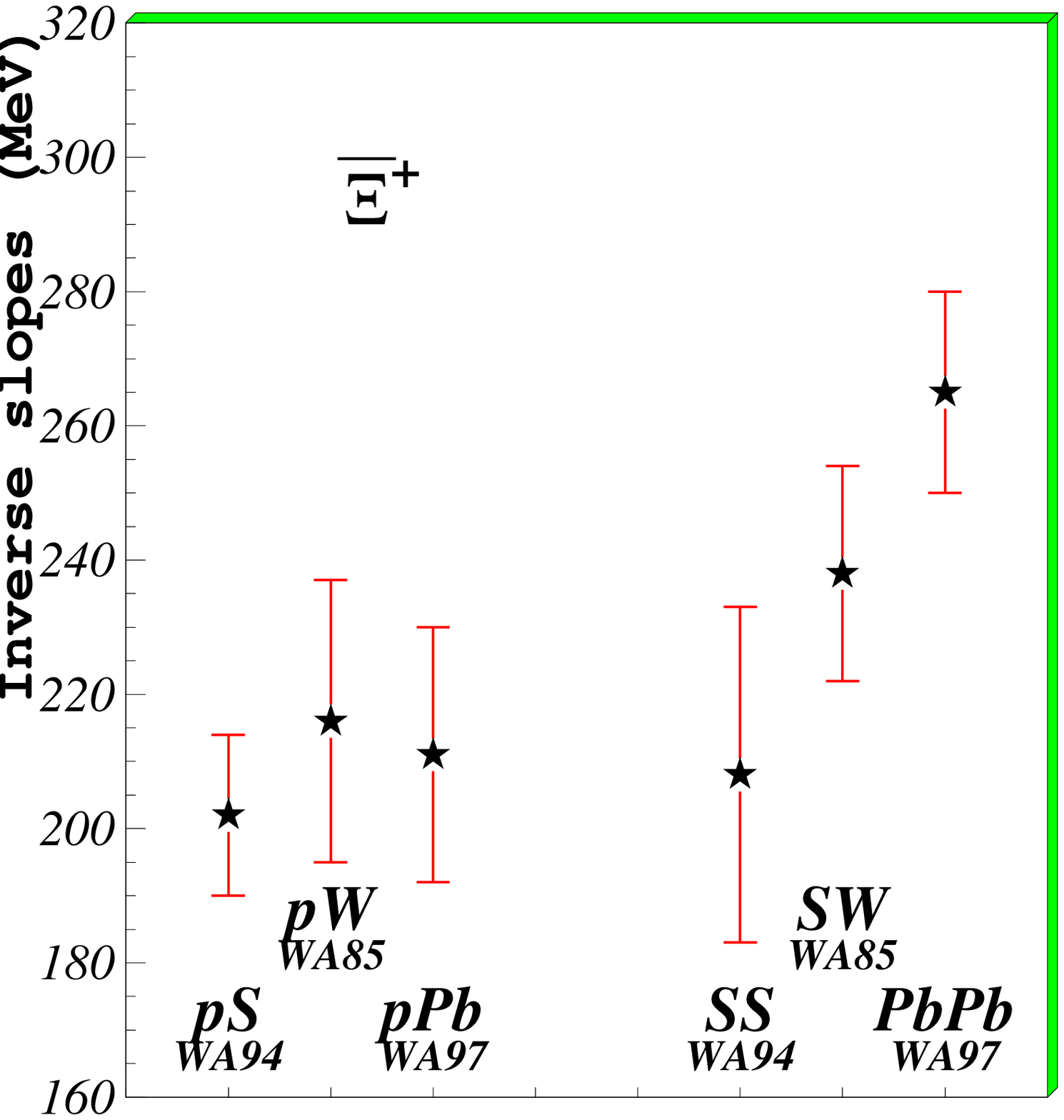}
\caption{\em Temperature apparenti di $\PgL$, $\PagL$, $\PgXm$ e
$\PagXp$  per differenti sistemi di collisione.}
\label{figuffafig}
\end{figure}

Al fine di individuare l'andamento dei parametri $T$ misurati per uno stesso
tipo di particella in diversi sistemi di collisione, le temperature 
apparenti riportate in tab.~\ref{tabtemp} sono state confrontate con quelle
determinate dagli esperimenti WA85 [Aba95] e WA94 [Abat95], rispettivamente in
interazioni S-W (p-W) e S-S (p-S) a $200~GeV/c$ per nucleone.
Le regioni cinematiche coperte da questi esperimenti
 sono leggermente differenti da quelle proprie
dell'esperimento WA97: le particelle rivelate hanno momento trasverso
$p_T>1~GeV/c$ e rapidit\`a nel laboratorio $2.3<y<2.8$ per
interazioni S-W (p-W) e $2.5<y<3.0$ per interazioni S-S (p-S).\footnote{
La rapidit\`a del centro di massa per collisioni S-W a $200~GeV/c$ per
nucleone \`e $y_{CM}\sim 2.54$, mentre quella per collisioni S-S e per il
sistema nucleone-nucleone alla stessa energia \`e $y_{CM}\sim 3.0$.}
In fig.~\ref{figuffafig} sono riportate le temperature apparenti, espresse
in $MeV$, di $\PgL$, $\PagL$, $\PgXm$ e $\PagXp$ al crescere della massa
atomica complessiva del sistema di collisione; i valori di $T$ sono  stati
ottenuti adoperando l'esponente $\alpha=\frac{3}{2}$ nell'espressione
(\ref{eqlikely}) e gli errori riportati sono solo di tipo
statistico.

\noindent
Si pu\`o notare che 
le temperature apparenti delle quattro specie di particelle esaminate assumono
valori simili tra loro in reazioni di tipo protone-nucleo, mentre un chiaro
incremento in funzione della complessit\`a del sistema di collisione \`e 
rilevabile per i valori relativi ad interazioni di tipo nucleo-nucleo.
Alla luce della (\ref{eqxu}), questo risultato \`e compatibile con la 
seguente interpretazione:
in collisioni tra nuclei di massa crescente  il
rilascio di energia iniziale nella regione centrale di interazione \`e
via via maggiore. La temperatura presente negli istanti iniziali della 
collisione ed il numero di interazioni tra le particelle prodotte ne risultano
cos\`{\i} incrementate, per cui nell'eq. (\ref{eqxu}) la temperatura apparente 
aumenta sia per effetto della pi\`u alta temperatura di {\em freeze-out}
 che per la
maggiore energia cinetica dovuta al flusso trasverso.
Per interazioni di tipo protone-nucleo, d'altra parte, l'evoluzione della
collisione non avviene secondo i modelli elaborati per quella tra nuclei
pesanti (par.~\ref{parmodel}) e la temperatura apparente non pu\`o essere
 parametrizzata secondo la (\ref{eqxu}).
Valori di temperatura apparente compatibili per le diverse interazioni
protone-nucleo indicano in ogni caso la presenza di fenomeni di 
{\em rescattering}
che non variano al variare della massa atomica del bersaglio, oppure la
completa assenza di fenomeni di questo tipo.
\par
Le temperature apparenti possono anche essere esaminate
 in funzione della massa
della particella esaminata, per un dato sistema di collisione.

\begin{figure}[htb]
\centering
\includegraphics[scale=0.6,bb=0 0 425 400,clip]%
                                {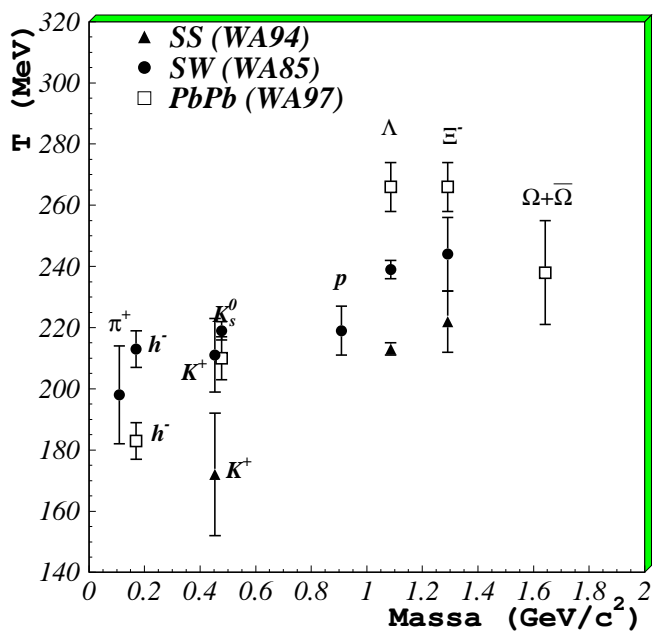}
\caption{\em Temperature apparenti degli adroni negativi ($h^-$) e delle diverse
particelle identificate in interazioni S-S (esperimento WA94), S-W
(esperimento WA85) e Pb-Pb (esperimento WA97).}
\label{fignnfig}
\end{figure}

\noindent
In fig.~\ref{fignnfig} sono mostrate le temperature apparenti misurate per
gli adroni negativi (trattati come pioni) 
e le diverse particelle identificate negli esperimenti WA94,
WA85 e WA97, rispettivamente in collisioni S-S, S-W e Pb-Pb 
[Aba95], [Aba96], [Ant97], [Abat97]. Per i
tre tipi di interazione (con qualche eccezione per Pb-Pb discussa pi\`u avanti)
si verifica un aumento del parametro $T$ in
funzione della massa della particella, attribuibile al termine di energia
cinetica nell'espressione ($\ref{eqxu}$).
Nell'esperimento WA85 \`e stato possibile misurare i parametri di temperatura
di $\Pgpp$ e $\Pp$ sfruttando le informazioni provenienti da un rivelatore
\v{C}erenkov multi-celle posto nella zona d'ombra delle camere proporzionali
 multifili (la relativa analisi \`e stata eseguita durante il primo anno di
dottorato ed \`e sintetizzata in appendice B) 
Dall'analisi della fig.~\ref{fignnfig} risulta che 
la crescita della temperatura
apparente con la massa della particella diventa sempre pi\`u rapida al 
crescere delle dimensioni del sistema interagente.
 Questa notevole
caratteristica emerge anche dall'analisi dei dati provenienti dagli
esperimenti NA35 [Alb94], [Alb96], NA44 [Bea96] e NA49 [Jac96] e,
interpretata in base alla relazione (\ref{eqxu}),
 indica una velocit\`a di flusso
 trasverso crescente con la complessit\`a del sistema interagente.
\par
\begin{figure}[htb]
\centering
\includegraphics[scale=0.68,bb=18 15 405 394,clip]%
                                {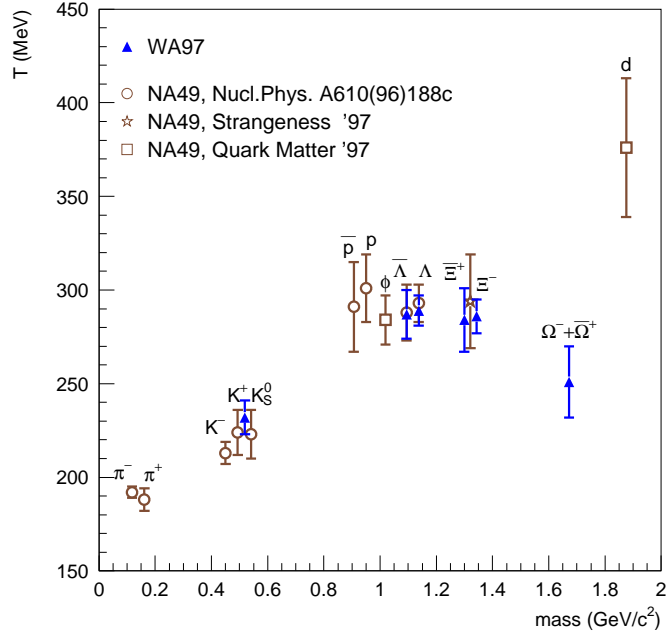}
\caption{\em Dipendenza della temperatura apparente dalla
massa della particella nel sistema di collisione Pb-Pb.}
\label{figt_vs_m}
\end{figure}

Per le $\Xi$ e le $\Omega$ identificate nell'esperimento WA97 si nota, tuttavia,
una deviazione dall'andamento crescente della temperatura apparente
 in funzione della massa. Per meglio caratterizzare questa osservazione si sono
considerate le temperature apparenti misurate nel solo sistema Pb-Pb, unendo
i risultati forniti in questa tesi a quelli dell'esperimento NA49
 [Jon96], [Rol97].
Questo confronto, riportato in fig.~\ref{figt_vs_m}, \`e stato ottenuto
 utilizzando il valore $\alpha=1$ nella funzione
(\ref{eqlikely}), essendo questo il valore usato dalla Collaborazione
NA49 nei dati pubblicati. 

\noindent
Gli intervalli di rapidit\`a coperti dall'esperimento NA49 sono molto
simili a quelli dell'esperimento WA97, ma una considerevole differenza 
risiede nella percentuale di sezione d'urto selezionata dal trigger
nei due esperimenti: $\sim 5\%$ in NA49 contro $\sim 40\%$ in WA97 [Fok97].
Una analisi delle distribuzioni di massa trasversa in funzione della
 centralit\`a dell'evento eseguita sui dati di WA97 ha tuttavia permesso
di escludere qualsiasi dipendenza dei valori di temperatura apparente
dalla percentuale di sezione d'urto selezionata, cio\`e dalla centralit\`a.
I risultati presentati in tab.~\ref{tabtemp}, infatti, si sovrappongono
perfettamente a quelli misurati da NA49 e contribuiscono ad evidenziare
la crescita del parametro $T$ con la massa della particella.
Tuttavia, mentre tale andamento risulta rafforzato dalla misura
della temperatura apparente del deuterone, i valori corrispondenti
alle particelle multi-strane se ne discostano, assumendo, come gi\`a rilevato,
valori simili, se non addirittura inferiori, a quelli delle $\PgL$ e $\PagL$.
L'assenza di una dipendenza univoca del parametro $T$ con la massa della
particella porta a scartare lo scenario di collisione comunemente proposto
in cui tutta la materia interagente si espande coerentemente, finch\`e
raggiunge un comune stato di {\em freeze-out}. \`E stata invece avanzata
l'ipotesi che gli adroni multi-strani possano raggiungere l'istante
di {\em freeze-out} prima delle altre particelle, risentendo solo
in minima parte del flusso trasverso collettivo [Hec98].
Il motivo per cui le $\Xi$ e le $\Omega$ si disaccoppiano prima dal resto
della materia interagente potrebbe risiedere nel numero limitato delle loro
risonanze. Infatti le interazioni nella fase di espansione sono dominate
dalla formazione di risonanze e, per il principio di indeterminazione,
una stima della frequenza di reazione \`e data dalla larghezza complessiva
delle  risonanze coinvolte.
Dai valori misurati [Bar96] si vede che tale larghezza \`e
correlata al contenuto di stranezza, 
%
%
e ci si aspetta che 
la frequenza di interazione per $\Xi$
ed $\Omega$ sia inferiore di circa $30\div 40\%$ rispetto a quella
delle $\Lambda$ o dei nucleoni.
Le interazioni delle $\Omega$, poi, potrebbero essere ancora pi\`u
soppresse a causa della natura della loro composizione a  quarks di valenza:
la creazione di risonanze nel sistema $\Omega\,\pi$ \`e proibita dalla
regola di OZI [Che76], mentre l'occorrenza di collisioni $\Omega + K$ o $\Omega + \eta$
risulta meno probabile a causa della minore presenza di partners mesonici
pi\`u pesanti dei pioni.
Per una verifica quantitativa di tale interpretazione \`e stato usato 
un modello teorico capace di descrivere l'intero processo di collisione
dal punto di vista microscopico. Tale modello, denominato RQMD,
riesce ad ottenere un buon accordo con i risultati sperimentali partendo 
dalle suddette ipotesi circa l'interazione della materia adronica, come
si vedr\`a nel prossimo capitolo.
\par
\begin{figure}[htb]
\centering
\includegraphics[scale=0.6,bb=0 0 425 400,clip]%
                                {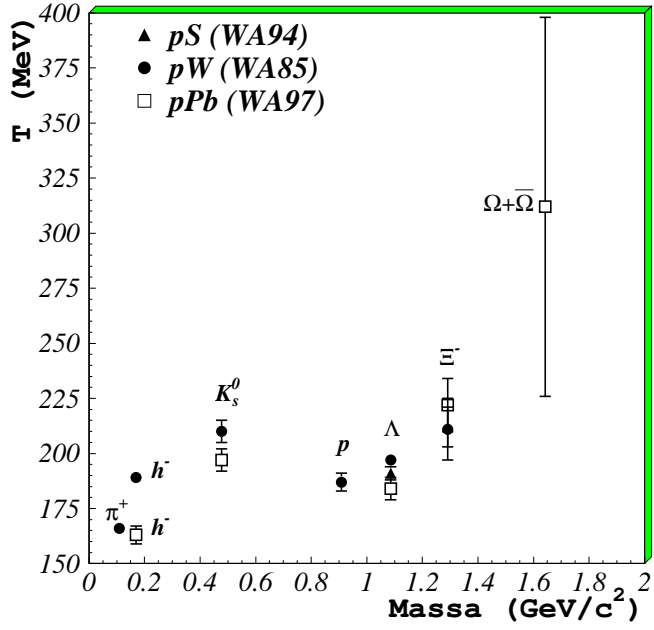}
\caption{\em Temperature apparenti delle negative ($h^-$) e delle diverse
particelle identificate in interazioni p-S (esperimento WA94), p-W
(esperimento WA85) e p-Pb (esperimento WA97).}
\label{figpnfig}
\end{figure}

In fig.~\ref{figpnfig} \`e presentata
una compilazione delle temperature apparenti misurate in interazioni di tipo
protone-nucleo dagli esperimenti WA94  (p-S), WA85 (p-W) e WA97 (p-Pb)
in funzione della massa della particella  [Aba97], [Abat97], [Ant97].

\noindent
In questo caso il parametro $T$  non sembra assumere un andamento definito
in funzione della massa della particella, 
tuttavia esso risulta analogo per i tre tipi di interazione.
Ci\`o avvalora l'ipotesi della presenza
di fenomeni di {\em rescattering} in misura uguale nei tre tipi di
collisioni, i quali non sono comunque in grado di innescare un processo
di espansione collettiva, poich\`e esso si manifesterebbe in
una dipendenza di $T$ dalla massa della particella coinvolta.
\par
\begin{figure}[htb]
\centering
\includegraphics[scale=0.52,clip]%
                                {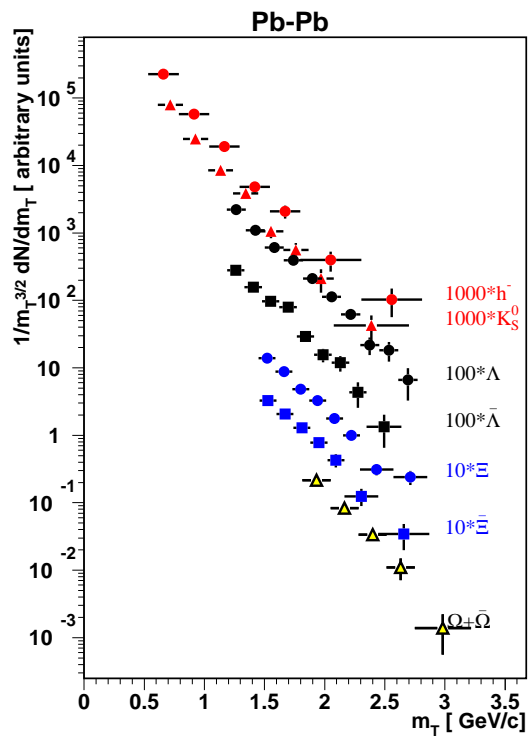}
\caption{\em Spettri di massa trasversa per tutte le particelle
identificate in interazioni Pb-Pb.}
\label{figmtscalefig}
\end{figure}

Si \`e finora osservato che gli spettri di massa trasversa 
delle particelle prodotte nelle 
collisioni tra nuclei pesanti hanno un andamento
compatibile con l'ipotesi di emissione da una sorgente termica in rapida
espansione. I relativi parametri di temperatura apparente possono variare da
un tipo di particella all'altro a causa delle loro differenti propriet\`a
di {\em freeze-out} e della dipendenza della massa dall'energia legata
al flusso trasverso. Questo risultato non \`e sufficiente da solo a provare il
raggiungimento di un equilibrio termico di tipo locale  anche negli
stadi precedenti al {\em freeze-out}, condizione
necessaria per giustificare l'approccio statistico alla dinamica della
 collisione e, ancor pi\`u, per provare l'esistenza di uno stato iniziale 
di QGP.
Tuttavia, studiando pi\`u in dettaglio le distribuzioni di massa trasversa delle particelle
 identificate nell'interazione Pb-Pb, mostrate tutte insieme  in
fig.~\ref{figmtscalefig}, si pu\`o notare che in corrispondenza degli
stessi intervalli di massa trasversa, particelle molto differenti mostrano
pendenze simili nei loro spettri.

\noindent
Questa propriet\`a, nota come ``{\em  $m_T$ - scaling}'' [Raf97], fa s\`{\i}
che la misura della temperatura apparente in uno stesso intervallo di $m_T$
fornisca lo stesso risultato per tutte le particelle studiate, come in 
effetti avviene per le coppie particella -- antiparticella. Ci\`o 
rappresenta una chiara indicazione del raggiungimento dell'equilibrio
 termico locale ed \`e 
compatibile con l'ipotesi di emissione da una sorgente in cui 
i gradi di libert\`a elementari (quarks e gluoni) abbiano raggiunto 
l'equilibrio termico [Raf97].

\section{Rapporti di produzione}
\index{Rapporti di produzione}

La misura dei rapporti di produzione fornisce
informazioni riguardanti l'equilibrio chimico
nella produzione delle diverse specie di particelle.
Nel paragrafo \ref{par_s} si \`e gi\`a accennato all'importanza di
stabilire il grado di saturazione nella produzione di stranezza, in quanto ci\`o
 potrebbe consentire di discriminare uno scenario di produzione adronica da
uno che prevede la formazione di QGP.
I rapporti di produzione, inoltre, vengono utilizzati nell'ambito di 
modelli termici contenenti l'intera evoluzione dinamica della collisione per
determinare i parametri caratteristici della sorgente negli  istanti
immediatamente successivi all'urto, secondo un approccio di tipo
statistico-termodinamico.
\par
I rapporti di produzione riguardanti le $\PgL$, $\Xi$ ed $\Omega$ prodotte
in collisioni Pb-Pb nell'esperimento WA97 sono riportati in tab.~\ref{tabratio}.

\begin{table}[ht]
\centering
\caption{\em Rapporti di produzione per le 
$\PgL$, $\Xi$ ed $\Omega$ prodotte in collisioni Pb-Pb.} 
 \begin{tabular}{|c|c|}
    \hline
    Rapporti &             Pb-Pb \\
   \hline
   \rule[2.50ex]{0mm}{1ex} 
  $\frac{\PgOm}{\PgXm}              $   & $0.18\pm0.02$ \\[0.3cm]
  $\frac{\PagOp}{\PagXp}            $   & $0.28\pm0.05$ \\[0.3cm]
  $\frac{\PgOm+\PagOp}{\PgXm+\PagXp}$   & $0.20\pm0.02$ \\[0.3cm]
 \hline\hline
  $\frac{\PgXm}{\PgL}               $  & $0.100\pm0.004$ \\[0.3cm]
  $\frac{\PagXp}{\PagL}             $  & $0.19\pm0.02$ \\[0.3cm]
  $\frac{\PgXm+\PagXp}{\PgL+\PagL}  $  & $0.110\pm0.004$ \\[0.3cm]
 \hline\hline
   \rule[2.50ex]{0mm}{1ex} 
   $\frac{\PagL}{\PgL}          $  & $0.133\pm0.007$ \\[0.3cm]
   $\frac{\PagXp}{\PgXm}        $  & $0.25\pm0.02$ \\[0.3cm]
   $\frac{\PagOp}{\PgOm}        $  & $0.38\pm0.08$ \\[0.3cm]
   \hline
  \end{tabular}
\label{tabratio}
\end{table}              

\begin{figure}[htb]
\centering
\includegraphics[scale=0.5,bb=0 0 470 500,clip]%
                                {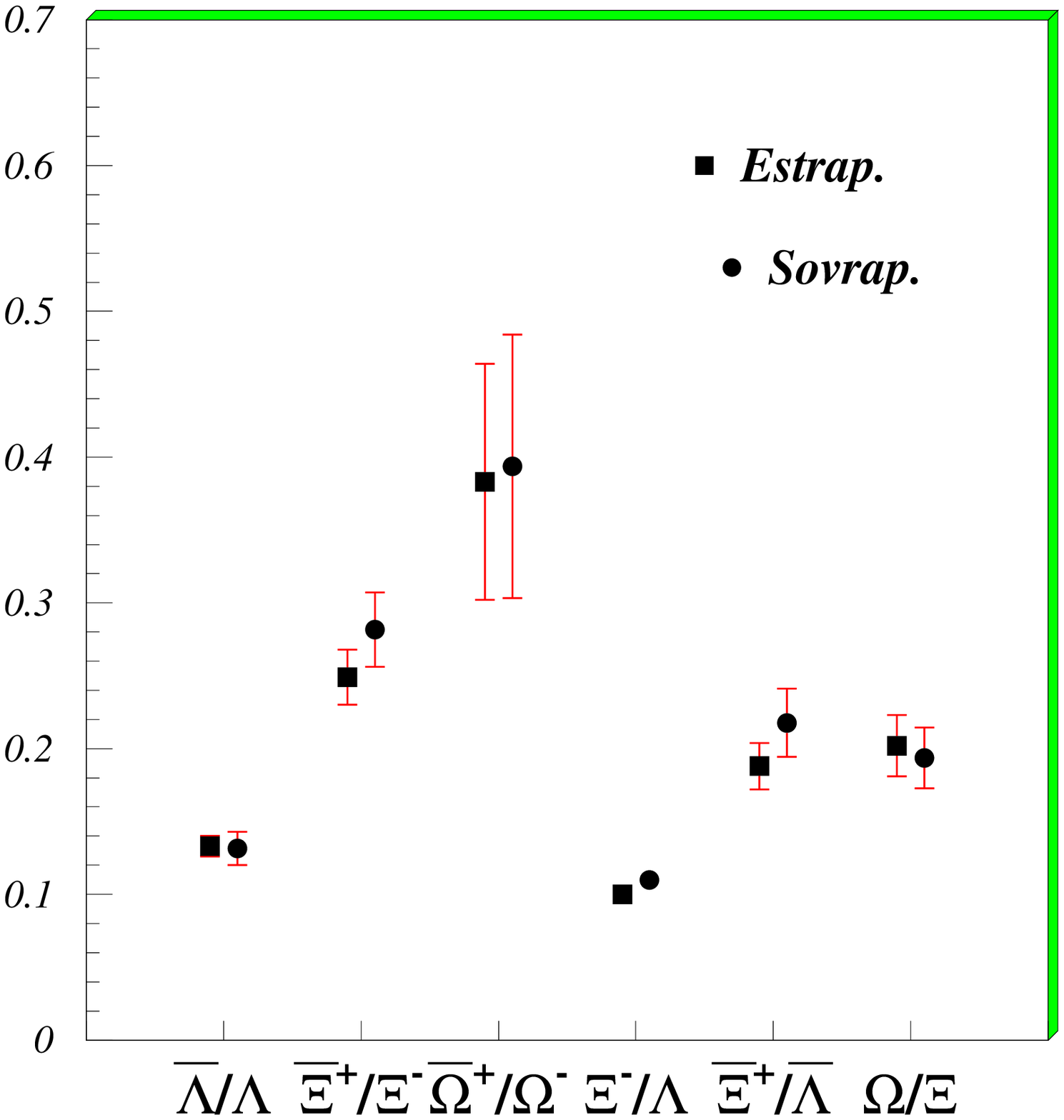}
\caption{\em Confronto tra i rapporti di produzione estrapolati e
quelli calcolati nelle regioni di sovrapposizione. Il rapporto $\Omega/\Xi$
si riferisce alla somma particella+antiparticella.}
\label{figratiover}
\end{figure}

\noindent
Essi sono stati ottenuti utilizzando la produzione di particelle calcolata
nella regione di estrapolazione $p_T>0~GeV/c$ e $|y-y_{CM}|<0.5$
(c.f.r. tab.~\ref{tabextra}). Gli errori riportati sono solo statistici
e non includono quelli sistematici di estrapolazione (tab.~\ref{taberrextra})
e quelli dovuti al {\em feed-down} (par.~\ref{par_feed}).
In fig.~\ref{figratiover}, tali rapporti sono confrontati con quelli calcolati
all'interno delle regioni di sovrapposizione tra le finestre cinematiche
della particella a numeratore e quella a denominatore.
La piena compatibilit\`a dei rapporti calcolati con i due metodi permette di
escludere eventuali dipendenze dei risultati dalla procedura di 
estrapolazione eseguita sui dati.
Dalla fig.~\ref{figratiover}, come dalla tab.~\ref{tabratio}, 
si osserva un andamento 
crescente dei rapporti del tipo antiparticella/particella, in funzione del
loro contenuto di stranezza.

I rapporti in tab.~\ref{tabratio} sono
 pienamente compatibili entro gli errori con quelli recentemente misurati
dall'esperimento NA49 [Jon96], [Gab98] e 
le leggere deviazioni sono da attribuirsi alle
differenze negli intervalli di centralit\`a selezionati e nelle
regioni cinematiche di rivelazione dei due esperimenti.
\par
Il calcolo della produzione di stranezza in interazioni tra nuclei pesanti
assume particolare rilevanza se riferita a quella dovuta ad interazioni 
pi\`u semplici, quali sono quelle di tipo protone-nucleo.
I rapporti di produzione delle particelle strane sono stati quindi
calcolati anche nel caso di collisioni p-Pb ed i risultati,
relativi alla stessa regione di estrapolazione considerata per le collisioni
Pb-Pb, sono riportati in tab.~\ref{tabratiop}.

\begin{table}[ht]
\centering
\caption{\em Rapporti di produzione per le 
$\PgL$, $\Xi$ ed $\Omega$ prodotte in collisioni p-Pb e loro variazioni 
nel confronto con gli analoghi rapporti calcolati in collisioni Pb-Pb.} 
 \begin{tabular}{|c|c|c|}
    \hline
    Rapporti &             p-Pb  & E($\frac{Pb-Pb}{p-Pb}$)\\
   \hline
   \rule[2.50ex]{0mm}{1ex} 
  $\frac{\PgOm}{\PgXm}              $   & $0.07\pm0.03$ &$2.5\pm1.0$\\[0.3cm]
  $\frac{\PagOp}{\PagXp}            $   & $0.04\pm0.02$ &$7\pm4$\\[0.3cm]
  $\frac{\PgOm+\PagOp}{\PgXm+\PagXp}$   & $0.06\pm0.02$ &$3.2\pm1.0$\\[0.3cm]
 \hline\hline
  $\frac{\PgXm}{\PgL}               $  & $0.051\pm0.004$ &$2.0\pm0.2$\\[0.3cm]
  $\frac{\PagXp}{\PagL}             $  & $0.08\pm0.01$ &$2.4\pm0.4$\\[0.3cm]
  $\frac{\PgXm+\PagXp}{\PgL+\PagL}  $  & $0.056\pm0.004$ &$2.0\pm0.2$\\[0.3cm]
 \hline\hline
   \rule[2.50ex]{0mm}{1ex} 
   $\frac{\PagL}{\PgL}          $  & $0.26\pm0.02$ &$0.52\pm0.05$\\[0.3cm]
   $\frac{\PagXp}{\PgXm}        $  & $0.38\pm0.05$ &$0.65\pm0.09$\\[0.3cm]
   $\frac{\PagOp}{\PgOm}        $  & $0.21\pm0.13$ &$1.8\pm1.2$\\[0.3cm]
   \hline
  \end{tabular}
\label{tabratiop}
\end{table}              

\noindent
Nella seconda colonna di tale tabella \`e riportata la variazione dei
 rispettivi rapporti nel passaggio da collisioni p-Pb a collisioni Pb-Pb,
ottenuta dividendo i rapporti in tab.~\ref{tabratio} con 
quelli della prima colonna.
Si pu\`o notare che, per quanto riguarda i rapporti che coinvolgono
particelle con diverso contenuto di stranezza, la variazione corrisponde
ad un incremento tanto maggiore quanto pi\`u grande
\`e il contenuto di stranezza delle particelle coinvolte.
L'incremento, inoltre,  risulta maggiore per le antiparticelle che per le
 particelle. Poich\`e le normali interazioni adroniche tendono ad attenuare
il contenuto di adroni multi-strani (ed ancor pi\`u quello di antibarioni
multi-strani) attraverso reazioni con scambio di {\em flavour}, 
l'incremento osservato non \`e compatibile con uno scenario di reazione
coinvolgente unicamente collisioni tra adroni. Tale risultato
\`e invece in linea con uno scenario che prevede la formazione
e l'emissione da una sorgente di materia deconfinata.
Per quanto riguarda i rapporti del tipo antiparticella/particella, le
variazioni riportate in tab.~\ref{tabratiop} corrispondono
ad una decrescita per le $\PagL$ e le $\PagXp$, attribuibile a fenomeni
di assorbimento degli antibarioni che si manifestano in maniera maggiore
 nel sistema di 
collisione pi\`u massivo.
Il rapporto relativo alle $\Omega$, al contrario, 
pur in presenza di un grande errore statistico, presenta un incremento
nel passaggio da interazioni p-Pb a quelle Pb-Pb, probabilmente in seguito alla 
concomitanza di due fattori: il modesto assorbimento di $\overline{\Omega}$
e $\Omega$ nel mezzo adronico (nel caso Pb-Pb) 
 e la soppressione della produzione di
barioni con tre unit\`a di stranezza in interazioni adroniche, a causa 
dell'alta soglia in massa (nel caso p-Pb).
Questa anomalia rafforza l'ipotesi che le $\Omega$ possano essere prodotte
in collisioni Pb-Pb secondo un meccanismo differente da quello presente in
interazioni p-Pb. 
\par
\begin{figure}[htb]
\centering
\includegraphics[scale=0.6,bb=0 0 425 400,clip]%
                                {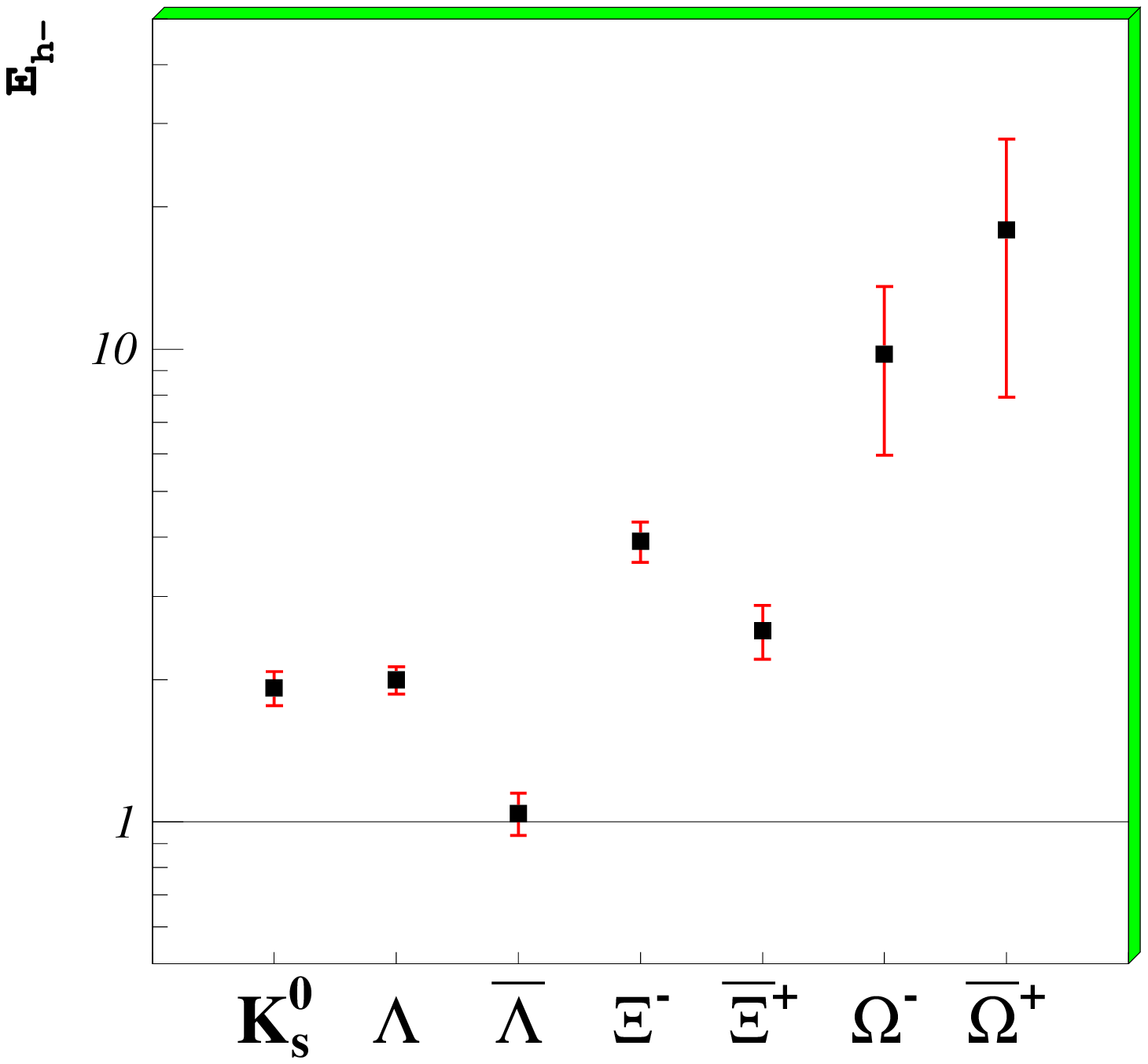}
\caption{\em Incremento di stranezza di $\PKzS$, $\PgL$, $\PagL$, $\PgXm$,
$\PagXp$, $\PgOm$ e $\PagOp$ in interazioni Pb-Pb rispetto ad interazioni
p-Pb, normalizzato alla produzione di particelle negative.}
\label{figenhfig}
\end{figure}

Come riferito nel par.~\ref{par_neg}, una maniera alternativa per valutare
l'incremento di stranezza nel passaggio da interazioni di tipo
protone-nucleo a quelle di tipo nucleo-nucleo consiste nel
normalizzare il numero di particelle strane di ciascun tipo 
al numero di particelle negative presenti nell'evento
(prevalentemente $\pi^-$).
Esso \`e stato allora calcolato mediante la formula

\begin{equation}
E_{h^-}=\left ( \frac{\overline{P}_Y}{\overline{P}_{h^-}} \right )_{Pb-Pb}
/
\left ( \frac{\overline{P}_Y}{\overline{P}_{h^-}} \right )_{p-Pb}
\label{eqenh1}
\end{equation}

\noindent
dove $\overline{P}_Y$ indica la produzione estrapolata delle varie particelle
 strane e $\overline{P}_{h^-}$ indica l'analoga quantit\`a per le negative.
I risultati sull'incremento di stranezza calcolato secondo la (\ref{eqenh1})
 sono mostrati in fig.~\ref{figenhfig} e riportati in
tab.~\ref{tabenh} per ciascun tipo di particella
strana identificata.

\begin{table}[htb]
\centering
\caption{\em Incrementi di stranezza nel passaggio da interazioni p-Pb a
             quelle Pb-Pb, normalizzati al numero di particelle negative 
            presenti nell'evento.}
\begin{tabular}{|c|c|}
    \hline
    Particelle &            $E_{h^-}$ \\
   \hline
 $\PKzS$    & $1.9\pm0.2$ \\
  $\PgL$     & $2.0\pm0.1$ \\
  $\PagL$    & $1.0\pm0.1$ \\
 $\PgXm$    & $3.9\pm0.4$ \\
  $\PagXp$   & $2.5\pm0.3$ \\
  $\PgOm$    & $10\pm4$  \\
 $\PagOp$   & $18\pm10$ \\
   \hline
  \end{tabular}
\label{tabenh}
\end{table}

\noindent
Tutti i punti rappresentativi della quantit\`a $E_{h^-}$ 
(ad eccezione della $\PagL$) si trovano al di 
sopra della linea continua orizzontale, 
corrispondente ad assenza di incremento.
Gli incrementi risultano crescenti con il contenuto di stranezza 
della particella 
anche se la si\-gnificativit\`a dell'incremento delle $\PgOm$
e $\PagOp$ \`e di sole 2 deviazioni standard, a causa principalmente
del piccolo numero di $\PgOm$ e $\PagOp$ osservate nell'interazione
p-Pb.  
Quantitativamente gli incrementi  seguono le seguenti proporzioni:

\begin{equation} 
\left\{
\begin{array}{rccccc}
E_{h^-}(\PKzS)\sim & E_{h^-}(\PgL) &:& E_{h^-}(\PgXm) &:& E_{h^-}(\PgOm) \\
                   &  1 &:& (2.0\pm0.2) &:& (5\pm 2) 
\end{array}
\right.
\label{line1}
\end{equation}
\vspace{0.8cm}
\begin{equation} 
\left\{
\begin{array}{ccccc}
E_{h^-}(\PagL) &:& E_{h^-}(\PagXp) &:& E_{h^-}(\PagOp)\\
1 &:& (2.4\pm 0.4) &:& (17\pm 10)
\end{array}
\right.
 \label{line2}
\end{equation}

\noindent
dove la prima relazione \`e stabilita tra particelle aventi rispettivamente
uno, due e tre quarks $s$ di valenza e la seconda riguarda antiparticelle con
uno, due e tre  quarks $\overline{s}$ di valenza.
Le stesse proporzioni sono ricavabili a partire dalle variazioni riportate
nella seconda colonna di tab.~\ref{tabratiop}.
\`E interessante notare come la crescita degli incrementi sia pi\`u sostenuta
per le antiparticelle (relazione \ref{line2}), pur non avendo esse alcun
{\em quark} di valenza in comune con i costituenti del bersaglio e del proiettile.
Questa caratteristica risulta difficilmente interpretabile in
un contesto di produzione adronica ed un incremento cos\`{\i} pronunciato
della produzione dello stato $\PagOp (\overline{s}\overline{s}\overline{s})$
nel passaggio da interazioni protone-nucleo a quelle nucleo-nucleo, presente
anche dopo aver considerato l'effetto di sovrapposizione dei nucleoni 
interagenti, indica chiaramente la possibilit\`a che le particelle possano
essere prodotte da una sorgente di  quarks e gluoni liberi e deconfinati.
\par
Una valutazione quantitativa dell'incremento di stranezza presenta, tuttavia,
alcune difficolt\`a dovute alla limitata regione dello spazio delle fasi
accessibile dai rivelatori. Si \`e gi\`a visto nel par.~\ref{par_rapid}
come le distribuzioni in rapidit\`a in sistemi di collisione del tipo 
protone-nucleo possano essere asimmetriche, con un picco pronunciato 
nella regione di frammentazione del bersaglio. Questa caratteristica accomuna
le particelle aventi almeno un {\em quark} di valenza in comune con il nucleone,
quali $h^-$, $\PKzS$, $\PgL$ e $\PgXm$; le antiparticelle e le $\Omega$ sono
invece prodotte in prevalenza nella regione centrale.
 In interazioni di tipo nucleo-nucleo,
invece, le distribuzioni di rapidit\`a appaiono pi\`u simmetriche, in
virt\`u della maggiore simmetria del sistema di collisione 
(cfr. fig.~\ref{figrapfig}).
Di conseguenza, il confronto tra la produzione di particelle strane e 
quella delle negative, eseguito in un ristretto intervallo centrale 
di rapidit\`a, pu\`o portare ad una dipendenza dei risultati dalla
forma della distribuzione di rapidit\`a della particella considerata,
specialmente nel caso di interazioni protone-nucleo.
Qualitativamente si pu\`o prevedere che gli incrementi calcolati per
 $\PagL$, $\PagXp$ e $\Omega$ possano essere sottostimati rispetto agli altri 
e questo potrebbe spiegare il  minore incremento misurato per $\PagL$
e $\PagXp$ rispetto  a quello delle rispettive particelle; la 
consistenza interna delle relazioni (\ref{line1}) e (\ref{line2})
\`e in ogni caso preservata.
\par
Una analoga misura dell'incremento di stranezza \`e stata recentemente 
ricalcolata dalla Collaborazione WA85 [Ant98], [Eva98] all'interno della regione
cinematica $2.5 < y < 3.0$ e $p_T > 1.4~GeV/c$ ed i risultati sono
riportati nella prima colonna di tab.~\ref{tabwa85}. Nella seconda
colonna sono riportati gli incrementi $E_{h^-}$ misurati nell'esperimento WA97
ed estrapolati nella regione $p_T > 1.4~GeV/c$, per facilitare il
confronto.

\begin{table}[ht]
\centering
\caption{\em Incremento di stranezza, normalizzato alle negative, 
misurato negli esperimenti WA85 (S-W e p-W)  e WA97 
(Pb-Pb e p-Pb) per $p_T > 1.4~GeV/c$.}
 \begin{tabular}{|c|c|c|}
    \hline
              &$E_{h^-}(\frac{S-W}{p-W})$&$E_{h^-}(\frac{Pb-Pb}{p-Pb})$ \\
   \hline
  $\PgL   $   & $2.76\pm0.22 $&$2.70\pm0.55$ \\
  $\PagL  $   & $2.69\pm0.38 $&$1.72\pm0.49$ \\
  $\PgXm  $   & $3.82\pm0.51 $&$2.77\pm0.60$ \\
  $\PagXp $   & $3.87\pm0.80 $&$1.98\pm0.60$ \\
   \hline
  \end{tabular}
\label{tabwa85}
\end{table}              

\noindent
Gli incrementi per le $\PgL$ e le $\PgXm$ risultano all'incirca 
compatibili nei
due esperimenti, mentre per le antiparticelle essi differiscono di
 pi\`u di una deviazione standard.
Una sovrastima degli incrementi delle antiparticelle nell'esperimento WA85
 \`e ipotizzabile considerando gli argomenti gi\`a esposti circa le loro
distribuzioni di rapidit\`a. I valori di rapidit\`a del centro di massa
per i sistemi S-W e p-W sono differenti e corrispondono ai due estremi
dell'intervallo $2.5<y<3.0$ di rivelazione (si veda la nota del 
par.~\ref{par_tempris}).
Il confronto tra la produzione di un
dato tipo di particella nelle due interazioni viene quindi eseguito in una
regione di rapidit\`a spostata verso la regione di frammentazione del bersaglio
nel caso di interazioni p-S ed in una 
spostata verso la regione di frammentazione del 
proiettile nel caso di interazioni S-W. Per particelle prodotte principalmente
a rapidit\`a centrale (quali le antiparticelle), ci\`o conduce ad
una sovrastima della quantit\`a $E_{h^-}$. Tale effetto, come precedentemente
esposto, si manifesta in senso opposto nell'esperimento WA97, per via della
simmetria del sistema di collisione Pb-Pb e poich\`e l'intervallo di rapidit\`a
coperto \`e centrato sul valore del centro di massa per entrambe le
interazioni p-Pb e Pb-Pb.

\begin{figure}[htb]
\centering
\includegraphics[scale=0.58,bb=0 5 425 400,clip]%
                                {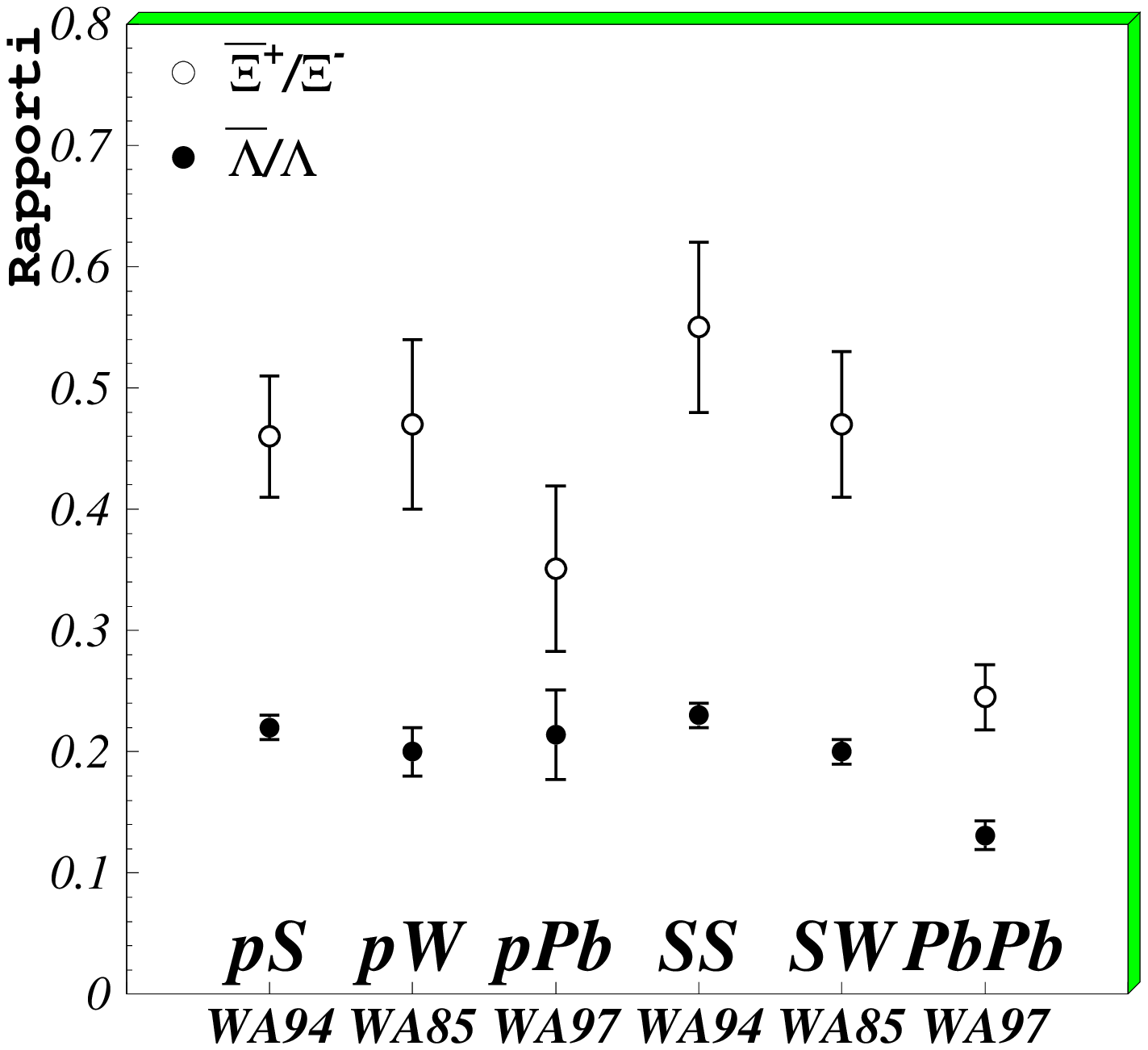}
\caption{\em Rapporti $\frac{\PagL}{\PgL}$ e
$\frac{\PagXp}{\PgXm}$ per sistemi di collisione di complessit\`a crescente.}
\label{confratio1}
\end{figure}

\begin{figure}[htb]
\centering
\includegraphics[scale=0.58,bb=0 5 425 400,clip]%
                                {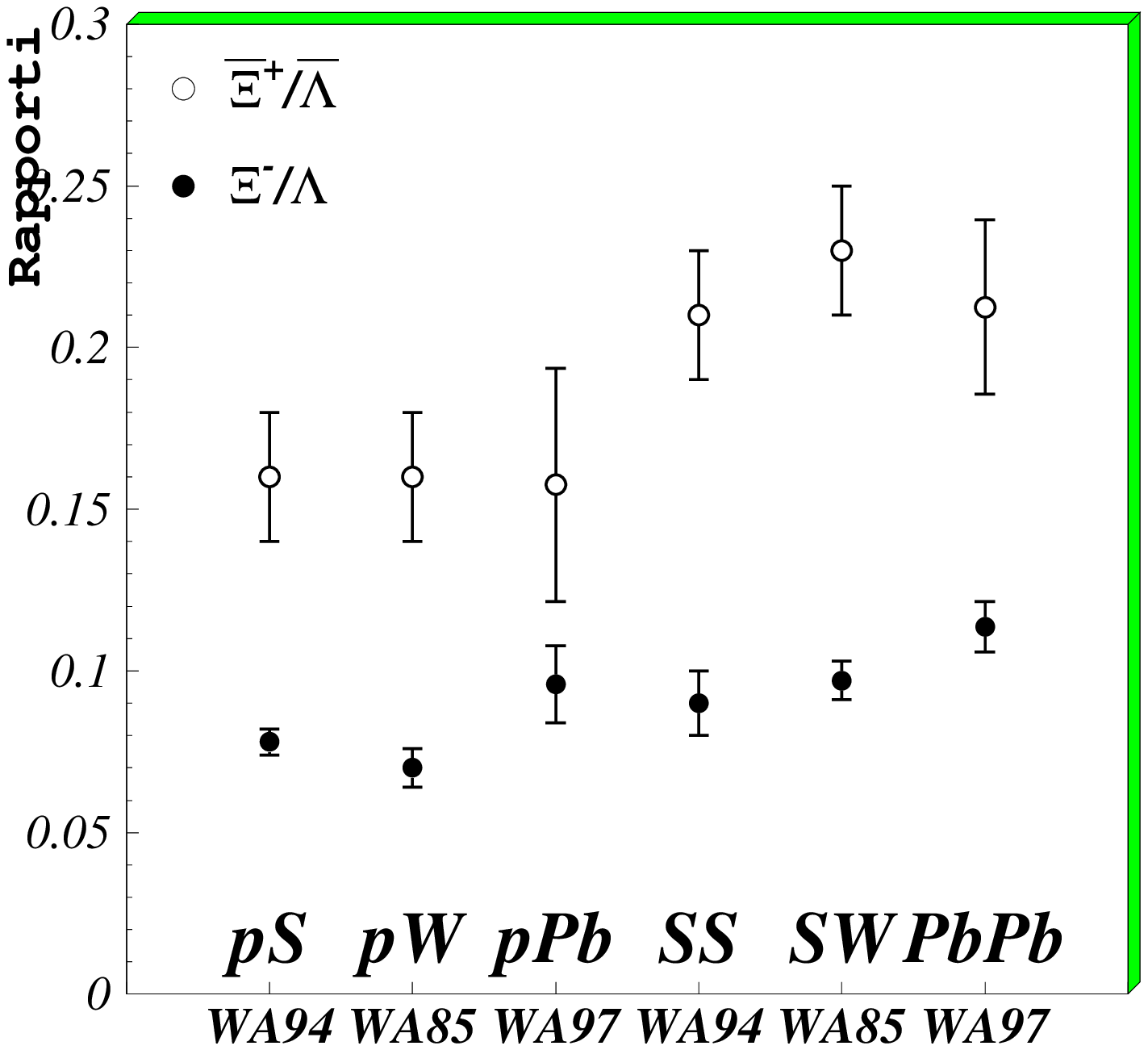}
\caption{\em Rapporti $\frac{\PgXm}{\PgL}$ e
$\frac{\PagXp}{\PagL}$ per sistemi di collisione di complessit\`a crescente.}
\label{confratio2}
\end{figure}

Un confronto dei rapporti di produzione in funzione del sistema di collisione
pu\`o essere effettuato utilizzando i risultati degli esperimenti
WA85 e WA94, forniti nella finestra cinematica $2.3 < y < 2.8$
e $p_T > 1.2~GeV/c$ per i dati p-W e S-W e  
$2.5 < y < 3.0$ e $p_T > 1.2~GeV/c$ per i dati p-S e S-S [Eva98].
Per facilitare il confronto, i dati dell'esperimento WA97 sono stati
estrapolati a $p_T > 1.2~GeV/c$.
La fig.~\ref{confratio1} mostra i rapporti $\frac{\PagL}{\PgL}$ e
$\frac{\PagXp}{\PgXm}$ per sistemi di collisione di complessit\`a crescente.
Entrambi i rapporti assumono valori compatibili in reazioni indotte da protoni
e ioni zolfo, mentre per collisioni tra ioni piombo essi sono considerevolmente
pi\`u bassi. Ancora una volta ci\`o potrebbe 
indicare la presenza di effetti di assorbimento,
dovuti al maggiore {\em rescattering} in questo tipo di collisioni.
In fig.~\ref{confratio2} sono invece riportati i rapporti $\frac{\PgXm}{\PgL}$
e $\frac{\PagXp}{\PagL}$ nei diversi sistemi di collisione: i valori calcolati
in collisioni p-Pb sono compatibili con quelli determinati in collisioni 
p-S e p-W e lo stesso vale per i rapporti in collisioni Pb-Pb rispetto
a quelli osservati in interazioni S-S ed S-W.

\section{Produzione di stranezza in funzione della centralit\`a}
\index{Produzione di stranezza in funzione della centralit\`a}
\label{par_mol}

L'esperimento WA97 consente di studiare la produzione di particelle strane 
in funzione della centralit\`a della collisione, sfruttando l'informazione
proveniente dai rivelatori di molteplicit\`a. Per avere una stima della
centralit\`a bisogna mettere in relazione la molteplicit\`a di particelle
cariche nell'evento, cio\`e l'osservabile fisica direttamente misurata,
con una variabile definita teoricamente che indica la centralit\`a della
collisione. Mediante il modello geometrico di collisione nucleare
descritto in appendice C, si pu\`o definire la variabile di centralit\`a
$N_p$ che indica il numero medio di nucleoni che partecipano alla
collisione tra due nuclei.
Si \`e poi assunto che la molteplicit\`a media di particelle cariche sia
proporzionale ad $N_p$, mediante una costante ($q$) che costituisce
l'unico parametro libero usato. Per tener conto delle fluttuazioni
intrinseche nel numero $N_p$ di nucleoni colpiti e nel numero di particelle
prodotte in seguito a tali collisioni, si assume che la molteplicit\`a
sia distribuita intorno al suo valor medio ($q\, N_p$) secondo una
gaussiana di larghezza $\sigma=q\,\sqrt{N_p}$
(ci\`o equivale ad assumere una dispersione poissoniana del numero
di partecipanti) [Kha97].
L'ulteriore fluttuazione connessa alla risoluzione sperimentale del
rivelatore di molteplicit\`a \`e stata stimata usando le informazioni
provenienti dalle due stazioni di microstrip al silicio (fig.~\ref{figmicro})
ed inserita nel modello.
Le molteplicit\`a fornite dalle due stazioni, una volta corrette per
accettanza, per efficienza del rivelatore e per i contributi
dovuti all'eventuale interazione fuori bersaglio, costituiscono
infatti due misure indipendenti di una stessa quantit\`a fisica.
La risoluzione sperimentale \`e stata allora determinata suddividendo 
l'intervallo di molteplicit\`a coperto in una serie di intervallini
e determinando in ciascuno
di essi la larghezza della gaussiana che descrive la
differenza tra le due misure di molteplicit\`a.

\begin{figure}[htb]
\centering
\includegraphics[scale=0.47,bb=3 28 520 562,clip]%
                            {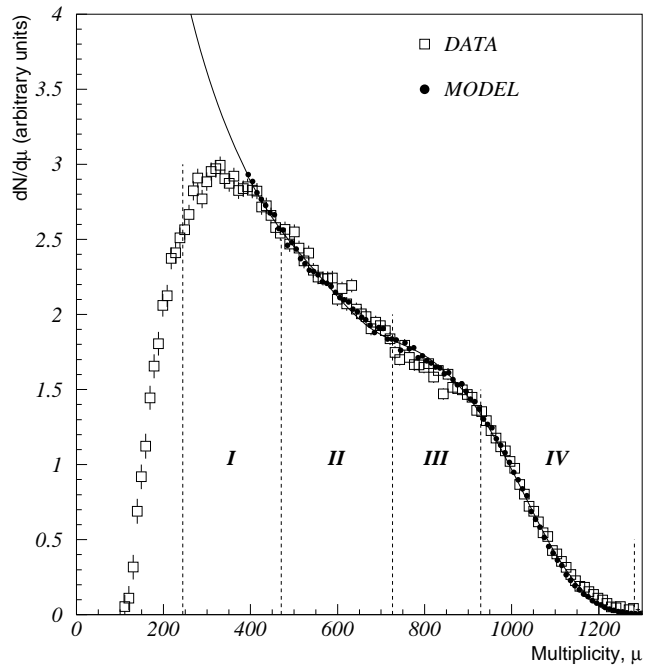}
\caption{\em Spettro di particelle cariche misurato (quadrati) confrontato
col modello sviluppato per descrivere geometricamente la collisione
tra nuclei (linea continua). I cerchi individuano la regione di spettro
utilizzata per la procedura di best fit.}
\label{multfig}
\end{figure}

\noindent
In fig.~\ref{multfig} \`e mostrato lo spettro di molteplicit\`a misurato,
con sovrapposto il risultato del modello elaborato.
Per molteplicit\`a al di sopra di $400$, il modello \`e in grado di riprodurre
fedelmente la molteplicit\`a misurata, utilizzando come valore
del parametro libero $q=2.7$ (che rappresenta il numero medio di
particelle cariche prodotte per nucleone partecipante).
La discrepanza a basse molteplicit\`a \`e dovuta alla soppressione di
collisioni periferiche operata dal trigger dell'esperimento.
\par
Per ottenere il numero di iperoni prodotti in 
funzione del numero di partecipanti,
la distribuzione di molteplicit\`a \`e stata divisa in quattro intervalli
(fig.~\ref{multfig}) ed in ognuno di essi
\`e stato calcolato il numero medio di partecipanti utilizzando il
modello sviluppato. L'estensione degli intervalli \`e stata scelta
in modo che in essi cadano lo stesso numero di $\Omega$  (la
particella con minore statistica) rivelate.
Per il primo intervallo, dove \`e presente la discrepanza col calcolo del 
modello, il numero medio di partecipanti \`e stato calcolato come media 
pesata, usando come peso il rapporto tra le molteplicit\`a misurate
e quelle risultanti dal modello. Questo accorgimento consente di
determinare il numero di partecipanti nelle interazioni Pb-Pb tenendo
conto della selezione operata dal trigger; per interazioni p-Pb, invece,
il numero medio di partecipanti \`e stato calcolato integrando su
tutti i possibili parametri di impatto.

\begin{figure}[htb]
\centering
\includegraphics[scale=0.55,bb=0 16 682 460,clip]%
                            {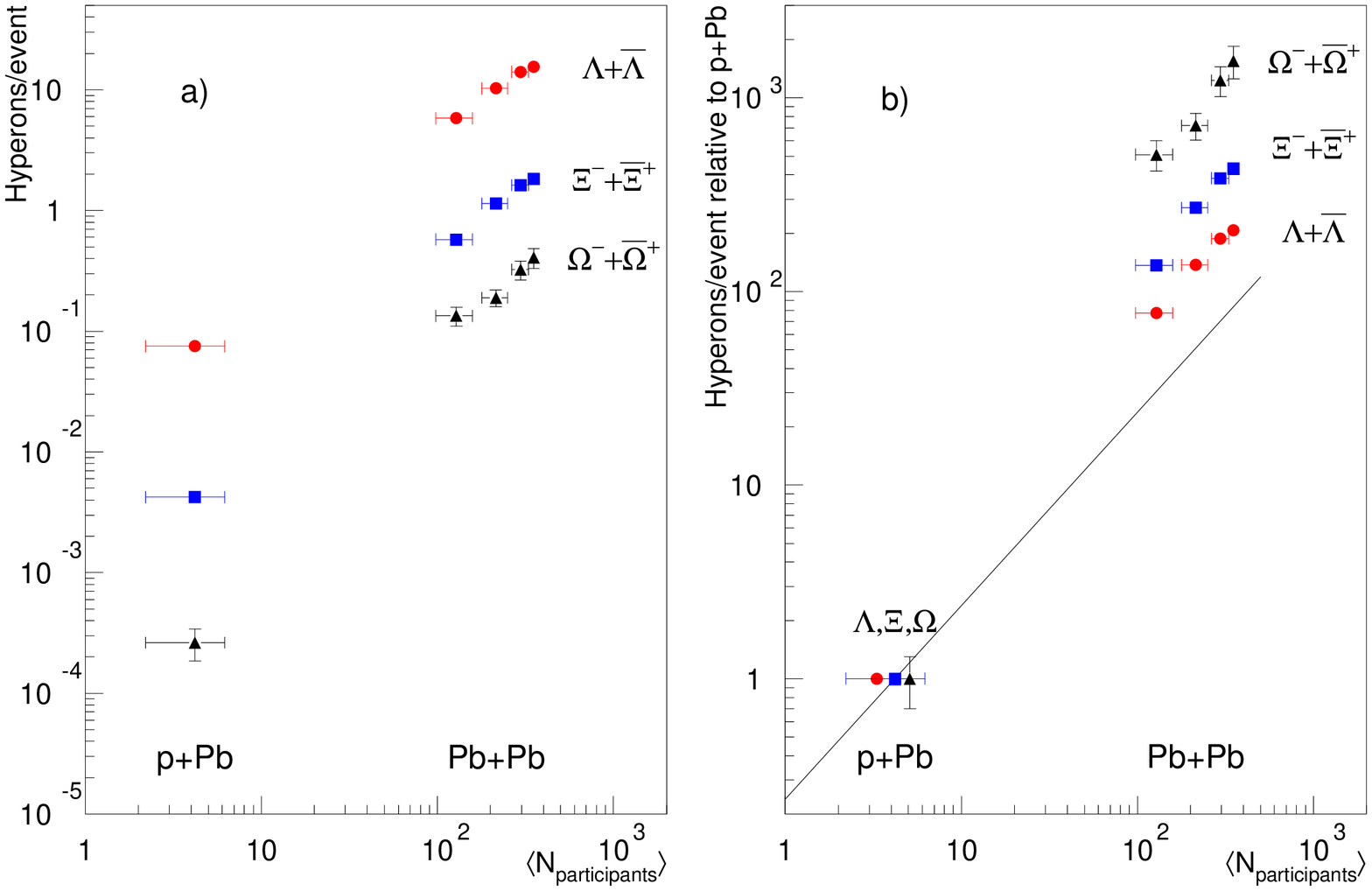}
\caption{\em a) Numero di $\Lambda$, $\Xi$ ed $\Omega$ prodotte per evento in
funzione del numero medio di partecipanti.  b) Numero di $\Lambda$,
$\Xi$ ed $\Omega$ prodotte per evento in unit\`a della corrispondente
produzione osservata in collisioni p-Pb. La linea continua indica la
proporzionalit\`a col numero di partecipanti.} 
\label{yields1}
\end{figure}

\noindent
In fig.~\ref{yields1}a \`e stato riportato il numero di $\Lambda$,
$\Xi$ ed $\Omega$ prodotte per evento nelle interazioni p-Pb e Pb-Pb
(cfr. tab.~\ref{tabextra}) in funzione del numero di partecipanti.
L'effetto del trigger di interazione presente nei dati p-Pb \`e stato preso
in considerazione nel calcolo della produzione di particelle, mentre
nel caso di interazioni Pb-Pb, l'effetto della selezione in centralit\`a
\`e presente coerentemente sia nel calcolo della produzione
di particelle che in quello del numero di partecipanti.
Le barre verticali di errore in fig.~\ref{yields1} mostrano le sole
incertezze statistiche e non includono i sistematici dovuti al
{\em feed-down} e quelli introdotti dalla procedura di estrapolazione.
Come discusso nel capitolo 4, essi sono comunque minori degli attuali errori
statistici.
Le barre orizzontali di errore includono invece le deviazioni standard del
numero di partecipanti in ciascuno dei quattro intervalli per
collisioni Pb-Pb e l'intero intervallo possibile per collisioni p-Pb.
In fig.~\ref{yields1}a \`e chiaramente visibile una crescita uniforme 
nella produzione di iperoni strani, a partire da collisioni p-Pb fino
agli urti Pb-Pb pi\`u centrali. In fig.~\ref{yields1}b la produzione
di ciascun tipo di particella \`e stata scalata in modo da porre
ad uno i relativi valori per interazioni p-Pb.
La produzione di iperoni risulta cos\`{\i} espressa in unit\`a della 
corrispondente produzione per interazioni p-Pb ed \`e confrontata con la
linea continua proporzionale al numero di partecipanti, passante per
il punto comune relativo all'interazione p-Pb.
Si pu\`o osservare che la produzione di tutti gli iperoni cresce con la
centralit\`a molto pi\`u velocemente rispetto a quanto previsto dalla
semplice proporziona-lit\`a col numero di
partecipanti (e quindi dal semplice effetto di sovrapposizione dei nucleoni
interagenti) e questo incremento aumenta in funzione del contenuto di
stranezza della particella, essendo maggiore per le $\Omega$ rispetto alle
 $\Xi$ e per le $\Xi$ rispetto alle $\Lambda$.
L'incremento delle $\Omega$ rispetto alla crescita col numero di partecipanti,
in particolare, si manifesta con pi\`u di un ordine di grandezza.

\begin{figure}[htb]
\centering
\includegraphics[scale=0.62,angle=-90,clip]%
                            {figure/t_pa1l.epsi}
\caption{\em Numero di particelle 
per evento, in unit\`a della corrispondente
produzione osservata in collisioni p-Pb, in funzione del
numero di partecipanti per a) $h^-$, $\PKzS$, $\PgL$ e $\PgXm$
 b) $\PagL$, $\PagXp$ e $\PgOm +\PagOp$.
La linea continua indica la
proporzionalit\`a col numero di partecipanti.} 
\label{yields2}
\end{figure}

In fig.~\ref{yields2} \`e riportata la produzione di tutte le particelle
e le antiparticelle identificate nell'analisi dei dati Pb-Pb,
 sempre espressa in unit\`a
della corrispondente produzione per interazioni p-Pb.
Le particelle sono state divise in due gruppi: quelle aventi almeno un 
{\em quark}
di valenza in comune con il nucleone, mostrate in fig.~\ref{yields2}a
(anche se in realt\`a $\PKzS$ contiene entrambi i contributi $\overline{d}s$
e $d\overline{s}$), e quelle che non hanno alcun {\em quark} in comune con il nucleone,
mostrate in fig.~\ref{yields2}b. L'analisi separata 
 dei due tipi di particelle rende ininfluenti
i possibili effetti dovuti alle loro 
diverse caratteristiche di produzione.
\par
\begin{figure}[htb]
\centering
\includegraphics[scale=0.5,clip]%
                            {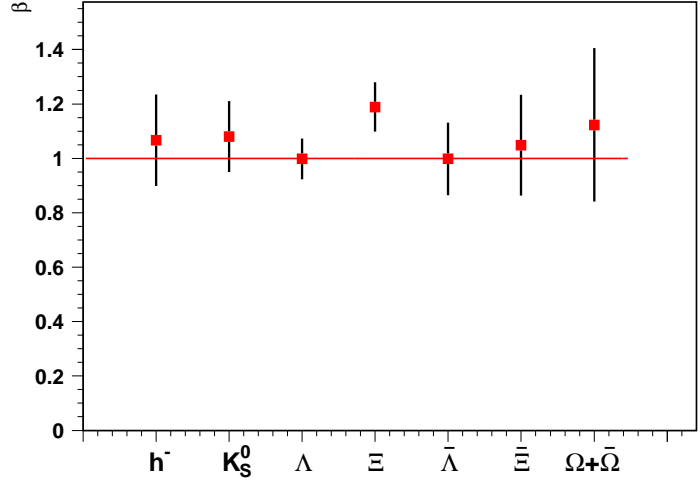}
\caption{\em Valori dell'esponente $\beta$ ricavati dai fit.}
\label{yields3}
\end{figure}

La statistica accumulata consente di studiare l'andamento della produzione
all'interno dell'intervallo di centralit\`a coperto dalle interazioni
Pb-Pb, corrispondente a $N_p > 100$. In tale intervallo, come risulta
dalla fig.~\ref{yields2}, la produzione di tutte le particelle segue
fedelmente le linee tratteggiate indicanti la proporzio\-nalit\`a col
numero di partecipanti. In maniera pi\`u quantitativa, una procedura
di {\em best fit}, applicata ai punti relativi alla sola produzione 
in collisioni Pb-Pb ed assumendo una dipendenza del tipo $N_p^\beta$, 
fornisce i valori dell'esponente $\beta$ riportati in 
fig.~\ref{yields3}: per tutte le particelle essi sono compatibili con 
l'unit\`a.

\begin{figure}[htb]
\centering
\includegraphics[scale=0.58,clip]%
                            {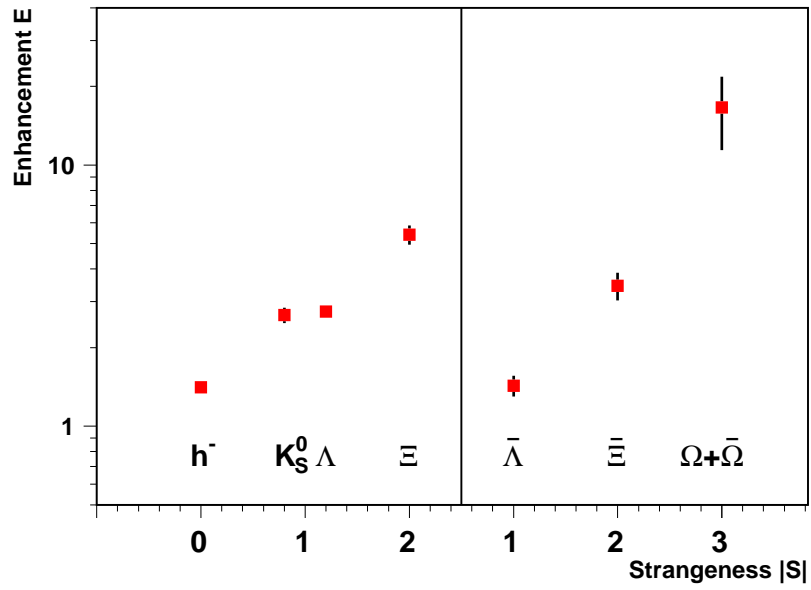}
\caption{\em Incremento di stranezza $E_{N_P}$ per le varie particelle
             identificate, in funzione del loro contenuto di stranezza.}
\label{yields4}
\end{figure}

\noindent
Ci\`o consente di dare una seconda definizione dell'incremento di stranezza
passando da interazioni p-Pb a quelle Pb-Pb, considerando il numero
di partecipanti come elemento di normalizzazione.
Secondo il modello geometrico descritto in appendice C, infatti, la
produzione di particelle per partecipante
rappresenta una quantit\`a totalmente svincolata dall'effetto di
sovrapposizione che si avrebbe se i nucleoni interagissero
 in maniera indipendente.
Si definisce allora l'incremento

\begin{equation}
E_{N_p}=\left ( \frac{\overline{P}_Y}{<{N_p}>} \right )_{Pb-Pb}
/
\left ( \frac{\overline{P}_Y}{<N_{p}>} \right )_{p-Pb}
\label{eqenh2}
\end{equation}

\noindent
dove $\overline{P}_Y$ indica la produzione estrapolata, riportata in 
tab.~\ref{tabextra} e $<N_{p}>$ \`e il numero medio di partecipanti 
relativo alla interazione considerata.
Graficamente, $E_{N_p}$ corrisponde alla distanza tra la linea continua
mostrata in fig.~\ref{yields2} e la linea tratteggiata corrispondente alla
particella considerata. I valori di $E_{N_p}$ calcolati
sono riportati in fig.~\ref{yields4} per i due tipi di particelle
prima definiti, in funzione del loro contenuto di stranezza.

\noindent
Confrontando tali valori con quelli riportati in fig.~\ref{figenhfig}, si nota
che l'andamento generale dell'incremento di stranezza \`e qualitativamente
riprodotto, anche se gli incrementi $E_{h^-}$ sono sistematicamente inferiori
a quelli definiti secondo la (\ref{eqenh2}).
Le negative, infatti, risultano crescere pi\`u velocemente del numero
di partecipanti, per cui l'incremento $E_{h^-}$ rappresenta una
sottostima  di quello calcolato rispetto al numero di
partecipanti, $E_{N_p}$. 
Le proporzioni \ref{line1} e \ref{line2} e la
discussione che le ha accompagnate continuano tuttavia a valere, anche
se riferite alla quantit\`a $E_{N_p}$.


%
%
%
%
%
%
\chapter{Confronto con modelli di produzione adronica}
\index{Confronto con modelli di produzione adronica}

\section{Introduzione}
\index{Introduzione}

L'aumento della produzione di particelle
strane e multi-strane nelle interazioni nucleo-nucleo,
riportato nel capitolo precedente, \`e stato valutato attraverso
il confronto con l'interazione protone-nucleo, cio\`e in
condizioni in cui non \`e prevista la formazione di QGP.
In questo tipo di analisi, la ricerca di fenomeni nuovi nelle
collisioni tra nuclei, indicati come caratteristici del QGP, si traduce come
ricerca di deviazioni rispetto alle osservazioni fatte su interazioni 
nucleone-nucleo; \`e perci\`o di fondamentale importanza una comprensione
il pi\`u possibile accurata 
della fisica che descrive l'interazione tra adroni, vista come
`` fondo convenzionale'' e presente anche nelle collisioni nucleo-nucleo.
\par
Ad alte energie, infatti, l'urto tra due nuclei non \`e direttamente
descrivibile come  semplice sovrapposizione di collisioni
nucleone-nucleone. Come gi\`a osservato in appendice C, per un nucleone
incidente che attraversa un nucleo bersaglio la scala di tempo tra due
collisioni  successive \`e troppo piccola rispetto a quella
richiesta dal processo di adronizzazione, per cui nell'interazione saranno
coinvolti ``oggetti intermedi'' risultanti dalle prime collisioni.
Questo nuovo tipo di dinamica potrebbe avere una influenza nelle
osservabili e determinare delle differenze tra le collisioni nucleo-nucleo
e protone-nucleo, senza dover necessariamente invocare una 
transizione di fase.
\par
D'altra parte, la formulazione di una teoria rigorosa, capace di descrivere
le collisioni protone-nucleo e gli eventi di fondo di tipo
convenzionale che hanno luogo in collisioni nucleo-nucleo, risulta ostacolata
dal carattere ``{\em soft}'' dei processi elementari coinvolti. La QCD 
perturbativa, infatti, non pu\`o essere applicata ad interazioni che
coinvolgono bassi momenti trasferiti, quali quelle coinvolgenti i 
nucleoni partecipanti.
Sono stati allora sviluppati modelli fenomenologici, basati sulla QCD, 
capaci di descrivere le collisioni tra nuclei estrapolando
i concetti sviluppati nello studio delle collisioni $e^+\,e^-$, adrone-adrone
e adrone-nucleo. Questi modelli di produzione adronica,  contenenti
la descrizione dell'intera evolu\-zione dinamica della collisione, sono in
grado di fornire predizioni quantitative e, in linea di principio,
potrebbero rivelare la presenza di fenomeni nuovi attraverso il
confronto con i dati sperimentali.
Essi, infatti, dovrebbero riprodurre fedelmente le osservabili fornite dallo
studio delle collisioni protone-nucleo e quindi le eventuali discrepanze tra le
loro predizioni ed i risultati relativi a collisioni nucleo-nucleo
 indicherebbero caratteristiche non interpretabili dalla fisica
convenzionale.
Questo progetto di ricerca appare subito molto arduo, anche
perch\`e la quantit\`a di fenomeni riproducibili attraverso
meccanismi derivati dalle cosiddette ``estrapo\-lazioni di concetti
di fisica adronica convenzionale'' si \`e ben presto dimostrata
superiore alle migliori previsioni. Il rischio \`e, quindi,
che questi modelli Monte Carlo, con l'introduzione di meccanismi
ad hoc, possano andare al di l\`a di una semplice
descrizione degli eventi di fondo nelle collisioni tra nuclei, 
nascondendo cos\`{\i} eventuali segnali di plasma, senza peraltro fornire
una precisa interpretazione teorica dei fenomeni riprodotti.
\par
Allo scopo di verificare la capacit\`a di tali modelli 
di riprodurre le caratteristiche generali delle interazioni
protone-piombo e piombo-piombo a 158~GeV/c per nucleone
ed in particolare la produzione di stranezza a rapidit\`a centrale, 
i risultati riportati nel capitolo 5 sono stati confrontati con
le previsioni fornite da due  Monte Carlo: VENUS 4.12 e RQMD 2.3.
\par
Nel seguito del capitolo verranno esposte le principali caratteristiche
dei due modelli, le modalit\`a usate nella generazione degli eventi
ed il metodo con il quale \`e stato condotto il confronto con i dati 
sperimentali. Seguiranno i risultati del confronto e le conclusioni che
\`e possibile trarre da esso.
\par
Questa analisi \`e stata oggetto di un poster presentato al
``LXXXXIV  Congresso Nazionale della Societ\`a Italiana di Fisica'',
tenutosi a Salerno dal 28 Settembre al 2 Ottobre 1998 ed una
sua versione preliminare \`e gi\`a stata pubblicata in [Vir98].

\section{I modelli VENUS e RQMD}
\index{I modelli VENUS e RQMD}
\label{par_mod}  

Due dei pi\`u accreditati e maggiormente usati modelli di produzione 
adronica in urti nucleari ad energie relativistiche sono il modello
VENUS ({\em Very Energetic NUclear Scattering}) [Wer93] e RQMD
({\em Relativistic Quantum Molecolar Dynamics}) [Sor95].
Essi rappresentano l'estensione a collisioni di tipo nucleo-nucleo
di due diversi modelli fenomenologici, ispirati alla QCD e sviluppati
per la descrizione dell'interazione tra adroni: rispettivamente 
 il {\em Dual parton model} [Cap87] e il {\em Lund model} [And87].
L'intera evoluzione dinamica della collisione tra nuclei \`e riprodotta 
all'interno di tali modelli attraverso meccanismi di interazione
che fanno uso del concetto di ``stringa'', intesa come tubo di flusso del 
campo di colore esistente tra  quarks.
Nel seguito del paragrafo verranno descritte le principali
caratteristiche di tali meccanismi, evidenziando le differenze
tra i due modelli e gli effetti che esse provocano sulle
osservabili fisiche da usare per il confronto.
\par
\begin{figure}[htb]
\centering
\includegraphics[scale=1.4,clip]%
                                {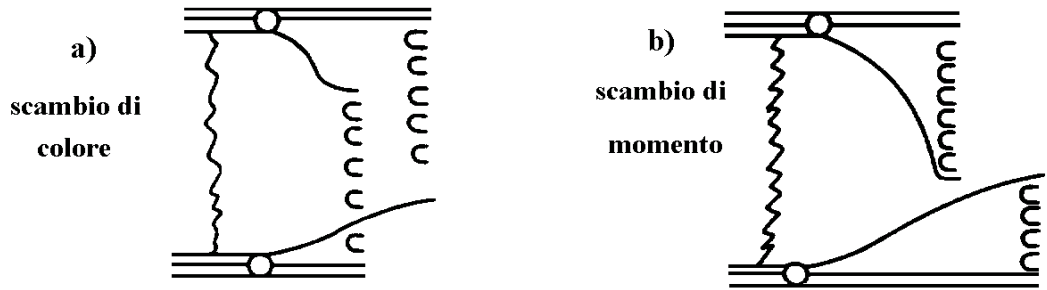}
\caption{\em Diagrammi che descrivono  la formazione di stringhe
in collisioni nucleari ad alta energia, dovute a) allo scambio
di colore in VENUS e b) allo scambio di momento in RQMD [Wah97].}
\label{figmod1}
\end{figure}

Entrambi i modelli trattano la collisione tra nuclei da un punto di vista
geometrico. I nucleoni dei nuclei proiettile e bersaglio si scontrano
viaggiando lungo traiettorie rettilinee e la probabilit\`a di interazione
\`e data dalla sezione d'urto anelastica $\sigma_{NN}$ definita in
appendice C. I nucleoni che subiscono almeno una interazione anelastica 
nel processo d'urto vengono definiti partecipanti,  i nucleoni
rimanenti vengono chiamati spettatori. In RQMD tale distinzione viene 
fatta anche a livello di  quarks, a seconda che essi interagiscano o meno
con gli altri quarks costituenti dei nucleoni.
Il risultato dell'interazione tra nucleoni \`e la formazione di una
stringa che lega due cariche di colore, costituite da una coppia
 quark - antiquark o quark - diquark. I meccanismi di
formazione delle stringhe, mostrati in fig.~\ref{figmod1}, sono
tuttavia differenti per i due modelli.

\noindent
In VENUS le stringhe sono formate attraverso lo ``scambio di colore''
(fig.~\ref{figmod1}a): un quark costituente di un nucleone bersaglio
\`e legato ad un  diquark appartenente ad un nucleone proiettile e
viceversa. Nell'interazione non vengono scambiate n\`e particelle n\`e momento,
 ma lo scambio di colore produce due nuove stringhe che acquistano una 
consistente energia grazie alla grande differenza di momento tra i partoni
legati. In RQMD, invece si ha uno ``scambio di momento'' tra i partoni
interagenti (fig.~\ref{figmod1}b) e le stringhe vengono formate tra il partone
che ha interagito  e i rispettivi spettatori 
dei nucleoni proiettile e bersaglio. La stringa viene eccitata
grazie al trasferimento di momento longitudinale tra gli adroni.
Nonostante le differenze nella loro formazione, le stringhe prodotte nei
due modelli hanno caratteristiche simili. La differenza
sostanziale nelle osservabili finali risiede nel diverso contenuto
a quark delle strighe: mentre nello scambio di momento le
stringhe contengono gli stessi quark dei nucleoni incidenti, nello scambio
di colore vengono formate stringhe utilizzando partoni provenienti da
differenti nucleoni; ci\`o pu\`o portare ad un incremento del contenuto
barionico nella regione centrale di rapidit\`a [Wer89].
\par
Seguendo l'evoluzione dinamica della collisione, la successiva 
produzione di particelle  \`e introdotta nei modelli attraverso
il processo di frammentazione delle stringhe. Il campo di colore induce
la produzione di una coppia $q - \overline{q}$, la quale scherma la forza
di colore tra le cariche originali e consente la formazione di due
sottostringhe. Queste continuano a dividersi secondo un processo
iterativo, finch\`e tutti i frammenti di stringa creati sono identificabili con
adroni stabili o con risonanze conosciute.
La procedura di frammentazione viene eseguita in maniera diversa nei due
 modelli,
con  differenze di natura principalmente  tecnica. Esse
 riguardano,
ad esempio, il modo di determinare i punti di rottura sulla superficie
delle stringhe. La differenza pi\`u sostanziale risiede nel fatto che in
RQMD una delle due sottostringhe prodotte per frammentazione della
stringa originaria \`e gi\`a identificato con un adrone
stabile e non frammenta pi\`u.
\par
Nelle collisioni tra nuclei ad alta energia le particelle prodotte nelle 
interazioni primarie possono interagire tra loro e con i nucleoni spettatori,
secondo un processo noto come {\em rescattering}. Questa eventualit\`a \`e
riprodotta nei modelli di produzione adronica introducendo una frammentazione
non indipendente delle stringhe. L'interazione pu\`o dunque avvenire
tra le stringhe originali, i frammenti di stringa intermedi, le risonanze
e gli adroni prodotti per frammentazione ed i nucleoni spettatori
del bersaglio e del proiettile. A tutti questi elementi, genericamente
indicati come ``oggetti'', vengono assegnate delle traiettorie nello
 spazio-tempo e l'interazione ha luogo se esse si avvicinano sufficientemente
l'una all'altra.
Nel modello VENUS viene associato il raggio $R_M$ agli oggetti con numero
barionico nullo ed il raggio $R_B$ agli oggetti rimanenti: l'interazione
avviene se la minima distanza tra le due traiettorie \`e minore della somma dei 
raggi dei relativi oggetti. La dimensione dei raggi \`e lasciata come 
parametro libero del modello. Il risultato dell'interazione tra oggetti
\`e la formazione di un ``agglomerato di quarks'' di {\em flavour}
 e momento somma di quelli degli oggetti costituenti. Successivamente
l'agglomerato decade in maniera iterativa, finch\`e non si riduce ad una 
risonanza o ad un adrone stabile.
\par
In RQMD, al contrario, la probabilit\`a di interazione \`e definita dalle
sezioni d'urto misurate sperimentalmente o, quando ci\`o non
\`e possibile, stimate attraverso 
vari modelli di interazione adronica.
In questo caso l'interazione coinvolge principalmente adroni,
in virt\`u del particolare meccanismo di frammentazione utilizzato, e procede
mediante collisioni binarie che portano alla formazione di risonanze,
le quali possono a loro volta interagire in base ai loro parametri
caratteristici misurati sperimentalmente. 
Questo secondo metodo ha il vantaggio di non utilizzare parametri liberi
e di fornire una interpretazione pi\`u realistica del processo di
 {\em rescattering}.
\par
In effetti,  la procedura di {\em rescattering} 
introduce le differenze pi\`u sostanziali nelle
predizioni dei due modelli. Essa influenza notevolmente le osservabili fisiche
ottenibili dai modelli e, tra l'altro, genera l'espansione trasversa delle 
particelle prodotte, dato che i momenti trasversi introdotti nella
frammentazione delle stringhe sono orientati a caso.
In RQMD le interazioni secondarie consentono al sistema interagente, 
costituito da un gas ideale di adroni e risonanze, di procedere verso uno stato
di equilibrio.
\par
I meccanismi introdotti all'interno dei modelli di produzione adronica
per simulare l'evoluzione dinamica della collisione sono regolati da un
certo numero di parametri interni, i cui valori vengono determinati attraverso
il confronto con i dati sperimentali. In VENUS, i parametri che regolano
la frammentazione delle stringhe sono individuati utilizzando dati
relativi agli urti $e^+\, e^-$ e leptone-nucleo ad alta energia; quelli
che regolano la formazione delle stringhe sono determinati
dall'accordo con i dati relativi agli urti adrone-adrone e, infine, i dati
sugli urti adrone-nucleo e nucleo-nucleo
permettono di fissare i parametri di geometria e di {\em rescattering}.
In tale modello i valori dei parametri cos\`{\i} determinati non vengono
cambiati a seconda del tipo di collisione studiata, per cui i meccani\-smi
introdotti sono intesi ad assicurare
 una interpretazione univoca e coerente di tutti 
i diversi processi di collisione ad alta energia.
RQMD, invece, \`e stato ideato appositamente per riprodurre le 
collisioni tra nuclei ed usa un modello (FRITIOF [Nil87]) gi\`a 
regolato e verificato  per lo studio dell'urto tra nucleoni.
Nello studio dell'urto tra nuclei, d'altra parte, l'introduzione in tale
modello delle 
sezioni d'urto e delle probabilit\`a di decadimento misurate 
consente di rinunciare all'uso di parametri liberi, permettendo
una verifica pi\`u diretta del meccanismo di interazione attraverso il 
confronto con i dati.

\begin{figure}[htb]
\centering
\includegraphics[scale=1.4,clip]%
                                {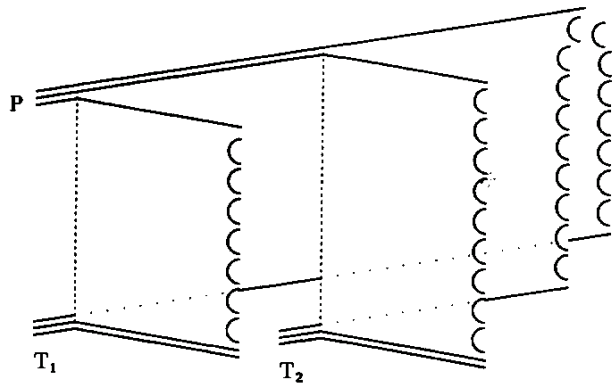}
\caption{\em Diagramma che descrive lo scambio multiplo di
colore tra nucleoni interagenti [Wer93].}
\label{venus1}
\end{figure}

Per quanto riguarda l'incremento nella produzione di stranezza nel passaggio
da collisioni nucleone-nucleone a interazioni tra nuclei, \`e importante
sottolineare che esso \`e in una certa misura contenuto nei modelli stessi,
perch\`e originato dalle
collisioni multiple dei nucleoni interagenti nella materia nucleare.
\par
\begin{figure}[htb]
\centering
\includegraphics[scale=1.3,angle=359.5,clip]%
                                {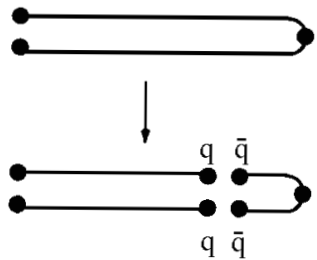}
\caption{\em Schema di frammentazione delle strighe doppie nel modello VENUS
[Wer93].}
\label{venus2}
\end{figure}

In VENUS, al crescere della densit\`a di materia nella regione di interazione,
 aumenta la probabilit\`a di scambi multipli di colore tra i nucleoni
interagenti, i quali avvengono secondo il diagramma illustrato in 
fig.~\ref{venus1}.
Procedendo da sinistra a destra, un nucleone del nucleo proiettile ($P$)
effettua uno scambio di colore con un nucleone del nucleo bersaglio ($T_1$).
Successivamente, l'interazione con un secondo nucleone del nucleo bersaglio 
($T_2$) determina un secondo scambio di colore con uno dei quarks
residui di $P$. L'ultimo quark di $P$ \`e legato mediante una doppia 
stringa ai quark rimanenti di $T_1$ e $T_2$. Le stringhe doppie
formate in scambi multipli di colore frammentano creando
coppie $q - \overline{q}$, secondo quanto mostrato in fig.~\ref{venus2}.

\noindent
La probabilit\`a di creare una coppia $s - \overline{s}$ \`e dunque doppia
rispetto alle stringhe ordinarie e ci\`o consente un aumento nella
produzione di stranezza in sistemi di collisione a densit\`a elevata.
Questo effetto risulta pi\`u evidente a rapidit\`a centrali, dato che la regione
centrale di interazione \`e quella a pi\`u alta concentrazione di 
materia interagente.
\par
In RQMD, le collisioni multiple subite dai quarks interagenti nello
stadio iniziale della collisione tra nuclei hanno l'effetto di modificare la
struttura a quark degli adroni formati per frammentazione.
I loro quark originali di valenza vengono rimpiazzati da quarks
provenienti dal mare e la probabilit\`a che un barione acquisti una o pi\`u
unit\`a di stranezza aumenta col numero di collisioni subite.
\par
Un ulteriore aumento di stranezza \`e generato nei modelli durante la
fase di {\em rescattering}. In VENUS, l'effetto delle interazioni secondarie
\`e quello di aumentare il momento trasverso delle particelle prodotte e
il numero di particelle strane nella regione di frammentazione del 
bersaglio e del proiettile, nelle quali \`e pi\`u alta la probabilit\`a
di interazione dei frammenti di stringa con i nucleoni spettatori.
In RQMD, la maggior parte dell'incremento di stranezza \`e prodotta durante
la fase adronica della collisione, dominata dalle interazioni tra risonanze.
Le coppie di quark non-strani sono preferenzialmente annichilate
in processi di formazione di risonanze e stringhe e vengono di tanto in
tanto sostituite da coppie di quarks strani nei corrispondenti
processi di decadimento.
\par
La necessit\`a di riprodurre i risultati sperimentali riguardanti la
produzione di stranezza e di coppie barioniche a rapidit\`a centrale ha
portato ad introdurre nel modello RQMD un ulteriore meccanismo che agisce
nello stadio di pre-equilibrio delle collisioni tra nuclei.
L'interazione tra stringhe in collisioni ad alta densit\`a nucleare \`e
modellata in termini di fusione di stringhe in ``funi di colore''.
Le funi possono essere considerate come un modello di materia
costituita da quarks deconfinati, dominata da eccitazioni
longitudinali. Il decadimento delle funi di colore, sempre descritto 
in modo fenomenologico, produce un aumento della molteplicit\`a
di particelle prodotte a rapidit\`a centrale e, in generale, porta ad una
modifica delle osservabili fisiche che tende inevitabilmente ad avvicinarle
alle previsioni fatte nell'ambito di modelli di QGP.
Le  modifiche apportate ai modelli per simulare le interazioni
che avvengono in ambienti ad alta densit\`a nucleare e di energia
si discostano chiaramente dall'originale progetto  che consiste
nella ricerca di fenomeni nuovi attraverso il confronto delle previsioni dei
modelli Monte Carlo con i risultati sperimentali. Infatti 
i fenomeni di tipo collettivo, quali l'agglomerato di quarks
introdotto in VENUS e le funi di colore introdotte in RQMD, 
rappresentano dei meccanismi
del tutto nuovi rispetto a quelli utilizzati nelle normali interazioni adroniche
meno complesse e tendono ad avvicinarsi alle idee che stanno alla base
dei fenomeni esotici che si vorrebbero evidenziare.

\section{Confronto con i risultati sperimentali}
\index{Confronto con i risultati sperimentali}

Allo scopo di confrontare i risultati sperimentali con le previsioni
dei due modelli sopra descritti, \`e stata eseguita una generazione in larga
scala di eventi p-Pb e Pb-Pb a 160~GeV/c per nucleone con i programmi VENUS
versione 4.12 e RQMD versione 2.3.
Gli eventi sono stati generati senza modificare i parametri interni dei
modelli ed usando le loro opzioni di base. 
 I numeri di eventi generati sono riportati nelle
prime due colonne di tab.~\ref{tabgen}, mentre la corrispondente statistica
accumulata dall'esperimento WA97 consiste di circa 120 milioni di 
eventi p-Pb e circa 200 milioni di eventi Pb-Pb.

\begin{table}[ht]
\centering
\caption{\em Statistica degli eventi generati con VENUS ed RQMD.}
 \begin{tabular}{|c|c|c|c|}
    \hline
            &    p-Pb    & Pb-Pb      &   Pb-Pb con trigger \\
   \hline
VENUS 4.12  &  500.000  &  20.000     &   6945                \\
RQMD 2.3    &  228.000  &  11.500     &   3825                \\
   \hline
  \end{tabular}
\label{tabgen}
\end{table}              

\noindent
Per riprodurre la selezione introdotta dal trigger di centralit\`a
presente nei dati Pb-Pb, le particelle prodotte nella simulazione 
sono state inseguite nell'apparato sperimentale per
mezzo del programma GEANT [Brun]. Questo programma, tra l'altro, \`e in grado
di simulare il rilascio di energia delle particelle che attraversano
i petali di scintillatore preposti alla selezione di eventi ad alta
moltepicit\`a.
La selezione  sperimentale di centralit\`a negli eventi generati
\`e stata riprodotta richiedendo che almeno 5 scintillatori su 6
 forniscano un segnale sopra una certa soglia e regolando questa in modo che
la distribuzione di molteplicit\`a di particelle cariche si avvicini
il pi\`u possibile a quella sperimentalmente misurata 
dai rivelatori di molteplicit\`a.
Il numero di tali eventi che soddisfano la condizione di trigger 
 \`e riportato nella terza colonna della tab.~\ref{tabgen} 
per entrambi i modelli e rappresenta il campione
statistico effettivamente utilizzato per il confronto con i dati sperimentali.

\begin{figure}[htb]
\centering
\includegraphics[scale=0.65,clip]%
                                {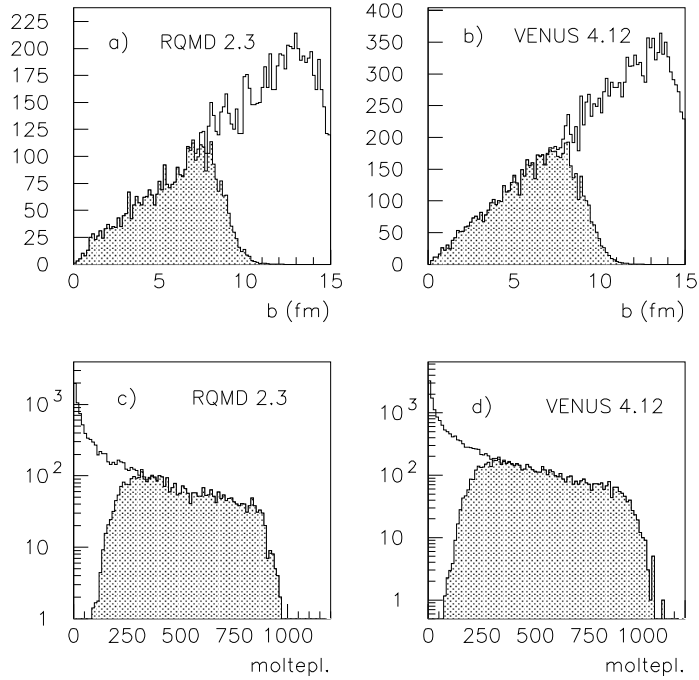}
\caption{\em Distribuzioni di parametro di impatto (a, b) e molteplicit\`a 
di particelle cariche nella regione
$2 < \eta < 4$ (c, d) per eventi Pb-Pb generati da RQMD e VENUS.
Le regioni in grigio indicano gli eventi selezionati dal trigger di
centralit\`a.}
\label{figul1}
\end{figure}

In fig.~\ref{figul1} \`e mostrata la distribuzione di parametro di impatto
(a, b) e di molteplicit\`a di particelle cariche nella regione
$2 < \eta < 4$ (c, d) per eventi Pb-Pb generati da RQMD e VENUS rispettivamente.
Le regioni in grigio individuano gli eventi selezionati dalla condizione di
 trigger riprodotta.

\begin{figure}[htb]
\centering
\includegraphics[scale=0.67,clip]%
                                {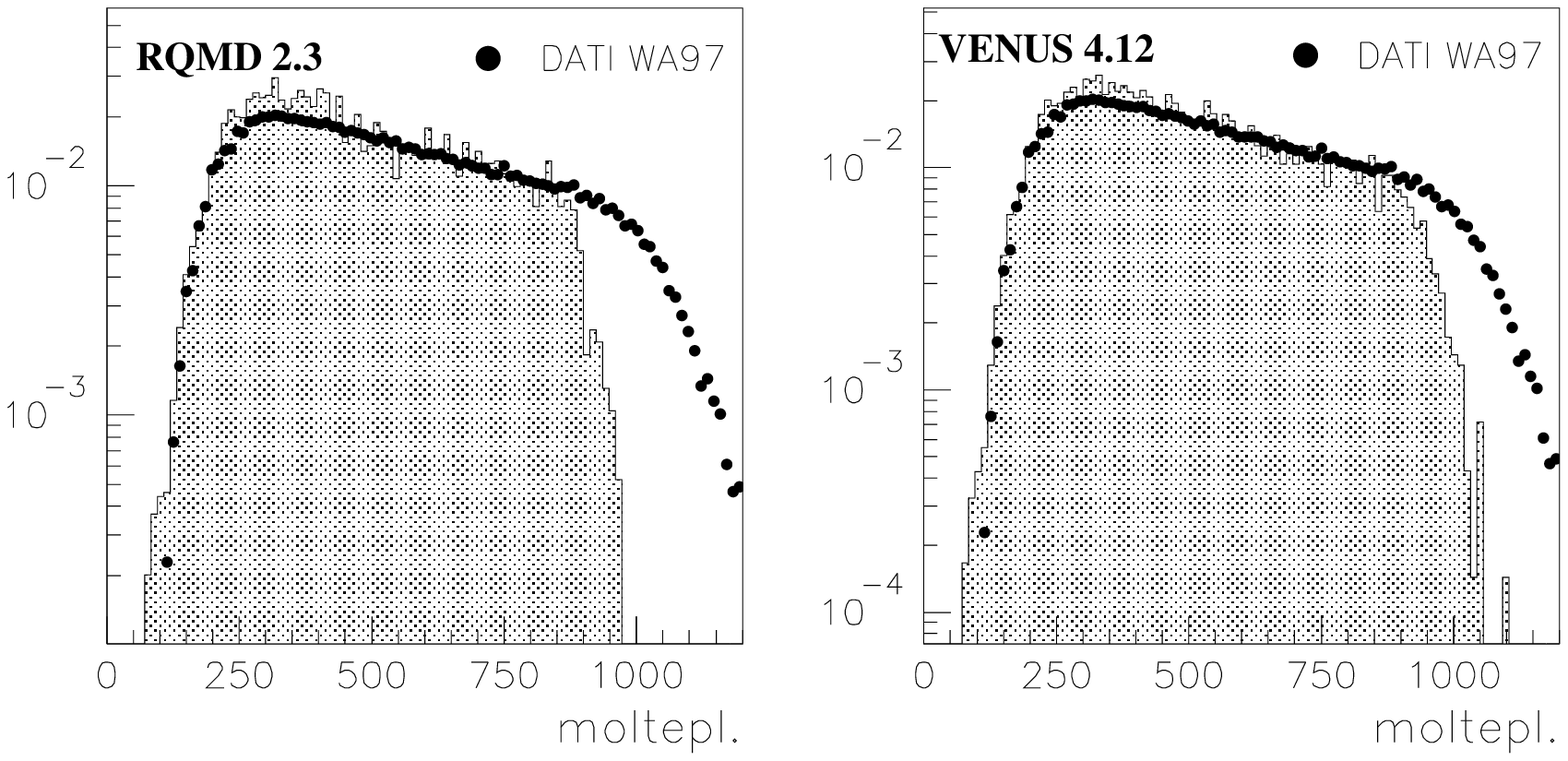}
\caption{\em Distribuzioni di molteplicit\`a prodotte da RQMD e VENUS
con la condizione di trigger. I pallini indicano l'analoga 
distribuzione sperimentalmente misurata.} 
\label{figul2}
\end{figure}

In fig.~\ref{figul2} le distribuzioni di molteplicit\`a prodotte dai due
modelli e seleziona\-te dal trigger sono confrontate con quella
misurata sperimentalmente. 
Entrambi i modelli riproducono in modo soddisfacente la distribuzione
misurata a basse molteplicit\`a, mentre la sottostimano ad alte molteplicit\`a;
la differenza \`e leggermente minore per gli eventi generati con VENUS.
\par
Le osservabili fisiche confrontate riguardano le particelle negative e
quelle strane
$\PKzS$, $\PgL$, $\PagL$, $\PgXm$, $\PagXp$, $\PgOm$ e $\PagOp$ e sono
\begin{description}
\item[a)] il numero di particelle prodotte 
    per interazione (nel caso di collisioni p-Pb) o per evento
    selezionato (nel caso di collisioni Pb-Pb); 
\item[b)]  le  temperature apparenti delle corrispondenti distribuzioni
            di massa trasversa.
\end{description}
La molteplicit\`a dei diversi tipi di particella \`e stata determinata negli 
eventi simulati semplicemente contando il loro numero all'interno delle
corrispondenti regioni  cinematiche di accettanza usate per i dati reali
e normalizzandolo al numero di eventi generati. La restrizione della
procedura di confronto all'interno delle  regioni cinematiche
di rivelazione delle diverse particelle consente di evitare eventuali
disaccordi dovuti alla estrapolazione effettuata sui dati sperimentali.
Per la determinazione delle temperature apparenti, \`e stato usato il metodo
della {\em Maximum Likelihood} con la funzione (\ref{eqlikely}), sempre
all'interno delle regioni di accettanza, secondo la procedura usata per lo
studio delle distribuzioni di massa trasversa  sperimentalmente misurate.
Nel confronto sono state considerate anche le particelle negative,
alle quali \`e stata assegnata la massa dei pioni, cos\`{\i} come fatto per i
dati reali; ci\`o ha consentito di verificare le previsioni
dei modelli su una osservabile
fisica comunemente accessibile in letteratura 
e legata alle caratteristiche globali della collisione.
Infine, i rapporti di produzione e gli incrementi di stranezza forniti dai
modelli Monte Carlo sono stati infine calcolati all'interno della regione
comune $|y-y_{CM}|<0.5$ e $p_T>0~GeV/c$, utilizzata per l'estrapolazione dei
risultati sperimentali.
\par
In tab.~\ref{tabgen1} sono riportati i numeri delle particelle generate
in interazioni p-Pb e Pb-Pb 
che cadono all'interno delle rispettive regioni di accettanza.

\begin{table}[ht]
\centering
\caption{\em Numero di particelle generate con VENUS ed RQMD in interazioni
p-Pb e Pb-Pb 
all'interno delle rispettive regioni di accettanza.}
 \begin{tabular}{|c|c|c|c|c|}
    \hline
        &\multicolumn{2}{|c|}{p-Pb} &\multicolumn{2}{|c|}{Pb-Pb}    \\\hline
        &  VENUS    &  RQMD     &  VENUS   &  RQMD  \\ 
   \hline 
$h^-$      & 268287    & 95408     & 509914   & 289796  \\
$\PKzS$    & 12039     & 4910      &  34562   & 23296   \\
$\PgL$     & 10673     & 4171      &  22157   & 17315   \\
$\PagL$    & 3559      & 1200      &  8572    & 2359    \\
$\PgXm$    & 882       & 131       &  3549    &  1480   \\
$\PagXp$   & 447       & 57        &  1964    &  392    \\
$\PgOm$    & 176       & 5         &  916     &  95     \\
$\PagOp$   & 123       & 4         &  768     &  40     \\
   \hline
  \end{tabular}
\label{tabgen1}
\end{table}              

\begin{figure}[htb]
\centering
\includegraphics[scale=0.7,clip]%
                                {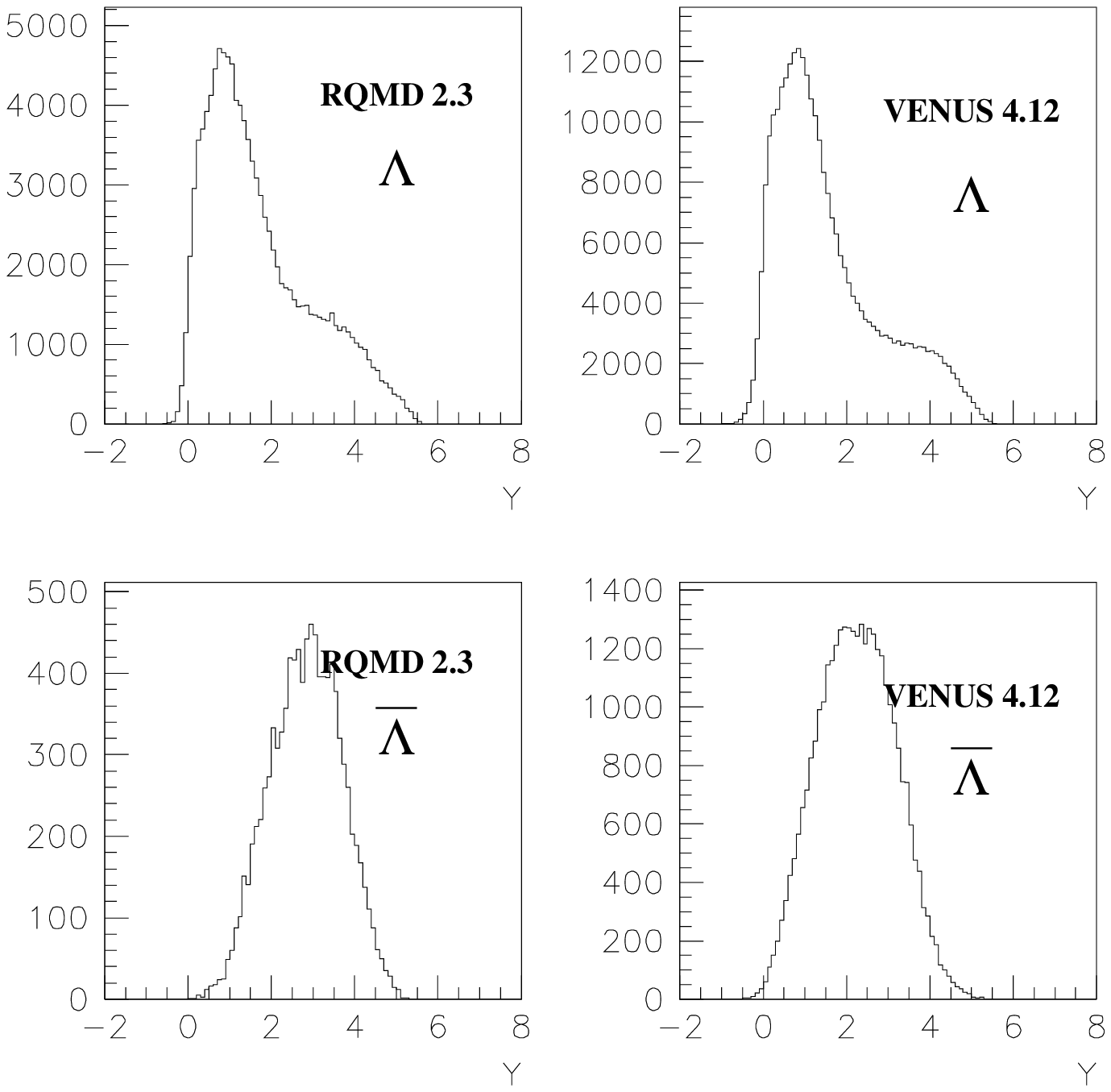}
\caption{\em Distribuzioni di rapidit\`a prodotte da RQMD e VENUS
             per $\PgL$ e $\PagL$ in interazioni p-Pb.}
\label{figul3}
\end{figure}

\begin{figure}[htb]
\centering
\includegraphics[scale=0.7,clip]%
                                {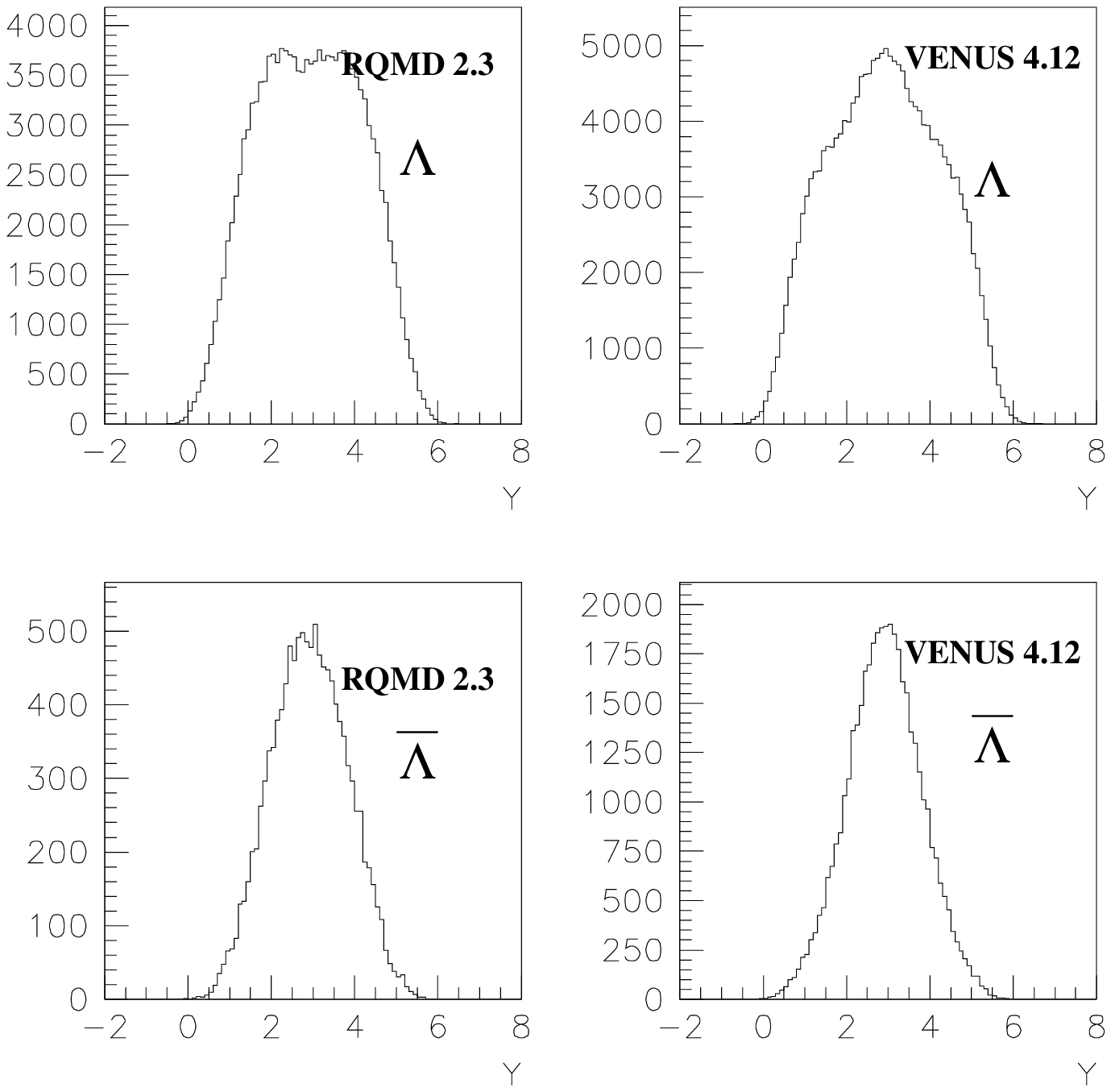}
\caption{\em Distribuzioni di rapidit\`a prodotte da RQMD e VENUS
             per $\PgL$ e $\PagL$ in interazioni Pb-Pb.}
\label{figul4}
\end{figure}

\noindent
In fig.~\ref{figul3} e fig.~\ref{figul4} sono invece riportate le
distribuzioni di rapidit\`a di $\PgL$ e $\PagL$ generate da VENUS e RQMD
rispettivamente per collisioni p-Pb e Pb-Pb.

\noindent
Si pu\`o notare come entrambi i modelli riproducano le principali
caratteristiche delle distribuzioni di $\PgL$ e $\PagL$ in collisioni
p-Pb, descritte nel par.~\ref{par_rapid}. VENUS, tuttavia, prevede una
produzione di $\PgL$ e $\PagL$ nella
regione di frammentazione del bersaglio maggiore di quella ottenuta
con RQMD, come si deduce soprattutto
 dallo spostamento a sinistra della relativa distribuzione di $\PagL$ rispetto a quella di RQMD.  Le distribuzioni
di $\PgL$ e $\PagL$ nel Pb-Pb appaiono invece simmetriche in entrambi
i modelli, in accordo con i risultati richiamati nel par.~\ref{par_rapid},
anche se sensibili differenze sono evidenti nella forma delle distribuzioni
 delle $\PgL$.  Queste differenze  sono attribuibili al
diverso meccanismo di {\em rescattering} presente nei due modelli [Wer93].

\subsection{Confronto per le interazioni p-Pb}
\index{Confronto per le interazioni p-Pb}

\begin{figure}[htb]
\centering
\includegraphics[scale=0.55,clip]%
                                {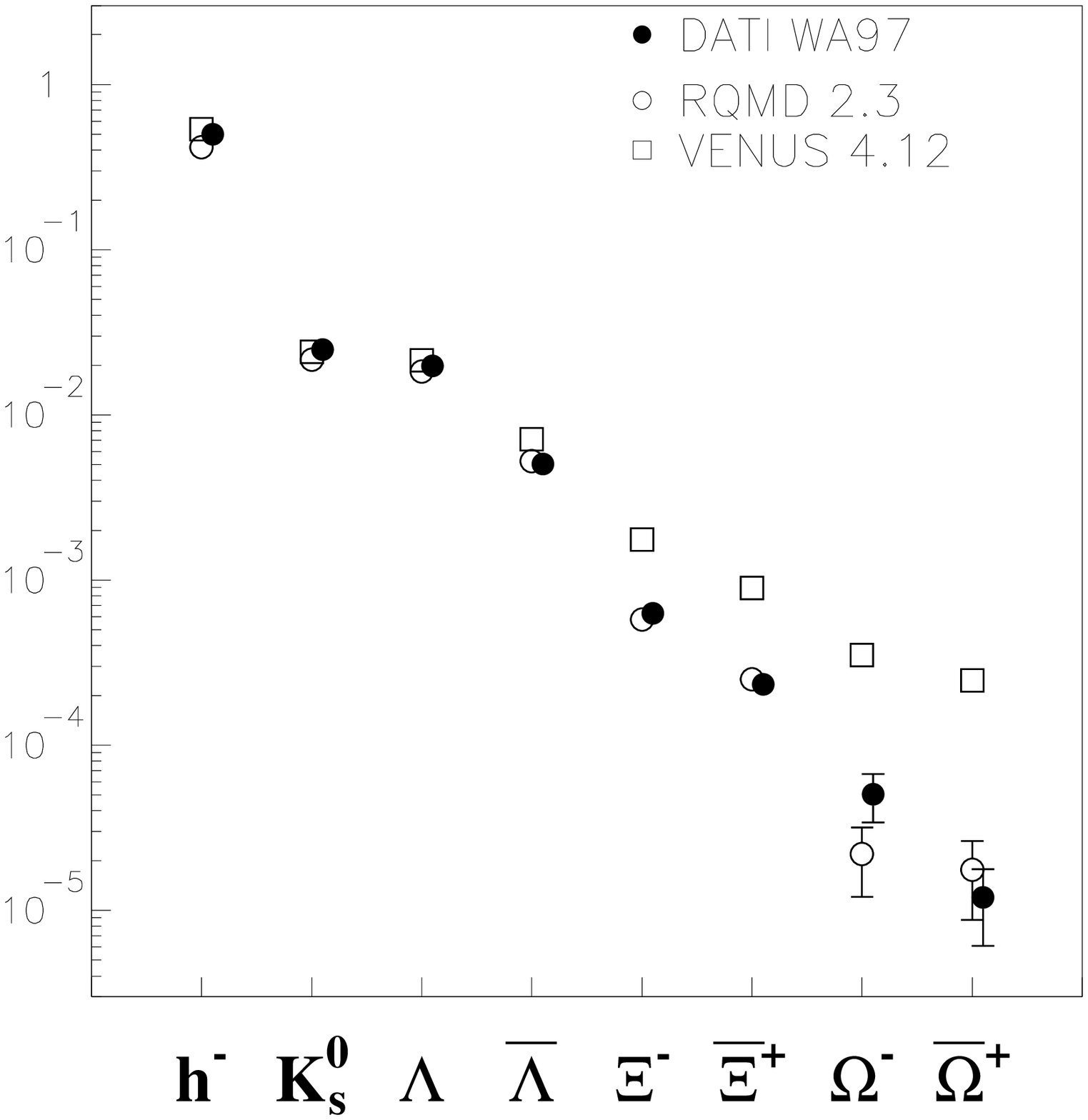}
\caption{\em Particelle per interazione p-Pb misurate dall'esperimento
WA97 nelle regioni di accettanza (cerchi neri), confrontate con le particelle per evento generate da VENUS (quadrati bianchi) ed RQMD (cerchi bianchi).}
 \label{figul5}
\end{figure}

In fig.~\ref{figul5} sono riportati i numeri di particelle per
interazione p-Pb, misurati dall'esperimento WA97 all'interno
delle rispettive regioni di accettanza, confrontati
con il numero di particelle per evento generate da VENUS ed RQMD.

\noindent
La scala orizzontale non ha significato, se non per
separare le differenti particelle. Nella maggior parte dei casi,
le barre di errore sono contenute nelle dimensioni dei simboli 
utilizzati. Si pu\`o notare che entrambi i modelli riproducono
bene la molteplicit\`a delle negative, $\PKzS$ e $\PgL$. Per le
successive particelle, VENUS mostra una significativa sovrastima rispetto
ai dati, che diventa sempre pi\`u consistente all'aumentare del
contenuto di stranezza della particella e passando da particelle 
ad antiparticelle.
Le discrepanze diventano di circa un ordine di grandezza in
corrispondenza delle $\PgOm$ e $\PagOp$. RQMD, al contrario, riproduce 
in maniera soddisfacente le molteplicit\`a di tutte le particelle
esaminate.
\par
\begin{figure}[htb]
\centering
\includegraphics[scale=0.55,clip]%
                                {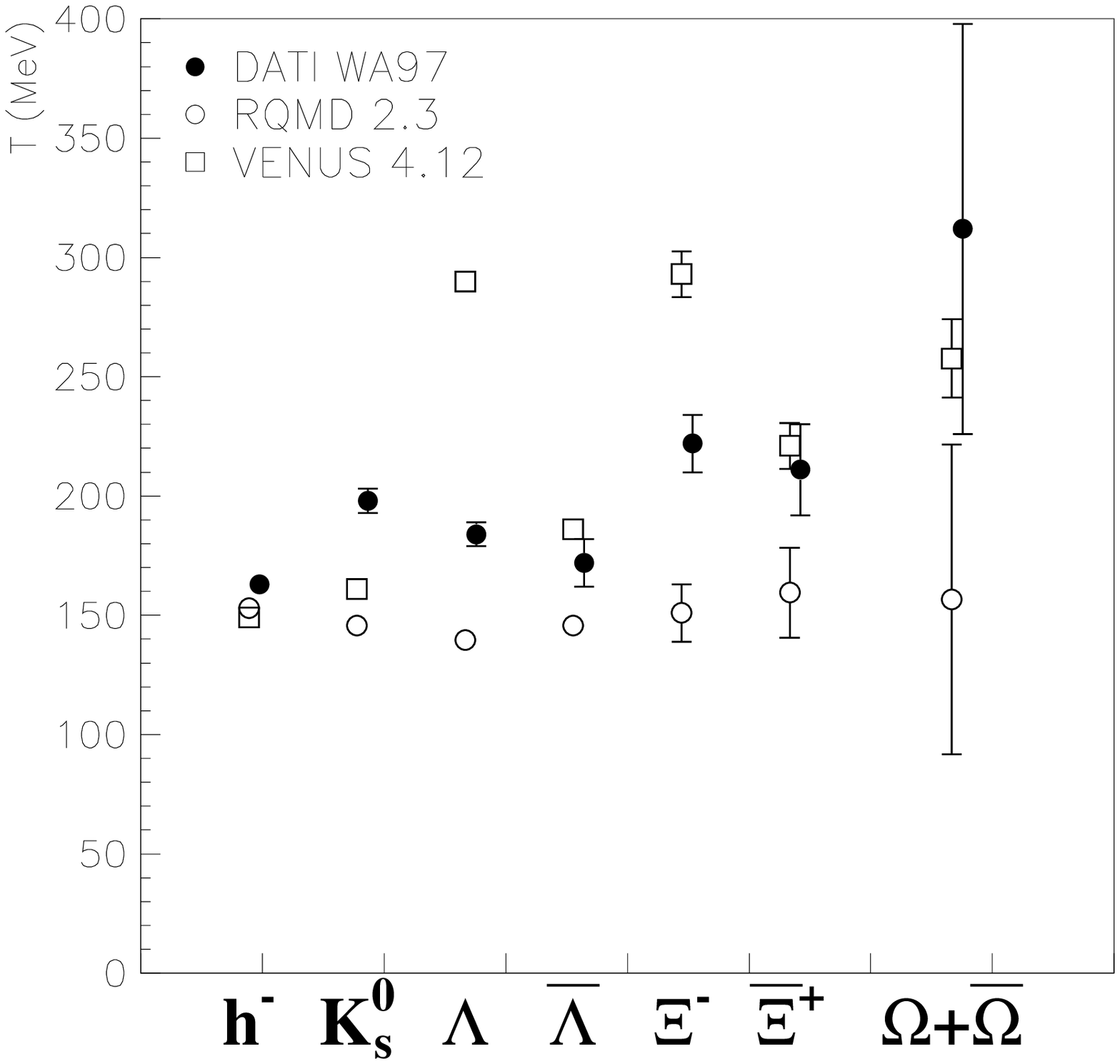}
\caption{\em Temperature apparenti misurate dall'esperimento
WA97 nelle regioni di accettanza per interazione p-Pb 
(cerchi neri), confrontate con quelle previste da VENUS
 (quadrati bianchi) ed RQMD (cerchi bianchi).}
 \label{figul6}
\end{figure}

Il confronto tra le temperature apparenti misurate e quelle previste dai
modelli \`e riportato in fig.~\ref{figul6}, nella quale sono stati
usati gli stessi simboli della figura precedente.
Le $\PgOm$ e $\PagOp$ sono state analizzate insieme a causa della bassa
statistica del campione sperimentale.

Si pu\`o osservare che RQMD prevede lo stesso valore
di temperatura apparente per tutte le particelle analizzate,
pari a circa $150~MeV$.
VENUS si discosta da questo comportamento e prevede temperature apparenti
che sembrano in graduale crescita con la massa della particella, 
tranne per le $\PgL$ e $\PgXm$, che mostrano valori sensibilmente pi\`u alti
e del tutto incompatibili con i dati sperimentali.
Questa discrepanza era stata gi\`a osservata in interazioni p-Ar e p-Xe per le 
$\PgL$ [Der91] ed ora la sua presenza \`e confermata anche per le $\PgXm$.
Essa pu\`o essere dovuta al meccanismo di {\em rescattering}, il quale aumenta 
la produzione di particelle ad alto momento trasverso nella regione di 
frammentazione del bersaglio, a causa delle interazioni dei frammenti
di stringa con i nucleoni spettatori.
Le $\PgL$ e  le $\PgXm$, infatti,
sono prodotte principalmente in tale regione ed \`e ragionevole supporre
che questo effetto si propaghi in qualche misura anche nella regione centrale
di rapidit\`a.
Le $\PagL$ e $\PagXp$ 
vengono prodotte pi\`u centralmente e dunque risentono meno di questo effetto.
\par
La regolazione dei paramentri di {\em rescattering} in VENUS potrebbe
essere anche alla base della discrepanza osservata per le molteplicit\`a
di particelle multi-strane in fig.~\ref{figul5}.
 Infatti, se si considera la struttura gerarchica
 dei suoi parametri, un cambiamento nei parametri introdotti per la
descrizione dei precedenti
stadi della collisione introdurebbe un disaccordo del modello con i 
numerosi dati relativi ad interazioni $e^+\, e^-$, leptone-nucleo
ed adrone-adrone.  D'altra parte, le osservabili comunemente misurate in
collisioni nucleo-nucleo, come le molteplicit\`a di
negative, $\PKzS$ e $\PgL$,  sono in buon accordo
con quelle dell'esperimento WA97 e di altri esperimenti [Too87], [Dem82];
perci\`o \`e possibile che i parametri del {\em rescattering} non
siano ancora stati opportunamente regolati per riprodurre le
caratteristiche di produzione delle particelle multi-strane.

\begin{figure}[htb]
\centering
\includegraphics[scale=0.53,clip]%
                                {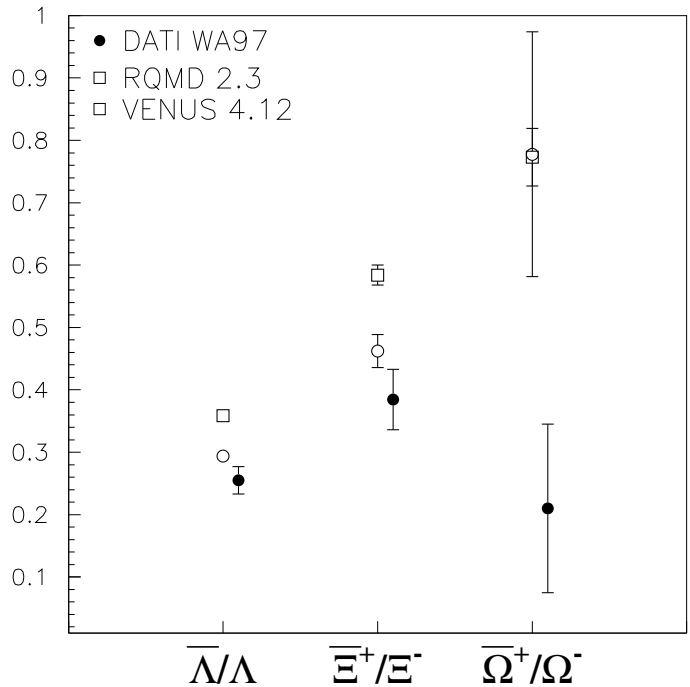}
\caption{\em Rapporti del tipo antiparticella/particella misurati
   da WA97 in interazioni p-Pb e confrontati con le previsioni di VENUS
   ed RQMD.}
 \label{figul7}
\end{figure}

In fig.~\ref{figul7}, infine, \`e mostrato il confronto dei
rapporti di produzione del tipo antiparticella/particella
per interazioni p-Pb. Entrambi i modelli concordano nel
prevedere un andamento crescente di tale rapporto col contenuto
di stranezza delle particelle coinvolte, in maniera simile a quanto
osservato per gli analoghi rapporti sperimentali in interazioni Pb-Pb 
(fig.~\ref{figratiover}).
Il rapporto $\frac{\PagOp}{\PgOm}$ misurato da WA97 sembra invece
discostarsi da questo andamento. I consistenti errori statistici  non
consentono tuttavia di trarre conclusioni definitive circa 
l'eventuale disaccordo con i modelli.

\subsection{Confronto per le interazioni Pb-Pb}
\index{Confronto per le interazioni Pb-Pb}

\begin{figure}[htb]
\centering
\includegraphics[scale=0.55,clip]%
                                {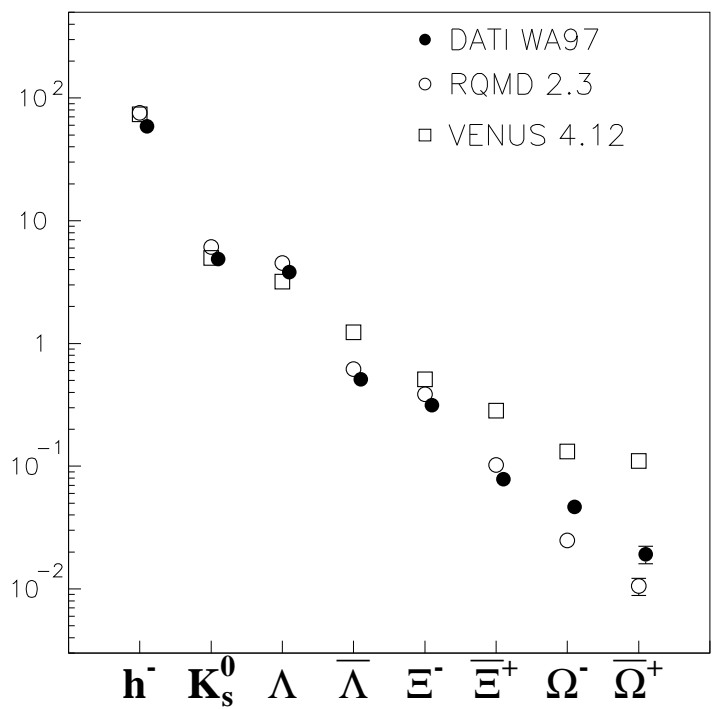}
\caption{\em Particelle per evento Pb-Pb misurate dall'esperimento
WA97 nelle regioni di accettanza (cerchi neri), confrontate con le 
particelle per evento generate da VENUS
 (quadrati bianchi) ed RQMD (cerchi bianchi).}
 \label{figul8}
\end{figure}

Il confronto della molteplicit\`a di particelle prodotte in collisioni Pb-Pb
nelle rispettive regioni di accettanza \`e riportato in fig.~\ref{figul8},
con le stesse notazioni delle figure precedenti.

\noindent
L'accordo con i dati sperimentali \`e soddisfacente per le molteplicit\`a
delle negative, $\PKzS$ e $\PgL$, mentre VENUS sovrastima nuovamente
le molteplicit\`a delle particelle successive, raggiungendo una
discrepanza di circa un ordine di grandezza per le $\PgOm$ e $\PagOp$.
RQMD mostra, invece, un discreto accordo per quanto riguarda le $\PagL$,
 $\PgXm$ e $\PagXp$, mentre sottostima la produzione di $\PgOm$ e $\PagOp$
di circa un fattore due.

\begin{figure}[htb]
\centering
\includegraphics[scale=0.55,clip]%
                                {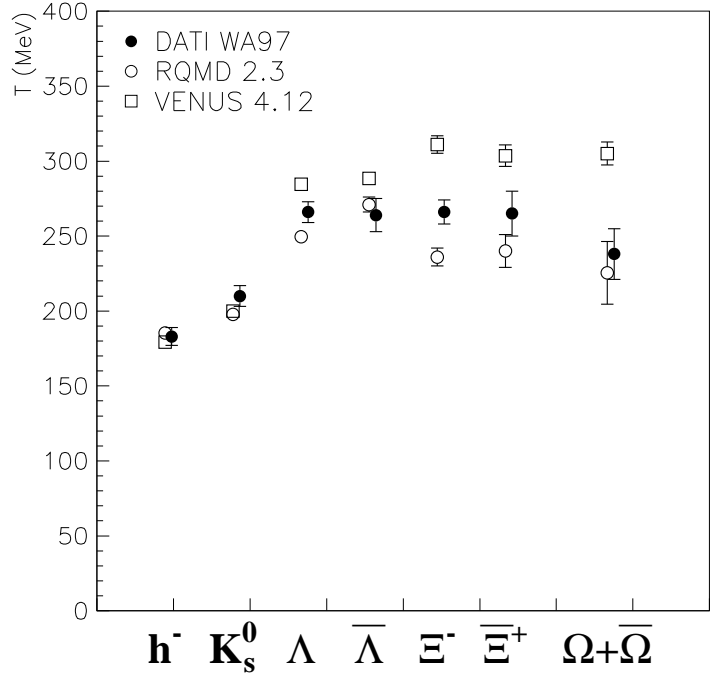}
\caption{\em Temperature apparenti misurate dall'esperimento
WA97 nelle regioni di accettanza per interazione Pb-Pb 
(cerchi neri), confrontate con quelle previste da VENUS
 (quadrati bianchi) ed RQMD (cerchi bianchi).}
 \label{figul9}
\end{figure}

Il confronto delle temperature apparenti, mostrato in fig.\ref{figul9},
rivela un generale accordo dei due modelli con i dati di WA97.
Le previsioni di RQMD, in particolare, riproducono fedelmente
le temperature apparenti misurate, secondo quanto \`e stato gi\`a
osservato in [Hec98] e discusso nel par.~\ref{par_tempris}.
La deviazione dei dati di WA97 dalla crescita lineare della 
temperatura apparente con la massa della particella introdotta dalla
relazione (\ref{eqxu}), sembra dunque essere confermata dalle previsioni di
entrambi i modelli. L'espansione trasversa  introdotta
all'interno dei modelli di produzione adronica \`e originata
nella fase di {\em rescattering}, determinata dalle interazioni delle particelle
 nello stato finale.

\begin{figure}[htb]
\centering
\includegraphics[scale=0.53,clip]%
                                {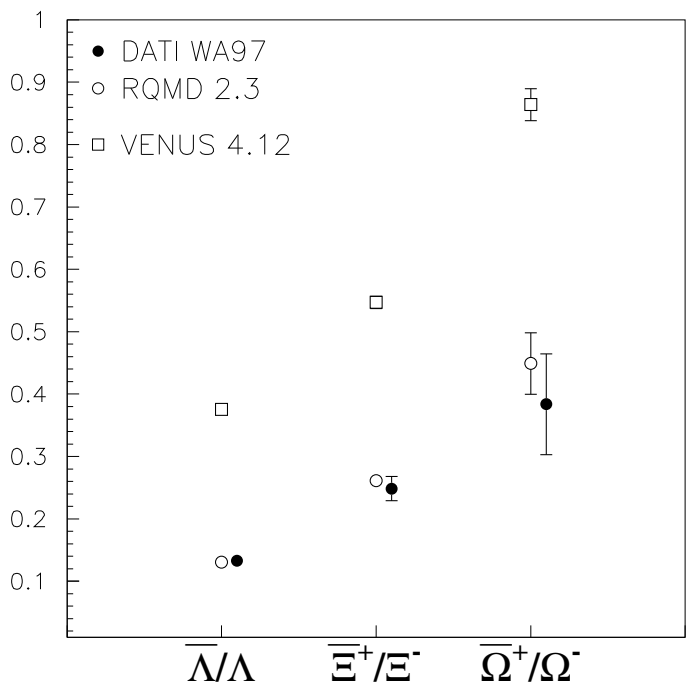}
\caption{\em Rapporti del tipo antiparticella/particella misurati
   da WA97 in interazioni Pb-Pb e confrontati con le previsioni di VENUS
   ed RQMD.}
 \label{figul10}
\end{figure}

In fig.~\ref{figul10} si pu\`o infine osservare come entrambi i modelli
 riproducano
l'andamento crescente dei rapporti del tipo antiparticella/particella
al crescere del loro contenuto di stranezza.
Tuttavia, mentre RQMD ne riproduce correttamente anche i valori, VENUS li
sovrastima di un fattore circa tre.

\subsection{Confronto dell'incremento di stranezza}
\index{Confronto dell'incremento di stranezza}

L'incremento di stranezza ($E_{h^-}$) nel 
passaggio da interazioni p-Pb a Pb-Pb,
valutato normalizzando la produzione di particelle strane a
quella delle negative, \`e stato calcolato nei modelli nella
regione $|y-y_{CM}|<0.5$ e $p_T>0~GeV/c$ e confrontato con quello
ottenuto estrapolando alla stessa regione i dati di WA97.

\begin{figure}[htb]
\centering
\includegraphics[scale=0.55,clip]%
                                {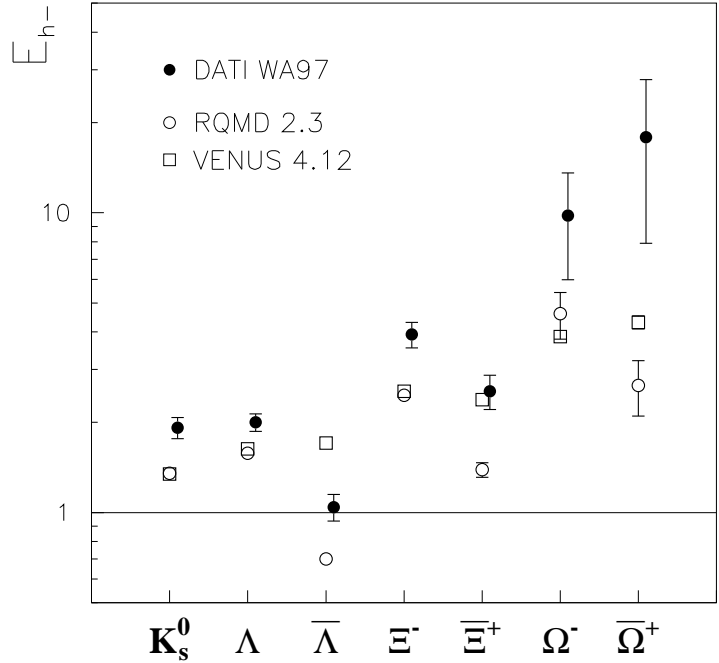}
\caption{\em Incremento di stranezza in interazioni Pb-Pb rispetto ad
 interazioni p-Pb, normalizzato alla produzione di particelle negative.
I cerchi neri indicano le misure di WA97, i quadrati ed i cerchi bianchi 
indicano le predizioni di VENUS ed RQMD.}
 \label{figul11}
\end{figure}

\noindent
In fig.\ref{figul11} \`e possibile notare come anche i modelli predicano
un incremento di stranezza ed il suo andamento in funzione del contenuto 
di stranezza risulti in generale compatibile con
quello osservato sperimentalmente. I risultati dei due modelli 
concordano per quanto concerne gli incrementi di $\PKzS$, $\PgL$,
$\PgXm$ e $\PgOm$, mentre VENUS prevede un incremento per le antiparticelle
maggiore di quello rilevato da RQMD, come gi\`a osservato in precedenza.
Dato che VENUS non \`e in grado di riprodurre le molteplicit\`a
di particelle multi-strane misurate nell'interazione p-Pb, usata
come riferimento per valutare gli incrementi di stranezza, \`e
preferibile discutere il confronto con i dati del solo modello RQMD.
Quest'ultimo sottostima tutti gli incrementi misurati ed in maniera
particolare quelli delle antiparticelle.
\`E opportuno osservare che se si introducono gli errori
sistematici di estrapolazione, riportati
in tab.~\ref{taberrextra},
si ha come effetto quello di diminuire la discrepanza negli 
incrementi di $\PKzS$ 
e di aumentarla per gli incrementi di $\PgOm$ e $\PagOp$, in entrambi i casi
 di circa il $15\%$.
Dal punto di vista numerico, gli incrementi di stranezza previsti da RQMD 
seguono le seguenti proporzioni

\begin{equation}
\left\{
\begin{array}{rccccc}
E_{h^-}(\PKzS)\sim & E_{h^-}(\PgL) &:& E_{h^-}(\PgXm) &:& E_{h^-}(\PgOm) \\
                      &    1      &:& (1.56\pm 0.06)  &:& (2.9\pm 0.5)  
\end{array}
\right.
\label{uline1}
\end{equation}
\vspace{0.5cm}
\begin{equation}
\left\{
\begin{array}{ccccc}
E_{h^-}(\PagL) &:& E_{h^-}(\PagXp) &:& E_{h^-}(\PagOp) \\
     1        &:& (2.0\pm 0.1)  &:& (4\pm 1)  
\end{array}
\right.
\label{uline2}
\end{equation}
\vspace{0.3cm}

\noindent
da confrontare con le (\ref{line1}) (\ref{line2}), valide per i dati di WA97.
Le proporzioni tra gli incrementi forniti da RQMD
risultano compatibili con quelle sperimentali entro due deviazioni standard.

\subsection{Confronto della produzione di stranezza in funzione della
centralit\`a}  
\index{Confronto della produzione di stranezza in funzione della
centralit\`a}  

\begin{figure}[htb]
\centering
\includegraphics[scale=0.6,clip]%
                                {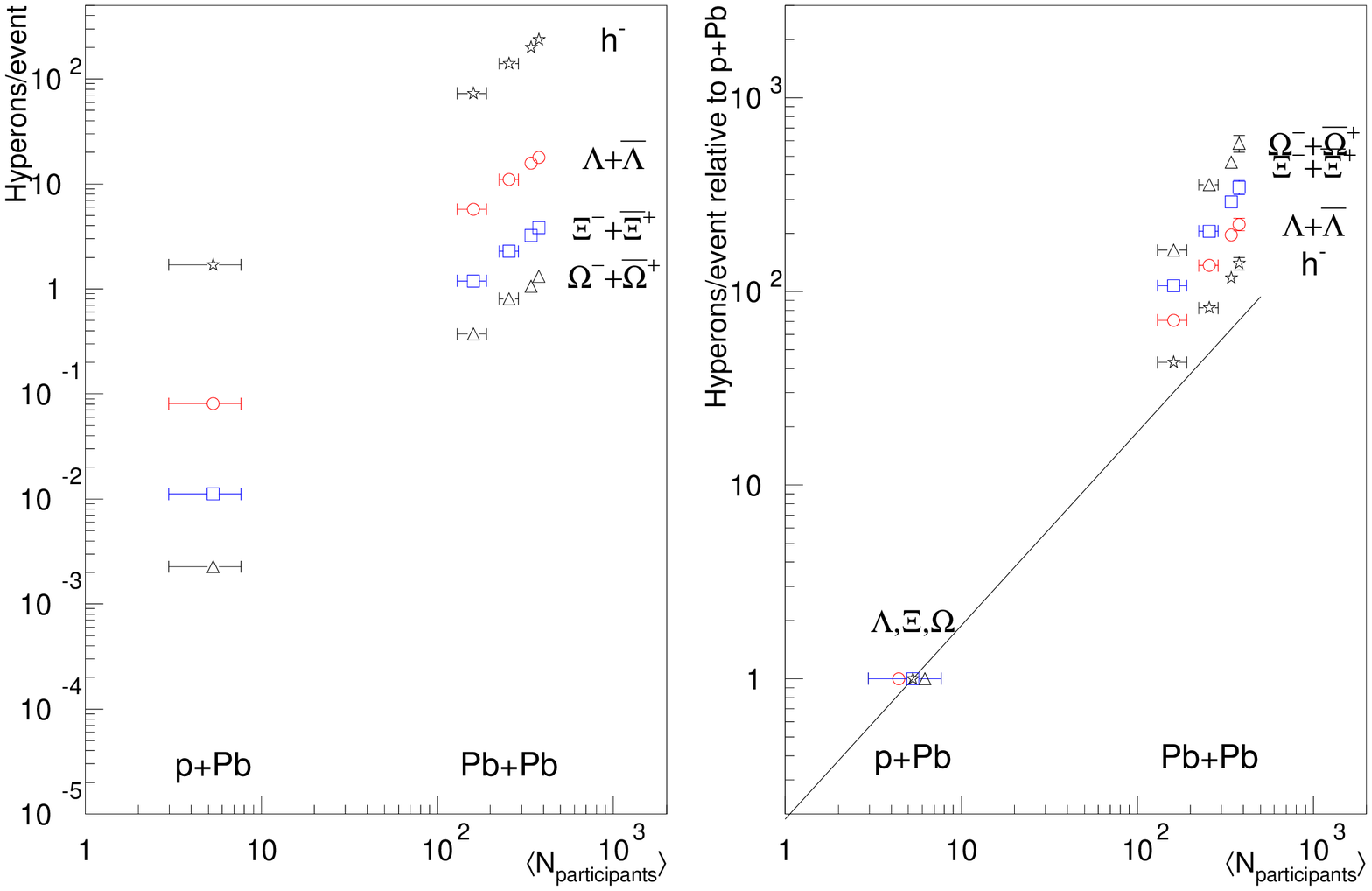}
\caption{\em a) Numero di $h^-$, $\PgL$, $\Xi$ e $\Omega$ 
  prodotte per evento generato da VENUS 
in funzione del numero medio di partecipanti.
b) Numero di $h^-$, $\PgL$, $\Xi$ e $\Omega$ per evento in unit\`a
della corrispondente produzione osservata in collisioni p-Pb.} 
 \label{figul12}
\end{figure}

\begin{figure}[htb]
\centering
\includegraphics[scale=0.6,clip]%
                                {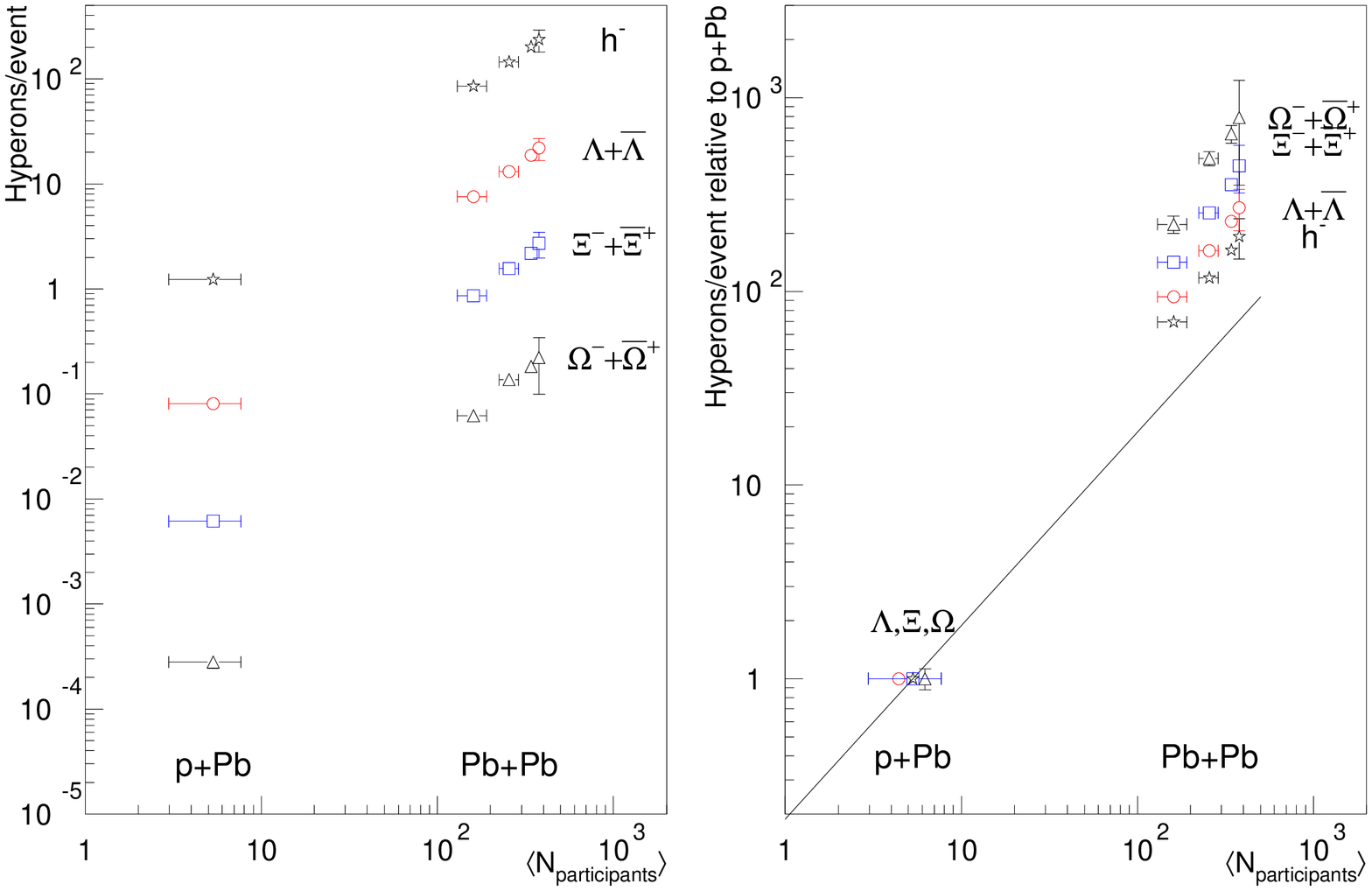}
\caption{\em a) Numero di $h^-$, $\PgL$, $\Xi$ e $\Omega$ 
  prodotte per evento generato da RQMD   
in funzione del numero medio di partecipanti.
b) Numero di $h^-$, $\PgL$, $\Xi$ e $\Omega$ per evento in unit\`a
della corrispondente produzione osservata in collisioni p-Pb.} 
 \label{figul13}
\end{figure}

L'analisi della produzione di stranezza in funzione 
della centralit\`a dell'evento pu\`o essere condotta in maniera
analoga a quanto fatto per i dati di WA97. Nelle distribuzioni di
 molteplicit\`a di particelle cariche mostrate in fig.~\ref{figul2} sono
stati individuati gli stessi intervalli scelti per la distribuzione
sperimentale e, all'interno di ciascuno di essi, \`e stato calcolato il
numero medio di partecipanti alla collisione, sfruttando l'informazione fornita
direttamente dal modello VENUS. 
I numeri cos\`{\i} ottenuti e quelli calcolati 
applicando  ai dati Monte Carlo il modello 
geometrico di collisione nucleare descritto in 
appendice C e nel par.~\ref{par_mol}, differiscono di non pi\`u del $10\%$ nel
primo intervallo e di circa il $4\%$ negli intervalli successivi.
La scala di centralit\`a  fornita dai modelli Monte Carlo
\`e dunque leggermente diversa da quella adoperata per i dati sperimentali.
La produzione di $h^-$, $\PgL$, $\Xi$ ed $\Omega$ in funzione del numero
di partecipanti, ottenuta da VENUS ed RQMD nella regione $|y-y_{CM}|<0.5$
e $p_T>0~GeV/c$ \`e mostrata rispettivamente in fig.~\ref{figul12} e
fig.~\ref{figul13}, secondo una rappresentazione grafica uguale a
quella utilizzata per i dati reali (fig.~\ref{yields1}).

\noindent
Si pu\`o osservare come entrambi i modelli riproducano qualitativamente sia
 l'incre\-mento di produzione di stranezza passando da collisioni p-Pb a
collisioni Pb-Pb (in misura maggiore rispetto a quanto previsto
dalla proporzionalit\`a  col numero di partecipanti), sia la gerarchia 
degli incrementi (crescenti con il contenuto di stranezza della particella)
gi\`a osservata sui dati. 
\par
Per quanto riguarda le negative, in particolare, anche i modelli prevedono
una crescita con la centralit\`a dell'evento pi\`u veloce di quella del
numero di partecipanti, cos\`{\i} come osservato nei dati reali. 
I modelli, d'altra parte,  consentono l'identificazione delle 
particelle negative all'interno 
della regione di accettanza sperimentale e la tab.~\ref{tabneg}
riassume le stime delle percentuali di particelle non strane
$h^-_{NS}$ ($e^-$, $\mu^-$, $\tau^-$, $\rho^-$, $\overline{p}$) e strane
$h^-_{S}$ ($K^-$, $\Sigma^-$, $\overline{\Sigma}^-$, $\PgXm$, $\PgOm$, ....)
presenti nelle negative, fornita da VENUS ed RQMD in collisioni p-Pb e Pb-Pb.

\begin{table}[ht]
\centering
\caption{\em Percentuali di particelle non strane
$h^-_{NS}$ ($e^-$, $\mu^-$, $\tau^-$, $\rho^-$, $\overline{p}$) e strane
$h^-_{S}$ ($K^-$, $\Sigma^-$, $\overline{\Sigma}^-$, $\PgXm$, $\PgOm$ ....)
presenti nell negative, fatta da VENUS ed RQMD in collisioni p-Pb e Pb-Pb.} 
 \begin{tabular}{|c|c|c|c|c|}
    \hline
            &   \multicolumn{2}{|c|}{p-Pb} & \multicolumn{2}{|c|}{Pb-Pb} \\     
\hline   
            &   VENUS    & RQMD       &   VENUS  &  RQMD  \\ 
   \hline
$h^-_{NS}/{h^-}$ &  89.8\%  & 86.2\%  &  86.4\%  & 83.1\% \\[0.25cm]   
$h^-_{S}/{h^-}$ &    10.2\%  & 13.8\%  &  13.6\%  & 16.9\% \\[0.25cm]   
   \hline
  \end{tabular}
\label{tabneg}
\end{table}              

\noindent
L'incremento delle negative rispetto al numero di partecipanti che si avrebbe
se fossero interamente costituite da particelle non strane
sarebbe, secondo la definizione (\ref{eqenh2}):

\begin{equation}
E_{N_p}(h^-_{NS})=\frac{\left(\frac{h^-_{NS}}{h^-}\right)_{Pb-Pb}}%
                       {\left(\frac{h^-_{NS}}{h^-}\right)_{p-Pb}}%
\times E_{N_p}(h^-)=0.96\times E_{N_p}(h^-)   
\label{eqneg}  
\end{equation} 

\noindent
dove il fattore moltiplicativo, calcolato sostituendo i valori di
 tab.~\ref{tabneg}, non cambia a seconda del modello usato. Gli
incrementi delle negative rispetto al numero di partecipanti 
osservati sperimentalmente e previsti dai modelli sono 
dunque influenzati dall'aumento di produzione delle particelle strane 
in esse contenute solo per un $4\%$, per cui essi sono quasi interamente
attribuibili ad un effetto intrinseco nella dinamica della collisione.

\begin{figure}[htb]
\centering
\includegraphics[scale=0.6,clip]%
                                {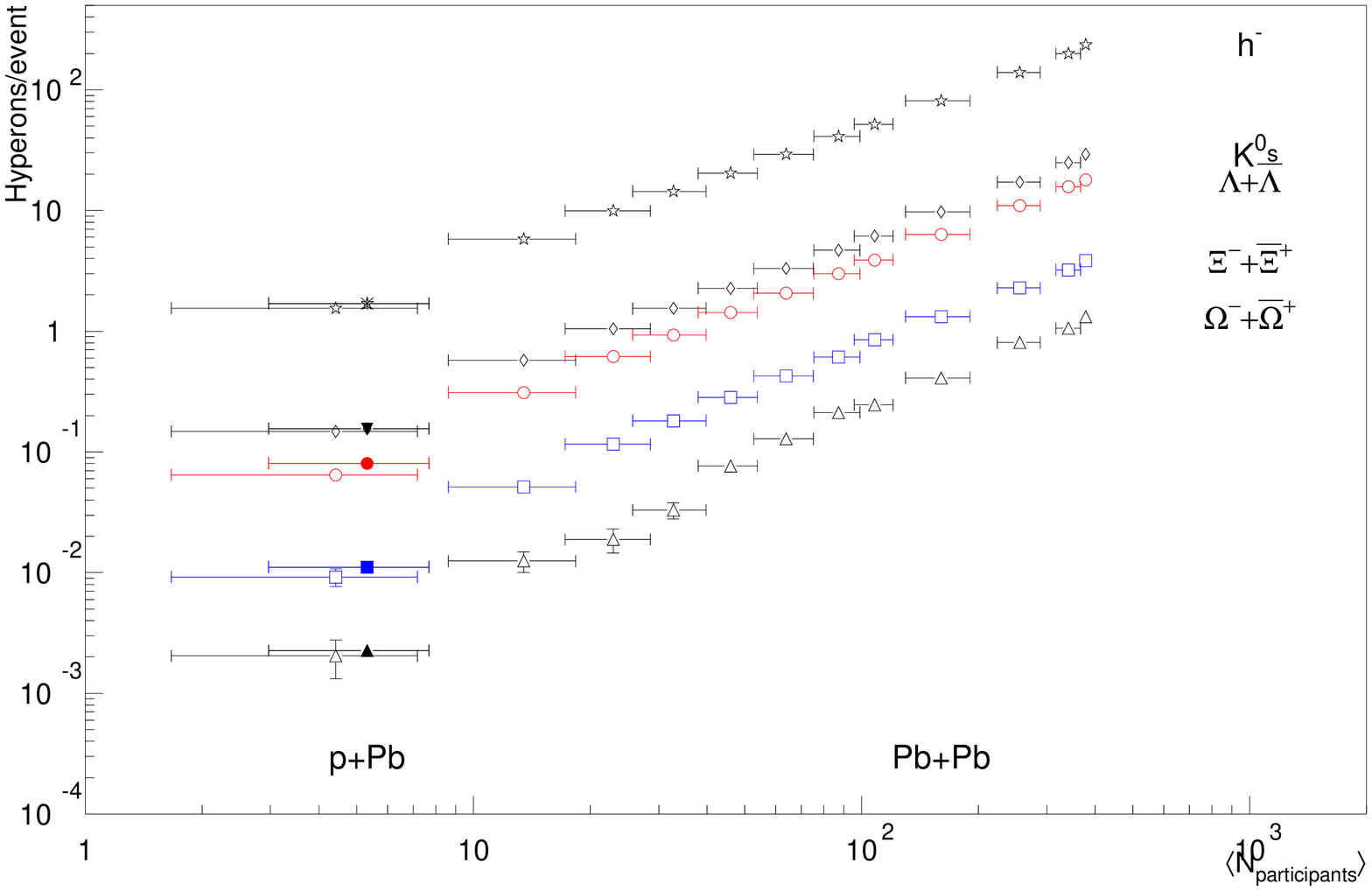}
\caption{\em Numero di $h^-$, $\PgL$, $\Xi$ e $\Omega$ prodotte per
evento generato da VENUS in funzione del numero medio di partecipanti.
La condizione di trigger sulle collisioni Pb-Pb \`e stata rilasciata
e in nero sono indicati i numeri relativi alle collisioni p-Pb.} 
 \label{figul14}
\end{figure}

\begin{figure}[htb]
\centering
\includegraphics[scale=0.55,clip]%
                                {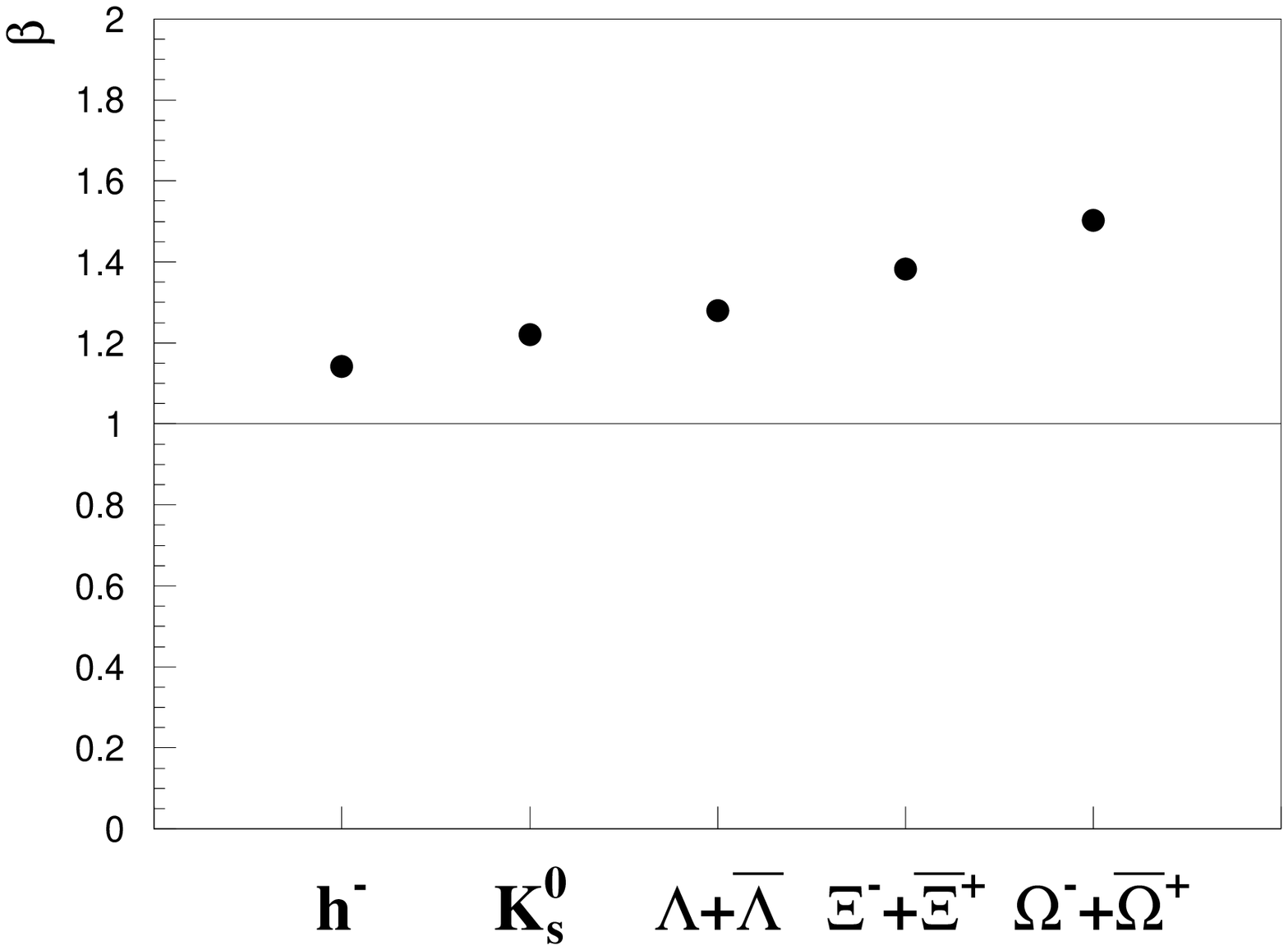}
\caption{\em Valori dell'esponente $\beta$ ricavati dai fit sui dati
di VENUS.}  
 \label{figul15}
\end{figure}

L'utilizzo  dei modelli consente di
 determinare il tipo di dipendenza della produzione delle
 particelle strane in funzione della centralit\`a. Per questo studio
si \`e ritenuto opportuno rilasciare la condizione di trigger riprodotta negli 
eventi simulati, onde sfruttare anche  l'informazione proveniente 
dall'intervallo di centralit\`a non coperto dall'esperimento WA97.
Data l'esigua statistica delle particelle multi-strane generate con RQMD,
\`e stato utilizzato per questo studio solamente il modello VENUS.
In fig.~\ref{figul14} \`e mostrato il numero di $h^-$, $\PgL$, $\Xi$ e
$\Omega$ prodotte da VENUS in collisioni Pb-Pb di qualsiasi parametro di 
impatto, in funzione del loro numero di partecipanti.  
I punti in nero  indicano la produzione delle stesse particelle prevista 
in interazioni p-Pb.
Il modello prevede quindi un incremento continuo della produzione di negative
e particelle strane in funzione della centralit\`a, passando da collisioni p-Pb a collisioni Pb-Pb. Assumendo una dipendenza del tipo $N_p^\beta$, la
procedura di {\em best fit} applicata ai punti di fig.~\ref{figul14} 
fornisce i valori dell'esponente $\beta$ riportati in fig.~\ref{figul15}
(da confrontare con  fig.~\ref{yields3}, relativa ai dati sperimentali).   
Per tutte le particelle, quindi, la crescita della produzione \`e pi\`u
 rapida del numero di partecipanti ed
avviene tanto pi\`u velocemente quanto maggiore \`e il contenuto di stranezza
della particella considerata.
Questo scenario \`e sostanzialmente differente da quello suggerito dai dati
sperimentali, per i quali la produzione di stranezza nell'intervallo
di centralit\`a coperto dall'esperimento WA97 
appare proporzionale al numero di partecipanti 
il parametro $\beta$ non dipende dal tipo di particella considerata.
Uno studio ana\-logo, condotto sulla statistica disponibile di eventi RQMD,
ha portato a risultati compatibili con quelli previsti dal modello VENUS.

\section{Discussione}
\index{Discussione}

Il confronto dei dati sperimentali con le previsioni di modelli Monte Carlo
consente una migliore comprensione dei processi
che hanno luogo durante le collisioni studiate, attraverso la
verifica dei meccanismi in essi introdotti per riprodurli.
Il disaccordo di VENUS con i dati sperimentali relativi sia al
numero di particelle multi-strane prodotte che alle distribuzioni
di massa trasversa di $\PgL$ e $\PgXm$ \`e attribuibile ad una inadeguata
descrizione dello stadio finale della collisione. I parametri liberi
$R_M$ e $R_B$ che regolano  il {\em rescattering} sono stati determinati
 dal confronto con i dati relativi ad interazioni protone-nucleo e 
nucleo-nucleo precedenti agli esperimenti con fasci di piombo all' SPS del CERN.
\`E dunque possibile che una loro migliore regolazione con dati pi\`u 
recenti possa ridurne il disaccordo. In ogni caso, il tentativo
di concentrare la mancanza di informazione circa il complicato processo
di {\em rescattering} entro pochi parametri si traduce in un modello molto
semplificato, quale  quello di ''agglomerato di quarks'' descritto nel 
par.~\ref{par_mod}, che pu\`o risultare intrinsecamente inadeguato.
\par
Il pi\`u complesso modello di {\em rescattering} introdotto in RQMD, che
peraltro fa riferimento a parametri misurati sperimentalmente, permette di
riprodurre meglio i dati relativi a collisioni p-Pb e Pb-Pb.
L'accordo tra le temperature apparenti di $\Xi$ e $\Omega$ misurate
in collisioni Pb-Pb e quelle previste dal modello rappresenta
una verifica dei meccanismi in esso presenti e consente di interpretare
il risultato sperimentale in uno scenario in cui gli adroni multi-strani
si disaccoppiano prematuramente dal resto del sistema interagente formato
in seguito alla collisione, prima che si sia sviluppato un apprezzabile 
flusso trasverso.  
\par  
Per quanto riguarda l'incremento della produzione di stranezza nel passaggio 
da collisioni p-Pb a Pb-Pb, l'accordo di massima del modello con i dati
sperimentali non consente una interpretazione altrettanto immediata.
I meccanismi collettivi introdotti in RQMD per riprodurre
tale incremento, ed in particolare il meccanismo della
fusione di stringhe in funi di colore, rappresentano concetti nuovi
rispetto alla sovrapposizione di collisioni adroniche elementari, quasi
quanto lo stato di QGP. Le modifiche che tali
meccanismi introducono nelle osservabili fisiche non permettono tuttavia una
loro differenziazione rispetto a quelle previste in caso di transizione
di fase di QCD, per cui la verifica o il
rigetto di una o dell'altra ipotesi
non possono aver luogo in queste condizioni.
La speranza \`e che l'osservazione di una
discontinuit\`a nella produzione di stranezza in funzione del
numero  di partecipanti,
possa fornire una osservazione
relativa al segnale di stranezza capace di discriminare una delle due 
ipotesi alternative.  

 
 \chapter*{Conclusioni}
\addcontentsline{toc}{chapter}{Conclusioni}

I risultati esposti in questa tesi hanno permesso di raggiungere
una pi\`u approfondita comprensione della dinamica delle collisioni tra nuclei
di piombo.
Lo studio della produzione di stranezza, in particolare, ha fornito nuove e 
pi\`u precise indicazioni circa la formazione di uno stato deconfinato di
quarks e gluoni durante i primi istanti della collisione.
\par
L'analisi delle distribuzioni di massa trasversa delle diverse particelle
identificate, soprattutto se eseguita in un intervallo comune di $m_T$,
ha evidenziato il raggiungimento di un equilibrio termico di tipo
locale della materia interagente formata in seguito alla collisione.
Le temperature apparenti estratte da tali distribuzioni permettono di
misurare l'energia delle particelle emesse, composta da un contributo dovuto al
moto termico (legato alla temperatura di {\em freeze-out}) e da uno
connesso all'espansione trasversa  del sistema interagente (legato alla
veloci\-t\`a di flusso trasverso).
Il confronto delle temperature apparenti in funzione del sistema 
di collisione ha rivelato una crescita di entrambi i contributi in funzione 
della massa complessiva del sistema interagente. Nelle collisioni
Pb-Pb, dunque, le temperature previste nello stadio iniziale della collisione
ed il numero di interazioni nello stadio finale risultano 
maggiori di quelle ottenute in seguito a collisioni di nuclei meno massivi.
\par
Il confronto delle temperature apparenti in funzione della massa delle 
particelle identificate, sempre nel caso di interazioni Pb-Pb, ha rivelato
una deviazione delle temperature apparenti delle $\Xi$ ed $\Omega$
dall'andamento crescente in funzione della massa della particella, seguito
dalle altre particelle. Questa caratteristica \`e stata interpretata 
ipotizzando un prematuro disaccoppiamento delle particelle multi-strane dal 
resto del sistema interagente, dovuto alla loro minore probabilit\`a di
interazione nello stato di gas adronico.
\par
Il calcolo dei rapporti di produzione di particelle strane a rapidit\`a 
centrale, inoltre, ha consentito di mettere in relazione la produzione di\
stranezza in collisioni Pb-Pb con quella misurata in collisioni p-Pb della
stessa energia.
Questo tipo di confronto, effettuato sia attraverso
i rapporti tra particelle con diverso contenuto di stranezza che attraverso
i rapporti delle particelle strane con le negative presenti nell'evento, ha
evidenziato un incremento nella produzione di particelle strane nel
passaggio dalle collisioni p-Pb a quelle Pb-Pb che risulta crescente con il
contenuto di stranezza della particella considerata.
Le proporzioni tra gli incrementi, riportate nelle relazioni
 (\ref{line1}) e (\ref{line2}),
mostrano che la loro crescita col contenuto di stranezza \`e pi\`u 
sostenuta per
le particelle strane 
che non hanno alcun quark di valenza in comune con i
costituenti dei nuclei bersaglio e proiettile. Queste caratteristiche sono
di difficile interpretazione in un contesto di
produzione adronica, data l'alta soglia in massa delle reazioni di produzione 
dei barioni multi-strani, e sono indicate come segnali 
della transizione di fase verso uno stato di QGP. 
\par
I risultati sulla produzione di stranezza in funzione della centralit\`a
(misurata dal numero di partecipanti), 
infine, suggeriscono il seguente scenario:
le particelle strane nella regione centrale di rapidit\`a  sono prodotte 
in maniera crescente con la centralit\`a 
e la crescita \`e pi\`u veloce di quanto previsto dalla proporzionalit\`a
col numero di partecipanti. Questo incremento risulta
pi\`u pronunciato all'aumentare del
contenuto di stranezza e raggiunge un ordine di grandezza nel caso di
$\Omega$. Limitatamente all'intervallo 
di centralit\`a coperto dalle interazioni Pb-Pb ($N_p~>~100$), 
tuttavia, la produzione di stranezza risulta proporzionale al numero
di partecipanti, per cui si pu\`o ipotizzare che tutti gli incrementi
raggiungano il loro massimo per valori di $N_p$ minori di $100$,
divenendo poi costanti per valori superiori.
\par
Questo scenario risulta compatibile con quanto misurato
dall'esperimento NA50 a proposito della soppressione del mesone
$J/\Psi$ in collisioni Pb-Pb a 158~GeV/c per nucleone [Abr96].
Come si pu\`o notare dalla fig.~\ref{fsupp2}, la produzione
di questo stato $c\overline{c}$
\`e soppressa in maniera anomala a partire da liberi cammini medi nella
materia nucleare maggiori di $10~fm$, corrispondenti, nell'ambito
del modello geometrico delle collisioni nucleari, ad un numero
di partecipanti $N_p$ maggiore di 160.
I risultati di entrambi gli esperimenti possono essere spiegati
assumendo la formazione di uno stato di QGP in collisioni Pb-Pb e la sua
assenza in collisioni p-Pb. 
Di conseguenza, diventa di primaria importanza
ricercare, nella regione di centralit\`a intermedia tra questi due
sistemi di collisione, una eventuale evidenza della 
transizione verso un nuovo stato di materia nucleare, che potrebbe
manifestarsi, ad esempio, in un repentino cambio nell'incremento della
produzione di particelle strane.
A questo scopo \`e stato progettato
l'esperimento NA57 [Cal96], naturale proseguimento di WA97, che 
 si propone di misurare la
produzione di stranezza in collisioni Pb-Pb estendendo l'intervallo di 
centralit\`a selezionato fino a $N_p\sim 50\div60$ ed \`e appena entrato
nella fase di presa dati.
\par


Il confronto dei dati sperimentali con due tra i modelli di produzione
adronica pi\`u accreditati ha mostrato che, a livello 
dell'interazione p-Pb, VENUS non \`e in grado di riprodurre la
 produzione di particelle multi-strane e sovrastima 
di circa un ordine di grandezza le $\PgOm$ e $\PagOp$.
RQMD, invece, mostra un complessivo accordo con i dati, tranne
probabilmente per quel che riguarda il rapporto $\frac{\PagOp}{\PgOm}$,
peraltro affetto dalla bassa significativit\`a statistica.
\par
Per quanto riguarda le collisioni Pb-Pb, RQMD \`e in grado di riprodurre 
sia le temperature apparenti, confermando 
la deviazione sperimentalmente osservata delle temperature apparenti
di $\Xi$ e $\Omega$ dalla proporzionalit\`a con la massa della particella,
sia la produzione di particelle strane, pur sottostimando quella
delle $\Omega$ di un fattore 2; l'incremento di stranezza per
particelle multi-strane misurato dal WA97 risulta leggermente sottostimato 
da RQMD, pur rimanendo confrontabile entro due deviazioni standard.
\par
Entrambi i modelli prevedono una crescita continua della produzione di 
particelle passando da interazioni p-Pb a interazioni Pb-Pb e questa 
caratteristica non sembra per il momento compatibile con i risultati
sperimentali, che invece propendono per una saturazione della 
produzione di stranezza ad alte centralit\`a, per le quali essa raggiunge la
proporzionalit\`a col numero di partecipanti.
Questa differenza nei possibili scenari di produzione rafforza la necessit\`a 
di una misura di particelle strane a rapidit\`a centrale in un pi\`u esteso
intervallo  di centralit\`a  delle collisioni tra nuclei di piombo, progettato
dal suddetto esperimento NA57.
In particolare, la rivelazione di un brusco salto nell'incremento della
produzione di stranezza per centralit\`a non coperte dall'esperimento WA97
sarebbe in completo disaccordo con le previsioni dei modelli
di produzione adronica analizzati e sarebbe difficilmente 
interpretabile nell'ambito di modelli che non contengono l'ipotesi di
formazione dello stato di QGP: la presenza di tale stato solo
in collisioni con sufficiente rilascio di energia nella regione centrale
di interazione tra i due nuclei fornirebbe una spiegazione pi\`u evidente
del fenomeno di soglia ipotizzato.  


\appendix

%
%
\chapter{Geometria del decadimento delle $V^0$}
\index{Geometria del decadimento delle $V^0$}

\begin{figure}[htb]
\centering
\includegraphics[scale=0.38,clip] 
                                {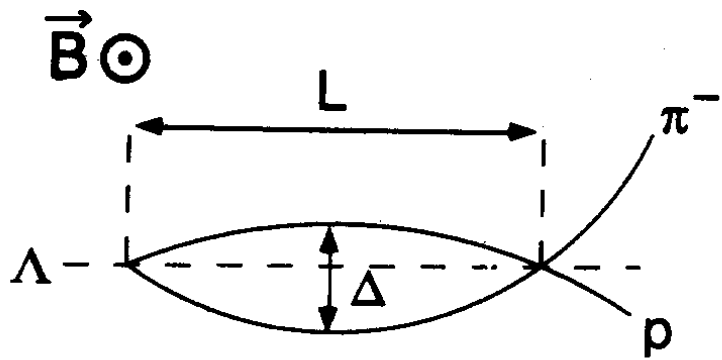}
\caption{\em Tipica geometria di decadimento  di una $\PgL$ nella
             configurazione cowboy.}
\label{figholmea}
\end{figure}

\begin{figure}[htb]
\centering
\includegraphics[scale=0.52,clip] 
                                {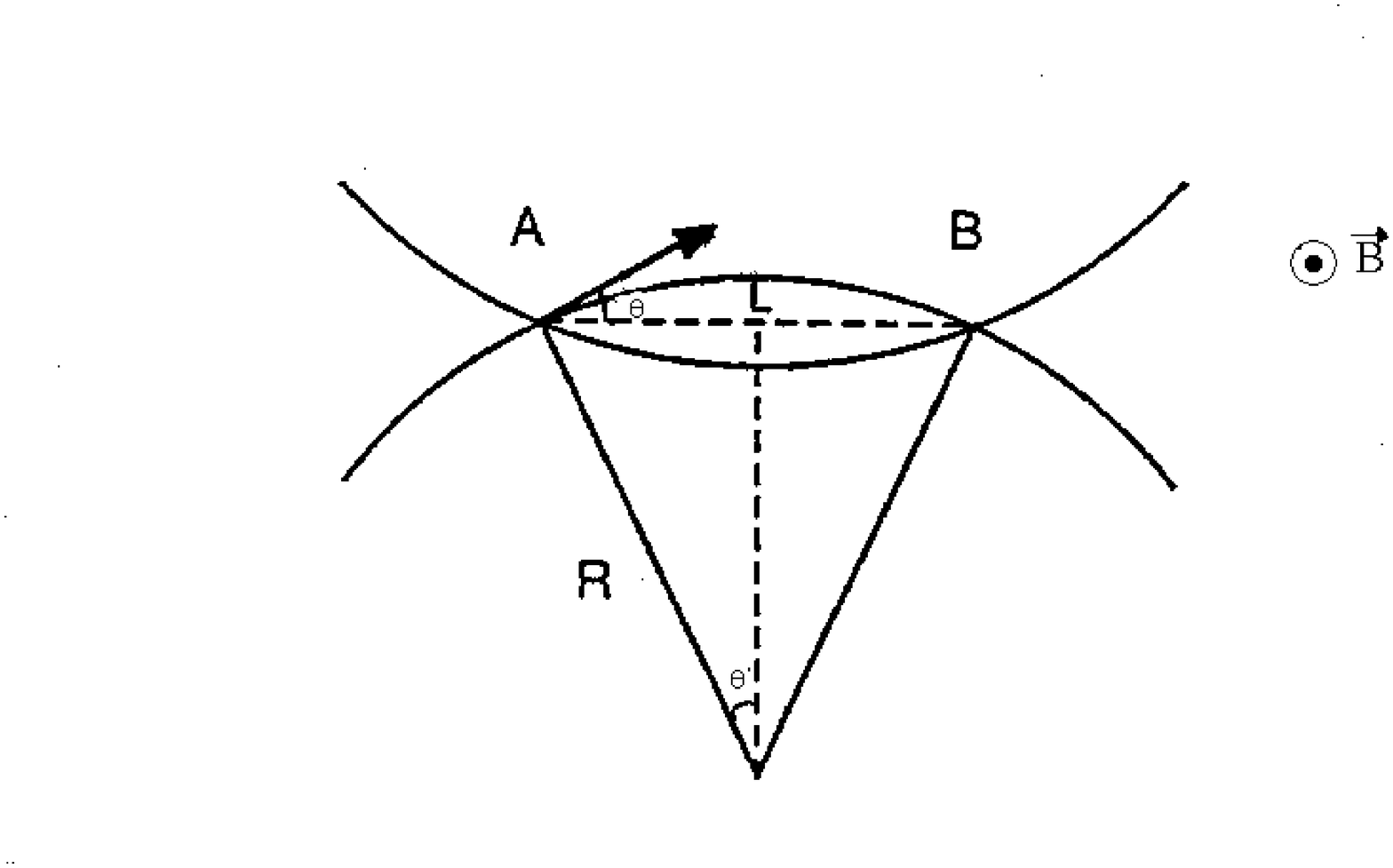}
\caption{\em Schema di decadimento di una $V^0$ nella configurazione cowboy.}
\label{figevans}
\end{figure}

In fig.~\ref{figholmea} \`e mostrata la proiezione nel piano $xy$
di un tipico decadimento di una $V^0$ (nel caso specifico di una $\PgL$)
 nella configurazione cowboy.
Si considerino le seguenti distanze: 
\begin{description}
\item[$L$] = distanza tra il vertice della $V^0$ ed 
             il secondo punto di intersezione delle tracce di decadimento
\item[$\Delta$] = massima distanza trasversa tra le due tracce
                   di decadimento lungo $L$.
\end{description}
Per il calcolo si far\`a riferimento alla fig.~\ref{figevans}.

Per una particella di carica $e$ e momento $p$ in moto entro un campo magnetico
$B$ ad essa perpendicolare, vale la relazione

\begin{equation}
R=\frac{p}{eB}
\label{eq_appa1}
\end{equation}

\noindent Dalla fig.~\ref{figevans} si ha

\begin{equation}
\sin\theta =\frac{L}{2R}
\label{eq_appa2}
\end{equation}

\noindent e, per piccoli $\theta$, risulta

\begin{equation}
\sin\theta \sim \tan\theta = \frac{q_T}{q_L}
\label{eq_appa3}
\end{equation}

\noindent 
dove $q_T$ e $q_L$ sono rispettivamente le componenti trasverse e longitudinali
del momento dei prodotti di decadimento.
Poich\`e le tracce di decadimento nell'esperimen\-to WA97 hanno momenti
dell'ordine di parecchi $GeV/c$ e il loro massimo momento trasverso
\`e $q_{T_{MAX}}=0.1~GeV/c$ per le $\PgL$ e $\PagL$ e $q_{T_{MAX}}=0.2~GeV/c$
per i $\PKzS$, risulta $p\gg q_T$, per cui $q_L\sim p$.
Sostituendo dunque le equazioni (\ref{eq_appa1}) e (\ref{eq_appa2})
nella (\ref{eq_appa3}), si ha:

\begin{equation}
L\sim 2\left( \frac{p}{eB} \right)\,\left( \frac{q_T}{p} \right)
 =\frac{2\, q_T}{eB}
\label{eq_appa4}
\end{equation}

\noindent
Tale distanza risulta quindi invariante per trasformazioni di Lorentz
longitudinali ed il suo valore massimo \`e caratteristico della
particella che decade. 
La massima distanza $L$, espressa in metri, risulta:

\begin{equation}
L_{MAX}\sim \frac{2\, {q_T}_{MAX}(GeV/c)}{e B(Tesla)}\,\,
          \left(\frac{10^9 e}{c}\right)
       = \frac{2\, {q_T}_{MAX}(GeV/c)}{0.2998\, B(Tesla)}
\end{equation}

\noindent
Considerando il campo magnetico $B=1.8~Tesla$ di OMEGA, si ha che

\vspace{0.7cm}
\begin{tabular}{ccl}
$L_{MAX}$ &$\sim $ & $40~cm$ per $\PgL$ e $\PagL$ \\
$L_{MAX}$ &$\sim $ & $80~cm$ per $\PKzS$.
\end{tabular}
\vspace{1.0cm}

Per quanto riguarda la distanza $\Delta$, dalla fig.~\ref{figevans} si ha:

\begin{equation}
\Delta=(R_1-R_1\cos\theta_1) + (R_2-R_2\cos\theta_2)
\label{eq_appa5}
\end{equation}

\noindent
dove gli indici $1$ e $2$ indicano le quantit\`a relative alle due tracce
di decadimento. Dalla relazione (\ref{eq_appa1}), considerando che
${\displaystyle \cos\theta=\frac{q_L}{p} }$, si ha:

\begin{equation}
(R_i-R_i\cos\theta_i) = \frac{p_i}{eB}\,{\left( 1-\frac{{q_L}_i}{p_i} \right)}=
\frac{1}{eB}\left( \sqrt{q_T^2+q_{L_i}^2}-q_{L_i}\right)
\label{eq_appa6}
\end{equation}

\noindent
Le relazioni che legano le componenti del momento delle particelle
di decadimento nel sistema del laboratorio alle quantit\`a $p^\star$,
$E^\star_i$ e $\theta^\star$ nel sistema del centro di massa sono:

\begin{equation}
\left\{ \begin{array}{ccl}
         q_{T_i} & = &(-1)^i\, p^\star\sin{\theta^\star} \\
         q_{L_i} & = &\gamma\:\:[\beta\, E_i^\star+(-1)^i\, p^\star\cos{\theta^\star}].
        \end{array}
\right.
\label{eq_appa7}
\end{equation}

\noindent
Considerando il caso ${\displaystyle \theta^\star=\frac{\pi}{2} }$ (decadimento con massimo
momento trasverso) e sostituendo le (\ref{eq_appa7}) nella 
(\ref{eq_appa6}), si ha:

\begin{equation}
(R_i-R_i\cos\theta_i) = \frac{1}{eB}\left( 
\sqrt{{p^\star}^2+(\gamma^2-1){E_i^\star}^2} - \sqrt{\gamma^2-1}\:\: E_i^\star
\right)
\label{eq_appa8}
\end{equation}

\noindent
per cui $\Delta$ dipende dal momento della particella che decade (attraverso
$\gamma$) e dalle quantit\`a $E_i^\star$ e $p^\star$, note a partire dalle
masse delle particelle coinvolte nel decadimento.
Esprimendo tali quantit\`a in $GeV/c$, si ha che

\begin{equation}
(R_i-R_i\cos\theta_i)(metri) = \frac{1}{0.2998\, B(Tesla)}\left( 
\sqrt{{p^\star}^2+(\gamma^2-1){E_i^\star}^2} - \sqrt{\gamma^2-1}\:\: E_i^\star
\right)
\end{equation}

\noindent
Facendo il calcolo per $\gamma=10$, corrispondente ai tipici momenti delle
particelle di decadimento, e considerando il campo magnetico di OMEGA, risulta:

\vspace{0.7cm}
\begin{tabular}{ccl}
$\Delta(\gamma=10)$ &=& $0.7~cm$ per $\PgL$ e $\PagL$ \\
$\Delta(\gamma=10)$ &=& $3.0~cm$ per $\PKzS$.
\end{tabular}

%
%

\chapter{Produzione di pioni e protoni in interazioni p-W e S-W a 200~GeV/c
per nucleone}
\index{Produzione di pioni e protoni in interazioni p-W e S-W\\
 a 200~GeV/c per nucleone}

\begin{figure}[htb]
\centering
\includegraphics[scale=0.57,clip] 
                                      {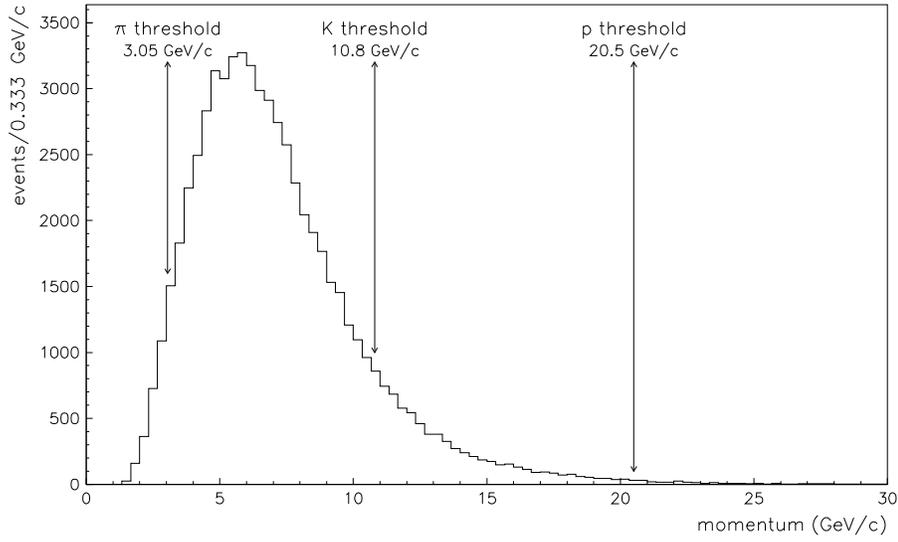}
\caption{\em Distribuzione in momento delle tracce che attraversano il 
        \v{C}erenkov; sono indicate le soglie per $\pi$, $K$ e $p$.}
\label{figsoglie}
\end{figure}

Nell'esperimento WA85 
un contatore \v{C}erenkov a soglia, costituito da 16 celle contenenti una 
mistura di freon e $C\,O_2$ a pressione atmosferica [WA85p]
permetteva  di identificare
le particelle cariche prodotte sia nelle interazioni p-W 
che in quelle S-W a 200 GeV/c per nucleone.
Le soglie per $\pi$, $K$ e $p$ sono indicate in fig.~\ref{figsoglie},
sovrapposte alla distribuzione in momento di un campione di tracce
cariche che attraversano il rivelatore.
Come si pu\`o notare, 
\`e possibile identificare come pioni le particelle di dato momento che
danno segnale nel \v{C}erenkov, mentre i protoni possono essere 
identificati soltanto in assenza di segnale;
l'identificazione anambigua dei kaoni non risulta praticabile.
L'identificazione di particelle di entrambi i segni di carica \`e consentita
invertendo la polarit\`a del campo magnetico di OMEGA.

\noindent
Come primo stadio della procedura di identificazione delle particelle, le
tracce cariche ricostruite nell'apparato sono state selezionate
richiedendo che esse abbiano segno compatibile con la polarit\`a del
campo magnetico adoperata, provenga\-no dal bersaglio e 
siano rivelate da almeno sei delle sette camere
proporzionali multifili usate come elementi traccianti.
Il secondo stadio consiste nella definizione delle matrici di correlazione
che determinano le corrispondenze tra le celle del \v{C}erenkov (C1)
e gli elementi di scintillatore appartenenti a due odoscopi
(HY1 e HY2) posti, rispettivamente, davanti e dietro il rivelatore.
Una particella \`e allora definita come un candidato pione se attraversa gli
elementi correlati del sistema HY1-C1-HY2 ed almeno una delle celle di C1
nella correlazione ha dato segnale (identificazione ``positiva'').
Un candidato protone (o antiprotone) \`e definito dall'assenza di segnale
nelle celle correlate (identificazione ``negativa'').
Il campione usato consiste di $\sim 3$ milioni di eventi p-W e di $\sim 1$
 milione di eventi S-W acquisiti con entrambe le polarit\`a del campo
magnetico ed i numeri di candidati identificati sono riportati in
tab.~\ref{tabstat}.

\begin{table}[ht]
\centering
\caption{\em Statistica delle particelle identificate.}
\begin{tabular}{|c|c|c|} \hline
      { }         &  $p$-W       & S-W       \\ \hline
{$\pi^+$}          & {232.962}   & {323.834}  \\ \hline
{$\pi^-$}          & {137.223}   & {185.891}  \\ \hline
{$p$}              & {11.768}    & {2.690}    \\ \hline
{$\overline{p}$}   & {1.457}     & {487}     \\ \hline
\end{tabular}
\label{tabstat}
\end{table}

\noindent
Come ultimo stadio della procedura di identificazione delle particelle, 
si sono considerati l'effetto dell'inefficienza del \v{C}erenkov e l'effetto
 del
fondo in esso presente, quest'ultimo dovuto a rumore elettronico,
ad interazioni secondarie nella sua struttura o a particelle sopra soglia
non ricostruite o rigettate dalla selezione sulle tracce.

\begin{figure}[htb]
\centering
\includegraphics[scale=0.5,clip] 
                                      {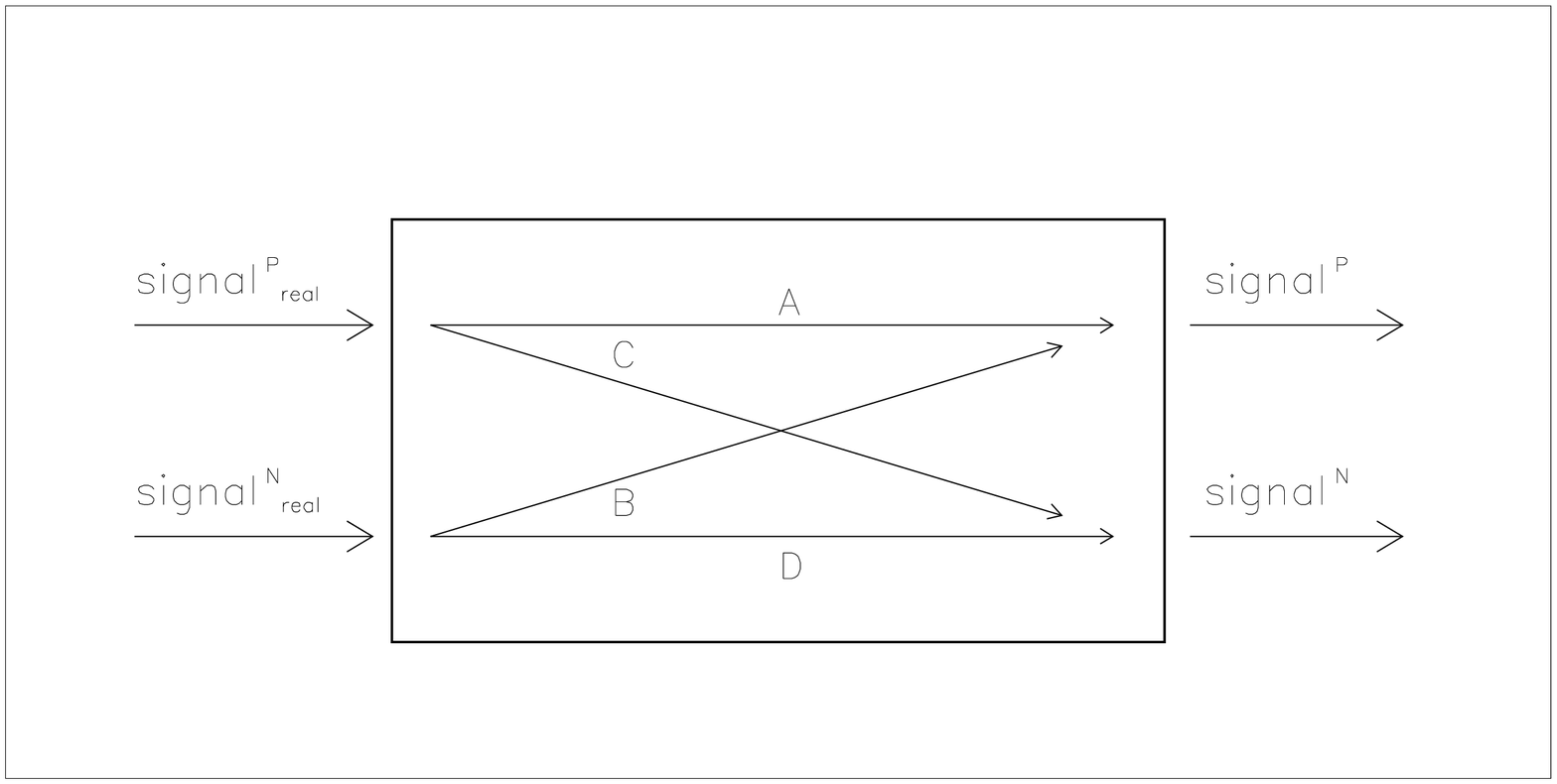}
\caption{\em Rappresentazione schematica dell'azione del \v{C}erenkov.}
\label{figmixing}
\end{figure}

\noindent
Il fondo e l'inefficienza del rivelatore danno origine ad un fenomeno di
mescolamento del segnale, in quanto una traccia sopra soglia pu\`o essere 
associata ad una cella inefficiente, mentre una cella associata ad una traccia
sotto soglia pu\`o dare segnale a causa del fondo.
L'azione del \v{C}erenkov pu\`o essere schematizzata come mostrato in 
fig.~\ref{figmixing}, nella quale $signal^{\rm P}_{\rm real}$ e
$signal^{\rm N}_{\rm real}$ rappresentano rispettivamente le effettive
particelle sopra e sotto soglia e
$signal^{\rm P}$ e $signal^{\rm N}$ indicano le particelle identificate,
come descritto in precedenza, rispettivamente in analisi positiva e negativa.
Il mescolamento pu\`o essere descritto dalla seguente relazione:

\begin{equation}
\left ( \begin{array}{c}
           { signal}^{\rm P} \\
           { signal}^{\rm N} 
        \end{array}
\right )
=
\left ( \begin{array}{cc}
               A & B \\
               C & D
        \end{array}
\right )
\times     
\left ( \begin{array}{c}
               { signal}^{\rm P}_{\rm real} \\
               { signal}^{\rm N}_{\rm real} \\
        \end{array}
\right )
\label{mix}
\end{equation}

\noindent
dove, in riferimento alla fig.~\ref{figmixing}, $A$ e $D$ rappresentano
le
efficienze di identificazione rispettivamente 
in analisi positiva e negativa e $B$ e $C$ 
denotano gli effetti di fondo ed inefficienza, rispettivamente.
Questi parametri sono stati determinati utilizzando le tracce di decadimento
delle $V^0$ ricostruite nello stesso esperimento e
identificate come $\PKzS$, $\PgL$ e $\PagL$ [Aba97], [Aba96].
\par
Le distribuzioni di massa trasversa sono state studiate nelle finestre 
cinematiche $2.5<y<3.0$ e $0.7<p_T <1.1~GeV/c$ per $\pi^{\pm}$ e
$2.75<y<3.0$ e $1.2<p_T <1.8~GeV/c$ per $p$ e $\overline{p}$.
Gli spettri di massa trasversa per ciascuna specie di particella
identificata sono stati considerati composti da due contributi, uno
dovuto alla effettiva particella in esame e l'altro dovuto
al fondo risultante da errate identificazioni, come spiegato in precedenza.
Inoltre, ad ogni spettro di massa trasversa \`e stato associato uno spettro
che rappresenta il corrispondente fondo:
sono state usate le particelle identificate in analisi negativa (i
candidati $K+p$ o $K+\overline{p}$, denominati $signal^{\rm N}$)
per lo spettro dei pioni e quelle identificate in analisi positiva
( i candidati $\pi + K$, denominati $signal^{\rm P}$) per lo spettro
dei protoni. Per ciascuna coppia di spettri, le particelle effettive
nel primo (ad esempio i $\Pgpp$) costituiscono il fondo nel secondo (per 
esempio i $K +p$) e viceversa. Ad essi \`e stata quindi applicata una procedura
di {\em best fit} congiunta, descrivendo entrambi i contributi con la funzione

\begin{figure}[htb]
\centering
\includegraphics[scale=0.85,clip] 
                                      {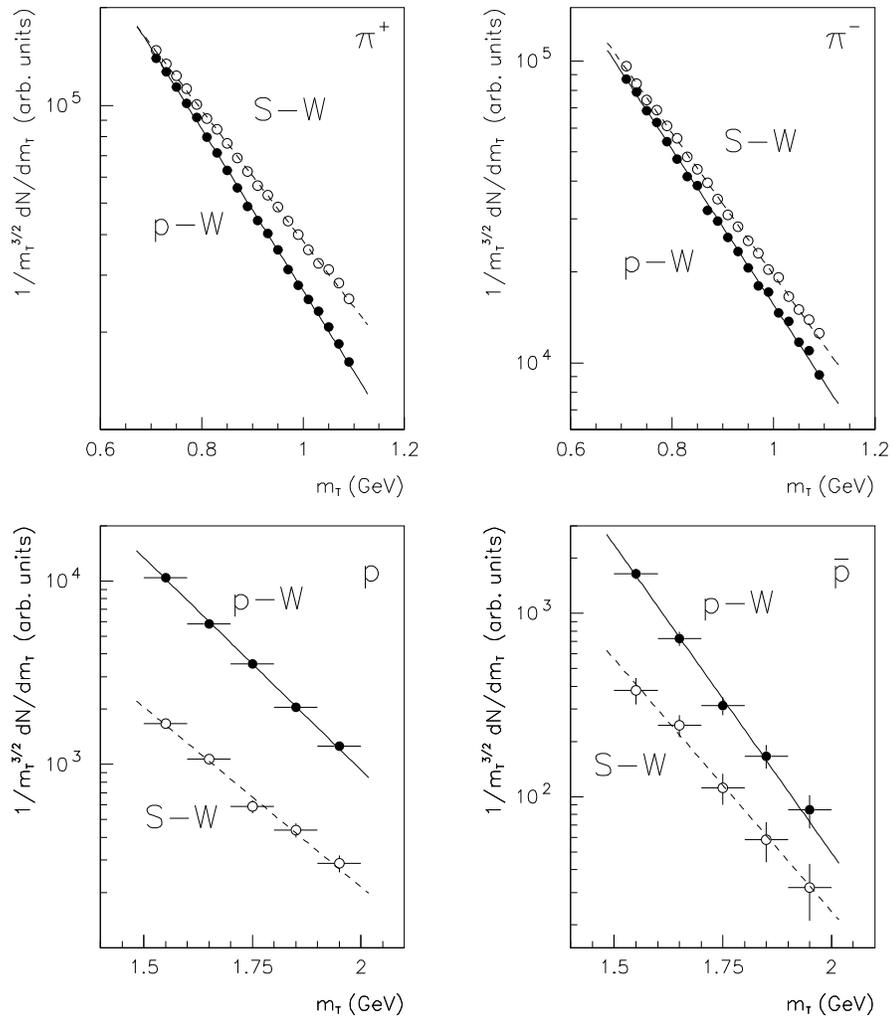}
\caption{\em Distribuzioni di massa trasversa per le particelle identificate.}
\label{figmtapp}
\end{figure}

\begin{center}
{\large $\frac{1}{m_T^{3/2}}\frac{dN}{dm_T}$} = $Ae^{-{\beta}m_T}$
\end{center}

\noindent
e richiedendo che i parametri $T$ dei contributi incrociati siano compatibili
entro gli errori. L'ambiguit\`a presente nei campioni \`e stata
calcolata attraverso la relazione (\ref{mix}) ed \`e stata mantenuta costante
nei {\em fit}.
I risultati della procedura di {\em best fit} sono mostrati in
 fig.~\ref{figmtapp} e le relative temperature apparenti sono riportate in 
tab.~\ref{tabmtapp}, insieme ai loro errori statistici.
\par
L'intera analisi \`e stata pubblicata in [Ant97].

\begin{table}[ht]
\centering
\caption{\em Temperature apparenti per le particelle identificate.}
\begin{tabular}{|c|c|c|} \hline
      { }         &  $p$-W       & S-W       \\ \hline
Particella          & {$1/\beta$ (MeV)} &  {$1/\beta$ (MeV)}\\ \hline
\hline
{$\pi^+$}         & {166 $\pm$ { 2}}  &  {198 $\pm$ {16}} \\ \hline
{$\pi^-$}         & {167 $\pm$ { 1}}  &  {186 $\pm$ {2 }} \\ \hline
{$p$}             & {187 $\pm$ { 4}}  &  {219 $\pm$ {8 }} \\ \hline
{$\overline{p}$}  & {136 $\pm$ { 7}}  &  {151 $\pm$ {28}} \\ \hline
\end{tabular}
\label{tabmtapp}
\end{table}

%
%

\chapter{Modello geometrico di collisione nucleare}
\index{Modello geometrico di collisione nucleare}

Il modello geometrico di collisione nucleare esposto nel seguito \`e stato
formulato in [Bia76] ed \`e sviluppato in analogia con il modello 
di Glauber delle interazioni nucleari.
L'assunzione di base del modello consiste nel fatto 
che la collisione anelastica tra due nuclei pu\`o essere descritta in
termini di
sovrapposizione incoerente delle collisioni dei singoli nucleoni costituenti.
\par
La variabile pi\`u indicata per caratterizzare la centralit\`a delle collisioni
tra nuclei \`e il numero di partecipanti $N_p$, inteso come il numero
di nucleoni che hanno subito almeno una collisione anelastica durante il 
processo d'urto. Nella collisione nucleare ad alta energia, infatti,
un nucleone del nucleo proiettile trapassa l'intero nucleo bersaglio
in un tempo molto minore di quello necessario alla produzione di adroni 
secondari per frammentazione del nucleone stesso.
La produzione di particelle secondarie
non dipende, quindi, dal numero di collisioni subite da un dato nucleone, ma
solo dal numero di essi che ha subito almeno una collisione.
La variabile $N_p$ risulta, dunque, pi\`u intimamente legata alla
molteplicit\`a di particelle secondarie prodotte nell'interazione.
In virt\`u dell'ipotesi di base del modello, \`e possibile determinare
$N_p$ usando il calcolo delle probabilit\`a.

\begin{figure}[htb]
\centering
\includegraphics[scale=1.75,clip] 
                                {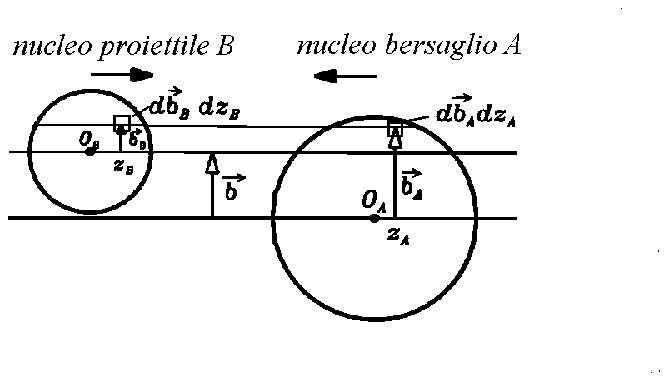}
\caption{\em Collisione di un nucleo proiettile B
(incidente lungo la direzione $\vec{z}$) con un nucleo bersaglio
            A con un parametro di impatto $\vec{b}$.}
\label{figmod}
\end{figure}

In fig.~\ref{figmod} \`e mostrata la geometria della collisione tra un nucleo
proiettile B ed un nucleo bersaglio A, con parametro
di impatto $\vec{b}$. 
Il numero di nucleoni partecipanti alla collisione \`e dato dalla somma
dei nucleoni partecipanti di $A$ e di $B$. Si consideri dapprima il
nucleo bersaglio $A$ ed un solo nucleone incidente di $B$: 
la probabilit\`a che tale nucleone
collida con uno qualsiasi dei nucleoni di $A$, 
avente parametro di impatto $\vec{b}_A$,
\`e $T_A (\vec{b}_A)\,\sigma_{NN}$, dove $\sigma_{NN}$ \`e la sezione d'urto 
anelastica per collisioni tra nucleoni, privata della componente riguardante
i fenomeni diffrattivi (il cui contributo alla
perdita di energia dei nucleoni collidenti \`e del tutto trascurabile).
Il suo valore, alle energie considerate, \`e $\sigma_{NN}=30~mb$. 
La funzione 
\[T_A(\vec{b})=\int dz\,\rho_A(z,\vec{b})\] 

\noindent
definisce il profilo del nucleo
$A$ lungo la direzione trasversa ed \`e normalizzata in modo che

\begin{equation}
\int d^2 \vec{b}\, T(\vec{b}) =1.
\label{eqc1}
\end{equation}

\noindent
$\rho_A$ descrive invece la distribuzione di densit\`a nucleare che, 
per nuclei sferici, pu\`o essere parametrizzata con la forma di
Woods-Saxon [deJ74]

\begin{equation}
\rho_A(r)=\frac{\rho_o}{1+e^{(r-R)/c}}, \;\;\;\;\; r=\sqrt{b^2+z^2},
\label{eqc2}
\end{equation}

\noindent
dove $R$ \`e il raggio nucleare, pari a $R=6.6~fm$ per il nucleo di piombo,
$c$ descrive lo spessore della corteccia nucleare ($c=0.54~fm$ per Pb) e
$\rho_o$ \`e la costante di normalizzazione che garantisce la (\ref{eqc1}).
\par
La probabilit\`a che il nucleone di B subisca un numero $n$ di urti 
con i nucleoni di A
aventi parametro di impatto $\vec{b}_A$, e solo con essi, \`e:

\begin{equation}
\left( \begin{array}{c}
          n_A\\
          n
       \end{array}
\right)\;\;
[T_A(\vec{b}_A)\,\sigma_{NN}]^n\;\;[1-T_A(\vec{b}_A)\,\sigma_{NN}]^{n_A-n},
\label{eqc3}
\end{equation}

\noindent
dove $n_A$ \`e il numero totale di nucleoni del nucleo $A$.
La probabilit\`a totale che il nucleone di B subisca almeno un urto
anelastico con i nucleoni di A aventi parametro di impatto $\vec{b}_A$
\`e ottenibile sommando la (\ref{eqc3}) su tutti i possibili $n$:

\begin{equation}
\sum_{n=1}^{n_A}
\left( \begin{array}{c}
          n_A\\
          n
       \end{array}
\right)\;\;
[T_A(\vec{b}_A)\,\sigma_{NN}]^n\;\;[1-T_A(\vec{b}_A)\,\sigma_{NN}]^{n_A-n}
=
1-[1-T_A(\vec{b}_A)\,\sigma_{NN}]^{n_A}.
\label{eqc4}
\end{equation}

\noindent
Integrando su tutti i possibili parametri di impatto dei nuclei di A, si 
ottiene la probabilit\`a che il nucleone di B collida con il nucleo A.
Questa \`e, in sostanza, la sezione d'urto anelastica nucleare di A:

\begin{equation}
\sigma_A =\int d^2 \vec{b}_A \{
1-[1-T_A(\vec{b}_A)\,\sigma_{NN}]^{n_A} \}
\label{eqc5}
\end{equation}

\noindent
Il numero medio di nucleoni di $B$ che partecipano alla collisione \`e dunque
proporzionale a 
$n_B\,\sigma_A$, se si indica con $n_B$ il numero totale di nucleoni
 del nucleo $B$.
Allo stesso modo, il numero medio 
di nucleoni di $A$ che partecipano alla collisione
\`e proporzionale a $n_A\,\sigma_B$,
per cui il numero medio di nucleoni partecipanti alla
collisione tra $A$ e $B$ risulta (a meno di una costante di normalizzazione)

\begin{equation}
N_p = n_A\,\sigma_B + n_B\,\sigma_A
\label{eqc6}
\end{equation}

\noindent
e sostituendo le formule per il calcolo della sezione 
d'urto anelastica dei nuclei
$A$ e $B$  (c.f.r. eq.~\ref{eqc5}) nella (\ref{eqc6}) si ottiene:

\begin{equation}
N_p  =n_A\,\int d^2\vec{b}_{B} \{ 1-[1-T_B(\vec{b}_B)\,\sigma_{NN}]^{n_B} \} +
n_B\,\int d^2\vec{b}_{A} \{ 1-[1-T_A(\vec{b}_A)\,\sigma_{NN}]^{n_A} \}
\label{eqc7}
\end{equation}

\noindent
La relazione (\ref{eqc7}) definisce la variabile ``numero di partecipanti''
a partire da quantit\`a note, quali la sezione d'urto nucleone-nucleone
ed i parametri della distribuzione di densit\`a nucleare.
 
%

%

%
\end{document}